\RequirePackage{lineno}
\documentclass[aps, prd, twocolumn,
superscriptaddress, nofootinbib]{revtex4-2}

\usepackage{amsmath}   
\usepackage{upgreek}
\usepackage{xspace}
\usepackage{graphicx}  

\usepackage{subcaption}  
\usepackage{float}
\usepackage[dvipsnames]{xcolor} 

\usepackage{units}
\usepackage{siunitx} 

\usepackage{multirow}

\usepackage{hyperref}
\usepackage{cleveref} 
\usepackage{placeins}

\newcommand{\sizecheck}{0} 
\newcommand{\PRLsupp}{0}   
\ifnum\PRLsupp=0
  
\else
  
\fi

\newif\ifpdf
\ifx\pdfoutput\undefined
   \pdffalse
\else
   \pdfoutput=1
   \pdftrue
\fi
\ifpdf
   \usepackage{graphicx}
   \usepackage{epstopdf}
\else
   \usepackage{graphicx}
\fi

\newcommand{\ssqthonethree}{\ensuremath{\sin^2\theta_{13}}\xspace}
\newcommand{\ssqthtwothree}{\ensuremath{\sin^2\theta_{23}}\xspace}
\newcommand{\ssqthonetwo}{\ensuremath{\sin^2\theta_{12}}\xspace}

\newcommand{\deltacp}{\ensuremath{\delta_{\scriptscriptstyle\mathrm{CP}}}\xspace}
\newcommand{\dmsq}{\ensuremath{\Delta{}m^2}\xspace}
\newcommand{\dmsqtwothree}{\ensuremath{\Delta{}m^2_{32}}\xspace}
\newcommand{\dmsqonetwo}{\ensuremath{\Delta{}m^2_{21}}\xspace}
\newcommand{\dmsqonethree}{\ensuremath{\Delta{}m^2_{13}}\xspace}

\newcommand{\appearance}{\deltacp vs \ssqthonethree\xspace}
\newcommand{\disappearance}{\dmsq vs \ssqthtwothree\xspace}

\newcommand{\NpointsFC}{\num{9}} 

\newcommand{\NtoysFakeDataNuebar}{$2 \times 10^4$}
\newcommand{\NtoysMargNuebar}{$2 \times 10^5$}
\newcommand{\pvalRateOnlyBetaZeroDataRunOneNineD}{$0.059$}
\newcommand{\sigmaRateOnlyBetaZeroDataRunOneNineD}{$1.89$}
\newcommand{\pvalRateOnlyBetaOneDataRunOneNineD}{$0.321$}
\newcommand{\sigmaRateOnlyBetaOneDataRunOneNineD}{$0.99$}
\newcommand{\pvalRateShapeBetaZeroDataRunOneNineD}{$0.016$}
\newcommand{\sigmaRateShapeBetaZeroDataRunOneNineD}{$2.40$}
\newcommand{\pvalRateShapeBetaOneDataRunOneNineD}{$0.300$}
\newcommand{\sigmaRateShapeBetaOneDataRunOneNineD}{$1.04$}

\newcommand{\pvalRateOnlyBetaZeroAsimovARunOneNineD}{$0.019$}
\newcommand{\sigmaRateOnlyBetaZeroAsimovARunOneNineD}{$2.36$}
\newcommand{\pvalRateOnlyBetaOneAsimovARunOneNineD}{$0.379$}
\newcommand{\sigmaRateOnlyBetaOneAsimovARunOneNineD}{$0.88$}
\newcommand{\pvalRateShapeBetaZeroAsimovARunOneNineD}{$0.006$}
\newcommand{\sigmaRateShapeBetaZeroAsimovARunOneNineD}{$2.76$}
\newcommand{\pvalRateShapeBetaOneAsimovARunOneNineD}{$0.409$}
\newcommand{\sigmaRateShapeBetaOneAsimovARunOneNineD}{$0.83$}

\newcommand{\nubar}{\ensuremath{\overline{\nu}}\xspace}
\newcommand{\nue}{\ensuremath{\nu_{e}}\xspace}
\newcommand{\numu}{\ensuremath{\nu_{\mu}}\xspace}
\newcommand{\nutau}{\ensuremath{\nu_{\tau}}\xspace}
\newcommand{\nueb}{\ensuremath{\nubar_{e}}\xspace}
\newcommand{\numub}{\ensuremath{\nubar_{\mu}}\xspace}

\newcommand{\nub}{\nubar}

\newlength{\parenbarKernelHeight}
\newcommand{\parenbar}[1]{%
    \settoheight{\parenbarKernelHeight}{\ensuremath{#1}}%
    \addtolength{\parenbarKernelHeight}{0.5pt}
    \llap{\raisebox{\parenbarKernelHeight}{\scalebox{0.5}[0.35]{(}}}%
    \overline{#1}%
    \rlap{\raisebox{\parenbarKernelHeight}{\scalebox{0.5}[0.35]{)}}}%
}

\newcommand{\nuany}{\ensuremath{\parenbar{\nu}}\xspace}
\newcommand{\numuany}{\ensuremath{\nuany_\mu}\xspace}
\newcommand{\nueany}{\ensuremath{\nuany_{e}}\xspace}

\newcommand{\nueCConepi}{\nue\text{ CC} $1\pi^{+}$\xspace}

\newcommand{\oscprob}[2]{\ensuremath{P\left(\nu_{#1}\rightarrow\nu_{#2}\right)}\xspace}

\newcommand{\oscprobany}[2]{\ensuremath{P\left({\nuany}_{#1}\rightarrow{\nuany}_{#2}\right)}\xspace}

\DeclareSIUnit\clight{\text{\ensuremath{c}}}
\DeclareSIUnit[per-mode=symbol]\eVperc{\eV\!\per\clight}

\DeclareSIUnit{\keV}{\kilo\eV}
\DeclareSIUnit{\keVperc}{\kilo\eVperc}
\DeclareSIUnit{\MeV}{\mega\eV}
\DeclareSIUnit{\MeVperc}{\mega\eVperc}
\DeclareSIUnit{\GeV}{\giga\eV}
\DeclareSIUnit{\GeVperc}{\giga\eVperc}

\newcommand*{\eV}{\,\si{\eV}\xspace}
\newcommand*{\eVmass}{\,\si{\eVperc^2}\xspace}

\newcommand*{\keV}{\,\si{\kilo\eV}\xspace}

\newcommand*{\MeV}{\,\si{\mega\eV}\xspace}

\newcommand*{\MeVmom}{\,\si{\MeVperc}\xspace}
\newcommand*{\GeV}{\,\si{\giga\eV}\xspace}
\newcommand*{\GeVmass}{\,\si{\GeVperc^2}\xspace}

\providecommand{\clap}[1] {\makebox[0pt][c]{#1}}

\renewcommand{\vec}[1]{\boldsymbol{#1}}
\newcommand*{\fabs}[1]{\ensuremath{\left\lvert#1\right\rvert}\xspace}

\newcommand*{\ue}{\mathrm{e}}  
\newcommand*{\ui}{\mathrm{i}}  
\newcommand*{\ud}{\mathrm{d}} 

\newcommand*{\SK}{SK\xspace}
\newcommand*{\CP}{\ensuremath{\mathrm{CP}}\xspace}
\newcommand*{\CLs}{\ensuremath{\mathrm{CL}_\mathrm{s}\xspace}}
\newcommand*{\NO}{\ensuremath{\mathrm{NO}\xspace}}
\newcommand*{\IO}{\ensuremath{\mathrm{IO}\xspace}}

\newcommand*{\nova}{$\text{NO}\upnu\text{A}$\xspace}
\newcommand*{\minerva}{$\text{MINER}\upnu\text{A}$\xspace}

\begin{document}

\title{Improved constraints on neutrino mixing from the T2K experiment with $\mathbf{3.13\times10^{21}}$ protons on target} 


\newcommand{\INSTHD}{\affiliation{University Autonoma Madrid, Department of Theoretical Physics, 28049 Madrid, Spain}}
\newcommand{\INSTEE}{\affiliation{University of Bern, Albert Einstein Center for Fundamental Physics, Laboratory for High Energy Physics (LHEP), Bern, Switzerland}}
\newcommand{\INSTFE}{\affiliation{Boston University, Department of Physics, Boston, Massachusetts, U.S.A.}}
\newcommand{\INSTGA}{\affiliation{University of California, Irvine, Department of Physics and Astronomy, Irvine, California, U.S.A.}}
\newcommand{\INSTI}{\affiliation{IRFU, CEA, Universit\'e Paris-Saclay, F-91191 Gif-sur-Yvette, France}}
\newcommand{\INSTGB}{\affiliation{University of Colorado at Boulder, Department of Physics, Boulder, Colorado, U.S.A.}}
\newcommand{\INSTFG}{\affiliation{Colorado State University, Department of Physics, Fort Collins, Colorado, U.S.A.}}
\newcommand{\INSTFH}{\affiliation{Duke University, Department of Physics, Durham, North Carolina, U.S.A.}}
\newcommand{\INSTBA}{\affiliation{Ecole Polytechnique, IN2P3-CNRS, Laboratoire Leprince-Ringuet, Palaiseau, France }}
\newcommand{\INSTEF}{\affiliation{ETH Zurich, Institute for Particle Physics and Astrophysics, Zurich, Switzerland}}
\newcommand{\INSTIE}{\affiliation{CERN European Organization for Nuclear Research, CH-1211 Genève 23, Switzerland}}
\newcommand{\INSTEG}{\affiliation{University of Geneva, Section de Physique, DPNC, Geneva, Switzerland}}
\newcommand{\INSTHJ}{\affiliation{University of Glasgow, School of Physics and Astronomy, Glasgow, United Kingdom}}
\newcommand{\INSTDG}{\affiliation{H. Niewodniczanski Institute of Nuclear Physics PAN, Cracow, Poland}}
\newcommand{\INSTCB}{\affiliation{High Energy Accelerator Research Organization (KEK), Tsukuba, Ibaraki, Japan}}
\newcommand{\INSTIB}{\affiliation{University of Houston, Department of Physics, Houston, Texas, U.S.A.}}
\newcommand{\INSTED}{\affiliation{Institut de Fisica d'Altes Energies (IFAE) - The Barcelona Institute of Science and Technology, Campus UAB, Bellaterra (Barcelona) Spain}}
\newcommand{\INSTEC}{\affiliation{IFIC (CSIC \& University of Valencia), Valencia, Spain}}
\newcommand{\INSTHH}{\affiliation{Institute For Interdisciplinary Research in Science and Education (IFIRSE), ICISE, Quy Nhon, Vietnam}}
\newcommand{\INSTEI}{\affiliation{Imperial College London, Department of Physics, London, United Kingdom}}
\newcommand{\INSTGF}{\affiliation{INFN Sezione di Bari and Universit\`a e Politecnico di Bari, Dipartimento Interuniversitario di Fisica, Bari, Italy}}
\newcommand{\INSTBE}{\affiliation{INFN Sezione di Napoli and Universit\`a di Napoli, Dipartimento di Fisica, Napoli, Italy}}
\newcommand{\INSTBF}{\affiliation{INFN Sezione di Padova and Universit\`a di Padova, Dipartimento di Fisica, Padova, Italy}}
\newcommand{\INSTBD}{\affiliation{INFN Sezione di Roma and Universit\`a di Roma ``La Sapienza'', Roma, Italy}}
\newcommand{\INSTEB}{\affiliation{Institute for Nuclear Research of the Russian Academy of Sciences, Moscow, Russia}}
\newcommand{\INSTHI}{\affiliation{International Centre of Physics, Institute of Physics (IOP), Vietnam Academy of Science and Technology (VAST), 10 Dao Tan, Ba Dinh, Hanoi, Vietnam}}
\newcommand{\INSTHA}{\affiliation{Kavli Institute for the Physics and Mathematics of the Universe (WPI), The University of Tokyo Institutes for Advanced Study, University of Tokyo, Kashiwa, Chiba, Japan}}
\newcommand{\INSTID}{\affiliation{Keio University, Department of Physics, Kanagawa, Japan}}
\newcommand{\INSTIF}{\affiliation{King's College London, Department of Physics, Strand, London WC2R 2LS, United Kingdom}}
\newcommand{\INSTCC}{\affiliation{Kobe University, Kobe, Japan}}
\newcommand{\INSTCD}{\affiliation{Kyoto University, Department of Physics, Kyoto, Japan}}
\newcommand{\INSTEJ}{\affiliation{Lancaster University, Physics Department, Lancaster, United Kingdom}}
\newcommand{\INSTFC}{\affiliation{University of Liverpool, Department of Physics, Liverpool, United Kingdom}}
\newcommand{\INSTFI}{\affiliation{Louisiana State University, Department of Physics and Astronomy, Baton Rouge, Louisiana, U.S.A.}}
\newcommand{\INSTHB}{\affiliation{Michigan State University, Department of Physics and Astronomy,  East Lansing, Michigan, U.S.A.}}
\newcommand{\INSTCE}{\affiliation{Miyagi University of Education, Department of Physics, Sendai, Japan}}
\newcommand{\INSTDF}{\affiliation{National Centre for Nuclear Research, Warsaw, Poland}}
\newcommand{\INSTFJ}{\affiliation{State University of New York at Stony Brook, Department of Physics and Astronomy, Stony Brook, New York, U.S.A.}}
\newcommand{\INSTGJ}{\affiliation{Okayama University, Department of Physics, Okayama, Japan}}
\newcommand{\INSTCF}{\affiliation{Osaka City University, Department of Physics, Osaka, Japan}}
\newcommand{\INSTGG}{\affiliation{Oxford University, Department of Physics, Oxford, United Kingdom}}
\newcommand{\INSTIC}{\affiliation{University of Pennsylvania, Department of Physics and Astronomy,  Philadelphia, PA, 19104, USA.}}
\newcommand{\INSTGC}{\affiliation{University of Pittsburgh, Department of Physics and Astronomy, Pittsburgh, Pennsylvania, U.S.A.}}
\newcommand{\INSTFA}{\affiliation{Queen Mary University of London, School of Physics and Astronomy, London, United Kingdom}}
\newcommand{\INSTGD}{\affiliation{University of Rochester, Department of Physics and Astronomy, Rochester, New York, U.S.A.}}
\newcommand{\INSTHC}{\affiliation{Royal Holloway University of London, Department of Physics, Egham, Surrey, United Kingdom}}
\newcommand{\INSTBC}{\affiliation{RWTH Aachen University, III. Physikalisches Institut, Aachen, Germany}}
\newcommand{\INSTFB}{\affiliation{University of Sheffield, Department of Physics and Astronomy, Sheffield, United Kingdom}}
\newcommand{\INSTDI}{\affiliation{University of Silesia, Institute of Physics, Katowice, Poland}}
\newcommand{\INSTBB}{\affiliation{Sorbonne Universit\'e, Universit\'e Paris Diderot, CNRS/IN2P3, Laboratoire de Physique Nucl\'eaire et de Hautes Energies (LPNHE), Paris, France}}
\newcommand{\INSTEH}{\affiliation{STFC, Rutherford Appleton Laboratory, Harwell Oxford,  and  Daresbury Laboratory, Warrington, United Kingdom}}
\newcommand{\INSTCH}{\affiliation{University of Tokyo, Department of Physics, Tokyo, Japan}}
\newcommand{\INSTBJ}{\affiliation{University of Tokyo, Institute for Cosmic Ray Research, Kamioka Observatory, Kamioka, Japan}}
\newcommand{\INSTCG}{\affiliation{University of Tokyo, Institute for Cosmic Ray Research, Research Center for Cosmic Neutrinos, Kashiwa, Japan}}
\newcommand{\INSTHF}{\affiliation{Tokyo Institute of Technology, Department of Physics, Tokyo, Japan}}
\newcommand{\INSTGI}{\affiliation{Tokyo Metropolitan University, Department of Physics, Tokyo, Japan}}
\newcommand{\INSTHG}{\affiliation{Tokyo University of Science, Faculty of Science and Technology, Department of Physics, Noda, Chiba, Japan}}
\newcommand{\INSTF}{\affiliation{University of Toronto, Department of Physics, Toronto, Ontario, Canada}}
\newcommand{\INSTB}{\affiliation{TRIUMF, Vancouver, British Columbia, Canada}}
\newcommand{\INSTDJ}{\affiliation{University of Warsaw, Faculty of Physics, Warsaw, Poland}}
\newcommand{\INSTDH}{\affiliation{Warsaw University of Technology, Institute of Radioelectronics and Multimedia Technology, Warsaw, Poland}}
\newcommand{\INSTFD}{\affiliation{University of Warwick, Department of Physics, Coventry, United Kingdom}}
\newcommand{\INSTGH}{\affiliation{University of Winnipeg, Department of Physics, Winnipeg, Manitoba, Canada}}
\newcommand{\INSTEA}{\affiliation{Wroclaw University, Faculty of Physics and Astronomy, Wroclaw, Poland}}
\newcommand{\INSTHE}{\affiliation{Yokohama National University, Department of Physics, Yokohama, Japan}}
\newcommand{\INSTH}{\affiliation{York University, Department of Physics and Astronomy, Toronto, Ontario, Canada}}

\INSTHD
\INSTEE
\INSTFE
\INSTGA
\INSTI
\INSTGB
\INSTFG
\INSTFH
\INSTBA
\INSTEF
\INSTIE
\INSTEG
\INSTHJ
\INSTDG
\INSTCB
\INSTIB
\INSTED
\INSTEC
\INSTHH
\INSTEI
\INSTGF
\INSTBE
\INSTBF
\INSTBD
\INSTEB
\INSTHI
\INSTHA
\INSTID
\INSTIF
\INSTCC
\INSTCD
\INSTEJ
\INSTFC
\INSTFI
\INSTHB
\INSTCE
\INSTDF
\INSTFJ
\INSTGJ
\INSTCF
\INSTGG
\INSTIC
\INSTGC
\INSTFA
\INSTGD
\INSTHC
\INSTBC
\INSTFB
\INSTDI
\INSTBB
\INSTEH
\INSTCH
\INSTBJ
\INSTCG
\INSTHF
\INSTGI
\INSTHG
\INSTF
\INSTB
\INSTDJ
\INSTDH
\INSTFD
\INSTGH
\INSTEA
\INSTHE
\INSTH

\author{K.\,Abe}\INSTBJ
\author{N.\,Akhlaq}\INSTFA
\author{R.\,Akutsu}\INSTHA
\author{A.\,Ali}\INSTCD
\author{C.\,Alt}\INSTEF
\author{C.\,Andreopoulos}\INSTEH\INSTFC
\author{M.\,Antonova}\INSTEC
\author{S.\,Aoki}\INSTCC
\author{T.\,Arihara}\INSTGI
\author{Y.\,Asada}\INSTHE
% ...

\collaboration{The T2K Collaboration}\noaffiliation

\date{\today}

\begin{abstract}
The T2K experiment reports updated measurements of neutrino and antineutrino oscillations using both appearance and disappearance channels.
This result comes from an exposure of $14.9~(16.4) \times 10^{20}$ protons on target in neutrino~(antineutrino) mode.
Significant improvements have been made to the neutrino interaction model and far detector reconstruction.  An extensive set of simulated data studies have also been performed to quantify the effect interaction model uncertainties have on the T2K oscillation parameter sensitivity.
T2K performs multiple oscillation analyses that present both frequentist and Bayesian intervals for the PMNS parameters.  For fits including a constraint on \ssqthonethree from reactor data and assuming normal mass ordering T2K measures \ssqthtwothree~$= 0.53^{+0.03}_{-0.04}$ and \dmsqtwothree~$= (2.45 \pm 0.07) \times 10^{-3}$~eV$^{2}$c$^{-4}$.
The Bayesian analyses show a weak preference for normal mass ordering (89\% posterior probability) and the upper \ssqthtwothree octant (80\% posterior probability), with a uniform prior probability assumed in both cases.
The T2K data exclude CP conservation in neutrino oscillations at the $2\sigma$ level.
\end{abstract}

\ifnum\sizecheck=0  
\maketitle
\fi

\newcommand*{\bra}[1]{\left<{#1}\right|}
\newcommand*{\ket}[1]{\left|{#1}\right>}
\newcommand*{\Ham}{\mathcal{H}}
\newcommand*{\Gfermi}{G_\mathrm{F}}
\newcommand*{\TrMat}[1]{\mathcal{T}_{#1}}
\newcommand*{\TrElmt}[2]{\mathcal{T}^{#2}_{#1}}
\newcommand*{\qq}[2]{\theta_{{#1}{#2}}}  
\newcommand*{\DD}[2]{\Delta_{{#1}{#2}}} 
\newcommand*{\DDAtm}{\DD{\mathrm{Atm}}{}} 
\newcommand*{\dmp}{\xi}  

\section{Introduction}
The fact that neutrino flavor mixing~\cite{Maki:1962mu} and oscillations~\cite{Pontecorvo:1967fh} account for the apparent depletion of neutrino fluxes from natural sources is now well-established by detailed observations of these sources~\cite{Aharmim:2011vm_sno,Abe:2017aap_skatm,Abe:2016nxk_sksol}, and verified by experiments using monitored artificial  sources~\cite{Ahn:2006zza_K2K, Adamson:2014vgd_MINOSlast, Gando:2013nba_kamland}.   Neutrino mixing requires that at least two of the neutrino masses ($m_1$, $m_2$ and $m_3$) be non-zero which, in turn requires expanding upon the Standard Model.  Masses require either new gauge singlets---right handed neutrinos---or a different mass generation mechanism from other Standard Model fermions, or a combination of both.  The observed pattern of neutrino masses and mixing is therefore of great interest as a window onto physics beyond the Standard Model. 

\subsection*{Unanswered questions in neutrino oscillations}
The generally accepted explanation of leptonic mixing and neutrino oscillation phenomena centers around the $3\times3$ PMNS mixing matrix, named for Pontecorvo,  Maki, Nakagawa and Sakata, which describes the neutrino mass eigenstates ($\nu_1$, $\nu_2$ and $\nu_3$) in terms of the weak flavor eigenstates ($\nue$, $\numu$, and $\nutau$). Under the assumption that neutrinos are Dirac particles, the matrix is conventionally parameterized as the product of three Tait--Bryan rotations (by $\theta_{ij}$) and a phase transformation (by $\deltacp$), as in Eq.~6 of~\cite{Chau:1984fp}:
\begin{equation}
U_\mathrm{PMNS} = \begin{pmatrix}
    U_{e1} & U_{e2} & U_{e3} \\
    U_{\mu1} & U_{\mu2} & U_{\mu3} \\
    U_{\tau1} & U_{\tau2} & U_{\tau3}     
\end{pmatrix} 
=     \small{ R(\qq23)U(\qq13,\deltacp)R(\qq12)}.
\end{equation}

It is well-established that all of these elements must be large, with the smallest $\fabs{U_{e3}}^2\sim1/45$ and the majority of elements having magnitude-squared of at least 1/4.   The top row elements are well-constrained by measurements of $\nue$ disappearance~\cite{Aharmim:2011vm_sno, Gando:2013nba_kamland,Adey:2018zwh, Bak:2018ydk, Abe:2015rcp}.  The $U_{\mu3}$ element is likewise constrained by disappearance of $\nu_\mu$~\cite{Adamson:2014vgd_MINOSlast, Acero:2019ksn_nova, Abe:2017aap_skatm, PhysRevLett.120.071801}, but the dependence is of the approximate form \mbox{$\fabs{U_{\mu3}}^2(1-\fabs{U_{\mu3}}^2)$}, leading to a degeneracy, commonly expressed in terms of the octant of the mixing angle $\theta_{23}$.

Since the matrix may be complex, a wide range of values for the the magnitudes of the four elements $U_{\mu1}$,$U_{\mu2}$,$U_{\tau1}$ and $U_{\tau2}$ are possible, depending on the phase parameter $\deltacp$.  A purely real matrix corresponds to \deltacp being an integer multiple of $\pi$; any other value is manifested as violation of Charge-Parity (\CP) symmetry in any neutrino \emph{appearance} channel, via the Kobayashi--Maskawa mechanism~\cite{Kobayashi:1973fv}.  The discovery of \CP violation in the lepton sector is of great interest and is a major focus of current experiments~\cite{Abe:2019vii_T2Knature, Acero:2019ksn_nova}.  The fact that \CP violation is controlled by a single parameter means that the rate of \nueb appearance is not independent of \nue appearance, so studying both channels provides a test of the standard PMNS picture. 

Another approximate degeneracy is in the ordering of neutrino masses.  It is known that $\Delta m^2_{21} = m^2_2 - m^2_1 > 0$, and that  $\fabs{\Delta m^2_{31}} > \Delta m^2_{21}$, but the sign of $\Delta m^2_{31}$ is as yet unknown. Neutrino masses (and the corresponding eigenstates) are numbered in order of decreasing $\nue$ content. In the case where $m^2_3 > m_1^2$, the predominant partner of the lightest charged lepton is the lightest neutrino. As the analogous pattern is seen in the quark sector, it is known as Normal Ordering (NO), with Inverted Ordering (IO) corresponding to a light $\nu_3$.  This Mass Ordering (MO) degeneracy is partially lifted at higher neutrino energies by the interactions of neutrinos with matter~\cite{Wolfenstein:1977ue,Mikheev:1986wj}. This matter effect changes both the propagation eigenstates (`effective masses') of the neutrinos and their mixing with the flavor states.  

These three remaining unknowns (octant, \deltacp, MO) are all accessible to current-generation long-baseline neutrino experiments such as T2K, through the $\numu\rightarrow\nue$ appearance channel and its \CP conjugate $\numub\rightarrow\nueb$.  Although \CP violation has the most obvious significance, the general structure of the matrix may give us a window into the deeper problem of neutrino mass and a broad range of new physics.  An inverted ordering would imply the lightest neutrino is relatively weakly coupled to the lightest charged lepton.  Similar in character would be resolving the octant degeneracy in favor of the upper octant, as this implies $\nu_3$ is not the predominant partner of the heaviest charged lepton.  More generally, the highly non-diagonal structure of the PMNS matrix is suggestive of an origin for neutrino masses that is separate from the electroweak physics that dominates the masses of the heavier fermions, and precision measurements of the $\numuany\rightarrow\nueany$ channel can help to determine the remaining elements. 

\section{The T2K experiment}
T2K is a second-generation accelerator neutrino oscillation experiment~\cite{T2K_NIM}, utilizing a narrow-band beam and a \SI{295}{\kilo\meter} baseline from J-PARC in T\=okai, Ibaraki to Super-Kamiokande (commonly Super-K or just SK) in Hida, Gifu.  The primary proton beam is accelerated to \SI{30}{\giga\eV} by J-PARC's Main Ring. In each cycle eight bunches are extracted in a single turn and directed due west at a downward angle of $-3.6^\circ$.  The intensity of the extracted beam is monitored by five current transformers that are also used to normalize the neutrino exposure between the various detectors of the experiment, while the beam profile is monitored with secondary emission monitors---most of which are removed during physics runs---and an optical transition radiation monitor.  The beam power has increased over time, reaching \SI{500}{\kilo\watt} by the end of May 2018.

The protons impinge on a \SI{91.4}{\centi\meter} graphite target to produce a secondary beam, primarily composed of pions and kaons that are focused (or defocused, according to charge) by a set of three magnetic horns~\cite{Sekiguchi:2015ghw_t2khorns} pulsed with a peak current of \SI{250}{\kilo\ampere}.  The focused secondary beam propagates in a \SI{96}{\meter} Helium-filled decay volume where the secondaries can decay to produce neutrinos, among other particles.  A beam dump sits at the end of the decay volume, followed by a muon monitor (MUMON) which is used to check the stability of the secondary beam.

If the horn current is run in the `Forward' direction, positively-charged secondaries are focused, which decay to produce a beam that is primarily \numu. This is referred to as `Forward Horn Current' (FHC) or neutrino mode.  Alternatively the horn current can be reversed, to give a beam of primarily \numub, which is referred to as `Reversed Horn Current' (RHC) or antineutrino mode.  In either case, secondary hadrons produced in the very forward direction pass through the field-free necks of the horns and contribute to a `wrong-sign' flux that is of order 1\% of the intended `right-sign' component in FHC mode, and order 10\% in RHC mode.    

The horn configuration is most effective at focusing pions with momenta between 2 and \SI{2.5}{\GeVperc}, resulting in a broadband neutrino flux along the beamline axis that peaks at around \SI{1}{\giga\eV}.  However, the beam is directed so that its center passes roughly \SI{4}{\kilo\meter} south of and \SI{12}{\kilo\meter} below the Super-Kamiokande detector, equivalent to an angle $2.5^\circ$ from the beam axis, as measured from the target.  This results in a narrow-band flux at the detector that peaks more strongly at \SI{.6}{\giga\eV}---roughly the energy of the first oscillation maximum for this baseline. This dramatically reduces the flux of higher-energy neutrinos that can mimic \nue signal events through neutral current interactions. 

The experiment utilizes a suite of Near Detectors (NDs) at a site \SI{280}{\meter} downstream of the target to characterize the initial flux of neutrinos and their interactions.  A high-mass monitoring detector, INGRID, is centered on the nominal beam axis and samples the beam out to about $1^\circ$ around the beam axis, as measured from the target. This allows the intensity and direction of the neutrino beam to be monitored on a daily basis, the latter to a precision of a few centimeters.

The second near detector, ND280, is optimized for measuring interaction rates and the properties of neutrino interactions.  The detector is positioned off-axis at the same azimuthal angle as Super-Kamiokande to minimize the effect of any directional shift of the beam, and at a distance from the beam axis to make the neutrino spectrum as similar as possible to what would be seen in SK in the absence of oscillations.  Residual differences in spectra exist because the decay volume covers a significant fraction of the \SI{280}{\meter} baseline; these are automatically accounted for in the way the ND280 data is used in the analysis.  The ND280 off-axis detector provides the most relevant information on the neutrino flux and interactions, but more importantly allows the analysis to form a multi-dimensional constraint, incorporating correlations between flux and interaction uncertainties.  The ND280 additionally has a \SI{0.2}{\tesla} magnetic field to provide charge identification of detected leptons and therefore separately constrain the neutrino and antineutrino components of the beam, improving sensitivity to \CP violation.  This is particularly important for the RHC data set which has a larger ($\sim30\%$) background of `wrong-sign' events.

Neutrinos that interact in SK produce signals that can be analyzed in one of five different event categories.  At T2K's beam energy, the lepton produced in a charged-current interaction will be either $\mu^\pm$ or $e^\pm$, so events are categorized as muon- or electron-like based on the pattern of Cherenkov light in the detector.  SK has no magnetic field, so the sign of the horn current is used to categorize events as a proxy for neutrino/antineutrino identification, and the kinematics give some discrimination between signal events and backgrounds.  Events without a lepton-like Cherenkov ring and events where there is evidence of additional hadronic activity are not used, except for a fifth class of events which selects FHC \mbox{\nue-like} events with evidence of a low energy positron from an ejected $\pi^+$.  These selections will be discussed in greater detail in Sec.~\ref{sec:sk_selection}.

\section{The Oscillation Probability}
Following~\cite{Nunokawa:2007qh}, and assuming three neutrinos masses labeled with $i,j \in \{1,2,3\}$, the vacuum oscillation probability between neutrino flavor states $\nu_\alpha$ and $\nu_\beta$ can be written
\begin{equation}
    \oscprob{\alpha}{\beta} =  \left|\sum_j{\bra{\nu_\beta}\ue^{-\ui \Ham L}\ket{\nu_\alpha}}\right|^2
\label{eq:osc_prob}
\end{equation}
where
\begin{equation}
    \label{eqn:vac_hamiltonian}
    \Ham L = 2\TrMat{2}\DD21 + 2\TrMat{3}\DD31 
    \text{,}
\end{equation}
and \mbox{$\DD{j}{i} = \Delta m^2_{ji} L / 4E$}.  For states $\nu_\alpha$ and $\nu_\beta$ the $\TrMat{i}$ take values $\TrElmt{i}{\alpha\beta} = U_{\alpha i}U^*_{\beta i}$ and, considering the specific case of \numu to \nue oscillations, the standard parameterization can be used to derive approximate forms for the two amplitudes: 
\begin{equation}
    \begin{gathered}
    \TrElmt{3}{\mu e} = \tfrac{1}{2} \sin2\qq13 \sin\qq23 \ue^{\ui\deltacp}
    = \fabs{\TrElmt{3}{\mu e}} \ue^{\ui\deltacp}\\
    \TrElmt{2}{\mu e} 
        \simeq \tfrac{1}{2}\sin2\qq12 \cos\qq13 \cos\qq23\text{.}
    \end{gathered}
\end{equation}
The appearance probability in vacuum can then be approximated as a sum of three terms,
\begin{multline}
\label{eqn:vac_prob}
    \oscprob{\mu}{e} \simeq~ 4\fabs{\TrElmt{3}{\mu e}}^2 \sin^2\DD31
    + 4\fabs{\TrElmt{2}{\mu e}}^2 \sin^2\DD21\\[0.5ex]
    + 8 \fabs{\TrElmt{2}{\mu e}} \fabs{\TrElmt{3}{\mu e}} \sin\DD31 \sin\DD21 \cos(\DD32+\deltacp), 
\end{multline}
which includes:
\begin{itemize}
    \item A dominant `atmospheric' ($\sin^2\DD31$) term, which is independent of the \CP-phase.
    \item A `solar' ($\sin^2\DD21$) term that is small at T2K's $L/E$.
    \item An `interference' term that depends on $\sin\DD31 \sin\DD21 \cos(\DD32+\deltacp)$.
\end{itemize}
Because the atmospheric term is proportional to $\fabs{U_{\mu3}}^2 \fabs{U_{e3}}^2$, the \nue appearance probability is free of the octant degeneracy seen in the \numu disappearance probability. In practice the \numu channel remains an important part of the fit, because there is a larger number of events, which gives higher precision on $\Delta m^2_{3i}$ and $\sin\theta_{23}$. The disappearance channel is also important because it is relatively insensitive to $\sin\theta_{13}$, $\deltacp$, and mass ordering, which helps to reduce the impact of degeneracies in the appearance channel.

The amplitude of the interference term is about 20\% of the atmospheric term, making it possible to measure the phase $\deltacp$. This interference term is of particular physical importance as the relative sign of $\deltacp$ and $\DD{j}{i}$ changes between neutrinos and antineutrinos, leading to \CP violation if $\deltacp$ is not an integer multiple of $\pi$.  Since T2K's event spectrum peaks very close to the oscillation maximum at $\DD32\sim \pi/2$, the contribution from the interference term is a direct measure of the amount of \CP violation in the neutrino sector.

The interference term also depends directly on $\mathrm{sign}(\Delta m^2_{31})$, through the $\sin\DD31$ part.  Up to a small difference between $\Delta m^2_{31}$ and $\Delta m^2_{32}$, this is degenerate with a substitution $\deltacp\rightarrow\pi-\deltacp$.  This means the same data will prefer opposite signs of \mbox{$(\deltacp - \pi/2)$} for normal and inverted orderings.  This degeneracy is lifted if the mass ordering is assumed, or if one or other ordering is strongly preferred in the fit.

\subsection{Matter effects} 
The (approximately constant) density of electrons in the Earth's crust along the T2K baseline changes~\cite{Barger} the propagation Hamiltonian (c.f. Eq.~\eqref{eqn:vac_hamiltonian}):
\begin{equation}
    \Ham L=   
    2\TrMat{2}\DD21 +
    2\TrMat{3}\DD31\pm
    \mathrm{diag}(1, 0, 0)\sqrt{2}\Gfermi n_e 
    \text{,}
\end{equation}
where $n_e$ is the number density of electrons, $\Gfermi$ is the Fermi coupling constant, and the minus sign applies for antineutrinos.  The impact can be understood using T2K's beam energy of $\sim\SI{0.6}{\giga\eV}$, and a crust density~\cite{Hagiwara:2011kw} of \SI{2.6}{\gram\per\cubic\centi\meter}, in which case:
\begin{equation}
    \underset{250\rule{0pt}{1.8ex}}{\Delta m^2_{31}} \gg
    \underset{11.9\rule{0pt}{1.8ex}}{2\sqrt{2}\Gfermi n_e E\vphantom{m^2_1}} >
    \underset{7.4\rule{0pt}{1.8ex}}{\Delta m^2_{21}}
    \text{.}\underset{(\times 10^{-5}\si{\eV^2})\rule{0pt}{1.8ex}}{\vphantom{m^2_1}}
\end{equation}
Because of the size ordering of terms in the Hamiltonian, the  matter effect dominates the solar term, but is only a small perturbation compared to the atmospheric and interference terms.  In practice, because the resulting oscillation probability is dominated by the atmospheric term, T2K's sensitivity to the mass ordering comes mostly from this perturbative effect on $\fabs{U_{e3}}$ and $\Delta m^2_{31}$.  In this regime the matter effect can be described with a dimensionless parameter
\begin{equation}
    \dmp(E) = \pm\frac{2\sqrt2 G_F n_e }{\Delta m^2_{31}}E \simeq \pm0.08\, E/\si{\GeV}\text{,}
\end{equation}
which is positive for neutrinos in the normal ordering and for antineutrinos in the inverted ordering.  The resulting probability for neutrinos or antineutrinos is
\begin{multline}
    \oscprobany{\mu}{e} \simeq\\
\begin{aligned}
    4\fabs{\TrElmt{3}{\mu e}}^2 
    \frac{\sin^2 \left([1\!-\!\dmp]\DD31\right) }{[1\!-\!\dmp]^2}
    + 4\fabs{\TrElmt{2}{\mu e}}^2
    \frac{\sin^2\left(\dmp\DD21\right)}{\dmp^2}&\\[0.5ex]
    +8 \fabs{\TrElmt{2}{\mu e}} \fabs{\TrElmt{3}{\mu e}} 
    \frac{\sin\left([1\!-\!\dmp]\DD31\right)}{1\!-\!\dmp} \frac{\sin\left(\dmp\DD21\right)}{\dmp}\cos\DD32 \cos\deltacp&\\[0.5ex] 
    \mp 8 \fabs{\TrElmt{2}{\mu e}} \fabs{\TrElmt{3}{\mu e}} 
    \frac{\sin\left([1\!-\!\dmp]\DD31\right)}{1\!-\!\dmp} \frac{\sin\left(\dmp\DD21\right)}{\dmp}
    \sin\DD32 \sin\deltacp&
    \text{,}
    \end{aligned}
\end{multline}
where the \CP-violating $\sin\deltacp$ term takes a negative sign for neutrinos and positive sign for antineutrinos.

For T2K, the largest observable effect from propagation in matter is the $[1\!-\!\dmp]^{-2}$ scaling of the atmospheric term.
This modification to the leading atmospheric term is about 5\%, leading to a roughly 10\% difference in the appearance probability for neutrino and antineutrinos.  Since this difference is about half the amplitude of the \CP-violating term, it is in general difficult to disentangle the two phenomena if they have opposite effects on the total number of events observed, or if the value of $\sin\deltacp$ is close to zero.
However if both phenomena enhance (or both suppress) the total number of events, then the net effect can be too large to attributable to either source alone, and there will be much less ambiguity.    

\subsection{The survival probability}
In the same notation, the \numuany survival probability at T2K is to a good approximation given by:
\begin{equation}
    P(\,\numuany\rightarrow\numuany)
    \simeq 1 - 4 \TrElmt{3}{\mu\mu}(1-\TrElmt{3}{\mu\mu})\sin^2\DDAtm
    \text{,}
    \label{equ:disappearance}
\end{equation}
where $\DDAtm = \DD32 +  \DD21 \times 
\TrElmt{1}{\mu\mu}/{(1-\TrElmt{3}{\mu\mu})}$. 
So although the observable survival probability is not sensitive to the mass ordering, the best-fit value of $\DD32$ is different for normal and inverted orderings.  As for the oscillation amplitude, in terms of the standard parameterization
\begin{equation}
    \fabs{\TrElmt{3}{\mu\mu}} = \sin^2\qq23\cos^2\qq13\text{,}
\end{equation}
so the amplitude reaches a maximum value around $\sin^2\qq23 = 1/(2\cos^2\qq13)$, and values on either side of this are degenerate.  

Propagation in matter does not change the survival probability by much; matter dependent effects are suppressed by a factor of $\TrElmt{3}{ee}(1-2\TrElmt{3}{\mu\mu})$~\cite{Hagiwara:2011kw}. Since the matter density can be approximated as symmetric, the probability also has no dependence on $\sin\deltacp$~\cite{Yokomakura:2002av}. However the relationship between $\DD32$ and the observable $\DDAtm$ does depend on $\cos\deltacp$ through $\TrElmt{1}{\mu\mu}$, which can for some parameter combinations give rise to a correlation between the measured $\DD32$ and $\deltacp$.

\section{Updates since the previous results}
This analysis uses a \SK data set collected up to the end of May 2018.  This corresponds to an exposure of \num{14.94e20}~Protons on Target (POT) in FHC mode and \SI{16.35e20}{POT} in RHC mode, the same as used to report indications of \CP violation in~\cite{Abe:2019vii_T2Knature}.  A detailed breakdown is given in Tab.~\ref{tab:POT_summary}.  Compared to the previous update~\cite{Abe:2018wpn_T2Krun8osc} this is a nominal increase of 1\% in FHC mode, but 116\% in RHC mode, which is particularly of interest for indications of \nueb appearance, described in Sec.~\ref{sec:nuebar}. 

In parallel with the statistical increase, our event selection has been refined since it was last described in detail~\cite{Abe:2017vif_T2Krun7osc}.  Event reconstruction is now based on an algorithm that matches the pattern of light observed in SK directly~\cite{Jiang:2019xwn_SKfitqun}. This makes use of more information about the event, providing better discrimination between event categories, and improving the resolution of the lepton momentum and vertex location.  As a result, the fiducial volume can also be expanded, roughly equivalent to a 20\% increase in statistics for the $\nue$ samples, as described in Sec.~\ref{sec:SKreco}.  The newer reconstruction algorithm has previously been used for rejecting neutral current $\pi^0$ events in the $\nue$ samples~\cite{Abe:2013hdq,Abe2015k} and for all aspects of event selection and reconstruction in more recent publications~\cite{Abe:2018wpn_T2Krun8osc, Abe:2019vii_T2Knature}.

A large fraction of the analysis development focuses on the interaction model, which incorporates constraints from a number of new external data sets and theoretical improvements.  Since reported in~\cite{Abe:2017vif_T2Krun7osc}, the dominant charged-current quasi-elastic (CCQE) models have been updated in various respects, including: the handling of weak charge screening in nuclei; the handling of nucleon removal energy and its effect on lepton kinematics; and additional freedom allowed in the kinematic dependence of interactions involving correlated nucleon pairs ($2p2h$).  Modeling of (and uncertainties assigned to) subdominant processes have also been improved, including coherent scattering and neutral current interactions. 

Constraints on neutrino oscillations and the associated parameters come from a combined analysis of disappearance and appearance channels across FHC and RHC configurations, using the same approach as in~\cite{Abe:2019vii_T2Knature}.  This paper provides a fuller description of the method and a broader range of results. 

\begin{table}[ht]
\sisetup{round-mode=places, round-precision=2}
    \newcommand{\ccol}[1]{\multicolumn{1}{c}{\text{#1}}}
    \centering
    \caption{T2K Run periods and exposure used in this analysis, for ND280 and SK.}
    \label{tab:POT_summary}
    \begin{tabular}{lr@{\quad}
    S[table-format=2.2]S[table-format=2.2]S[table-format=2.2]S[table-format=2.2]}
    \hline\hline
    \multirow{2}{*}{T2K run} & \multirow{2}{*}{End date} 
    \rule{0pt}{2.5ex}& \multicolumn{2}{c}{SK POT $\scriptsize{/10^{20}}$} & \multicolumn{2}{c}{ND280 POT $\scriptsize{/10^{20}}$}\\  
    & & FHC & RHC & FHC & RHC\\
         \hline
         Run 1 & Jun. 2010 & 0.326 & {---} & {---} & {---}\\
         Run 2 & Mar. 2011 & 1.122 & {---} & 0.78 & {---}\\
         Run 3 & Jun. 2012 & 1.599 & {---} & 1.56 & {---}\\
         Run 4 & May\; 2013 & 3.597 & {---} & 3.47 & {---}\\
         Run 5 & Jun. 2014 & 0.244 & 0.512 & {---} & 0.43 \\
         Run 6 & Jun. 2015 & 0.192 & 3.546 & {---} & 3.43 \\
         Run 7 & May\; 2016 & 0.484 & 3.498 & {---} & {---}\\
         Run 8 & Apr. 2017 & 7.169 & {---} & {---} & {---}\\
         Run 9 & May\; 2018 & 0.204 & 8.788 & {---} & {---}\\
         \hline
         \multicolumn{2}{c}{Total} & 14.938 & 16.346 & {5.80} & {3.86}\\
         \hline \hline
    \end{tabular}
\end{table}

\section{Analysis Overview}
The T2K near and far detectors have different target nuclei and are based on different particle detection techniques.
The T2K oscillation analysis therefore uses parameterized models of the neutrino beam flux and the neutrino interaction cross section to propagate near detector information to predict the far detector event rate.

The neutrino flux prediction has been described in detail in Ref.~\cite{t2k_flux}.
The collision of 30~GeV protons from the J-PARC main ring with the T2K neutrino production target is simulated using FLUKA~\cite{Ferrari:2005zk, BATTISTONI201510, BOHLEN2014211}.
The resultant secondary particles are passed to a GEANT3~\cite{Brun:1082634} simulation of the magnetic focusing horns and decay volume downstream of the target.
GCALOR~\cite{Zeitnitz:1992vw, Fasso:1993kr} is used to model hadronic interactions of the secondary particles as they traverse the focusing horns and decay volume.
Particles are then allowed to decay to produce neutrinos.

Data from proton beam monitors is used to tune the initial proton beam parameters in the simulation.   NA61/SHINE, a fixed-target experiment at CERN's Super Proton Synchrotron, measures particle production in nucleus and hadron collisions with a large acceptance spectrometer.  This includes measurements of the collisions of 30~GeV protons with graphite. 
Data from the NA61/SHINE~\cite{na61_thinpi,na61_thin_k,na61_2007_k0,na61_2009_thin} experiment are then used to tune the secondary particles produced from the target.   
Finally, the INGRID~\cite{Abe_INGRID} on-axis near detector is used to monitor the neutrino beam direction.
The uncertainty from each of these measurements are combined with uncertainties from the beam simulation to give the final flux uncertainty.
This is parameterized as a function of neutrino energy, neutrino species and whether the beam is 
operating in FHC or RHC mode.

\begin{figure}[htbp]
\centering
\begin{subfigure}{0.47\textwidth}
  \includegraphics[width=0.98\columnwidth]{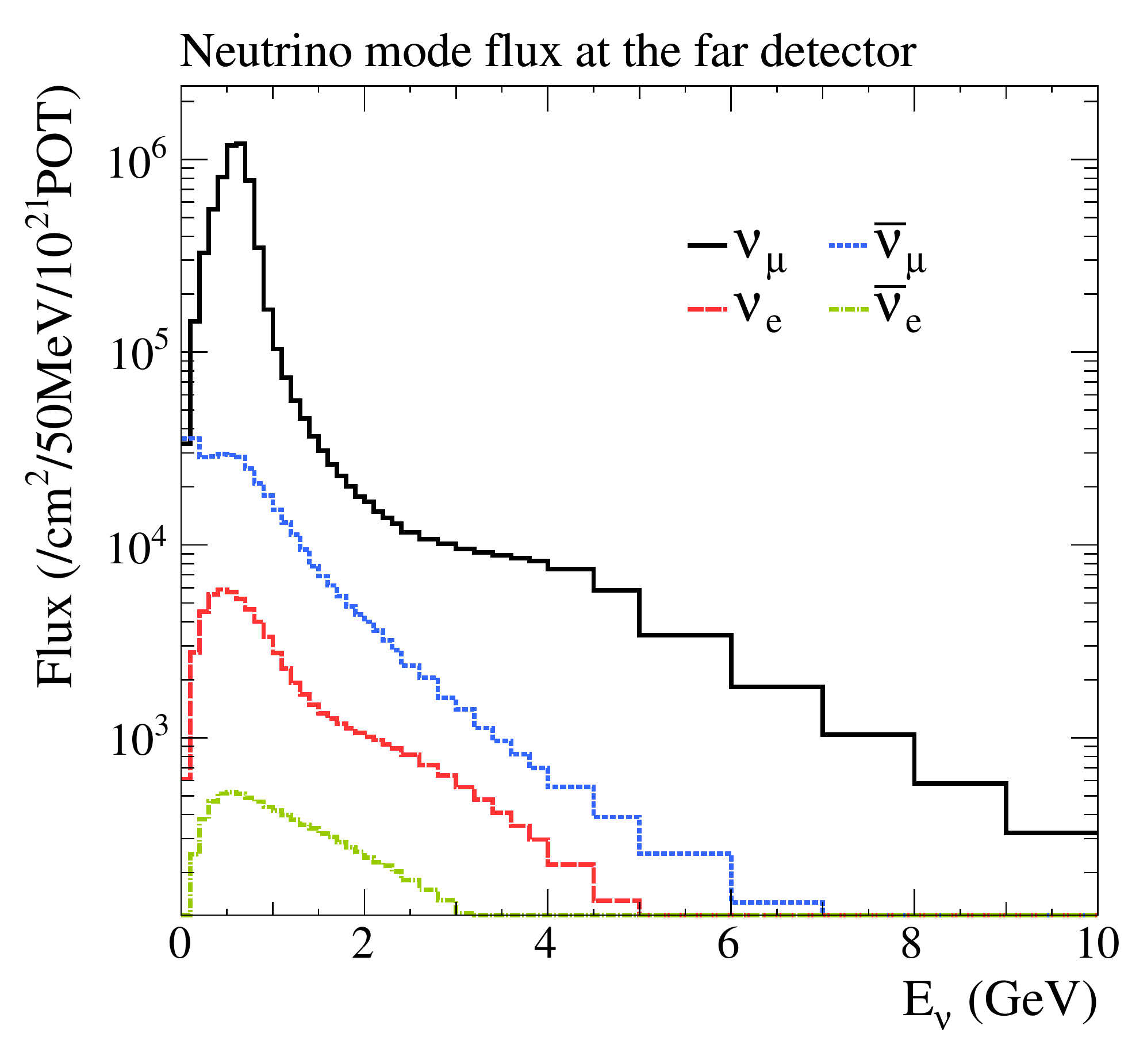} 
\end{subfigure}
\begin{subfigure}{0.47\textwidth}
  \includegraphics[width=0.98\columnwidth]{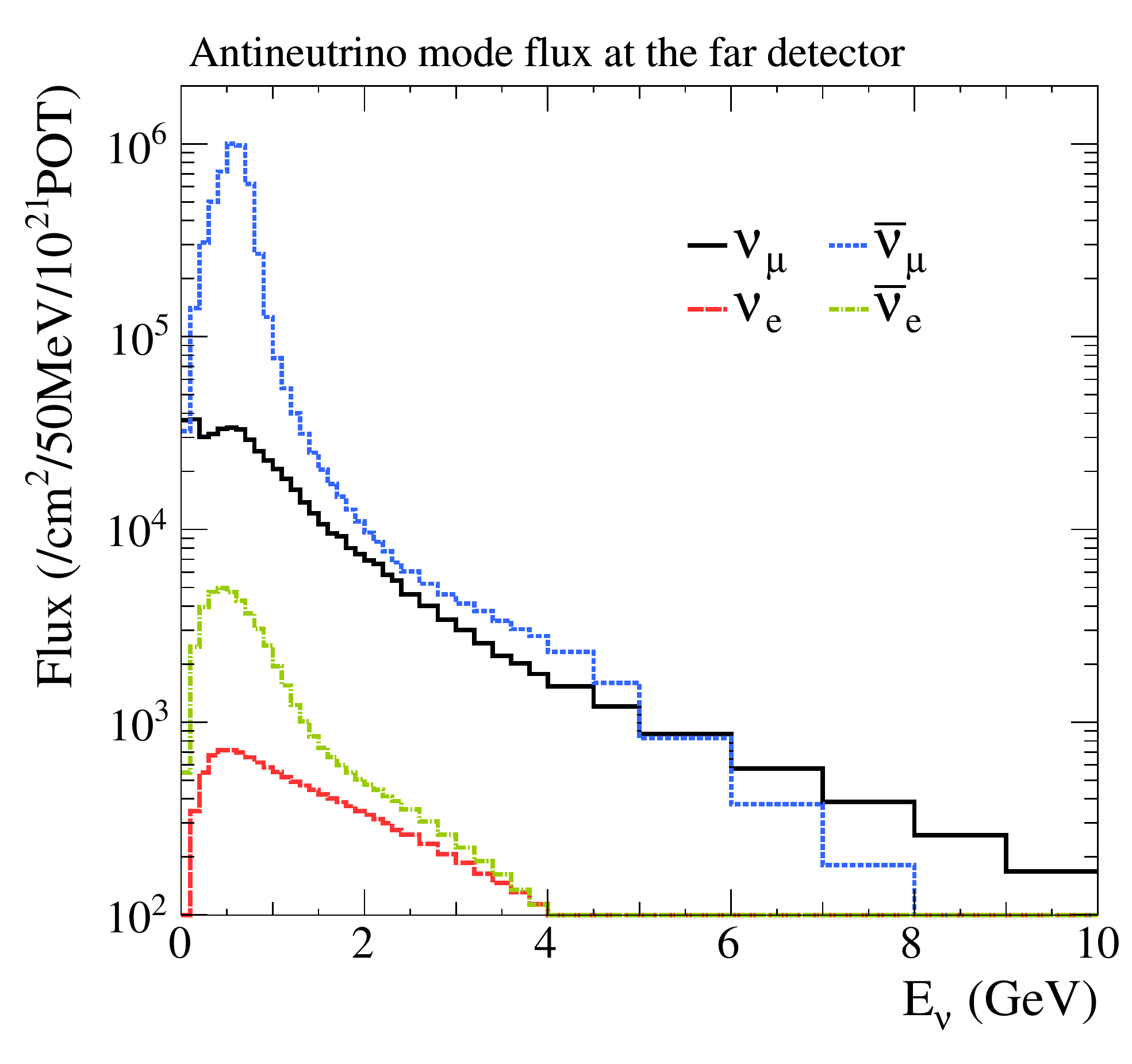} 
\end{subfigure}
\caption{The SK flux prediction for Runs 1--9a with horns operating in FHC ($250$\,kA) mode (upper) and RHC ($-250$\,kA) mode (lower).}
\label{fig:SK_flux}
\end{figure}

Figure~\ref{fig:SK_flux} shows the predicted neutrino fluxes at SK for both the FHC and RHC modes.   Previous T2K flux estimates~\cite{t2k_flux} used thin target hadron production data collected in the NA61/SHINE experiment in 2007~\cite{na61_thinpi, na61_thin_k,na61_2007_k0}, where a 31~GeV proton beam impinged upon a graphite target with a thickness of 4\% of a nuclear interaction length (the so-called \textit{thin} target).
The work presented here uses an updated flux prediction based on higher statistics thin target hadron production data collected in the NA61/SHINE experiment in 2009~\cite{na61_2009_thin}, including the yields of $\pi^{+}\!$, $\pi^{-}\!$, $K^{+}\!$, $K^{-}\!$, $K^{0}_{s}$, $\Lambda$ and $p$.  Future analyses will include NA61/SHINE hadron production measurements on a replica of the T2K neutrino production target.

\begin{figure}[htbp]
\centering
\begin{subfigure}{0.47\textwidth}
\includegraphics[width=0.98\columnwidth]{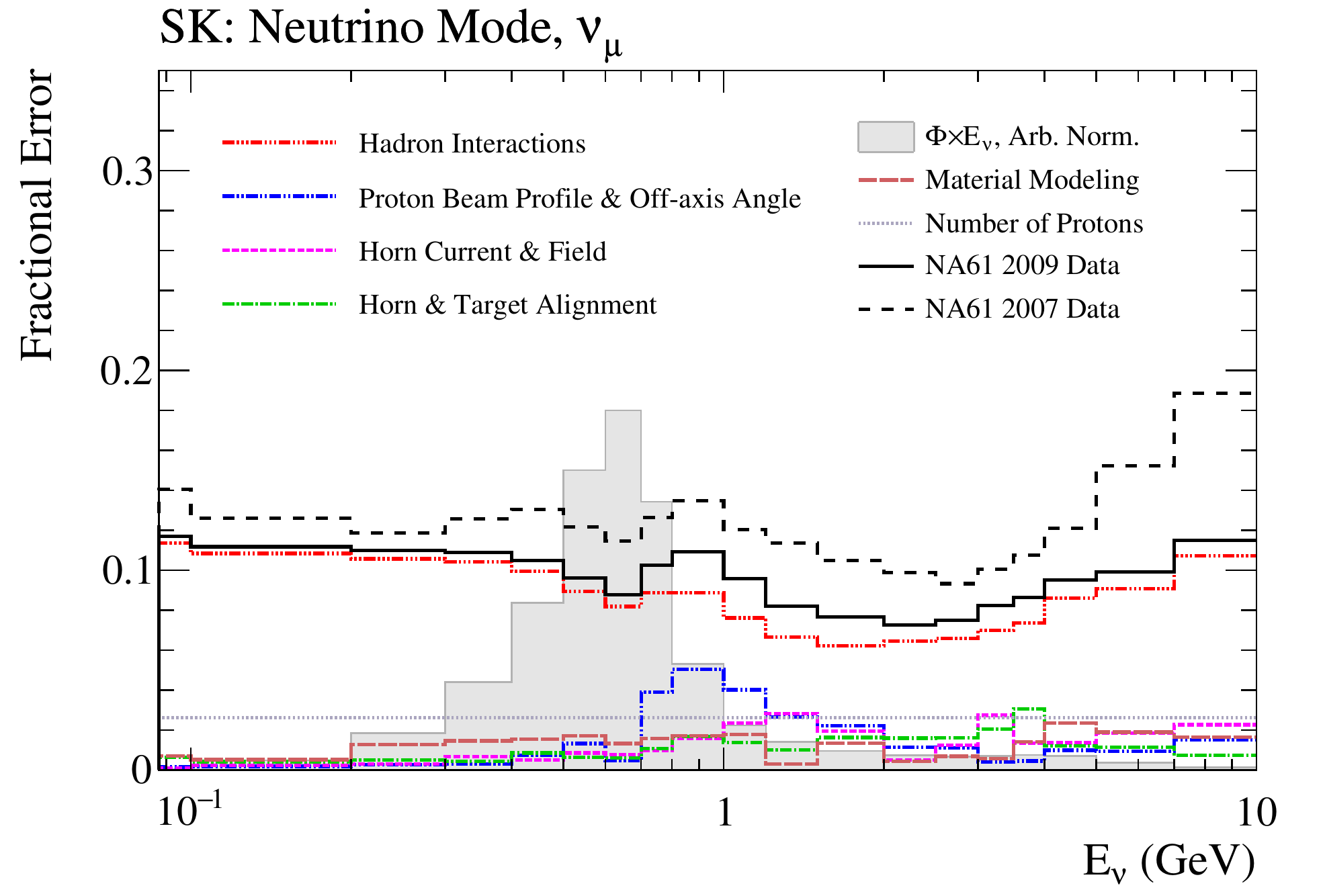} 
\end{subfigure}
\begin{subfigure}{0.47\textwidth}
\includegraphics[width=0.98\columnwidth]{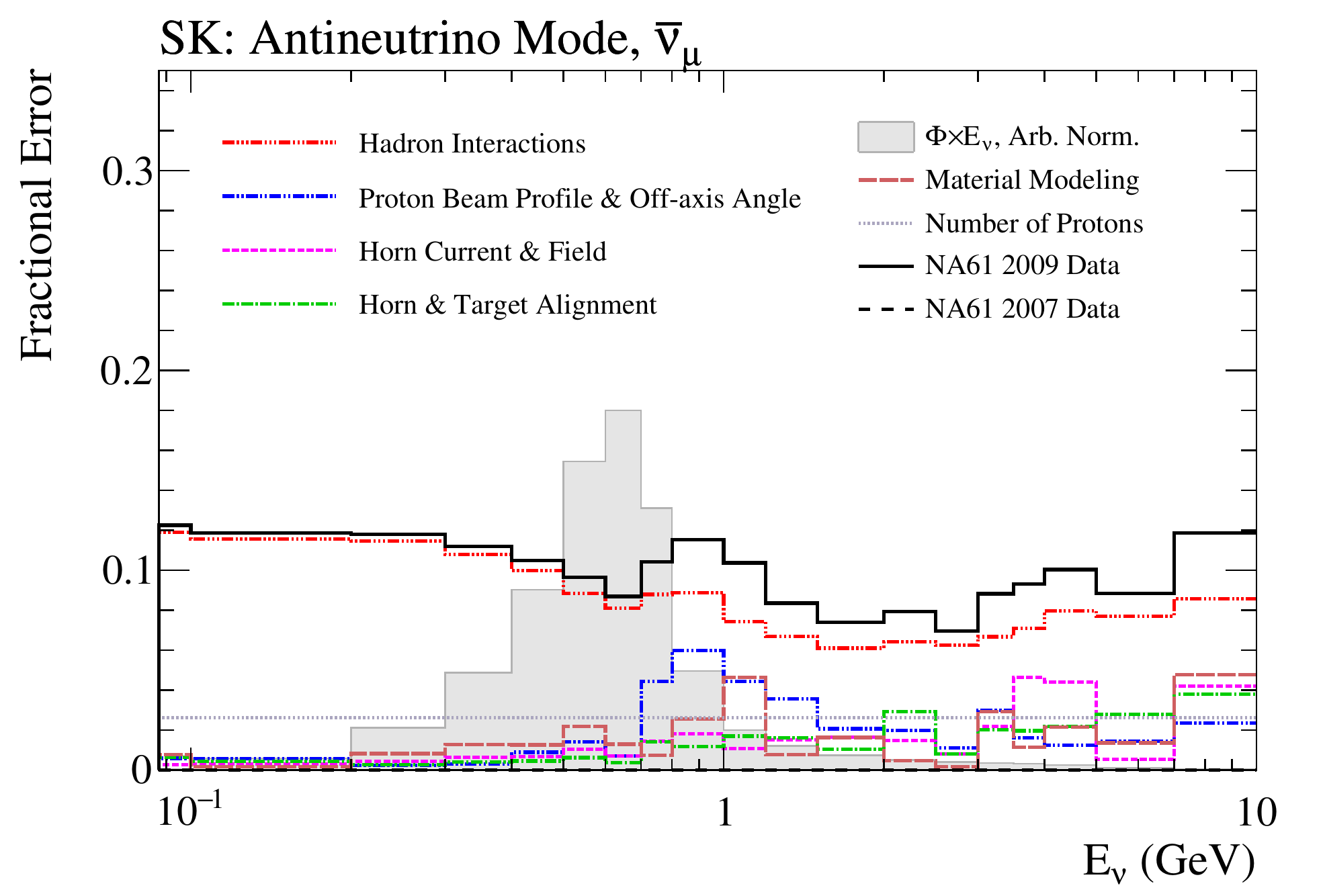} 
\end{subfigure}
\begin{subfigure}{0.47\textwidth}
\includegraphics[width=0.98\columnwidth]{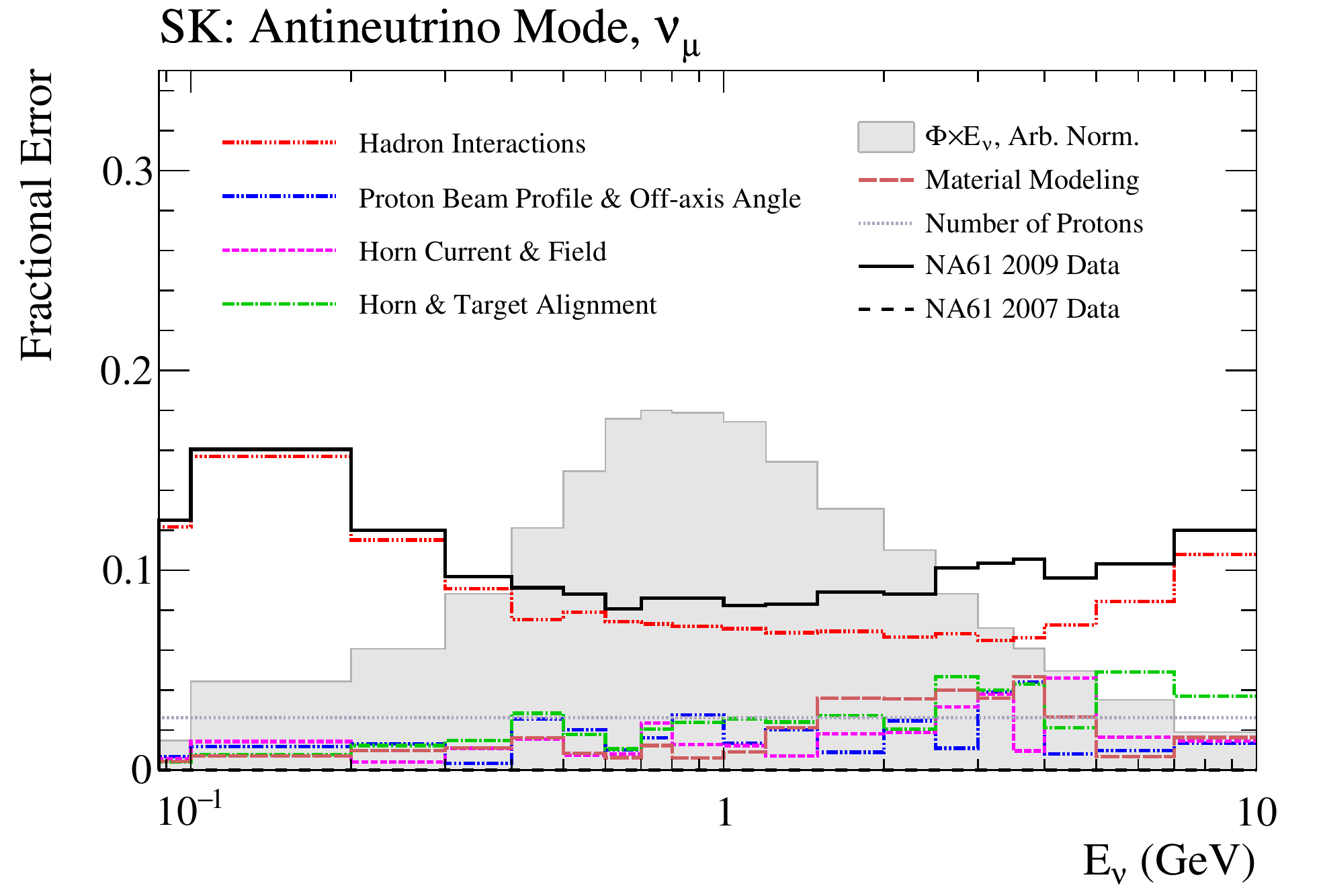} 
\end{subfigure}
\caption{The fractional systematic uncertainty on the $\numu$ flux at SK in FHC mode (top), on the right-sign $\numub$ flux at SK in RHC mode (middle), and on the wrong-sign $\numu$ flux at SK in RHC mode (bottom). The solid black line shows the current total fractional uncertainty (NA61/SHINE 2009 Data), while the dashed black line in the top panel shows the fractional uncertainty from an earlier flux prediction (NA61/SHINE 2007 Data).}
\label{fig:flux_err}
\end{figure}

Figure~\ref{fig:flux_err} shows the fractional uncertainty on the $\numu$ flux at SK in FHC mode, on the wrong-sign $\numub$ flux at SK in FHC mode and on the right-sign $\numub$ flux at SK in RHC mode. The improvement obtained by including the 2009 NA61/SHINE thin target data is also indicated.

The T2K neutrino interaction model is described extensively in Section~\ref{sec:interaction_model}. 
The model incorporates a number of tunable parameters whose prior uncertainties and nominal values come from comparisons to electron and neutrino scattering data sets.
There are a number of models that agree with existing data equally well, and it is not always possible to parameterize the differences between these models.
In this case simulated data studies have been performed using these alternate models, as described in Section~\ref{sec:fake_data_studies}.

The neutrino flux and interaction models are fit to data collected by the T2K off-axis near detector, ND280~\cite{T2K_NIM}.
The fit varies the parameters within both models simultaneously to best match the ND280 data, as described in Section~\ref{sec:ndfit}.
This results in tuned flux and interaction models with correlated uncertainties, providing a more accurate and precise prediction of the event rate at Super-Kamiokande. 

At the far detector a simultaneous fit of muon-like and electron-like samples from both neutrino and antineutrino beams is used to constrain the PMNS oscillation parameters.
The conventional $\{\theta_{ij},\deltacp\}$ parameterization is used, enforcing unitarity, and the effect of propagation in matter is included.
Data from \nue and \nueb disappearance experiments are used to constrain the parameters ($\theta_{12}$ and $\Delta m^2_{21}$)~\cite{pdg_2014} that T2K has little sensitivity to.
Fits are performed both with and without an external constraint on $\theta_{13}$~\cite{pdg_2018}.
Systematic uncertainties are treated by a numerical marginalization technique: all parameterized uncertainties are randomly sampled many times according to prior constraints, including ND280 data, and the likelihoods averaged over the ensemble.
This process is described in detail in Section~\ref{sec:fitters}, with the results described in Section~\ref{sec:results}.

\section{Neutrino Interaction Modeling}
\label{sec:interaction_model}

\newcommand{\NEUT}[0]{\textsc{neut}\xspace}

\newcommand{\figref}[1]{Fig.~\ref{fig:#1}\xspace}
\newcommand{\Figref}[1]{Figure.~\ref{fig:#1}\xspace}

\newcommand{\miniboone}[0]{MiniBooNE\xspace}
\newcommand{\Qsq}[0]{\ensuremath{Q^{2}}\xspace}
\newcommand{\eb}[0]{\ensuremath{E_\textsc{b}}\xspace}
\newcommand{\maqe}[0]{\ensuremath{M_{\textsc{A}}^{\textsc{QE}}}\xspace}
\newcommand{\erecqe}[0]{\ensuremath{E^{\textrm{Rec}}_{\textsc{QE}}}\xspace}

Oscillation parameter values are inferred from spectra of observable quantities, herein either reconstructed charged-lepton kinematics, $\left(p_{\ell},\theta_{\ell}\right)$, or reconstructed neutrino energy. The reconstructed neutrino energy is estimated from final-state charged-lepton kinematics only as
\begin{equation}
\erecqe\left(p_{\ell},\theta_{\ell}\right)=\frac{2M_{N,i}E_{\ell} - M_{\ell}^{2} + M_{N,f}^{2} - M_{N,i}^{2}}{2\left(M_{N,i}-E_{\ell}+p_{\ell}\cos{\theta_{\ell}}\right)},
\end{equation}
where $M_{N,i}$, $M_{N,f}$, and $M_{\ell}$ are the mass of the initial-state nucleon (an effective, off-shell mass that includes the `nucleon removal energy' is often used), final-state nucleon, and final-state charged lepton respectively; $E_{\ell}$, $p_{\ell}$, and $\theta_{\ell}$ are the energy, three-momentum, and angle of the final-state charged-lepton respectively. \erecqe provides a smeared but minimally-biased estimate of the neutrino energy for quasi-elastic neutrino scattering off bound nucleons (CCQE). For other interaction channels, such as those that produce extra hadrons, \erecqe underestimates the energy of the interacting neutrino.

The procedure of inferring oscillation parameter values from observable quantities implicitly relies on an accurate understanding of the rate of background processes and a mapping between true energy and observable quantities, \emph{e.g.} $\erecqe\left(E_\nu\right)$, both of which are derived from simulation.
As a result, accurate neutrino interaction modeling is critical. Event selections are trained on the simulated distribution of final-state particles and the predicted rate of various signal and background processes. The predicted rate of a number of neutrino interaction processes, which exhibit different true-to-observable mappings, is constrained by near detector data and then used to interpret the observed far detector data.
This section briefly describes the neutrino interaction model, accounted-for freedom within the model, and specific studies used to test resilience to known weaknesses of the model.

\subsection{The Base Interaction Model}
The samples of simulated neutrino interactions used in this analysis were made with version 5.3.3 of the \NEUT interaction generator~\cite{Hayato:2009zz}. \NEUT simulates known neutrino interaction channels relevant for few GeV neutrinos, these channels are broadly categorized as: 1p1h, 2p2h, single pion production, and deep inelastic scattering (DIS). In addition to the `primary' interaction channels, the effect of using nuclear targets, where the struck nucleons are bound within a nuclear potential, needs to be modeled well. These effects can be separated into initial-state and final-state effects. Most updates to the interaction model since the previous analysis~\cite{Abe:2017vif_T2Krun7osc}, are in the treatment of systematic uncertainties; however, a short description of the whole model is included here for completeness. As the `base' model has not changed, the interested reader is directed to Ref.~\cite{Abe:2017vif_T2Krun7osc} and Ref.~\cite{PhysRevD.93.072010} for a discussion of the motivations behind any specific model choices.

\paragraph{Initial-state nuclear effects:} Nucleons bound within a nuclear potential undergo non-negligible `Fermi motion'. For carbon, this means bound nucleons have a momentum of $p_{\mathrm{f}} \lesssim 217\MeVmom$, or equivalently, a Fermi energy of $E_{\mathrm{f}} \lesssim 25\MeV$. A Global Relativistic Fermi Gas (GRFG) is used to model the initial-state nucleon momentum distribution in this analysis. Neutrino interactions with bound nucleons are largely handled under the impulse approximation, whereby a single `struck' nucleon receives a four-momentum kick while the rest of the target nucleus acts as a group of non-interacting `spectator' nucleons. This rudimentary nuclear model is a simple approximation for the correct modeling of the initial nucleon momentum distribution and nucleon removal energy, a study, presented below, accounts for the effect of this approximation.

\begin{figure}[htbp]
    \centering
    \includegraphics[width=0.47\textwidth]{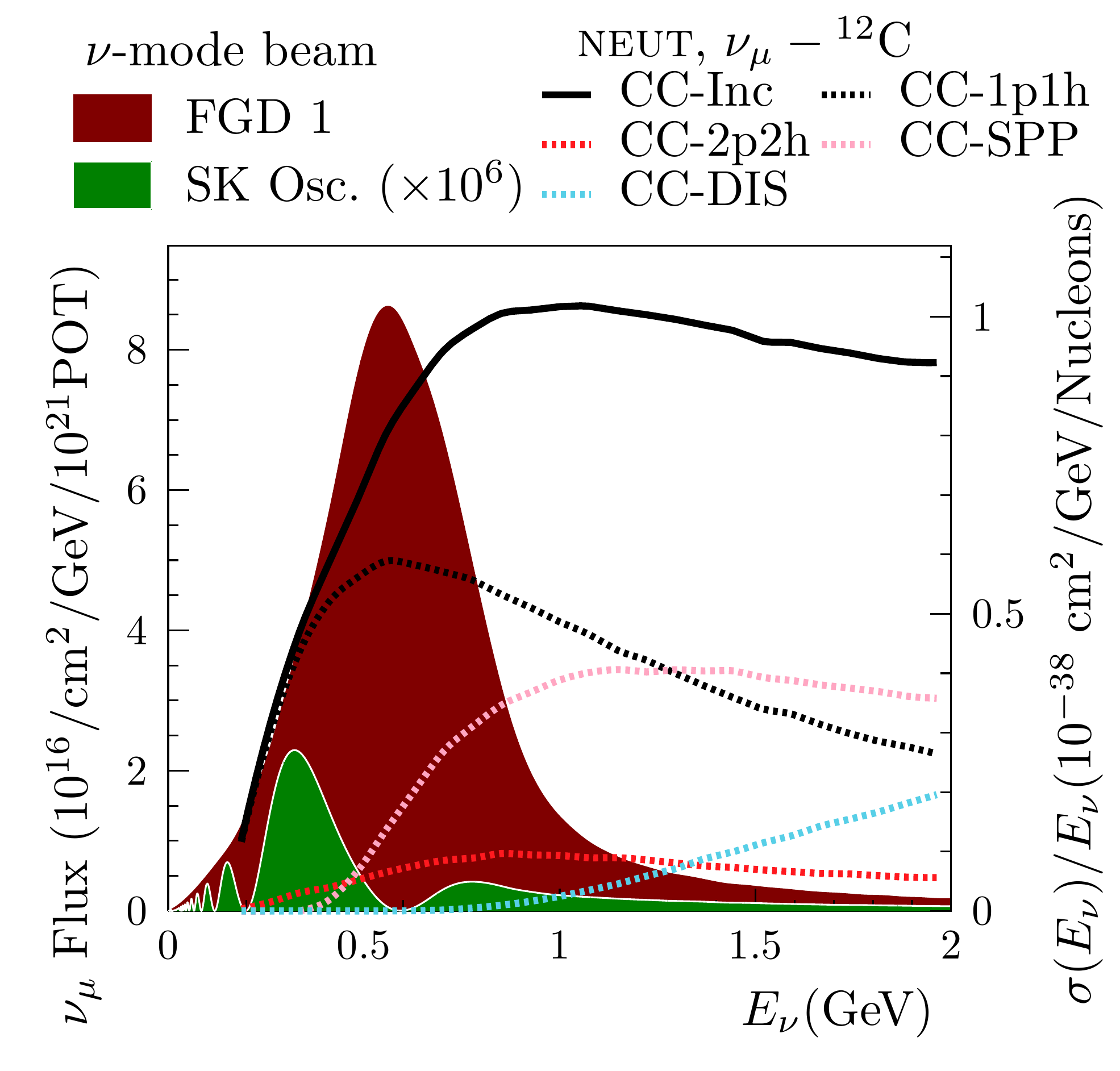}
    \caption{The total charged-current cross section for muon neutrinos interacting with a carbon nucleus, as predicted by \NEUT, overlaid on the ND280 muon neutrino flux, and an example oscillated muon neutrino flux at SK. The oscillation parameters used here are the best fit from the previous analysis ~\cite{Abe:2017vif_T2Krun7osc}. The total (Inc) cross section is separated into 1p1h, 2p2h, single pion production (SPP), and deep inelastic scattering (DIS) channels.}
    \label{fig:intmodel_sigenu}
\end{figure}

\paragraph{1p1h:} One particle, one hole interactions are those where the neutrino interacts quasi-elastically with a single bound nucleon---the interaction is only \emph{quasi}-elastic because of the bound nature of the target nucleon and, for charged current events (CCQE), the initial-to-final-state charged-lepton and nucleon rest mass difference. Such interactions are modeled in the Lewellyn--Smith formalism ~\cite{llewellyn_smith_neutrino_1972}, using the BBBA05~\cite{Bradford_2006} description for the vector part of the nucleon form factors, and a simple dipole form for the axial part. The \NEUT model includes two additional features of note: the nucleon removal energy, `NRE', and in-medium modifications to the W boson propagator via the Random Phase Approximation  (`RPA'). Variations in the average nucleon removal energy modify the predicted kinematics of final-state particles, most importantly charged leptons. 
When comparing predictions based on Fermi Gas nuclear models to 1p1h-like cross-section data, a suppression at low four-momentum transfer is favored relative to the free-nucleon-target calculation~\cite{morfin2012recent}. This is often attributed to a weak-charge screening effect as a result of the nuclear medium~\cite{NIEVES20161830}. The effect is termed `RPA' after the `Random Phase Approximation' technique used to sum up the series of contributing W-boson self-energy diagrams. Here, the distribution of four-momentum transfer is modified by the RPA calculation from Nieves~\emph{et.\,al.}~\cite{NIEVES20161830}. As can be seen in \figref{intmodel_sigenu}, 1p1h is the dominant interaction channel at T2K energies.

{}

\begin{figure}[htbp]
    \centering
    \begin{subfigure}{0.47\textwidth}
        \hspace{0.25cm}\includegraphics[width=0.98\columnwidth]{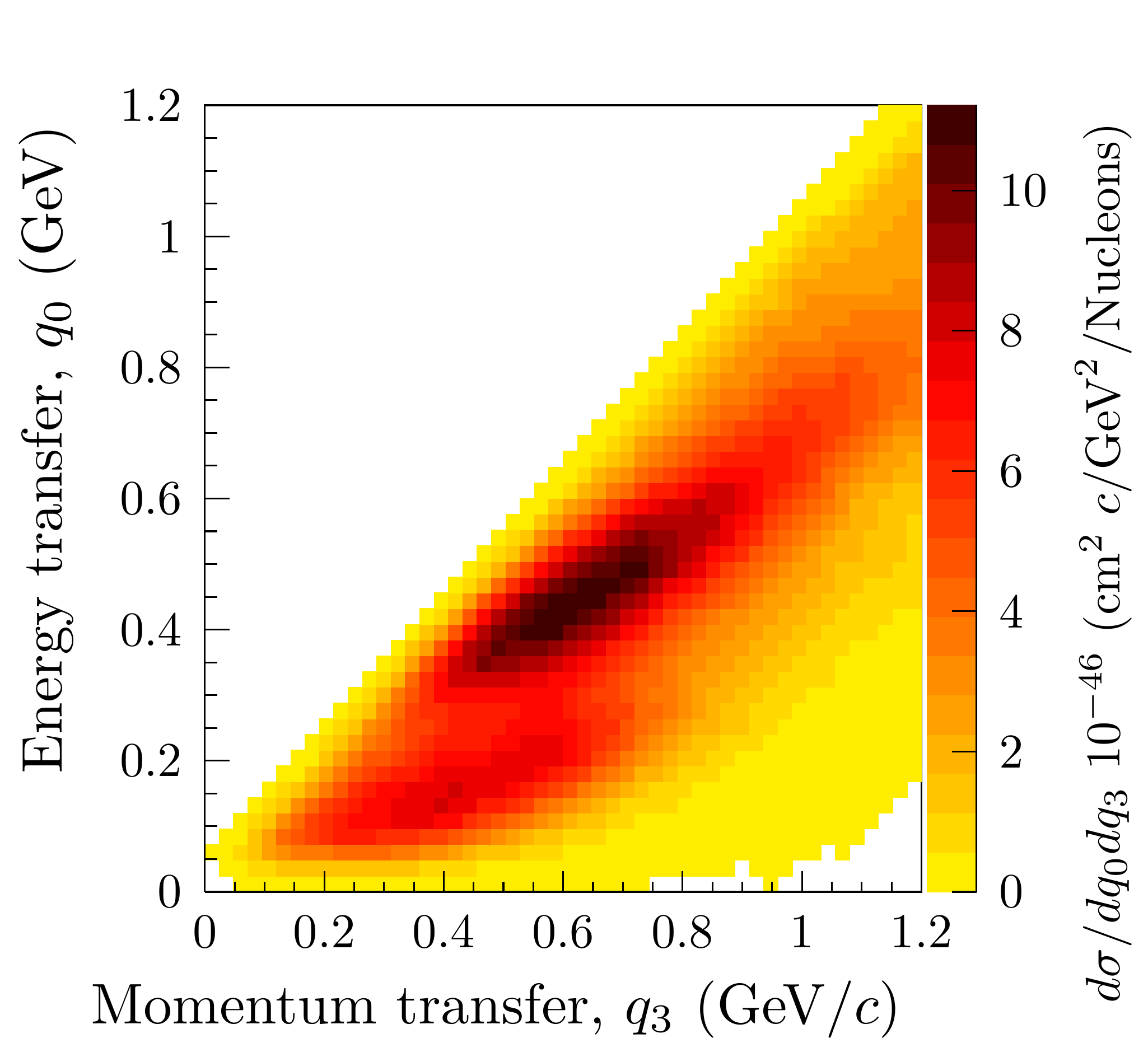}
    \end{subfigure}
    \begin{subfigure}{0.47\textwidth}
        \hspace{-0.5cm}\includegraphics[width=0.98\columnwidth]{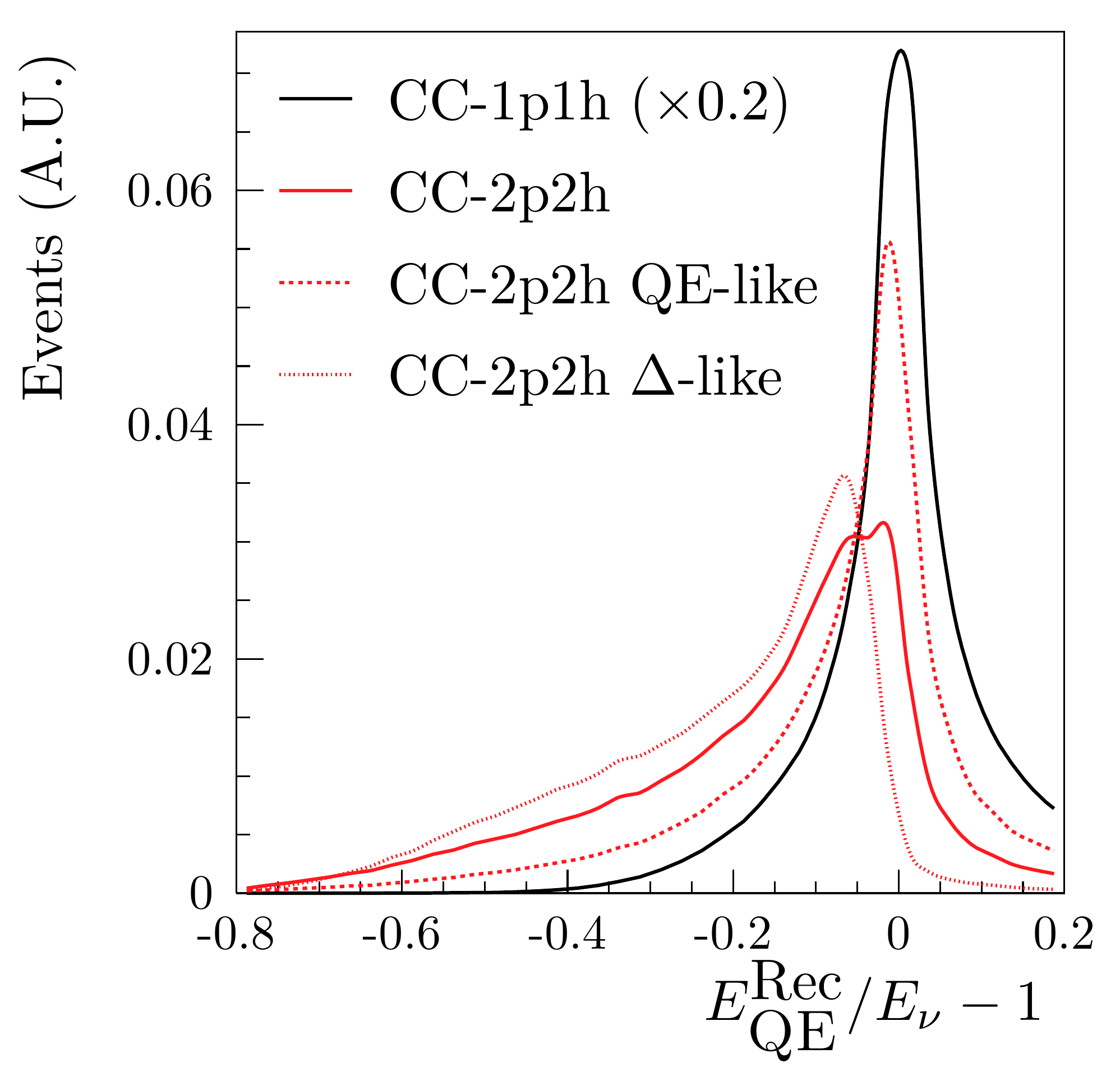}
    \end{subfigure}
    \caption{Top: The energy and momentum transfer distribution for the Nieves~\emph{et.\,al.} 2p2h model in \NEUT. The two peak structure is clear, with QE-like kinematics corresponding to the lower left peak, and Delta-like kinematics to the stronger central peak. Bottom: The reconstructed energy bias at SK is shown for 1p1h and 2p2h events for an oscillated muon neutrino flux. The different reconstructed energy smearing for 2p2h events with QE-like and Delta-like interaction kinematics can be seen.}
    \label{fig:qel_erec_bias}
\end{figure}

\paragraph{2p2h:} Two particle, two hole interactions are an inherently nuclear-target process, whereby the incoming neutrino interacts with a bound pair of nucleons, knocking both out of the nuclear potential. The Nieves~\emph{et.\,al.} model~\cite{PhysRevD.88.113007} is used to predict the cross-section as a function of lepton kinematics. While this process is sub-dominant, it produces observable final states that are indistinguishable from 1p1h interactions in the T2K detectors, but with different observed lepton kinematics as a function of neutrino energy. In the Nieves~\emph{et.\,al.} 2p2h model, there are two distinct regions of strength in the energy and momentum transfer space: the quasi-elastic-like (energy transfer, $q_{0}\lesssim0.3$), and Delta-like regions ($q_{0}\gtrsim0.3$). The energy-momentum transfer distribution and the corresponding \erecqe biases can be seen in \figref{qel_erec_bias}.

{}

\paragraph{Single pion production:} Single pion production can be separated into three sub-processes: resonant, non-resonant, and coherent single pion production. The resonant and non-resonant processes describe the production of a pion involving neutrino scattering of a single nucleon, either via an intermediate baryon resonance (resonant), or not (non-resonant). These processes are modeled in the Rein--Sehgal formalism~\cite{Rein:1980wg}, with an improvement that includes the effect of the final-state charged-lepton mass~\cite{PhysRevD.76.113004}, and updated nucleon axial form factors from Graczyk \& Sobczyk~\cite{PhysRevD.90.093001}. The contributions from 17 baryon resonances are considered, with the $\Delta(1232)$  being dominant, and interference terms between the resonances are taken into account. The non-resonant channel augments the production of half-unit isospin final states (\emph{e.g.} $\nu + n \rightarrow \ell^{-} + p^{+} \pi^{0}$ and $\nu + n \rightarrow \ell^{-} + n \pi^{+}$); any interference between the resonant and non-resonant contributions is ignored. These processes are used to model final states with an invariant hadronic mass of $W \leq 2.0\GeVmass$. The modeling of the so-called 'transition region' between single pion production off a nucleon and shallow- and deep-inelastic scattering is an unsolved theoretical problem~\cite{ALVAREZRUSO20181}. In the \NEUT model, the region $1.3 \leq W \leq 2.0\GeVmass$ contains contributions from both the Rein--Sehgal single pion model, described above, and the deep-inelastic-scattering model, described below. For higher invariant masses the deep inelastic scattering model is used. 

The axial form factors and the strength of the non-resonant contribution in the Rein--Sehgal model were tuned to published cross-section data using the NUISANCE framework~\cite{Stowell_2017}. As these parameters control only the nucleon-level interaction, the central values were determined from fits to deuterium-target bubble chamber data, which is largely free from nuclear effects. Data from ANL~\cite{PhysRevD.25.1161} (with some reanalyzed distributions taken from Ref.~\cite{PhysRevD.90.112017}) and BNL~\cite{PhysRevD.34.2554} was used. The uncertainties determined from the fits to bubble chamber data were then inflated to approximately cover cross-section data from \miniboone~\cite{PhysRevD.83.052007} and \minerva~\cite{PhysRevD.92.092008,PhysRevD.94.052005}.

Coherent single pion production describes the interaction of a neutrino coherently with a whole nucleus. This is a sub-dominant pion production process, observed at low energy for the first time by the \minerva experiment~\cite{PhysRevD.97.032014} and is characterized by very little four momentum transfer to the struck nucleus. We follow the preference of the \minerva data and use the Bergher--Sehgal model~\cite{PhysRevD.79.053003}. 

\paragraph{Deep inelastic scattering:} For interactions producing hadronic systems with two or more pions and invariant hadronic masses of $W > 1.3\GeVmass$, the cross section is constructed from nucleon structure functions that depend on the Bjorken scaling variables $x$ and $y$. The structure functions are calculated from the GRV98~\cite{gluck_dynamical_1998} parton distribution functions, with modifications from Bodek~\emph{et.\,al.}~\cite{bodek2003modeling} to account for the relatively low momentum transfers involved. For interactions with  $1.3 < W \leq 2.0\GeVmass$ the hadronic state is generated by a custom multi-pion production model, above $W = 2.0\GeVmass$ PYTHIA 5.72 is used~\cite{sjostrand_high-energy-physics_1994}.

\paragraph{Final-state nuclear effects:} After the \emph{primary} neutrino interaction has been simulated, a number of additional `nuclear effects' are included. For interactions that produce a final-state proton or neutron, the Pauli exclusion principle is applied, rejecting any events that produce a nucleon below the Fermi energy. This results in a suppression at low four-momentum transfer for 1p1h events. Final-state hadrons produced at the neutrino interaction vertex are stepped through the nuclear medium in a classical cascade, in which they may: interact and produce secondary particles, be absorbed, or undergo charge exchange (\emph{e.g.} $\pi^{+} + n \rightarrow \pi^{0} + p$). Such re-interactions are often called `Final-State Interactions', or FSIs. 
Finally, after the primary interaction and hadronic cascade, the remnant nucleus can be left in an excited state that will subsequently decay. For interactions on oxygen, nuclear de-excitations that result in secondary, low energy photons ($\mathcal{O}\left(1-10\right)\,\textrm{MeV}$) are modeled following Ref.~\cite{PhysRevC.48.1442}.

With the exception of 2p2h interactions, these channels and effects are also implemented for neutral current interactions, but the details are not repeated here for brevity. The total charged-current cross-sections, broken down by interaction channel, are shown in \figref{intmodel_sigenu}.

\subsection{The Uncertainty Model}

As the number of observed events included in the analysis grows with exposure, a robust interaction uncertainty model is required to assess the significance of the results. The uncertainty model for 1p1h and 2p2h interactions have seen recent improvements and will be discussed in detail here. For details on other, unchanged sources of interaction uncertainty see Ref.~\cite{Abe:2017vif_T2Krun7osc} Section III.

\paragraph{1p1h} The \NEUT 1p1h interaction model implements three main sources of uncertainty: the mass used in the dipole axial form factor (\maqe), the effect of RPA on the cross section as a function of four-momentum transfer, and the Nucleon Removal Energy associated with scattering off a bound nucleons.

\begin{figure}[htbp]
    \centering
    \includegraphics[width=0.47\textwidth]{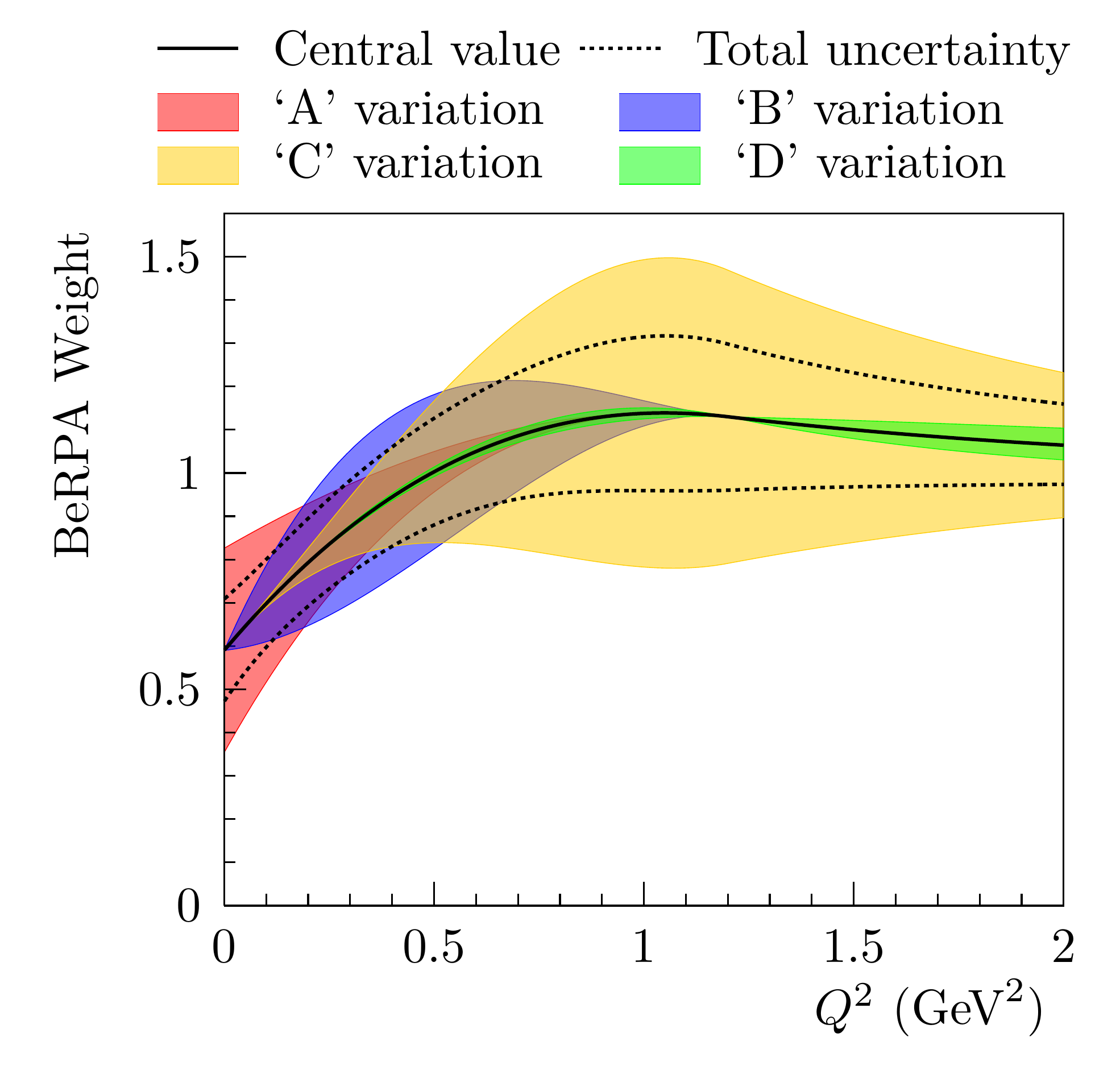}
    \caption{The central value BeRPA suppression factor (solid) and the total prior uncertainty (dashed) determined from $10^{5}$ uncorrelated throws of the free parameters, overlaid on the corresponding envelopes generated by varying each parameter. The continuity conditions result in a perhaps unintuitive total uncertainty envelope given the individual parameter variations.}
    \label{fig:intmodel_berpa}
\end{figure}

In this analysis, \maqe does not have a prior uncertainty and is constrained by near detector data alone. The parameterization of the uncertainty on the RPA suppression has been updated in this analysis; the previous implementation proved problematic because variations of different free parameters effected a similar response in \Qsq. For this analysis, Bernstein polynomials were used to model the shape below some \Qsq cutoff, $U$, above which an exponential decay form is used: 
\begin{equation*}
  \label{eq:berpa}
  f(x) = 
  \begin{cases}
    A(1-x')^{3} + 3B(1-x')^{2}x'\\
    \hspace{10pt}+~3p_{1}(1-x')x'^{2} + Cx'^{3}, & x < U \\[5pt]
    1 + p_{2}\exp(-D(x-U)), & x > U
  \end{cases},
\end{equation*}
where $x=Q^{2}$ and $x^\prime = \nicefrac{Q^{2}}{U}$. To ensure continuity at $Q^{2} = U$, the conditions:
\begin{align*}
  p_{1} &= C + \frac{UD(C-1)}{3}\\
  p_{2} &= C - 1
\end{align*}
are enforced, leaving four free parameters, $A, B, C, D$. The fifth, $U$, is kept fixed at $\SI{1.2}{\giga\eV\squared}$.
This parameterization has been termed `BeRPA' after the Bernstein polynomials on which it is based.
The effect of varying each of the four free parameters relative to the theoretical uncertainty calculated by following Ref.~\cite{VALVERDE2006325} can be seen in \figref{intmodel_berpa}. Together, the four free BeRPA parameters and the \maqe parameter give effective freedom over a range of \Qsq. The \Qsq distribution is then largely constrained by the fit to near detector samples.
{}

\paragraph{2p2h}
The details of the 2p2h process are highly uncertain. The total cross-section, evolution with neutrino energy, and energy and momentum transfer characteristics of the process are all predicted differently by the available models (\emph{e.g.} Nieves~\emph{et.\,al.}~\cite{PhysRevD.88.113007}, Martini~\emph{et.\,al.}~\cite{PhysRevC.80.065501}, SUSAv2-MEC~\cite{Simo_2017}, GiBUU~\cite{PhysRevC.94.035502}). While data from the T2K near detector~\cite{PhysRevD.93.112012,PhysRevD.98.032003,PhysRevD.101.112001}, \minerva~\cite{PhysRevLett.116.071802,PhysRevLett.120.221805}, and \nova~\cite{PhysRevD.98.032012} favor a process with similar interaction kinematics to such a multi-nucleon process\footnote{\emph{i.e.} at fixed momentum transfer, a process that occurs between the quasi-elastic and pion production peaks in energy transfer.}, experimental sensitivity to this exclusive channel is weak. As a result, significant freedom is afforded to the 2p2h process in this analysis.

\begin{figure}[htbp]
    \centering
    \begin{subfigure}{0.47\textwidth}
        \includegraphics[width=0.98\columnwidth]{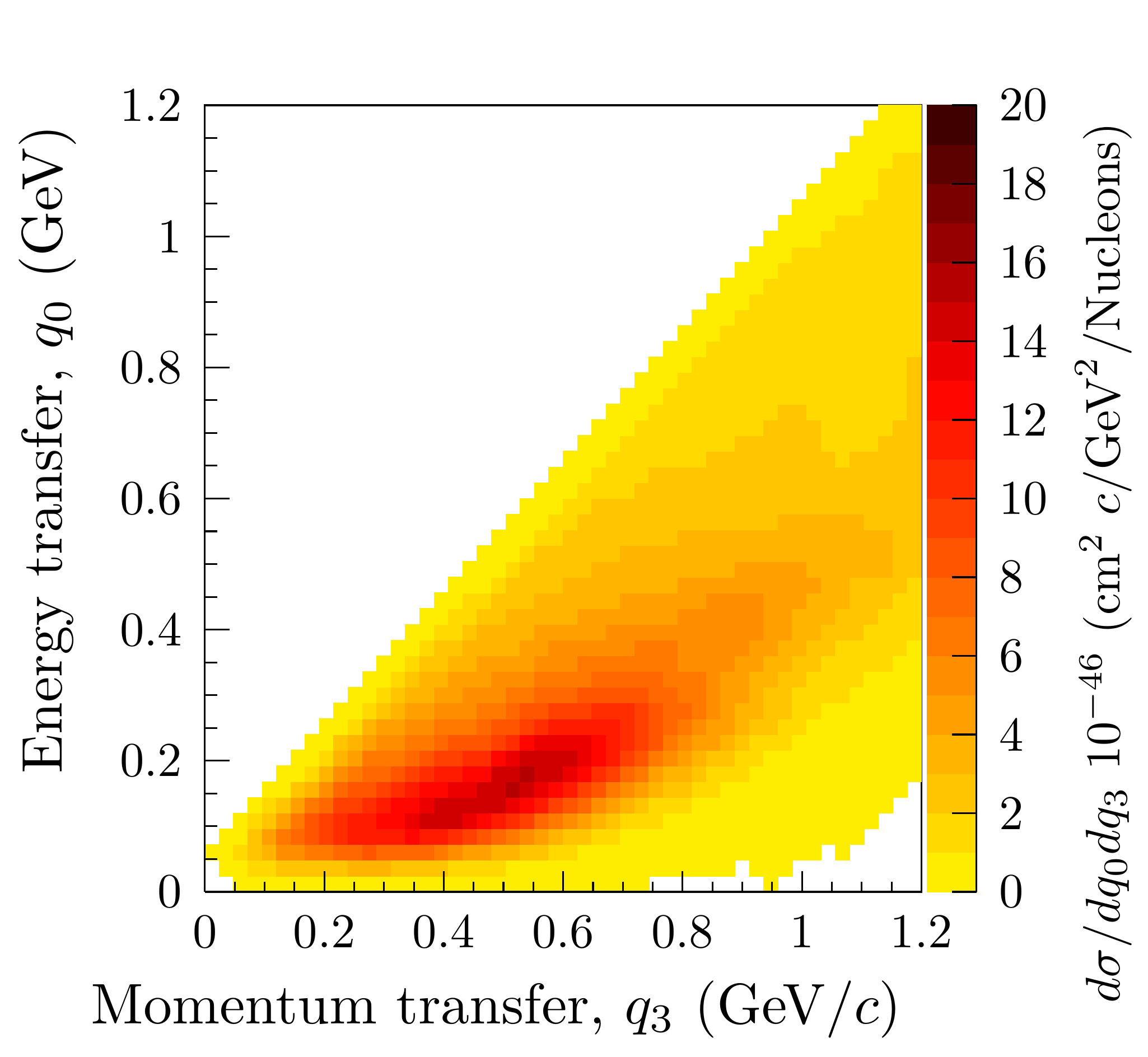}
    \end{subfigure}
    \begin{subfigure}{0.47\textwidth}
        \includegraphics[width=0.98\columnwidth]{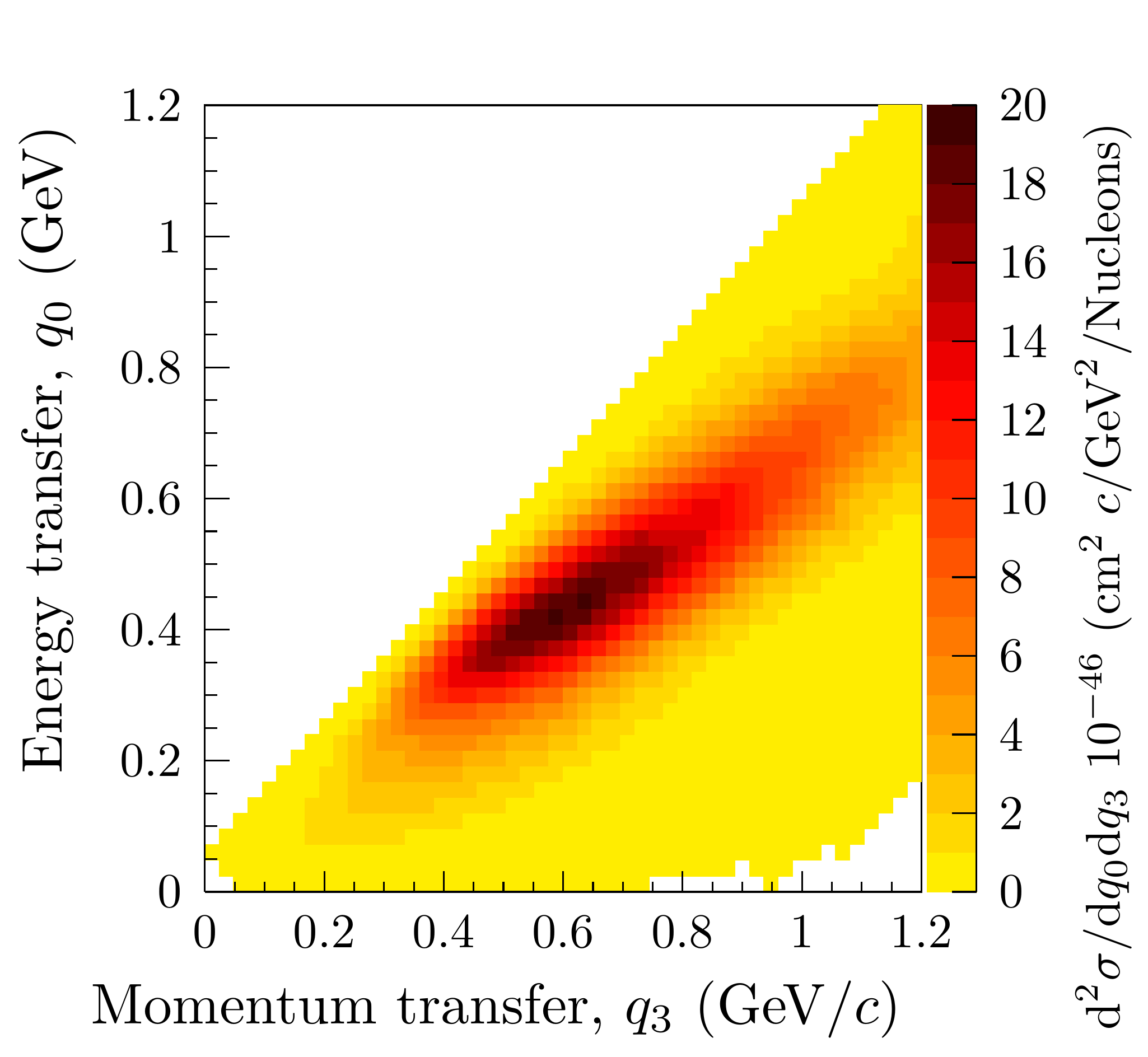}
    \end{subfigure}
    \caption{The differential cross section for the two extreme variations of the 2p2h `shape' parameter---QE-like (top) and Delta-like (bottom)---\emph{c.f.} \figref{qel_erec_bias} (top) for the central-value-predicted cross section.}
    \label{fig:intmodel_q0q3}
\end{figure}

The uncertainty on the 2p2h process is separated into normalization and shape components. An overall 100\% normalization uncertainty is separately assigned to interactions involving neutrinos and those involving antineutrinos. An additional parameter that introduces freedom in the relative normalization of 2p2h interactions with carbon and oxygen nuclei is used.
Finally, a parameter that varies the relative strength of the QE-like and Delta-like components, while keeping the total cross-section for 2p2h events constant, is introduced; the effect of extreme variations of this parameter are seen in \figref{intmodel_q0q3}. 

{}

\subsection{Simulated data studies}

It is strongly suspected that the described uncertainty model may not cover all differences between nature and the interaction model described above. To begin to address this, we perform fits of the model to targeted `simulated data sets' that test the robustness of the model and associated uncertainties to known missing features. In some cases the results of these studies are used to motivate additional uncertainties. This section introduces the simulated data sets that were analyzed to address specific concerns, the results of the fits will be discussed in Section~\ref{sec:fake_data_studies}.

\begin{figure}[htbp]
    \centering
    \includegraphics[width=0.47\textwidth]{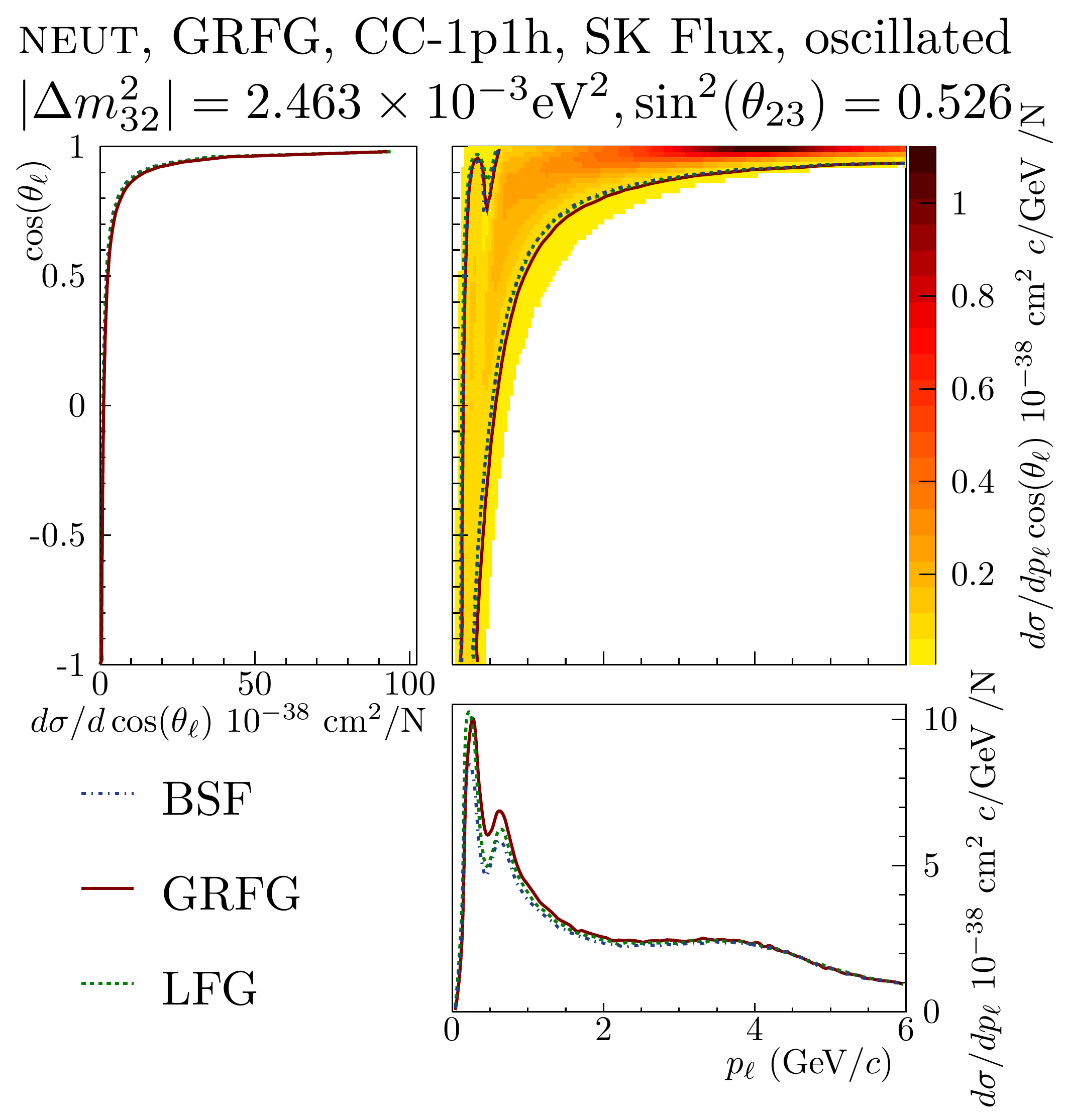}
    \caption{The predicted flux-averaged cross section for three different nuclear response models. The different models result in different predictions of the observed final-state lepton kinematics for the same oscillated neutrino flux. If such variations are not accounted for in the uncertainty model, extracted oscillation parameters may be biased. The contours contain the region of phase space with a differential cross section
    $\ud\sigma/\ud p_{\ell}\;\!\cos(\theta_{\ell}) > 0.05 \times 10 ^{-38} \textrm{cm}^{2}\ c/\textrm{GeV}\ /\textrm{N}$.}
    \label{fig:nucmodels}
\end{figure}

\paragraph{Alternative 1p1h Nuclear Models} The GRFG used to model the nuclear initial state is a simple model that contains no correlations between initial momentum and Nucleon Removal Energy ($\textsc{nre}$). Such correlations may be important for correctly modeling the observed charged lepton spectrum~\cite{PhysRevD.72.053005} and are seen in nuclear response measurements from electron scattering experiments. To test the robustness of the implemented uncertainty model to such details, two simulated data sets are used: the Nieves~\emph{et.\,al.} 1p1h model, and the spectral function model of Benhar~\emph{et.\,al.}~\cite{PhysRevD.72.053005} (BSF). Both contain some correlation between the initial momentum and the $\textsc{nre}$. The Nieves~\emph{et.\,al.} model differs from the base model by implementing a local, rather than a global, Fermi Gas (LFG), in which the concept of a radially-dependent nuclear density profile introduces such correlations. In the BSF model, initial nucleons are chosen from a full, two dimensional nuclear response distribution, which is constructed from ($e$,$e^{\prime}p$) data~\cite{PhysRevD.72.053005}. It should be noted that the BSF model contains no `RPA'-like, low four-momentum-transfer suppression effect. In constructing the simulated data, only the 1p1h cross section is modified. The predicted final-state muon kinematics for each model are shown in \figref{nucmodels}.

{}

\begin{figure}[htbp]
    \centering
    \begin{subfigure}{0.47\textwidth}
        \includegraphics[width=0.98\columnwidth]{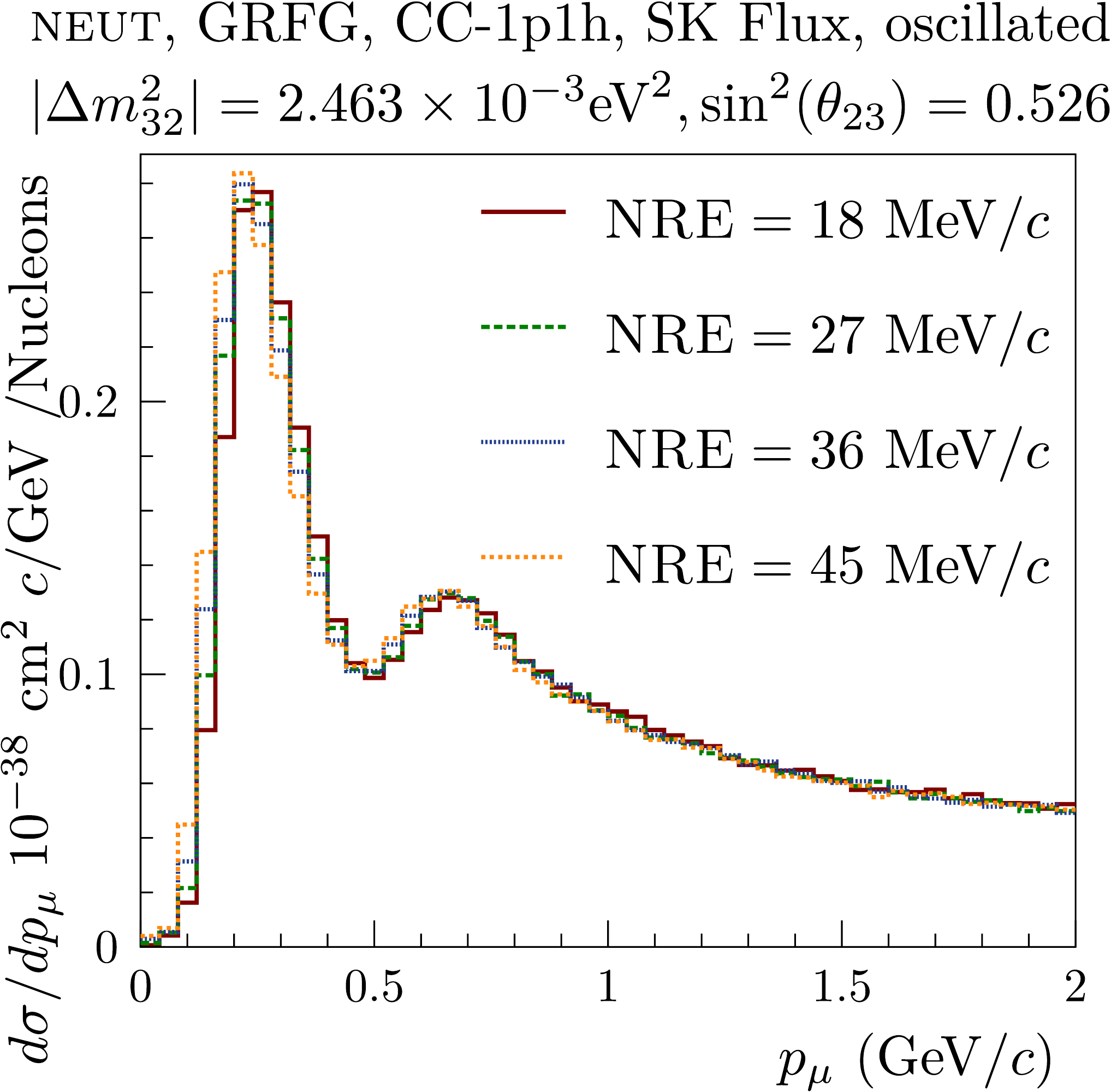}
    \end{subfigure}
    \begin{subfigure}{0.47\textwidth}
        \includegraphics[width=0.98\columnwidth]{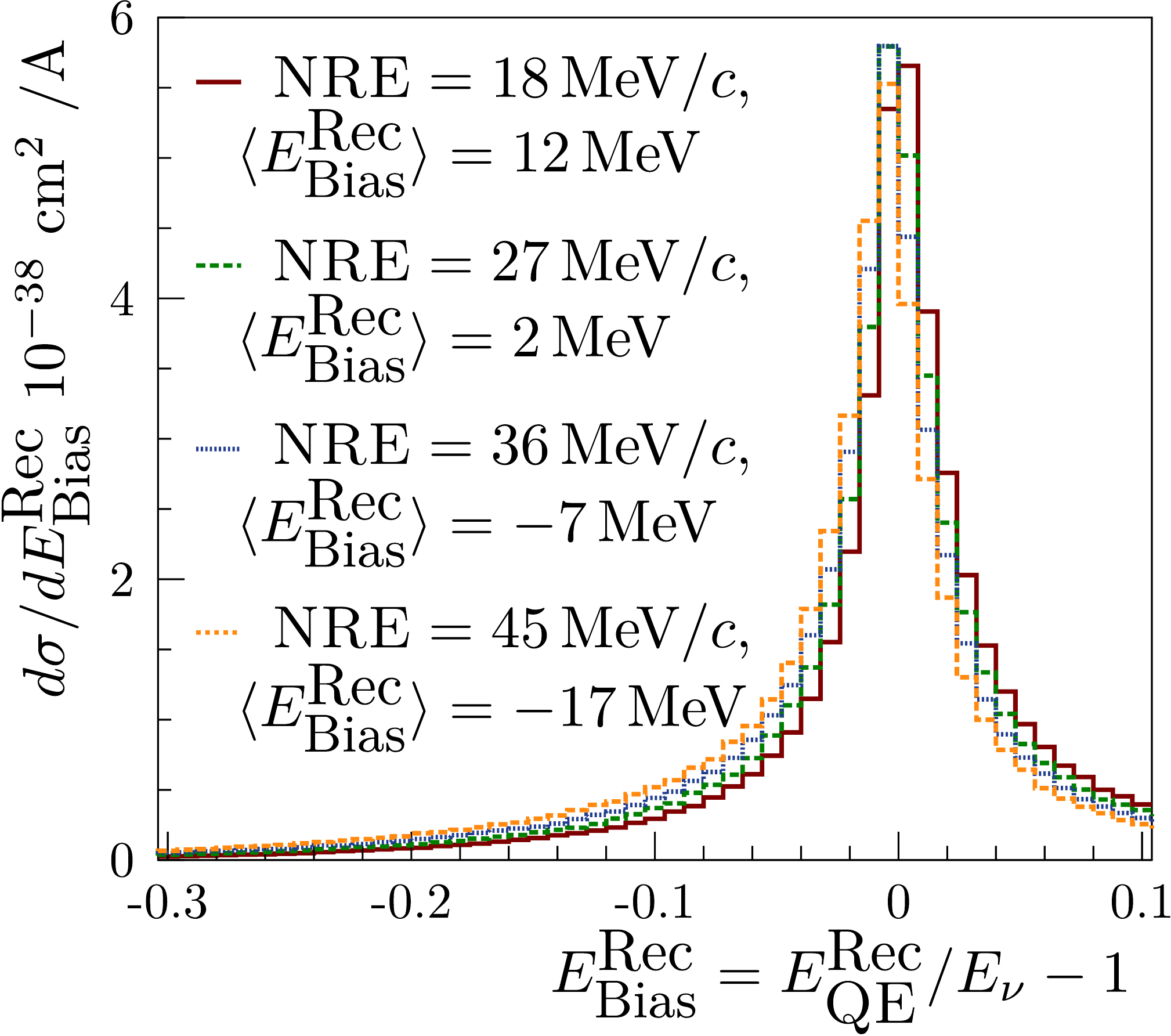}
    \end{subfigure}
    \caption{Top: The predicted muon momentum spectrum for a number of different values of $\textsc{nre}$ in an oscillated SK flux. The shift towards lower momentum may be confused for a shift in $|\Delta{}m_{32}^2|$. Bottom: The effect of varying $\textsc{nre}$ on the reconstructed energy bias; higher values result in more energy `feed down'.}
    \label{fig:intmodel_emiss}
\end{figure}

\paragraph{Nucleon Removal Energy} 
The implementation of the $\textsc{nre}$ parameter was revised for this analysis. The previous implementation relied on calculating the change in the predicted differential cross-section for a variation $\textsc{nre}\rightarrow \textsc{nre}^\prime$, $\sigma^{\textsc{CCQE}}\left(E_\nu,p_\ell,\theta_\ell,\textsc{nre}\right)$; this proved problematic as variations of $\textsc{nre}$ modify the available kinematic phase space for the production of final-state muons. For some simulated interaction, $\left(E_\nu,p_\ell,\theta_\ell\right)$, and binding-energy variation, $\textsc{nre}^\prime$, the variation weight, \(w = \nicefrac{\sigma^{\textsc{CCQE}}\left(E_\nu,p_\ell,\theta_\ell,\textsc{nre}^\prime\right)}{\sigma^{\textsc{CCQE}}\left(E_\nu,p_\ell,\theta_\ell,\textsc{nre}\right)}\), will be ill-defined when the denominator is vanishingly small. Instead, an effective implementation was used that shifts the final-state charged-lepton momentum in response to variations of $\textsc{nre}$. The momentum shifts were calculated in bins of true neutrino energy and true final-state charged-lepton polar angle---in the \NEUT implementation, variations of the binding energy effect only small changes in the final-state lepton angular distribution. An example of such a variation can be seen in \figref{intmodel_emiss}. The prior uncertainty on the new $\textsc{nre}$ parameterization was taken as 18 MeV---this large uncertainty is motivated in part because of implementation choices in \NEUT~\cite{Bodek2019} and in part because of uncertainties on the analyzed electron-scattering measurements.
{}

The uncertain $\textsc{nre}$ parameter was not included as a free parameter in the near detector fit for this analysis. Instead, an extremal variation based on the results in Ref.~\cite{Bodek2019} was included as a simulated data set.

\paragraph{Martini~\emph{et.\,al.} 2p2h} As previously mentioned, the modeling of 2p2h interactions is highly uncertain. We include a simulated data set based on an alternate 2p2h calculation by Martini~\emph{et.\,al.}~\cite{PhysRevC.80.065501}. This calculation predicts a larger inclusive 2p2h cross section than the Nieves~\emph{et.\,al.} calculation---importantly increasing the relative neutrino/antineutrino 2p2h strength---and is thus an instructive alternate model.

\paragraph{Kabirnezhad Single Pion Production} The Rein--Sehgal model accounts for interference between pion production channels that include a baryon resonance, however, interference between resonance and non-resonance channels is neglected. A new model, developed by Kabirnezhad~\cite{PhysRevD.97.013002}, overcomes this limitation and is used to build a simulated data set.

\paragraph{Data-driven CC $0\pi$ $E_{\nu}-Q^2$ dependence} The untuned ND280 CC $0\pi$ sample prediction underestimates the data by approximately 5\%.  This sample is largely composed of 1p1h and 2p2h interactions but with a significant contribution from interactions that produce a pion which is then absorbed before leaving the nucleus.  The 2p2h interaction can be further classified into events with and without a virtual $\Delta$(1232) particle. Simulated data sets are created by assigning the observed CC $0\pi$ data--simulation discrepancy to either the 1p1h or 2p2h event categories.  At the near detector the event category is weighted in bins of lepton momentum and angle so that the simulation matches the data.  This weighting is then projected as a function of neutrino energy and $Q^2$ and applied to the far detector simulation to create the simulated far detector data.

\paragraph{Coulomb Correction} As the final-state charged lepton leaves the nuclear potential, it undergoes a small momentum shift because of interaction with the Coulomb field of the nucleus. In addition, the Coulomb potential results in a small variation in the relative neutrino/antineutrino cross section. The effect of the Coulomb potential was not included in the base model, and thus a simulated data set was included in which a momentum shift was applied to final-state (anti-) muons and (anti-) electrons, following Ref.~\cite{PhysRevC.60.044308}, and the relative charged-current cross section for neutrinos and antineutrinos was varied by 3\%.

\section{Near detector data}

\begin{figure}[htbp]
\centering
  \includegraphics[width=0.47\textwidth]{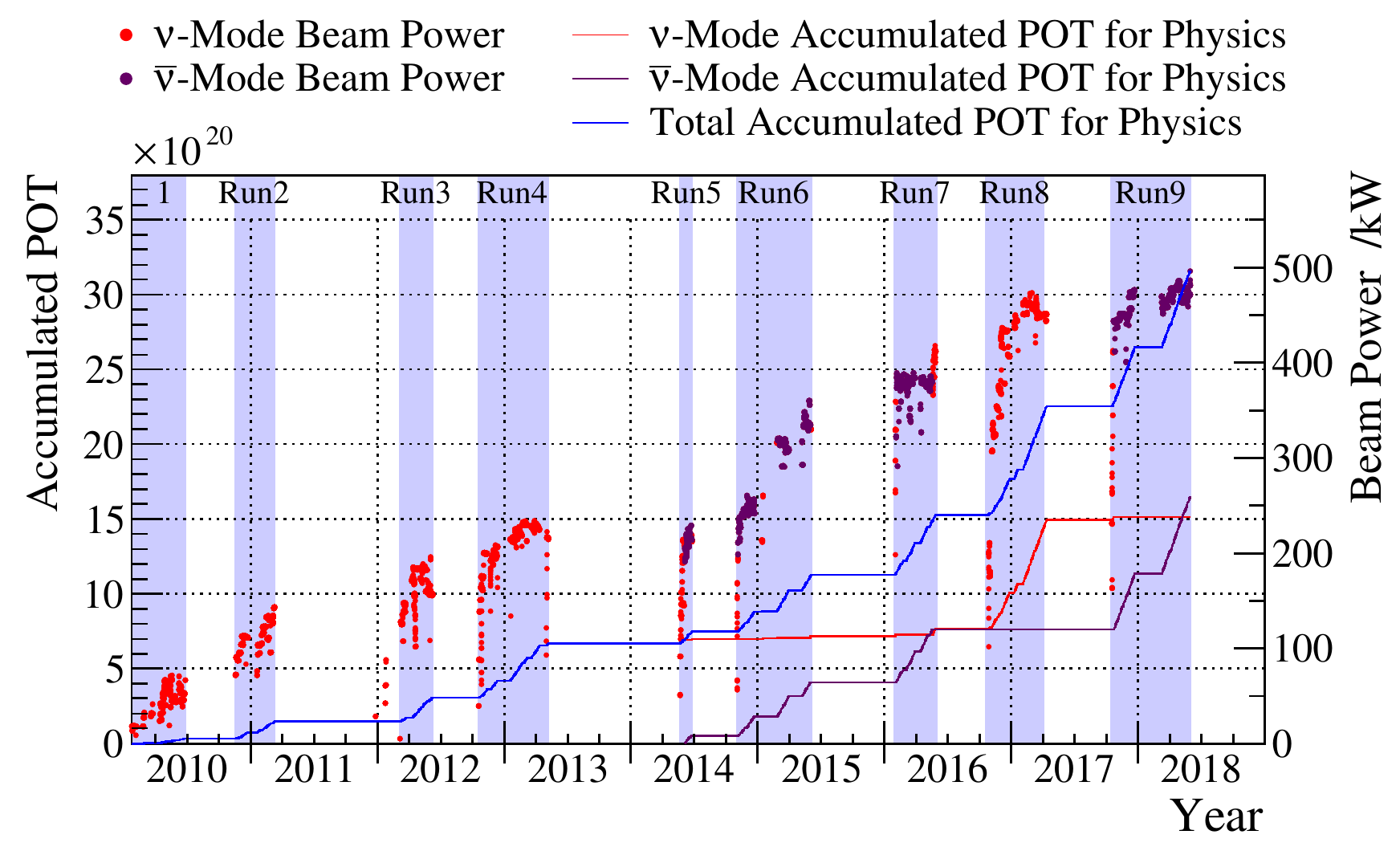}
\caption{The T2K data-taking periods, showing the accumulated beam protons on target delivered as a function of time.}
\label{fig:PoT_run1-9}
\end{figure}

\begin{figure}[htbp]
\centering
  \includegraphics[width=0.49\textwidth] {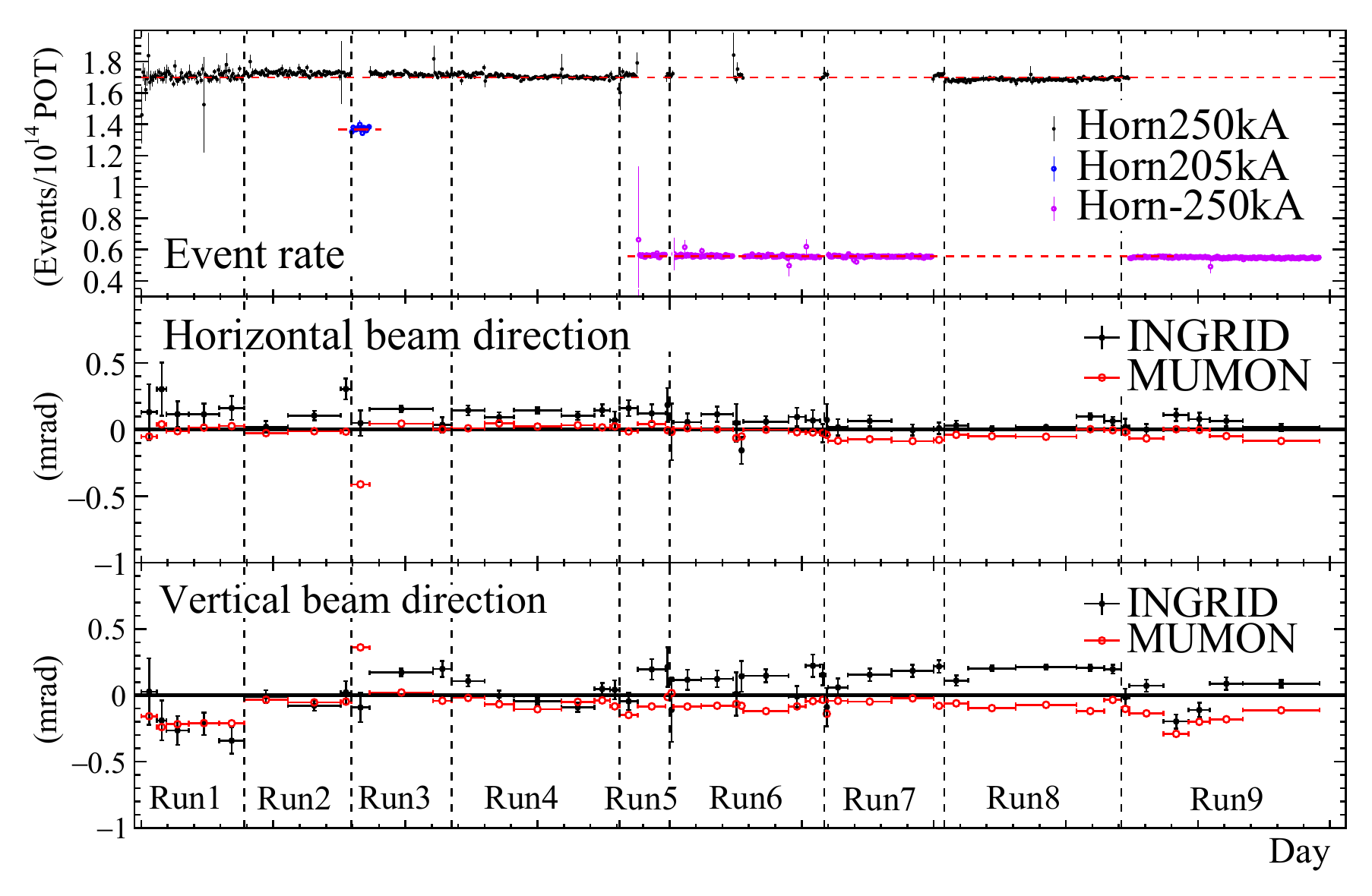} 
\caption{Data in the muon monitors (MUMON)  and INGRID on-axis detector for runs 1--9.  The top row shows the event rate in both detectors, which is reduced as expected for RHC mode. The middle and bottom rows show the horizontal and vertical beam direction in both detectors.}
\label{fig:INGRID_stability}
\end{figure}

The history of protons on target delivered to the T2K beamline until the end of May 2018 is shown in Fig.~\ref{fig:PoT_run1-9}.  
Data for runs 1--9 in the muon monitors and the on-axis INGRID detector are shown in  Fig.~\ref{fig:INGRID_stability}.  The rate is stable throughout the run periods, and the horizontal and vertical beam positions are stable to less than 1\,mrad throughout all of the run periods.

The off-axis near detector, ND280, is located 280~m upstream of the beam source. It consists of several sub-detectors inside a 0.2~T magnet. Charged current (CC) \numu and \numub neutrino interactions are selected in the tracker region, which is composed of two fine-grained detectors (FGD1 and FGD2)~\cite{amaudruz2012t2k} interleaving three time projection chambers (TPC)~\cite{abgrall2011time}. 
The FGDs provide the target mass for neutrino interactions. The first FGD consists of 30 layers each composed of 192 plastic scintillator bars.  The bars in each layer are oriented perpendicularly to the neutrino beam direction and to the bars in the preceding layer. Each pair of layers forms a single module providing a three-dimensional position for charged particles passing through it.  The second FGD consists of alternating plastic scintillator modules and water panels. There are seven scintillator modules interspersed with six water panels, providing a water target for neutrinos to interact within. This allows effects relating to neutrino interactions on water to be isolated from those on carbon, reducing the uncertainty in extrapolating the event rate measurement from ND280 to SK.
The TPCs measure both the curvature of charged particles in the magnetic field of ND280 and the energy lost by the particles as they travel through the TPC gas.  The curvature of the particles provides a precise measurement of their momentum and charge, while the energy loss allows the particle species to be identified.

Only ND280 data from runs 2--6 are used in this analysis, a smaller sample than for SK. Data quality is assessed weekly, and the total ND280 data taking efficiency across runs 2--6 was $\sim93\%$.
The near detector analysis uses a total exposure of 5.81$\times$10$^{20}$\,POT in FHC and 2.84$\times$10$^{20}$\,POT in RHC, as shown in Tab.~\ref{tab:POT_summary}.

The event selection for FHC is unchanged since the analysis described in Ref.~\cite{Abe:2017vif_T2Krun7osc}. 
The highest momentum, negatively charged track in each event is selected as the lepton candidate.
The candidate track must start within the fiducial volume of FGD1 or FGD2 and be identified as muon-like by the TPC.
This produces a selection of charged-current (CC) \numu interactions.
The selected events are divided into three samples for each FGD, based on the reconstructed pion multiplicity. 
Positive pions are identified in three ways: a positive charged FGD-TPC track with energy loss consistent with a pion; a positively charged FGD-contained track with charge deposition consistent with a pion; or a delayed energy deposit in the FGD due to stopped $\pi^{+}\rightarrow\mu^{+}\rightarrow e^{+}$ decays.
Negatively charged, minimally ionising TPC tracks are identified as negatively charged pions.
Neutral pions decay instantaneously to pairs of photons, which can then convert to electron-positron pairs.
TPC tracks with charge depositions consistent with an electron are used to identify these decays.

\begin{figure*}[htbp]
\centering
\begin{subfigure}{0.47\textwidth}
  \includegraphics[width=0.98\columnwidth]{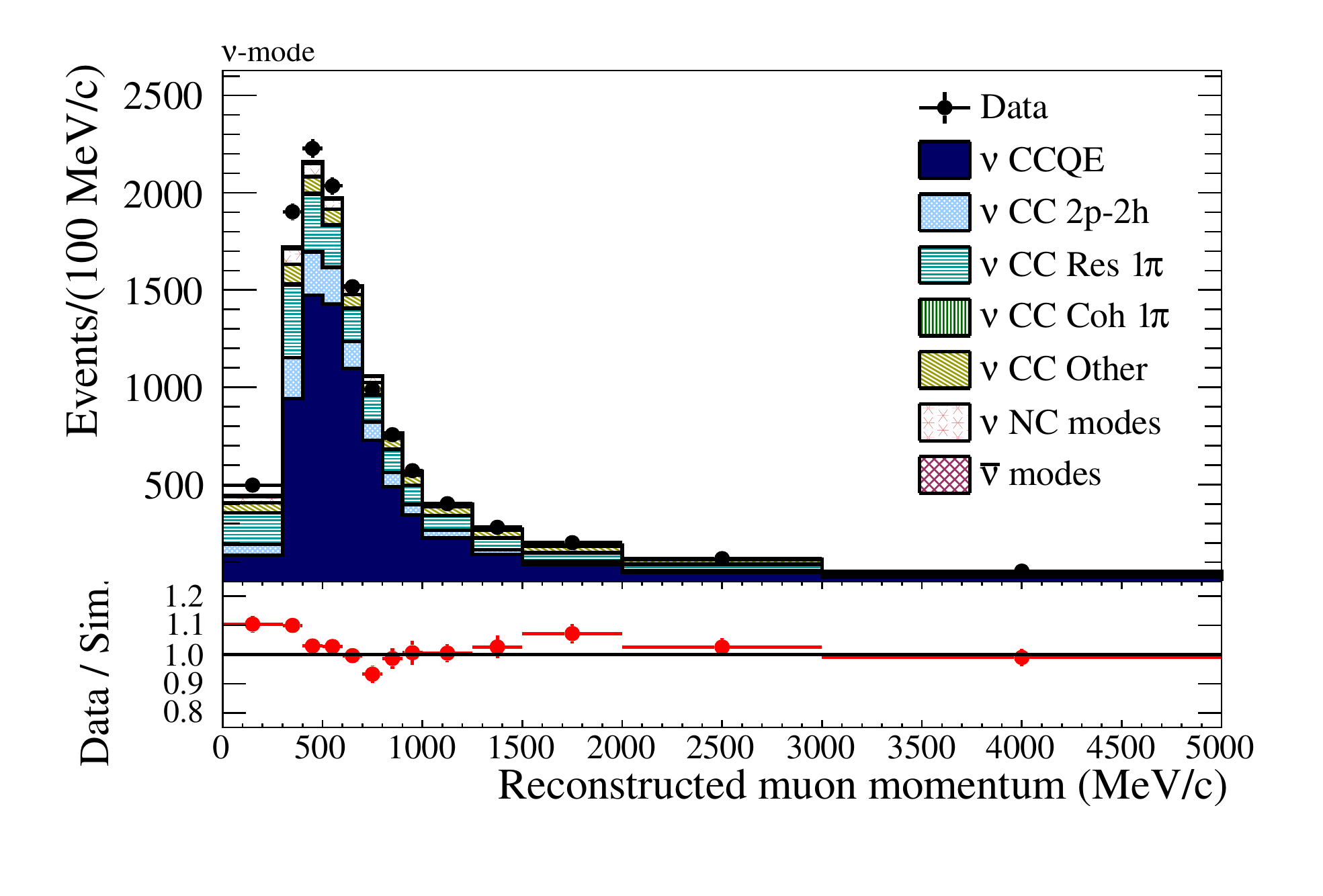} 
\end{subfigure}
\begin{subfigure}{0.47\textwidth}
  \includegraphics[width=0.98\columnwidth]{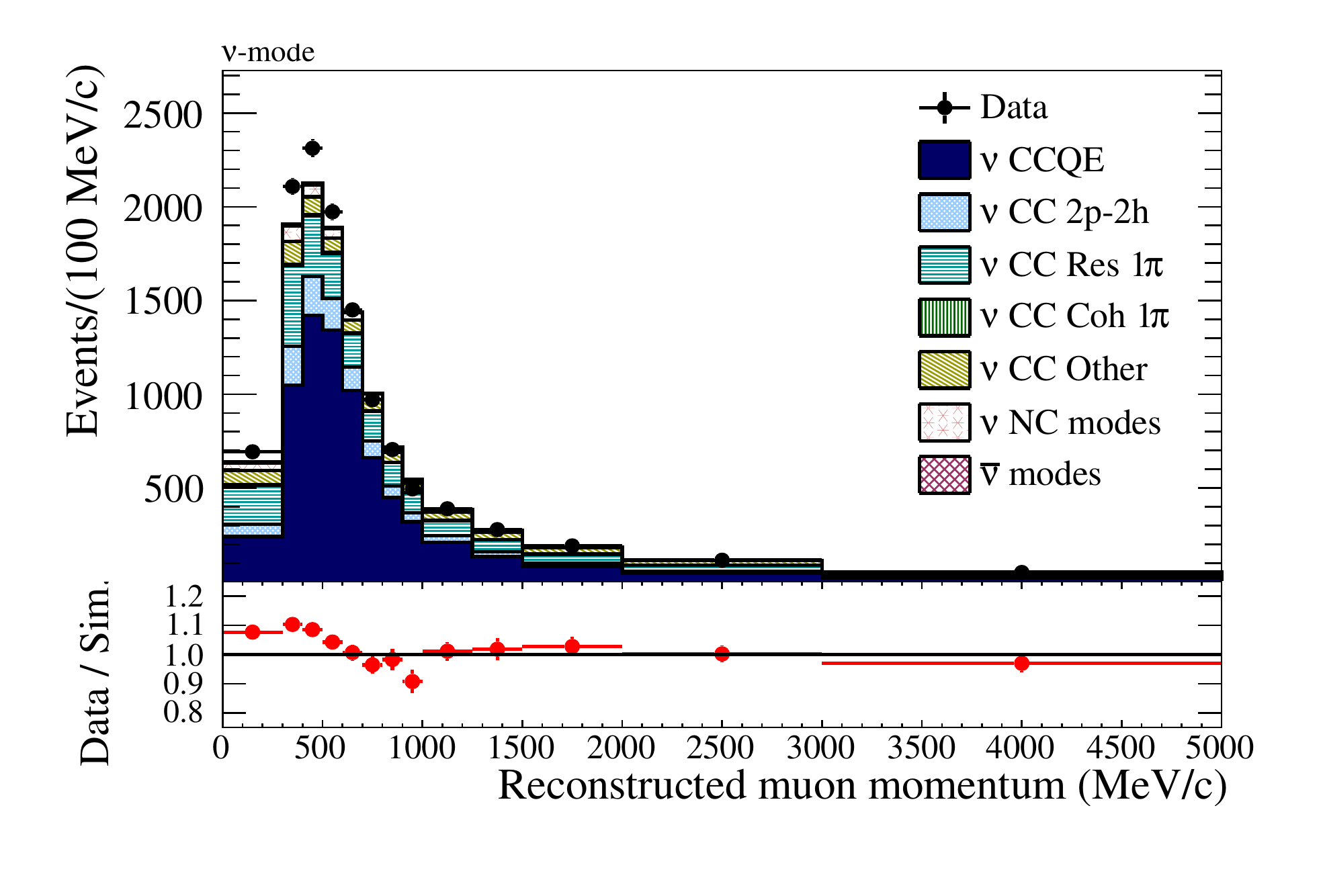}
\end{subfigure}
\begin{subfigure}{0.47\textwidth}
  \includegraphics[width=0.98\columnwidth]{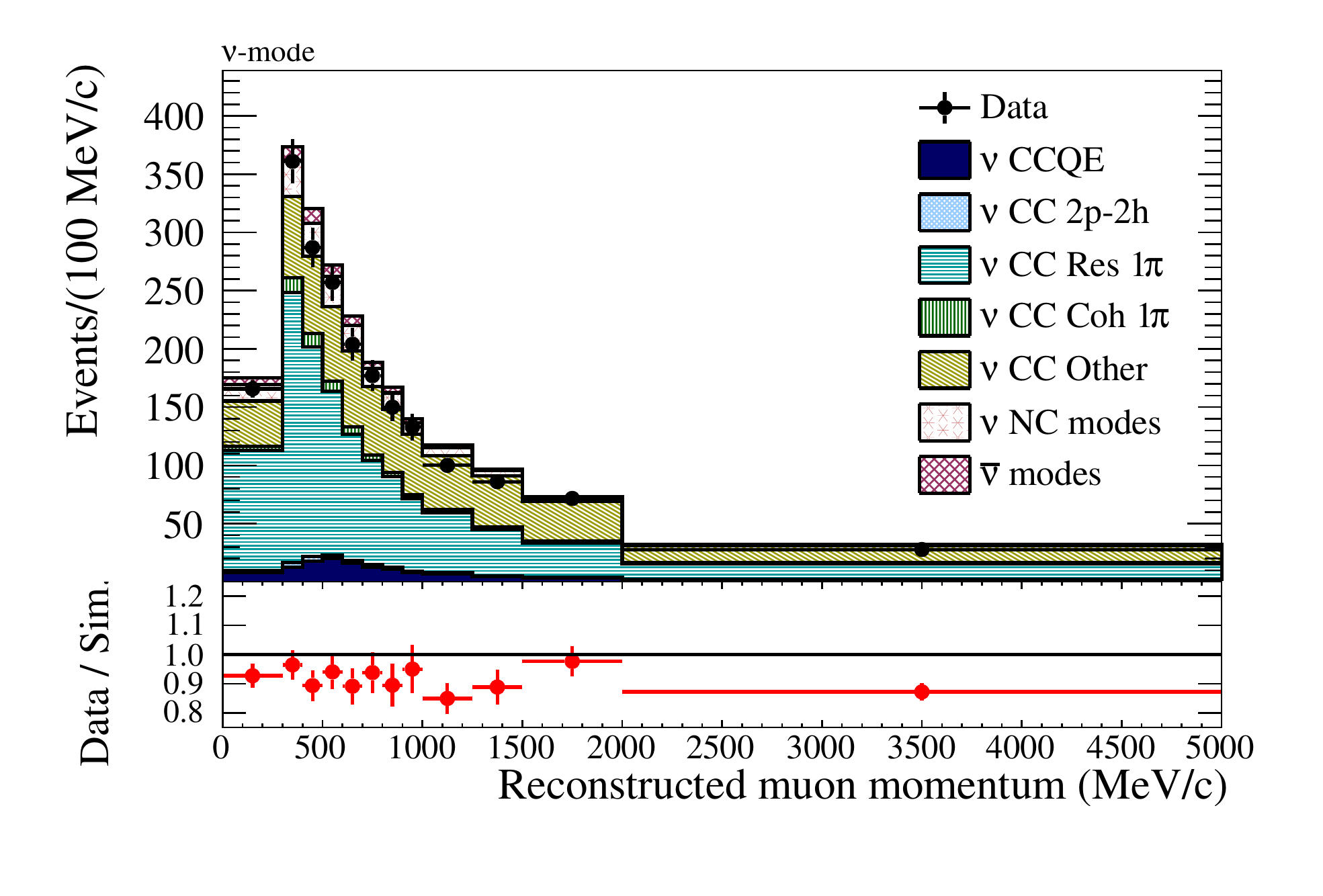} 
\end{subfigure}
\begin{subfigure}{0.47\textwidth}
  \includegraphics[width=0.98\columnwidth]{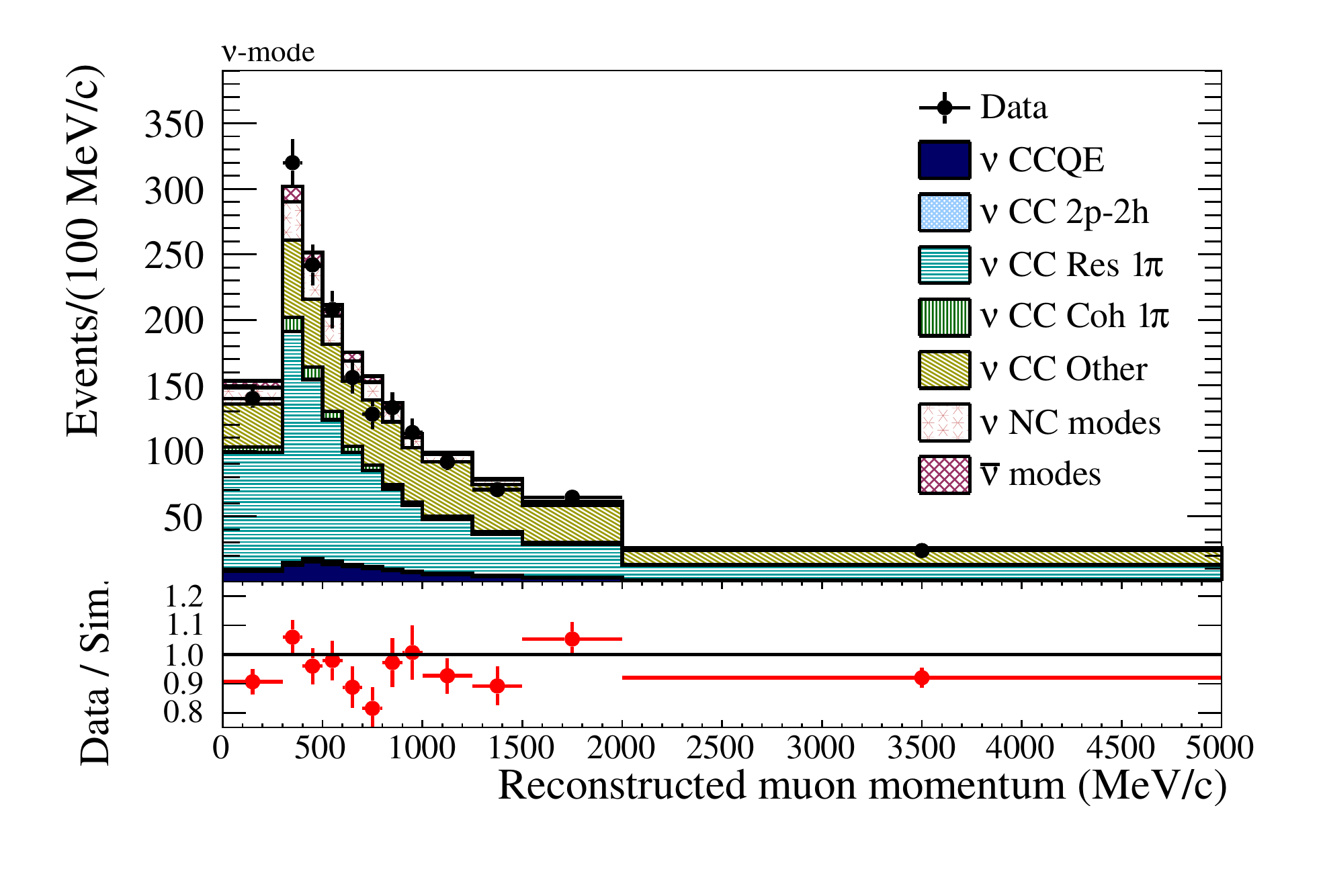}
\end{subfigure}
\caption{Final state muon momentum distributions of the FHC $\nu_{\mu}$ CC 0$\pi$ (top) and $\nu_{\mu}$ CC 1$\pi^+$ (bottom) data and simulation samples in FGD1 (left) and FGD2 (right).}
\label{fig:FHCNumuPrefit}
\end{figure*}

\begin{figure*}[htbp]
\centering
\begin{subfigure}{0.47\textwidth}
  \includegraphics[width=0.98\columnwidth]{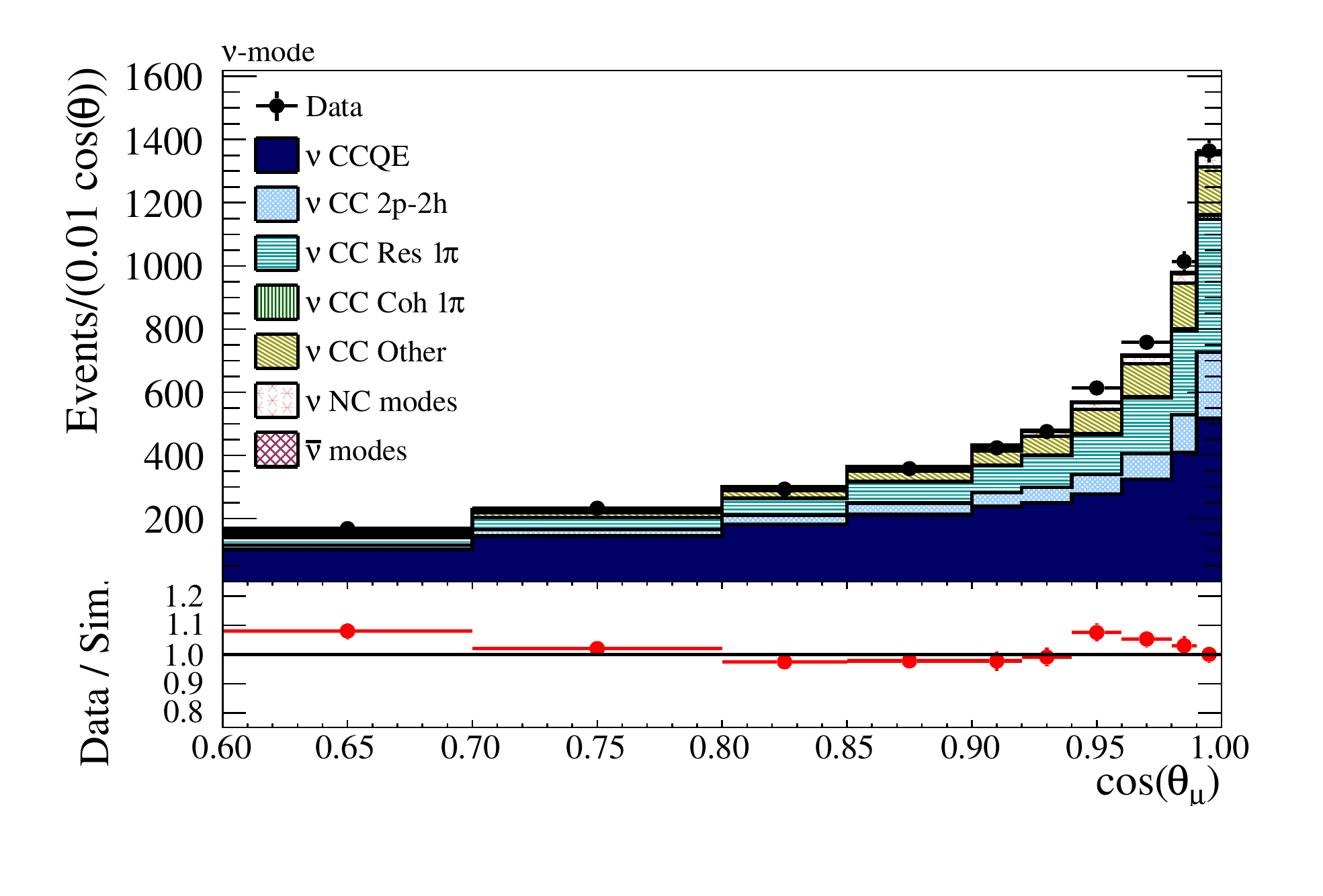} 
\end{subfigure}
\begin{subfigure}{0.47\textwidth}
  \includegraphics[width=0.98\columnwidth]{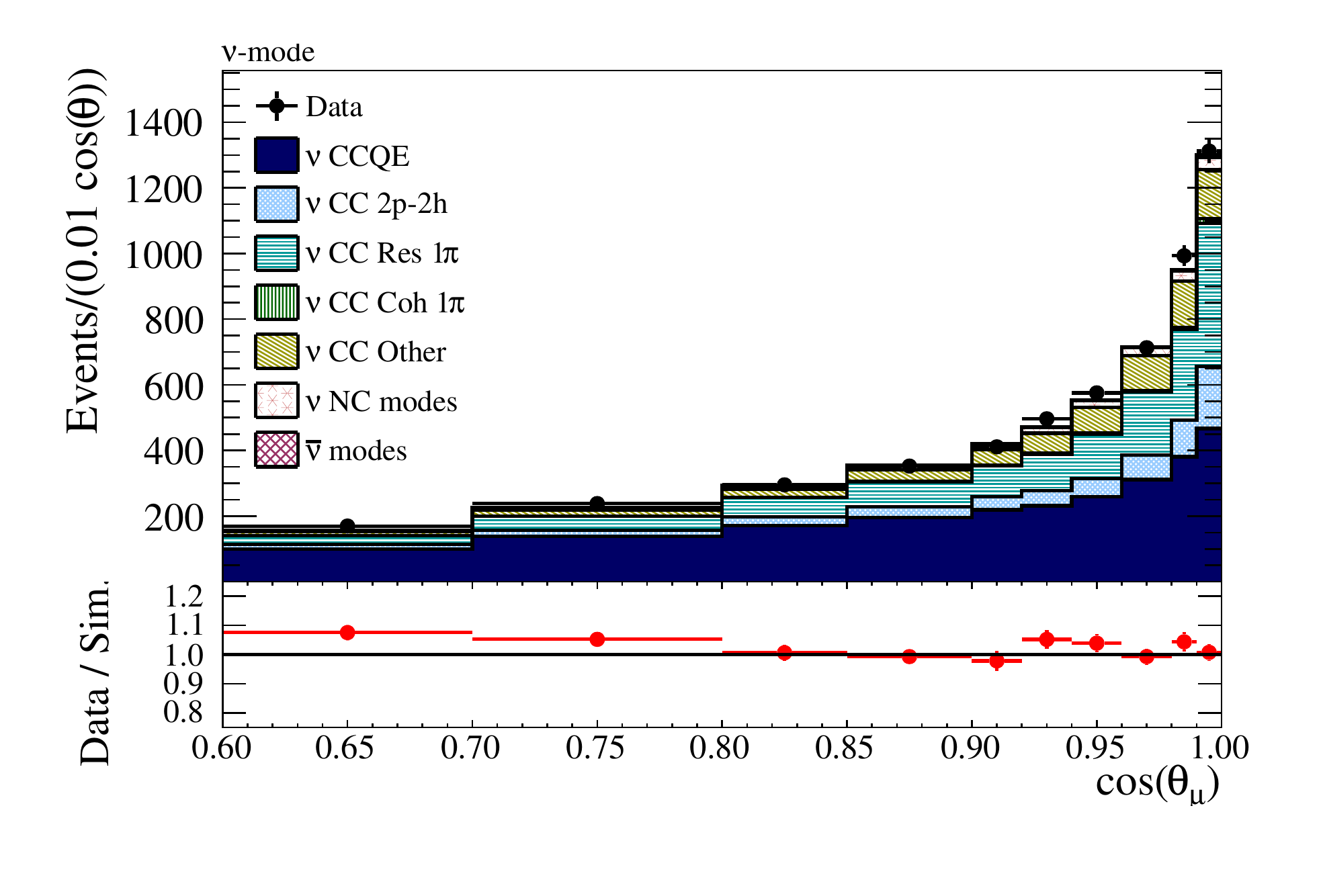}
\end{subfigure}
\begin{subfigure}{0.47\textwidth}
  \includegraphics[width=0.98\columnwidth]{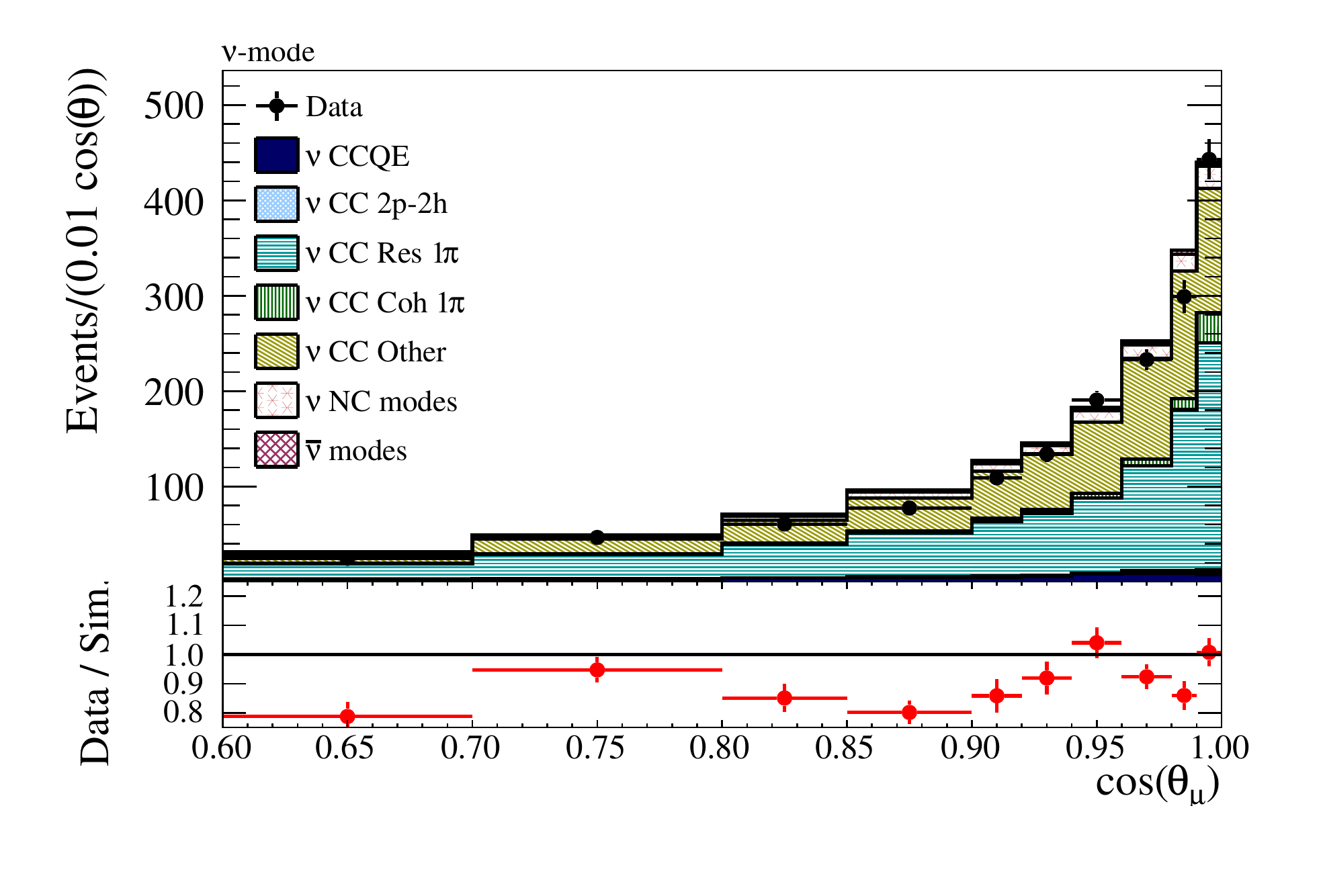} 
\end{subfigure}
\begin{subfigure}{0.47\textwidth}
  \includegraphics[width=0.98\columnwidth]{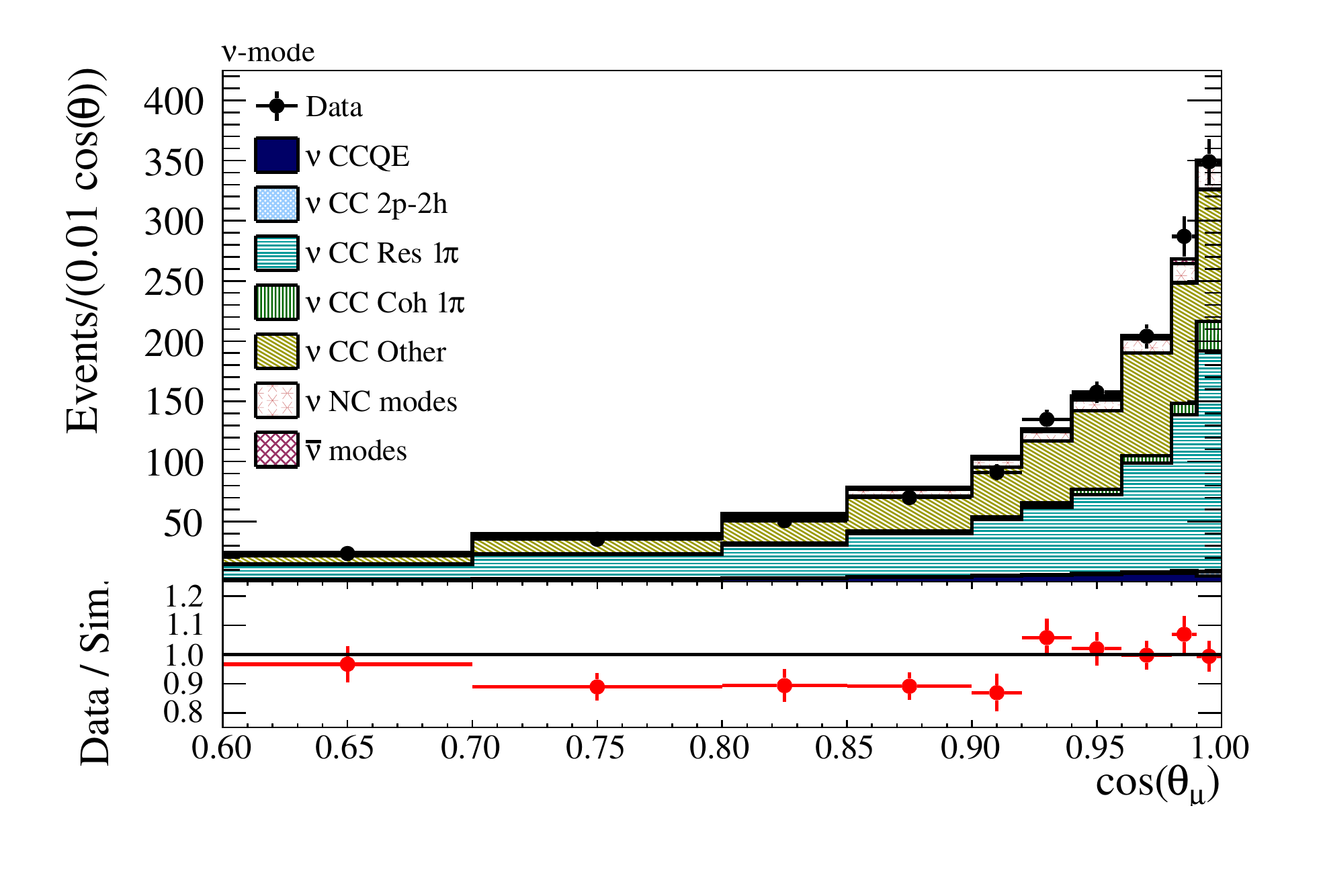}
\end{subfigure}
\caption{Distributions of the final state muon angle of the FHC $\nu_{\mu}$ CC 0$\pi$ (top) and $\nu_{\mu}$ CC 1$\pi^+$ (bottom) data and simulation samples in FGD1 (left) and FGD2 (right).}
\label{fig:FHCNumuPrefit_theta}
\end{figure*}

The three FHC CC sub-samples are CC 0$\pi$, which is dominated by CCQE interactions, CC 1$\pi^+$, which is dominated by CC resonant single pion production, and CC Other, which is dominated by interactions producing multiple pions. The reconstructed muon momentum and angle of the selected data and simulation events in the FHC CC 0$\pi$ and CC 1$\pi^+$ samples are shown in Fig.~\ref{fig:FHCNumuPrefit} and Fig.~\ref{fig:FHCNumuPrefit_theta}, for both FGD1 and FGD2. The numbers of events recorded in each sample and the expectation prior to the ND280 fit are shown in Tab.~\ref{tab:ndrates}.
\begin{table}[htbp]
\newcommand{\FGD}[1]{\footnotesize{FGD}$\,#1$}
\sisetup{table-format=5.0}
\caption{ND280 samples, with the observed and expected numbers of events (before and after fitting at ND280).}
\label{tab:ndrates}
\begin{tabular}{llcSSS} 
\hline
\hline
Beam & \multicolumn{1}{c}{Topology} & Target & \multicolumn{1}{c}{Data} & \multicolumn{1}{c}{Prediction} & \multicolumn{1}{c}{Postfit}\\
\hline
&\multirow{2}{*}{\numu CC 0$\pi$}      
     & \FGD1 & 17136 & 16724 & 17122 \\
    && \FGD2 & 17443 & 16959 & 17495 \\[1ex]
\multirow{2}{*}{FHC} 
&\multirow{2}{*}{\numu CC 1$\pi^{+}$}  
     & \FGD1 & 3954 & 4381 & 4062\\
    && \FGD2 & 3366 & 3564 & 3416\\[1ex]
&\multirow{2}{*}{\numu CC other}
     & \FGD1 & 4149 & 3944 & 4096\\
    && \FGD2 & 4075 & 3571 & 3915\\[0.5ex]
\hline
\rule{0pt}{3ex}
&\multirow{2}{*}{\numub CC 1-track}
     & \FGD1 & 3527 & 3588 & 3504\\
    && \FGD2 & 3732 & 3618 & 3685\\[1ex]
&\multirow{2}{*}{\numub CC $N$-track}
     & \FGD1 & 1054 & 1067 & 1053\\
\multirow{2}{*}[-0.5ex]{RHC}
    && \FGD2 & 1026 & 1077 & 1097\\[1ex]
&\multirow{2}{*}{\numu CC 1-track} 
     & \FGD1 & 1363 & 1272 & 1353\\
    && \FGD2 & 1320 & 1263 & 1330\\[1ex]
&\multirow{2}{*}{\numu CC $N$-track}
     & \FGD1 & 1370 & 1357 & 1354\\
    && \FGD2 & 1253 & 1247 & 1263\\[1ex]
\hline
\hline
\end{tabular}
\end{table}

The event selection for $\numu$ and $\numub$ interactions in the RHC beam mode is unchanged since the previous analysis~\cite{Abe:2017vif_T2Krun7osc}. These selections differ from the FHC selections described above in two important ways. As a larger number of interactions are produced by ``wrong-sign" neutrinos, selections of both $\numu$ and $\numub$ interactions are used in the RHC beam. Taking into account differences in the flux and cross-section, the wrong-sign contamination is approximately 30$\%$ in the selected RHC samples compared to 4$\%$ in the FHC samples. 

\begin{figure*}[htbp]
\centering
\begin{subfigure}{0.47\textwidth}
  \includegraphics[width=0.98\columnwidth]{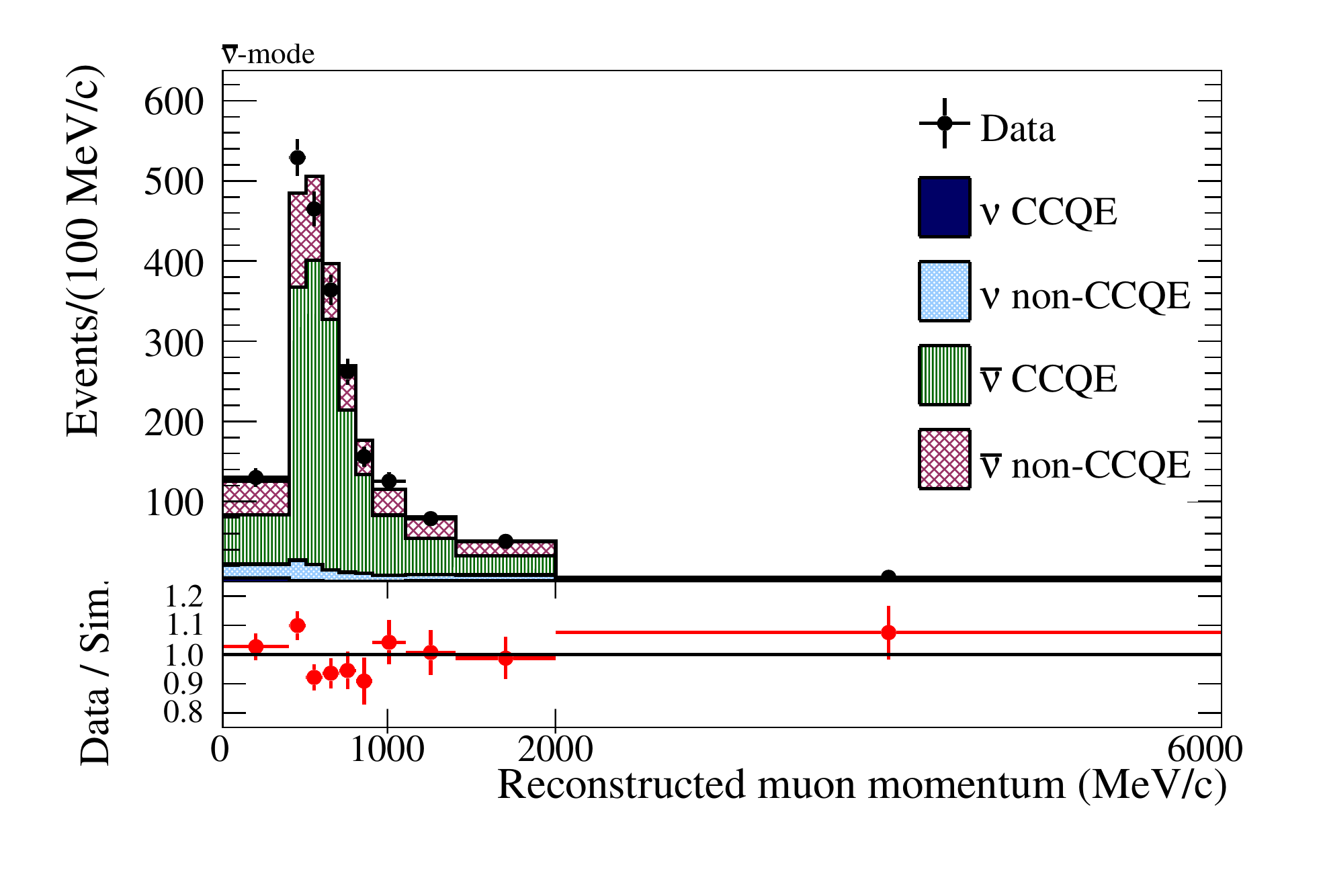} 
\end{subfigure}
\begin{subfigure}{0.47\textwidth}
  \includegraphics[width=0.98\columnwidth]{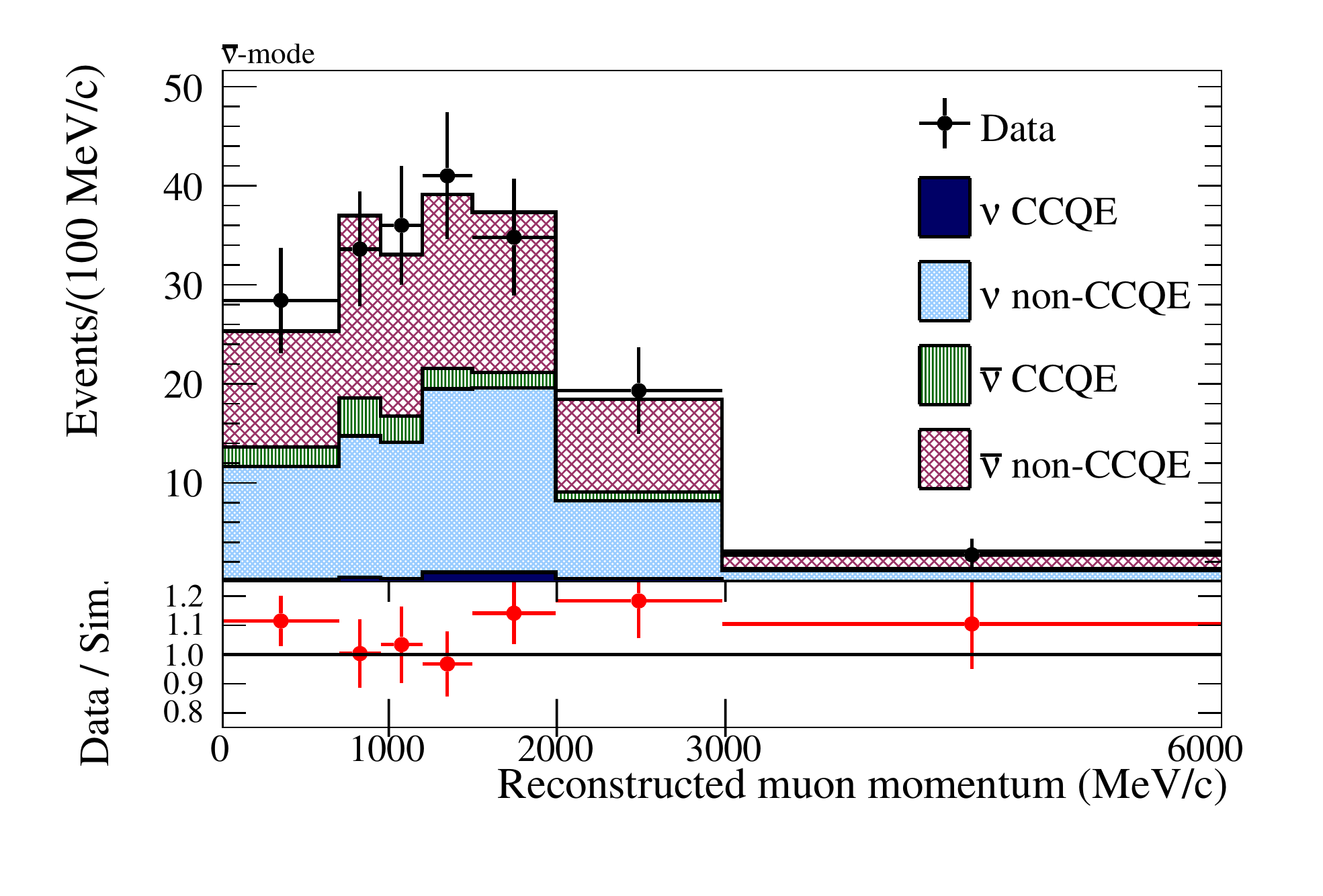}
\end{subfigure}
\begin{subfigure}{0.47\textwidth}
  \includegraphics[width=0.98\columnwidth]{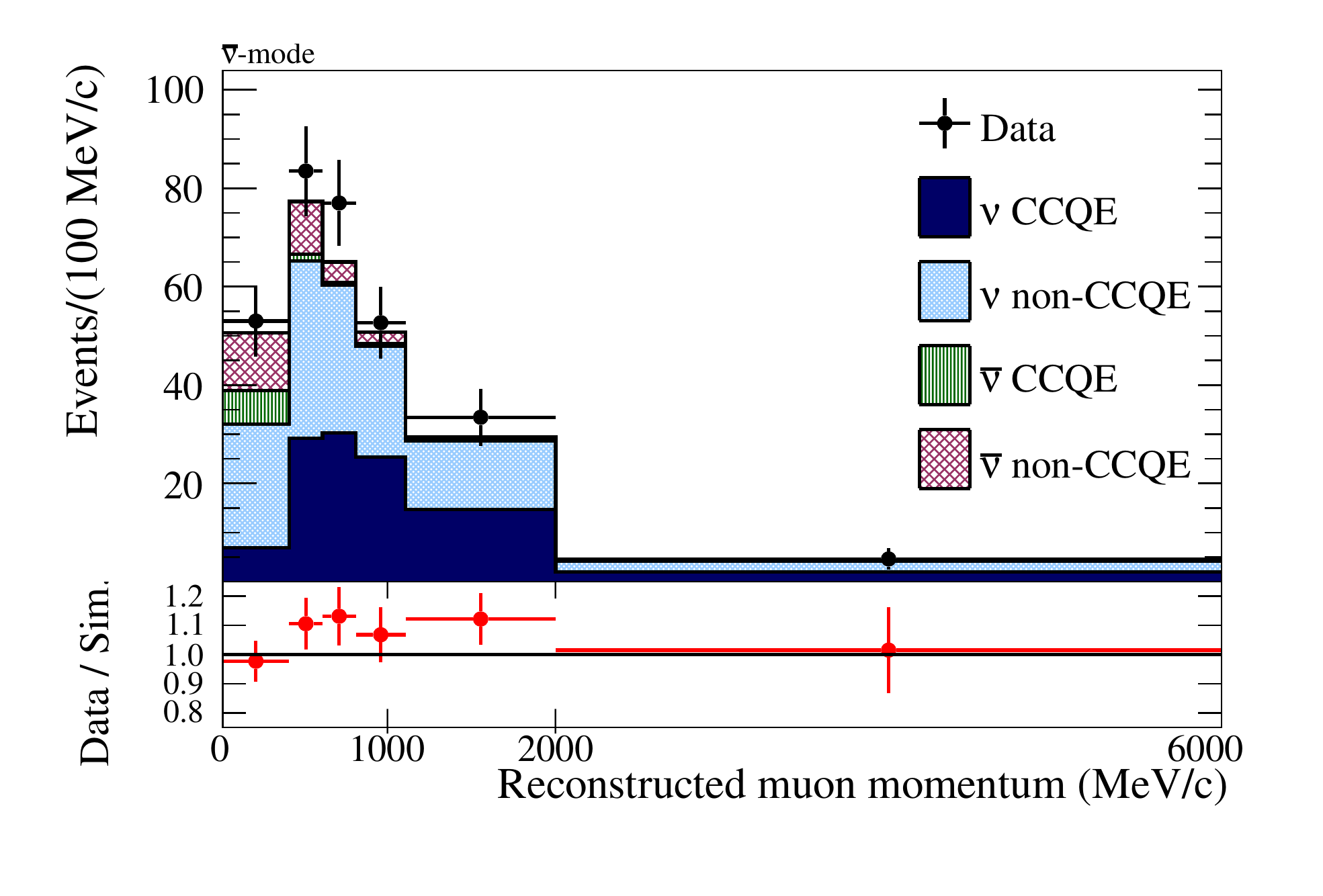} 
\end{subfigure}
\begin{subfigure}{0.47\textwidth}
  \includegraphics[width=0.98\columnwidth]{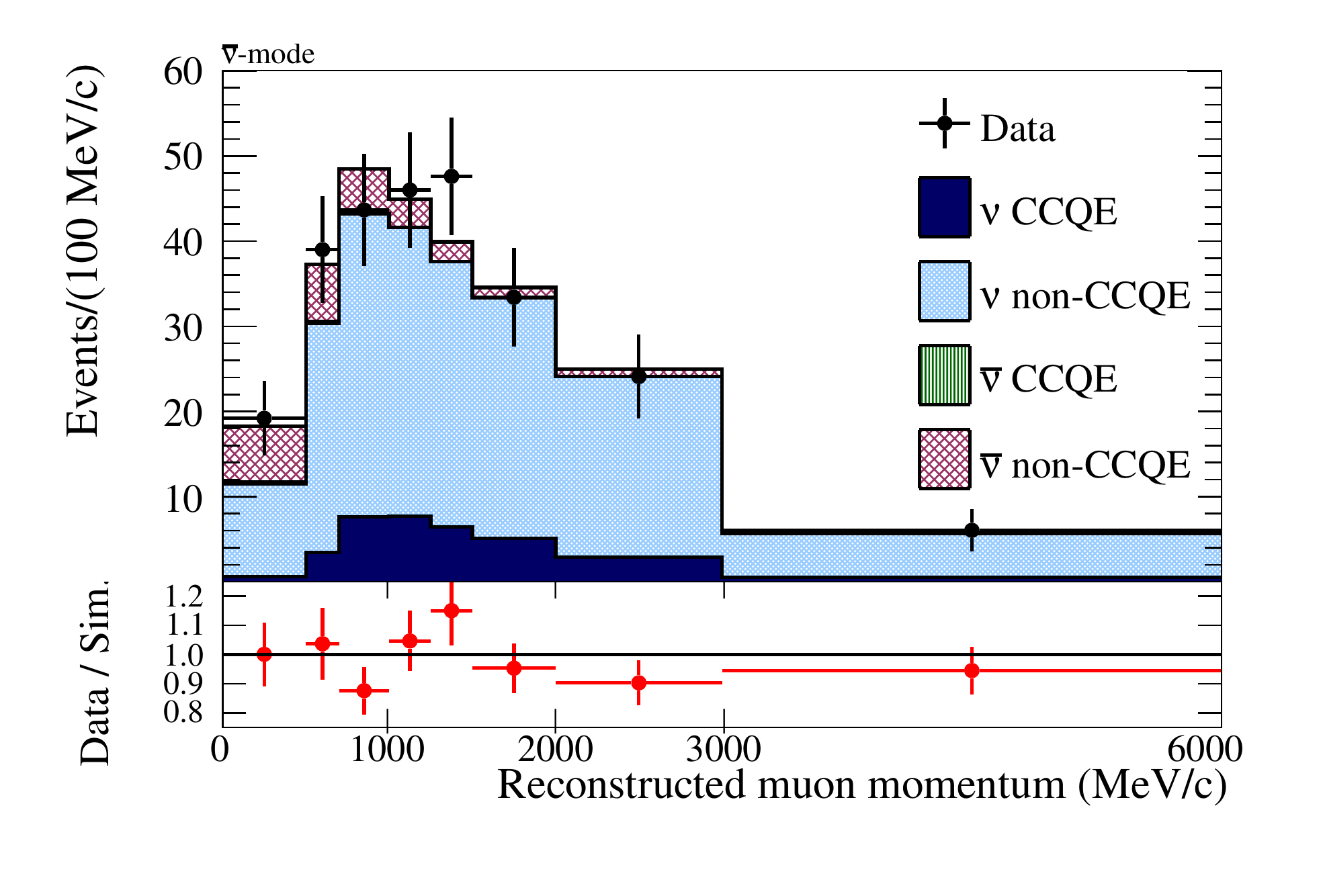}
\end{subfigure}
\caption{Final state muon momentum distributions for the RHC $\numub$ (top) and $\numu$ (bottom) CC 1-track (left) and CC $N$-track (right) FGD1 simulation samples. These distributions are before the ND280 fit.}
\label{fig:RHCNumuPrefit}
\end{figure*}

\begin{figure*}[htbp]
\centering
\begin{subfigure}{0.47\textwidth}
  \includegraphics[width=0.98\columnwidth]{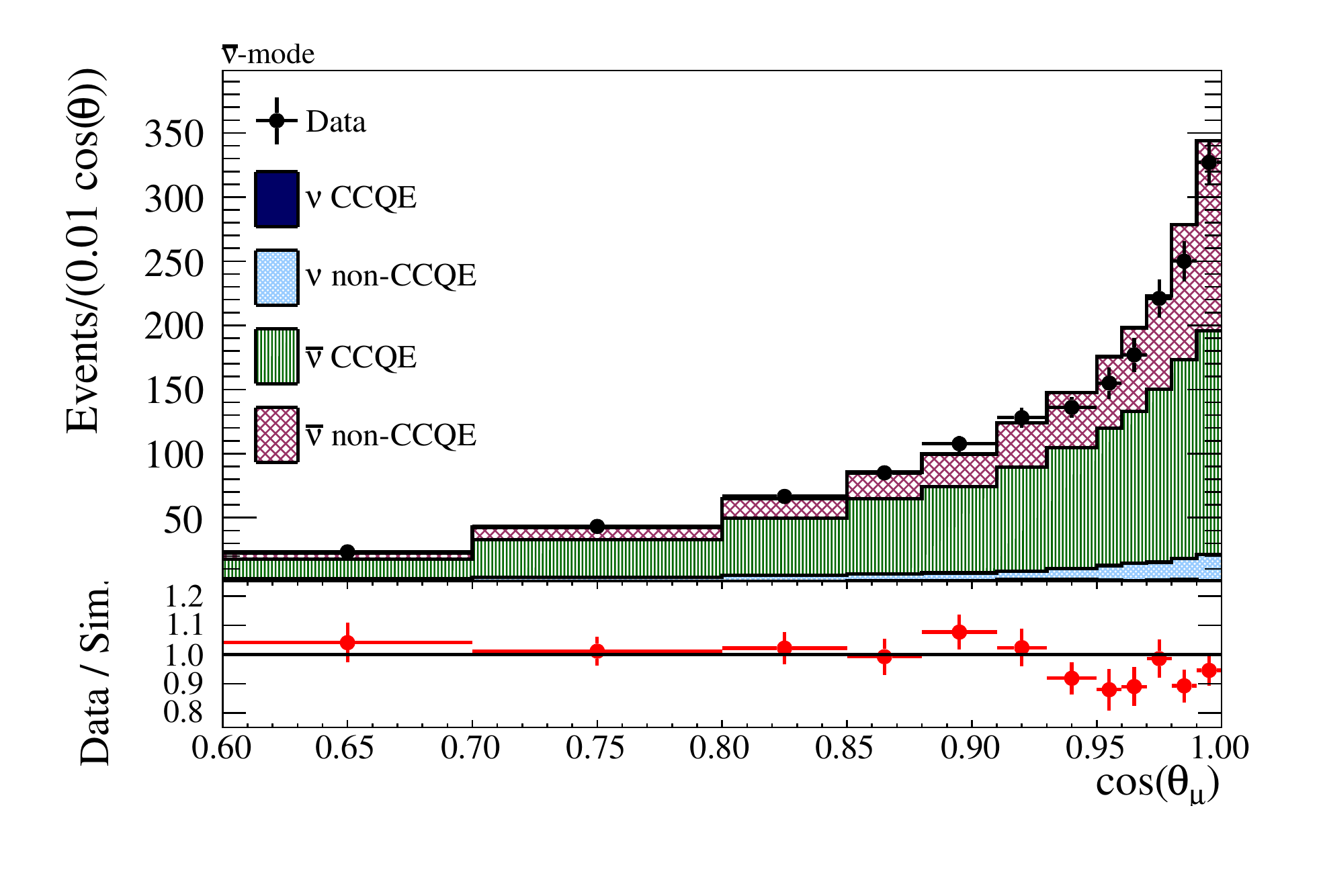} 
\end{subfigure}
\begin{subfigure}{0.47\textwidth}
  \includegraphics[width=0.98\columnwidth]{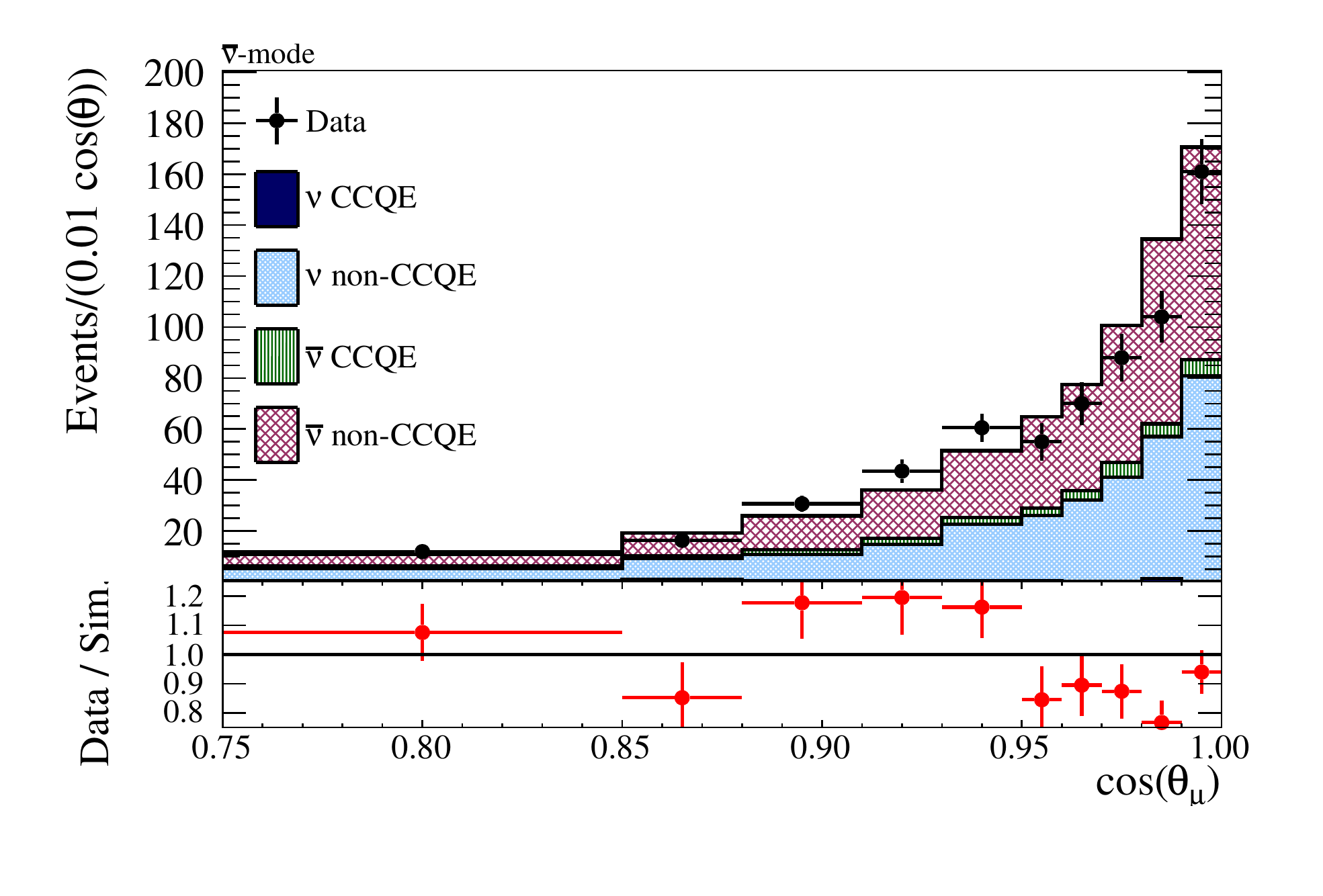}
\end{subfigure}
\begin{subfigure}{0.47\textwidth}
  \includegraphics[width=0.98\columnwidth]{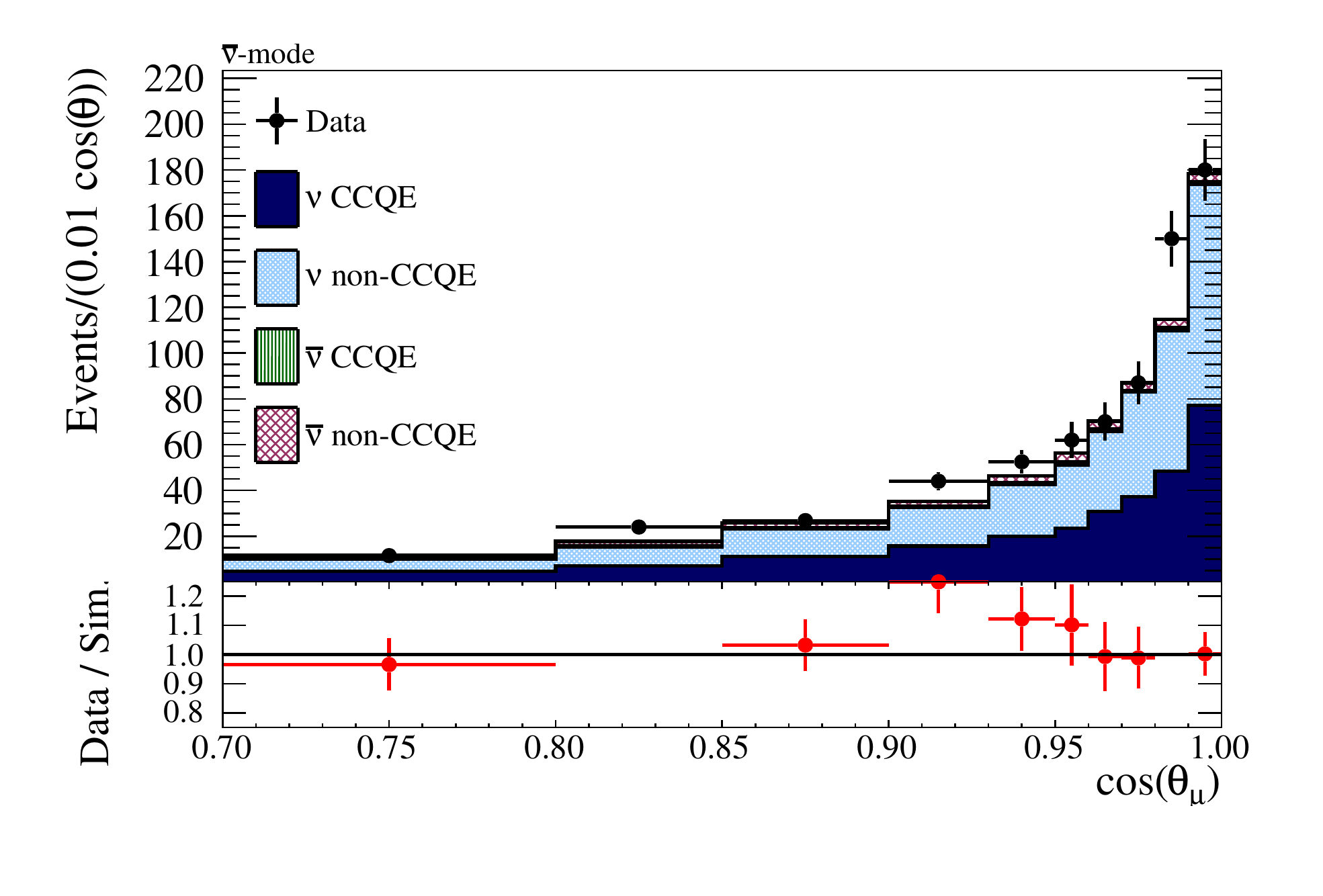} 
\end{subfigure}
\begin{subfigure}{0.47\textwidth}
  \includegraphics[width=0.98\columnwidth]{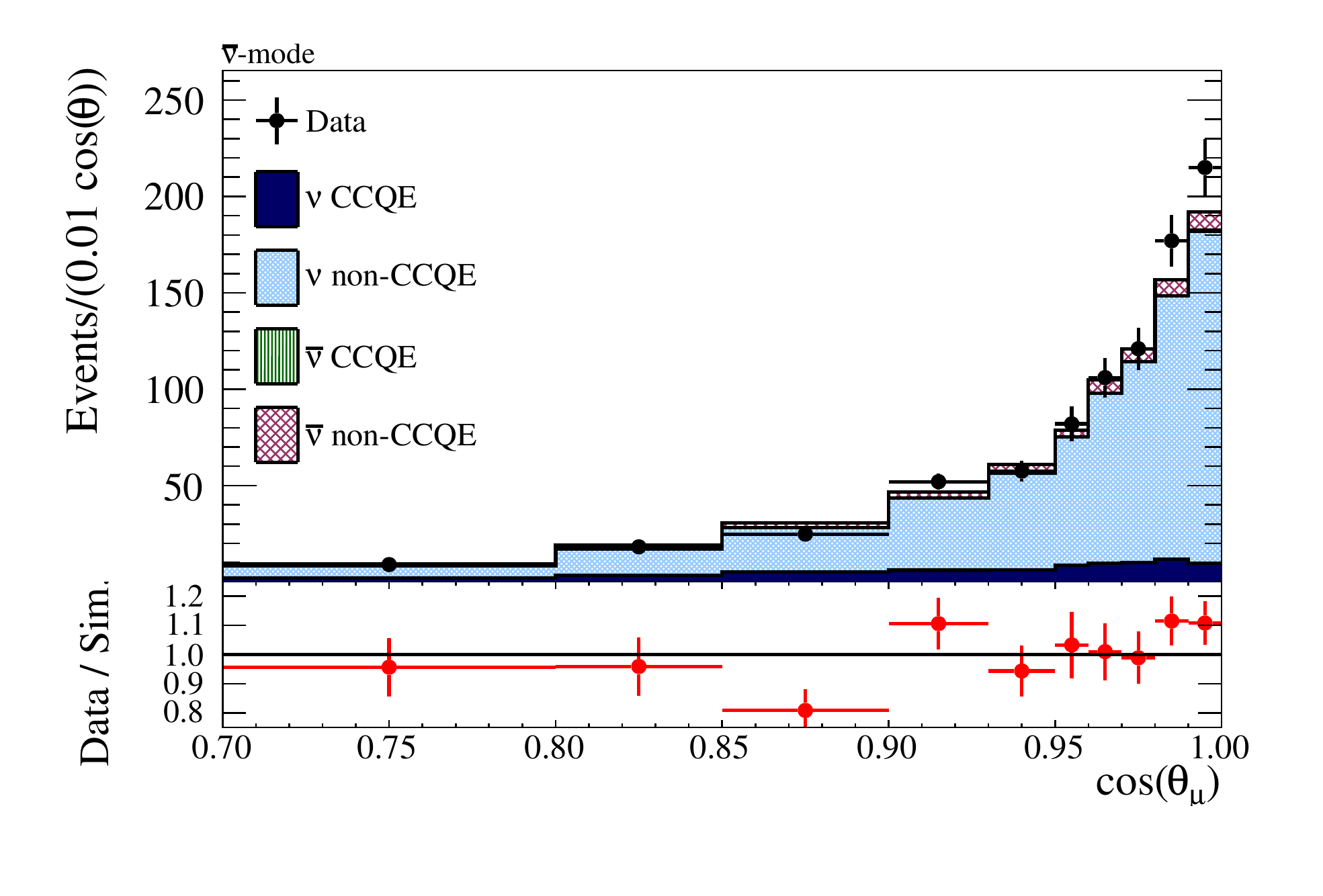}
\end{subfigure}
\caption{Distributions of the final state muon angle for the RHC $\numub$ (top) and $\numu$ (bottom) CC 1-track (left) and CC $N$-track (right) FGD1 simulation samples. These distributions are before the ND280 fit.}
\label{fig:RHCNumuPrefit_theta}
\end{figure*}

The selected $\numub$ ($\numu$) CC candidate events are divided into two samples for each FGD, based on the number of reconstructed tracks crossing a TPC. These are CC 1-track, which is dominated by CCQE-like interactions, and CC $N$-track, which is dominated by interactions producing pions.
The events are not divided according to the number of observed pions, unlike the FHC selections, due to the lower interaction rate for antineutrinos. The reconstructed muon momentum and angle of the selected events in these samples for FGD1 are shown in Fig.~\ref{fig:RHCNumuPrefit} and Fig.~\ref{fig:RHCNumuPrefit_theta} respectively.

In total there are 14 ND280 event samples: six for FHC (CC 0$\pi$, 1$\pi^+$ and Other, for FGD1 and FGD2), four for right-sign RHC (CC 1-track and CC $N$-Track, for FGD1 and FGD2) and four for wrong sign RHC (CC 1-Track and CC $N$-Track, for FGD1 and FGD2). The number of observed and predicted events for each sample are shown in Tab.~\ref{tab:ndrates}.

\section{Super-Kamiokande data and simulation}
\label{sec:sk_data_reduction}
The Super-Kamiokande detector~\cite{Fukuda:2002uc} consists of a cylindrical tank filled with 50~kilotonnes of pure water, located in the Mozumi mine in Hida, Gifu. An overburden of 2700~meter-water-equivalent provided by Mount Ikeno suppresses the cosmic ray muon flux by five orders of magnitude. Photo-multiplier tubes (PMTs) are supported by a 55\,cm wide steel structure, placed 2\,m away from the tank walls, which divides the detector into two optically separated regions. The outer detector (OD) region, used to identify events with entering particles, is lined with reflective material and viewed by 1885 $8''$ PMTs. The inner detector (ID) region contains 32 kilotons of water and is instrumented with 11146 $20''$ PMTs which make up 40\% of the detector's inner surface. The high density of PMTs in the ID allows for the imaging of the ring-like light patterns projected on the detector walls by particles traveling above the Cherenkov threshold in the water.

\subsection{Super-Kamiokande data}
Pulses on PMTs exceeding a charge threshold corresponding to roughly 0.1 photo-electrons are registered as hits, all of which are processed by a software trigger system~\cite{4436298}. For T2K analyses, all hits occurring in the 1\,ms windows centered on each beam spill arrival are written to disk. Beam spills are excluded from the analysis if they coincide with problems in the data acquisition system or the GPS system used to synchronize SK with the accelerator at J-PARC. Additionally, spills that occur within $100~\mu$s of a beam-unrelated event are rejected to reduce the contamination of T2K data with cosmic ray muon decay electrons. The beam spill selection introduces an inefficiency of 1\%, with roughly half of this being due to the preceding detector activity criterion.

For the analysis presented here, events associated with accepted spills are further required to have a reconstructed energy corresponding to an electron of at least 30~MeV and no more than 15 hits in the largest OD hit cluster. Additional criteria are used to reject spurious events that originate from spontaneous corona discharges in PMTs. Only events reconstructed in the $[-2, 10]\,\si{\micro\second}$ window around the leading edge of the beam spill are used in the analysis.

\begin{figure}[htbp]
  \centering
  \begin{subfigure}[b]{0.47\textwidth}
    \includegraphics[width=0.98\columnwidth]{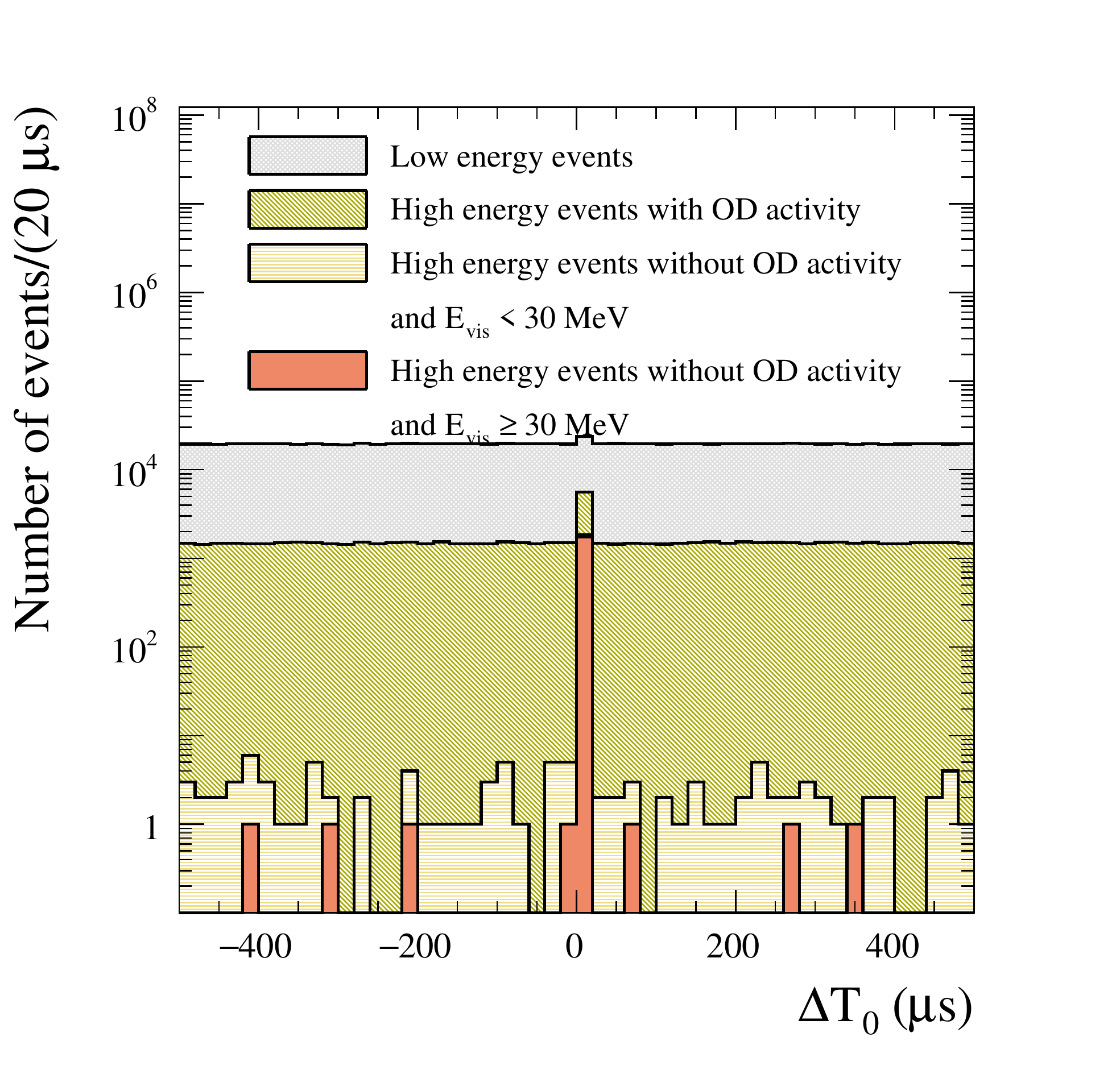}
  \end{subfigure}
  \begin{subfigure}[b]{0.47\textwidth}
    \includegraphics[width=0.98\columnwidth]{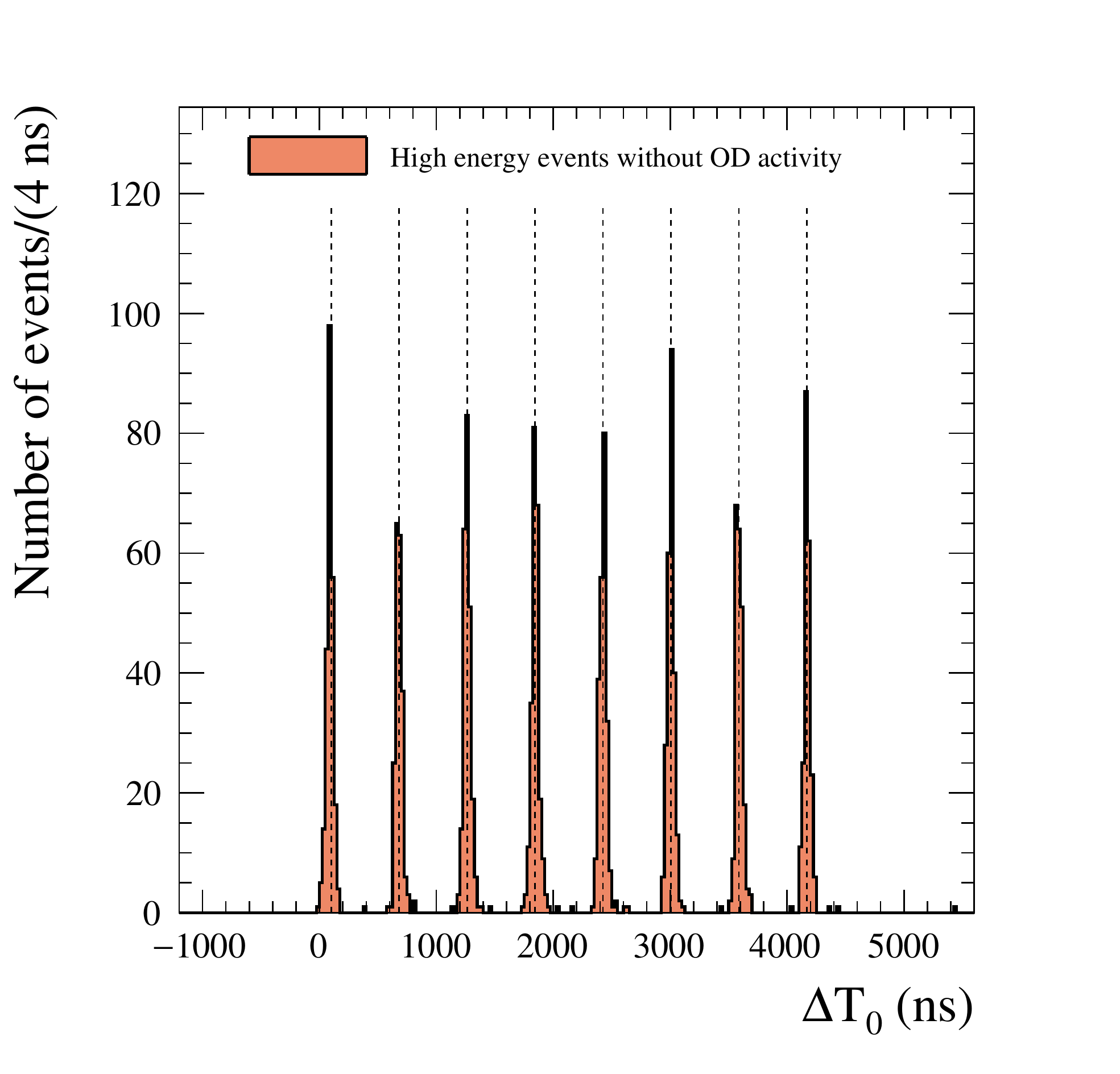}
  \end{subfigure}
  \caption {Reconstructed event time distributions in the 1\,ms (top) and $[-1.2, 5.6]\,\mu\mathrm{s}$ (bottom) windows around the beam spill leading edge.}
  \label{fig:SK_timing}
\end{figure}

Distributions of the reconstructed times for events in both 1~ms and $[-1.2, 5.6]\,\si{\micro\second}$ windows around the beam spill arrival are shown in Fig.~\ref{fig:SK_timing}. In the 1~ms window, a peak of events coincident with the beam arrival is clearly seen; after applying the OD and minimum energy criteria very few events remain outside this peak. The eight-bunch structure of the J-PARC beam is clearly seen in the narrow window event time distribution.

\subsection{Super-Kamiokande event simulation}
Events at SK are simulated using J-PARC (anti-) neutrino flux predictions and the neutrino interaction generator \textsc{NEUT}, which implements the neutrino interaction model described in detail in Section~\ref{sec:interaction_model}. Particles resulting from the neutrino interactions are propagated through an SK detector model using the same Geant3-based~\cite{Brun:1082634} SKDETSIM 13.90 package as in~\cite{Abe:2017vif_T2Krun7osc}. The detector model, including the optical properties of the ultra-pure water and detector materials, is tuned to calibration data~\cite{Abe:2013gga}.

\section{Super-Kamiokande event reconstruction and selection}
\label{sec:SKreco}
\newcommand{\plaincol}[1]{\multicolumn{1}{c}{#1}}
\newcommand{\fixcol}[2]{\multicolumn{1}{>{\centering}p{#1}}{#2}}

Events at SK are reconstructed with the FiTQun maximum likelihood estimation algorithm~\cite{Jiang:2019xwn_SKfitqun}. While this algorithm was initially used exclusively for NC$\pi^0$ background suppression in the $\nu_e$ appearance channel~\cite{Abe:2013hdq, Abe:2017vif_T2Krun7osc}, in recent T2K publications~\cite{Abe:2018wpn_T2Krun8osc, Abe:2019vii_T2Knature} FiTQun was used for all aspects of event reconstruction. Updating the reconstruction tools prompted a re-optimization of the event selection criteria, including an expansion of the fiducial volume (FV). 

In this section, the reconstruction algorithm is briefly described, as well as the updated event selection criteria and the procedure for their optimization. A discussion of the systematic uncertainties related to the SK detector concludes the section.

\subsection{Event reconstruction algorithm}
The FiTQun likelihood function consists of the probability of each PMT registering a hit in a given event, and for hit PMTs, the probability density functions for the charge and time of the hit. Particles in an event are described by tracks (or track segments) parameterized by particle type, momentum, direction and initial position. The FiTQun likelihood is a function of these track parameters and multiple tracks can be combined to form complex event hypotheses.

In an initial pre-fitting stage, the approximate location of the neutrino interaction is found with a simplified likelihood using only the time of the PMT hits. A residual time is calculated for each PMT hit by subtracting the Cherenkov photon time-of-flight, calculated using the straight-line distance from the vertex position to the PMT, from the hit time. Hits are associated to one or more clusters in residual time, with the initial cluster containing hits due to particles produced in the neutrino interaction and subsequent clusters containing hits due to products of weakly decaying prompt particles. Each hit cluster is then reconstructed separately by maximizing the likelihood function for the $e$, $\mu$, $\pi^+$ and $p$ single-particle hypotheses. 

For the earliest hit cluster only, multiple-track event hypotheses are also reconstructed using the results of the single-particle fits as the starting point. A multi-particle search algorithm is used to determine the number of particles observed in the event. This algorithm proceeds by iteratively adding a new electron-like or $\pi^+$-like track to the event until the best-fit likelihood after adding the new track fails to improve beyond a set threshold. In the analysis described here, additional event hypotheses targeting neutral current backgrounds are used: a $\pi^0$ hypothesis consisting of two electron-like tracks consistent with a $\pi^0\rightarrow\gamma\gamma$ decay, and a $\pi^+$ hypothesis with two track segments compatible with a $\pi^+$ undergoing a hard scatter.

\subsection{Event selection}
\label{sec:sk_selection}
Events are selected into samples using cuts on best-fit likelihood ratios between signal-like and background-like hypotheses: $\Lambda^{\alpha}_{\beta} \stackrel{\mathrm{def}}{=} \log\frac{L_{\alpha}}{L_{\beta}}$, where $\alpha$ and $\beta$ are competing hypotheses.  The cut points are typically parameterized as a function of reconstructed kinematics, such as the best-fit electron momentum or the reconstructed invariant mass obtained from the $\pi^0$ hypothesis best-fit kinematics.\\

Five signal-enriched SK samples are used in the analysis. Samples of events containing a single reconstructed $\mu$-like ring (1R$_\mu$) and a single reconstructed $e$-like ring (1R$_e$) target $\nu_{\mu}$ and $\nu_{e}$ CCQE interactions in both FHC and RHC beam modes. An additional sample, used in FHC data only, targets CC 1$\pi^+$ interactions where the $\pi^+$ is below Cherenkov threshold. The $\pi^+$ is identified by the detection of a delayed $\mu$-decay electron following the single prompt electron which results from the CC interaction (1R$_{e}$ $+$ 1 d.e). The CCQE-like selection criteria are the same for FHC and RHC samples.\\

Events in all samples are required to be fully contained (FC) in the ID using the cut on OD activity described in Sec.~\ref{sec:sk_data_reduction} above and to have only one prompt reconstructed particle identified by the multi-particle iterative search algorithm. Events are separated into $e$-like and $\mu$-like with a criterion based on the likelihood ratio of the best-fit $e$-like to $\mu$-like hypothesis ($\Lambda^{e}_{\mu}$) and the reconstructed momentum for the $e$ hypothesis ($p_{e}$).

The FV criteria are defined in terms of the distance from the event vertex to its closest point on the detector walls ($wall$) and the distance from the event vertex to the detector wall along the track direction ($towall$). This parameterization of the FV allows for a larger volume of the detector to be used by reducing the $wall$ threshold compared to previous T2K neutrino oscillation analyses, while ensuring that Cherenkov rings projected on the detector walls illuminate a large number of PMTs with the $towall$ criterion, introduced for the first time in the analysis described here. The $wall$ and $towall$ criteria are chosen separately for each sample to maximize the sensitivity to $\theta_{23}$ and $\deltacp$, as described in Sec.~\ref{sec:sk_fv_opt}. For the $\mu$-like samples, a minimum $wall$ of 50~cm is required, with a minimum $towall$ of 250~cm. The requirements for the $e$-like samples with no decay-$e$ are $wall>80$~cm and $towall>170$~cm, while for the sample with one decay-$e$ $wall>50$~cm and $towall>270$~cm are required.\\

\begin{figure}[htbp]
  \centering
  \begin{subfigure}[b]{0.47\textwidth}
    \includegraphics[width=0.98\columnwidth]{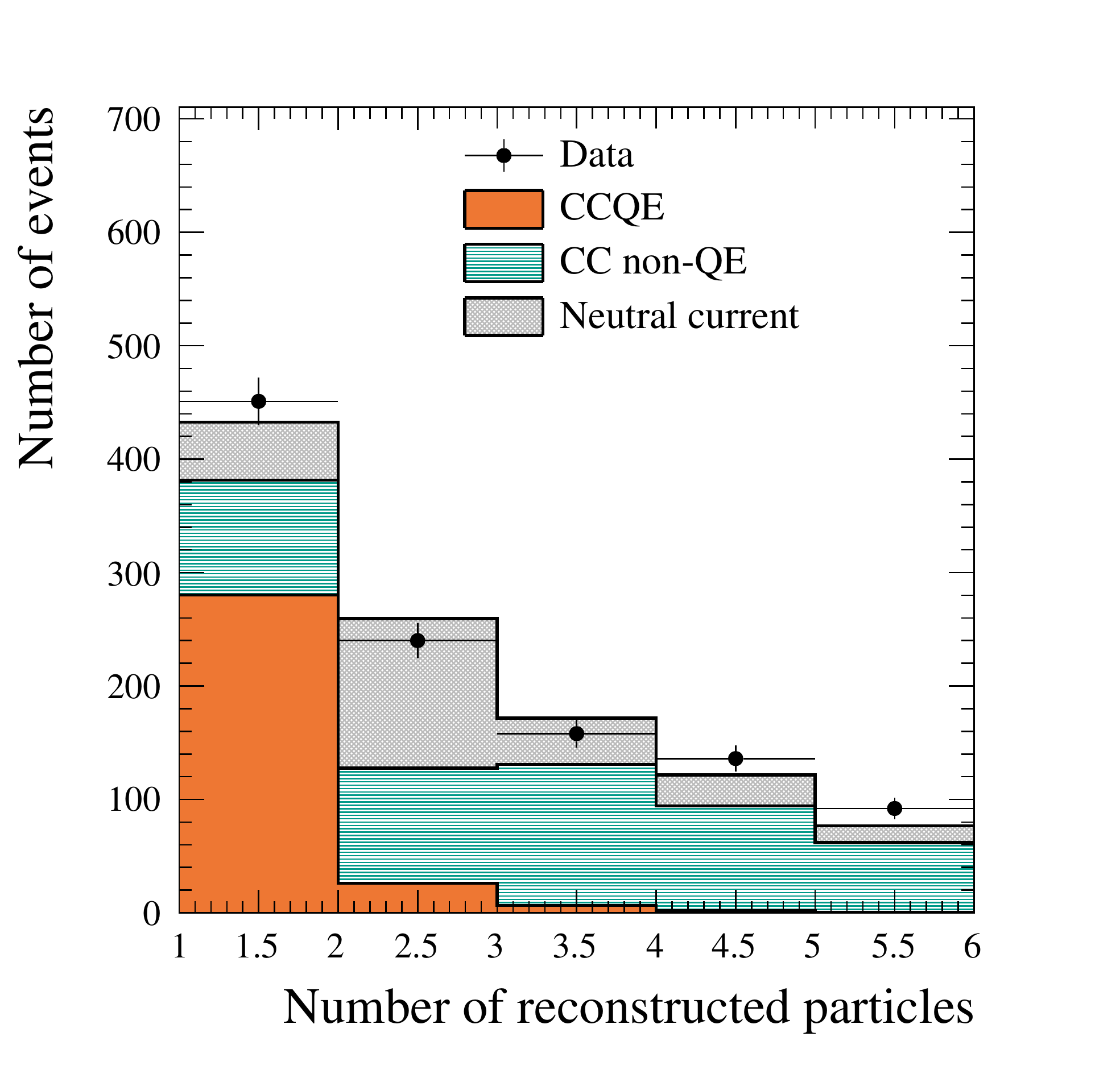}
  \end{subfigure}
  \begin{subfigure}[b]{0.47\textwidth}
    \includegraphics[width=0.98\columnwidth]{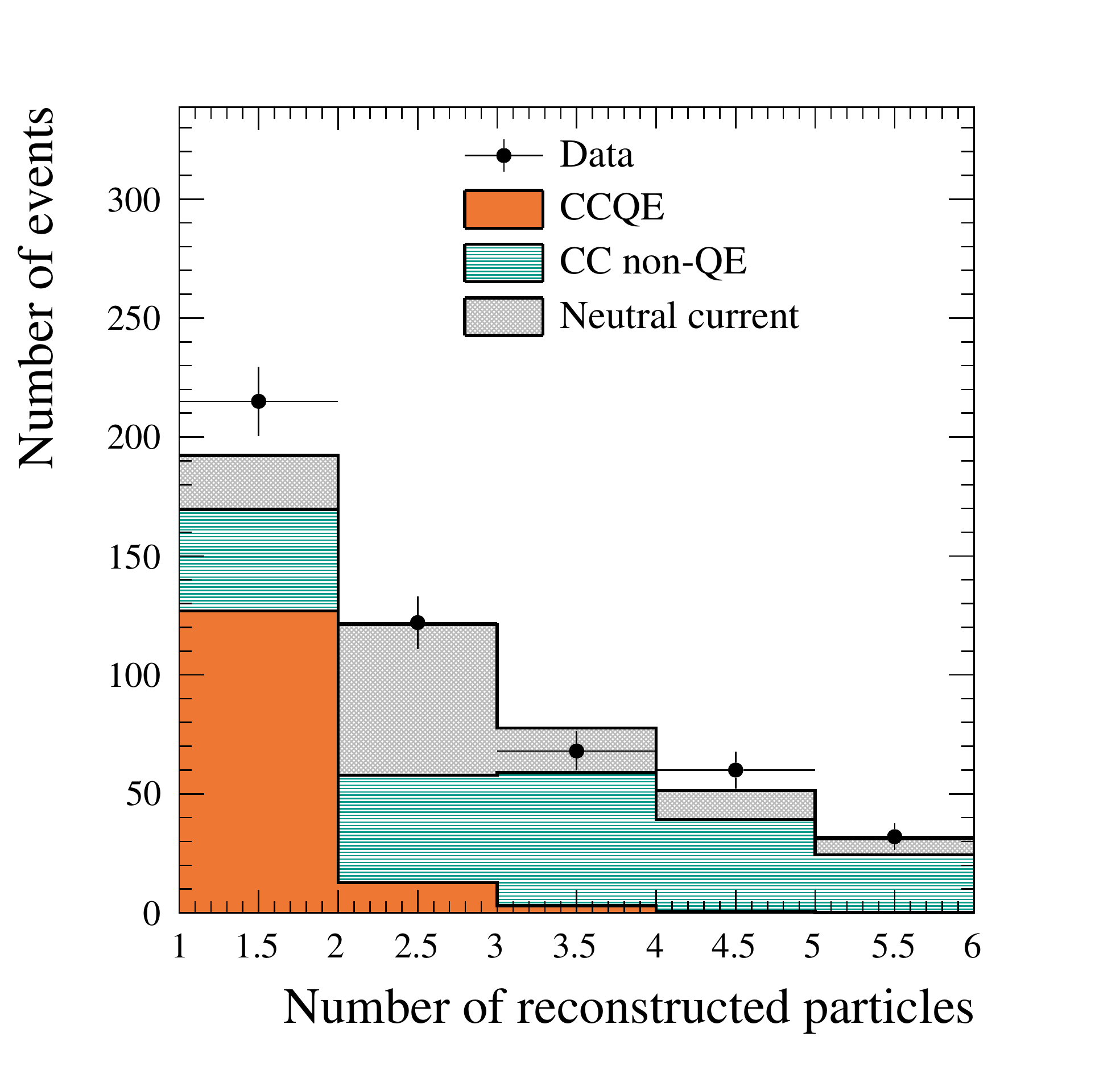}
  \end{subfigure}
  \caption {Distribution of number of reconstructed particles for events passing the $wall>80$~cm and $towall>170$~cm FV criteria in FHC (top) and RHC (bottom) data.}
  \label{fig:SK_nring}
\end{figure}

\begin{figure}[htbp]
  \centering
  \begin{subfigure}[b]{0.47\textwidth}
    \includegraphics[width=0.98\columnwidth]{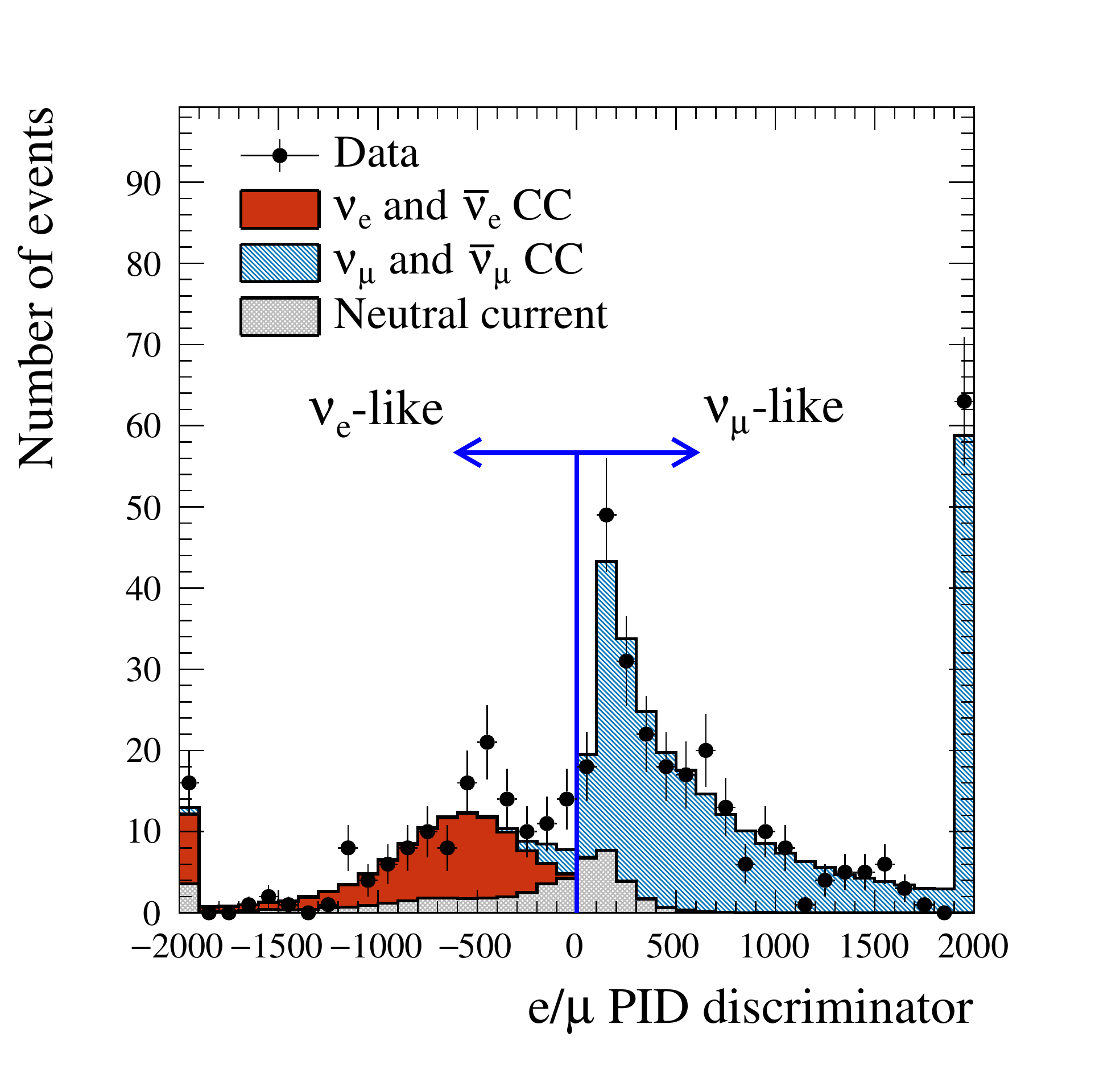}
  \end{subfigure}
  \begin{subfigure}[b]{0.47\textwidth}
    \includegraphics[width=0.98\columnwidth]{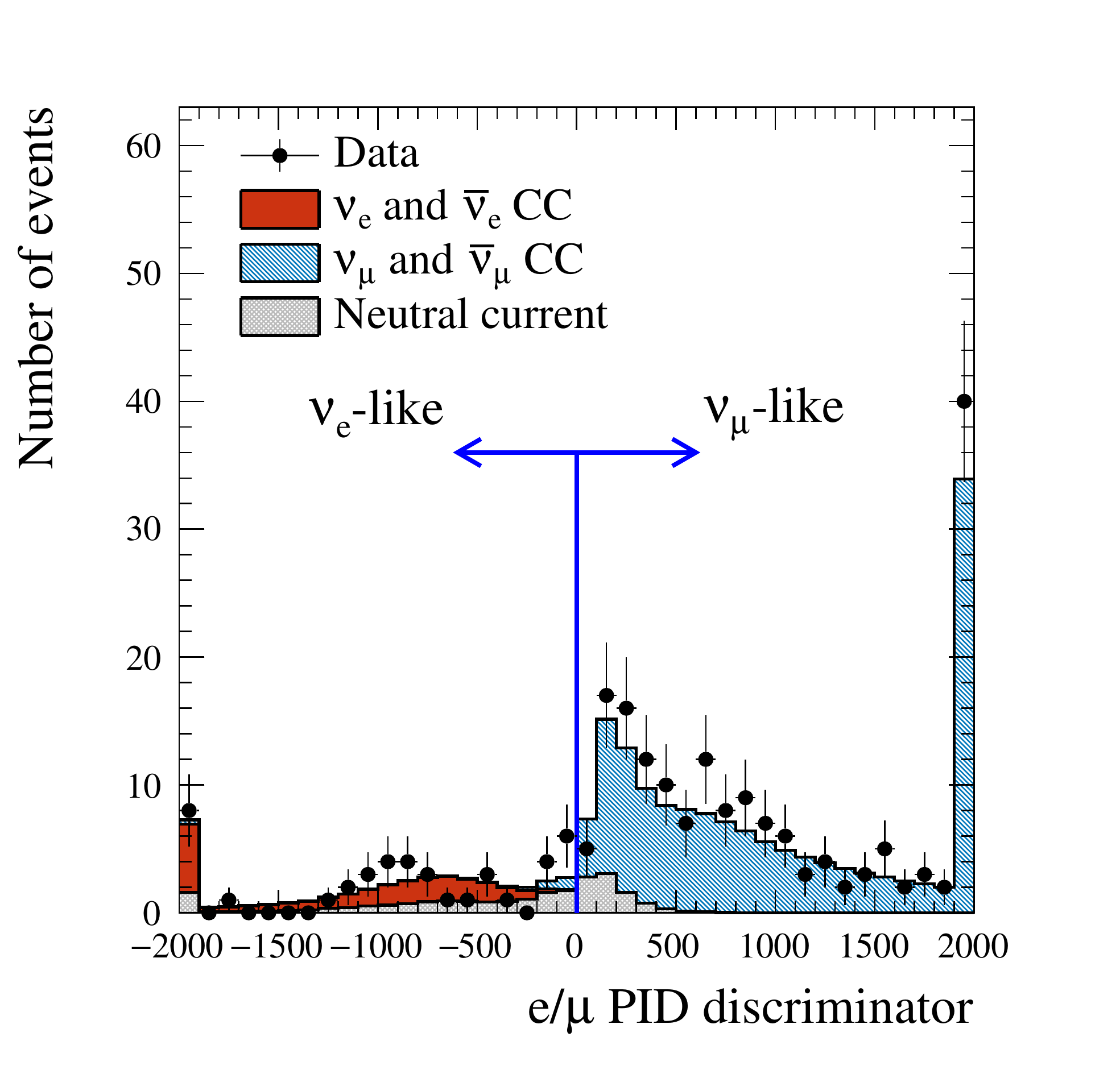}
  \end{subfigure}
  \caption {Distribution of the $e$/$\mu$ PID discriminator for single-particle events passing the $wall>80$~cm and $towall>170$~cm FV criteria in FHC (top) and RHC (bottom) data.}
  \label{fig:SK_emu_PID}
\end{figure}

\begin{figure}[htbp]
  \centering
  \begin{subfigure}[b]{0.47\textwidth}
    \includegraphics[width=0.98\columnwidth]{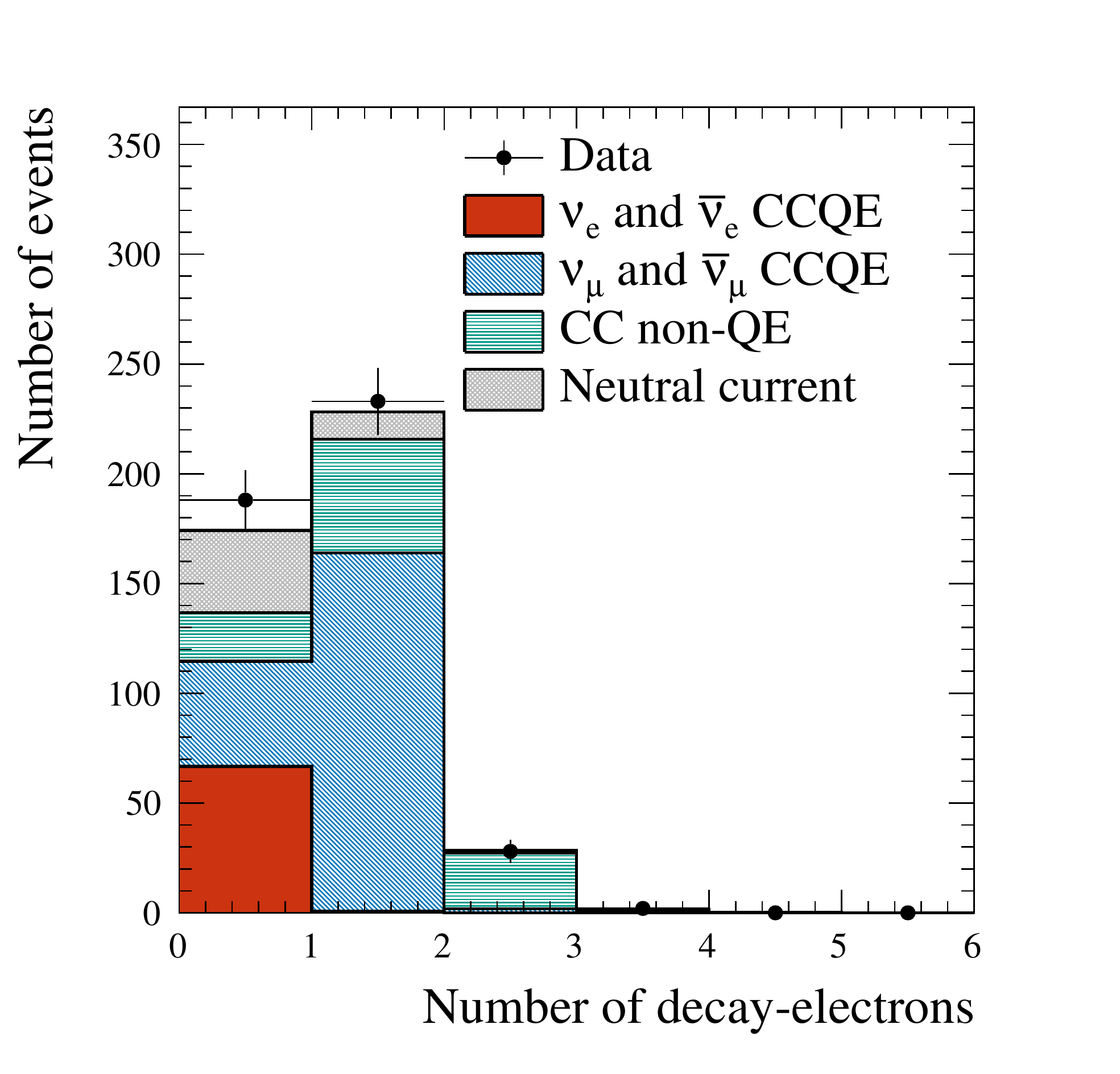}
  \end{subfigure}
  \begin{subfigure}[b]{0.47\textwidth}
    \includegraphics[width=0.98\columnwidth]{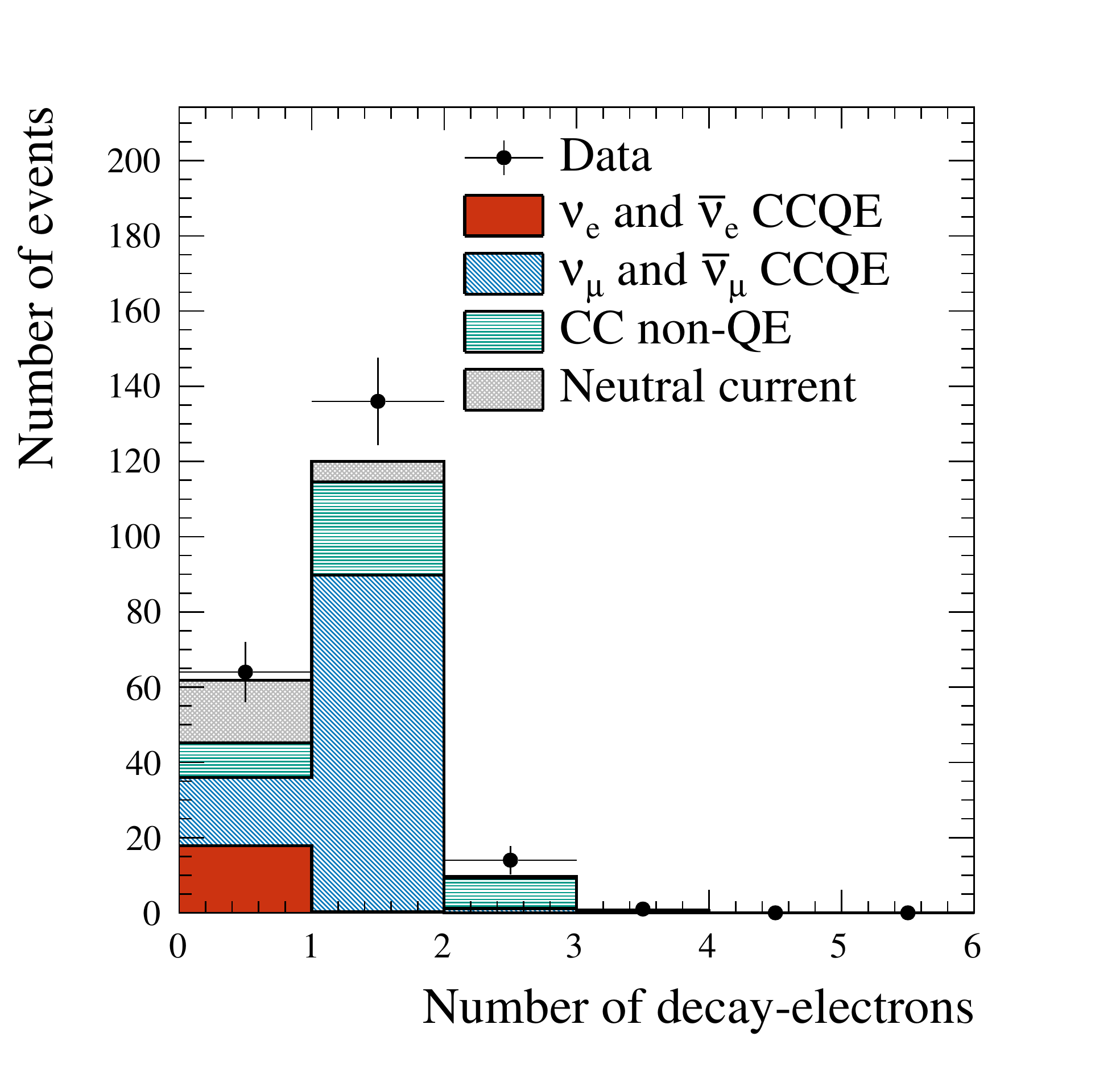}
  \end{subfigure}
  \caption {Distribution of the number of identified decay-electrons for single-particle events passing the $wall>80$~cm and $towall>170$~cm FV criteria in FHC (top) and RHC (bottom) data.} \label{fig:SK_ndecaye}
\end{figure}

For both FHC and RHC FC events, the distributions of the number of reconstructed particle tracks are shown in Fig.~\ref{fig:SK_nring}. For events with a single reconstructed track, the distributions of the $e$/$\mu$ discriminator and number of identified $\mu$-decay electrons are shown in Figs.~\ref{fig:SK_emu_PID} and~\ref{fig:SK_ndecaye} respectively. In these figures, and throughout this section, the MC predictions are produced with the neutrino mixing parameters given in Tab.~\ref{tab:asimova_params}, and the flux and cross-section parameters set to the best-fit value resulting from the near detector analysis described in Sec.~\ref{sec:fitters}.

\begin{table}[htbp]
\centering
\caption{
	Reference set of values of the oscillation parameters used to evaluate the expected sensitivity, number of events and effect of systematic uncertainties. The values of $\sin^2 \theta_{12}$ and $\Delta m^2_{21}$  are taken from \cite{pdg_2014}, the value of $\sin^2 \theta_{13}$  from \cite{pdg_2018}, and all the other values are set to the most probable value found in a previous T2K measurements \cite{Abe2015k}.
}
\label{tab:asimova_params}
\newcommand{\massmag}[1]{{\times}10^{#1}\si{\eV^2\!/\clight^4}} 
\newcommand{\point}{$&$}
\newcommand{\nopoint}{\multicolumn{1}{l@{}}{}}
\begin{tabular}{lr@{.}l}
\hline
\hline
Parameter            & \nopoint & Value  \\
\hline
$\sin^{2}\theta_{23}$   & $0 \point 528$  \\
$\sin^{2} \theta_{13}$  & $0 \point 0212$\\
$\sin^{2} \theta_{12}$	& $0 \point 304$\\[0.25ex]
$\deltacp$           & $-1\point 601$ \\[0.25ex]
\begin{tabular}{@{}ll@{}}
$\Delta m^{2}_{32}$ & (NO)\\
$\Delta m^{2}_{13}$ & (IO) 
\end{tabular}
    & $2 \point 509\massmag{-3}$ \\
$\Delta m^{2}_{21}$     
    & $7 \point 53\phantom{0} \massmag{-5}$	\\[0.25ex]
Mass Ordering           & \nopoint &  Normal \\
\hline
\hline
\end{tabular}
\end{table}

The reconstructed momentum is required to be larger than 100~MeV/$c$ for the $e$-like samples to reduce contamination from below-threshold-$\mu$ decays, and larger than 200~MeV/$c$ for the $\mu$-like samples.

Events in the $\mu$-like samples can have up to one reconstructed decay-$e$, while the $e$-like samples are required to have zero and one decay-$e$ for the samples targeting CCQE and CC1$\pi^{+}$ interactions, respectively.

Neutral current $\pi$ production events are a background in both $e$-like and $\mu$-like samples. In the former, electromagnetic showers resulting from $\pi^0 \rightarrow \gamma \gamma$ decays can mimic an electron-like event; while in the latter, charged pion detector signatures are only significantly different from those of muons through their hadronic interactions, which are not always present. Additional criteria are used to remove these backgrounds in all analysis samples.

For the $\mu$-like samples a cut is applied on the likelihood ratio of the best-fit single-$\pi^{+}$ to the single-$\mu$ hypotheses ($\Lambda^{\pi^{+}}_{\mu}$) as a function of reconstructed $\mu$ momentum ($p_{\mu}$). This selection criterion is only available in the FiTQun reconstruction algorithm and is deployed for the first time in the current analysis.

\begin{figure}[htbp]
  \centering
  \begin{subfigure}[b]{0.47\textwidth}
    \includegraphics[width=0.98\columnwidth]{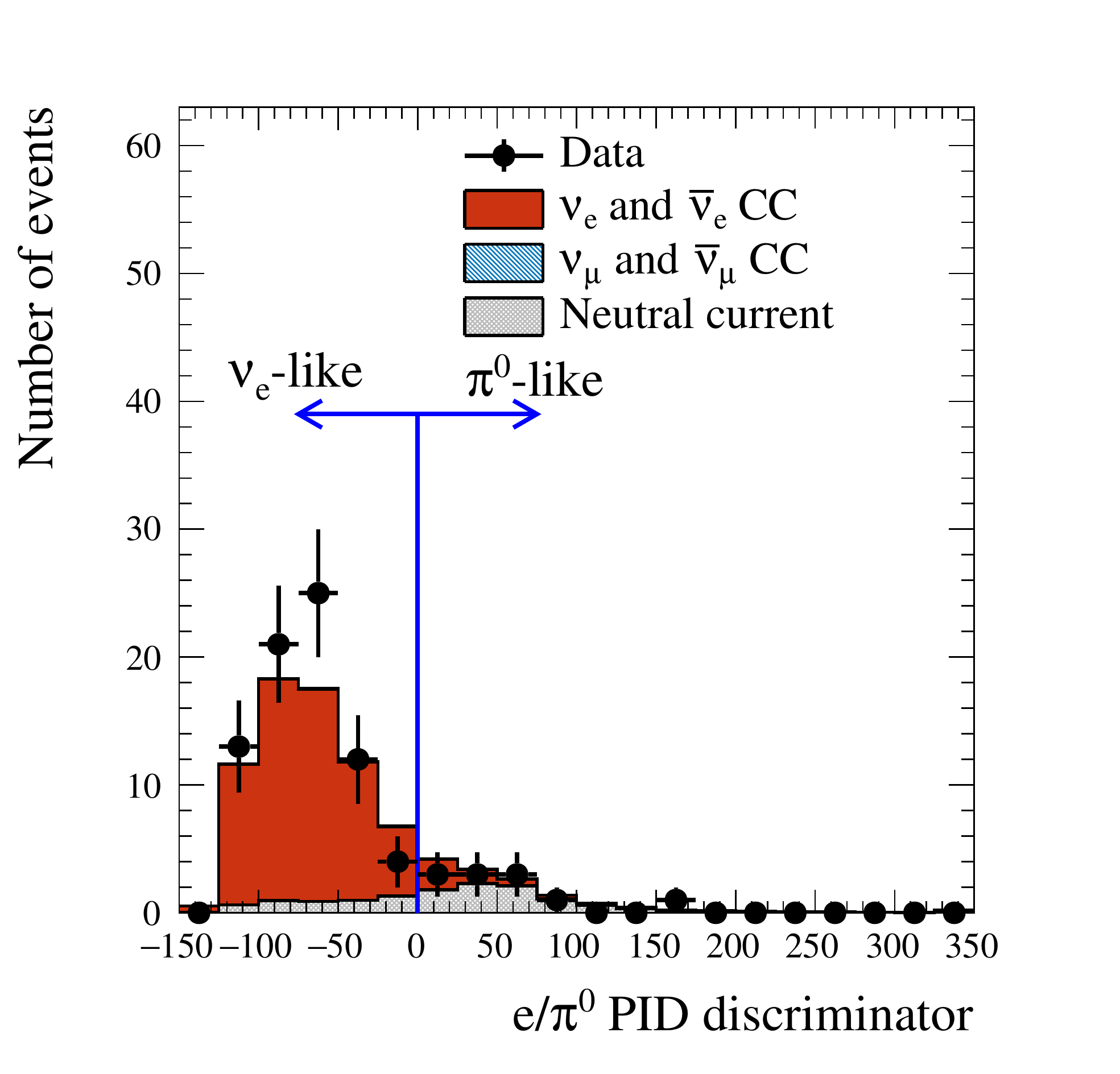}
  \end{subfigure}
  \begin{subfigure}[b]{0.47\textwidth}
    \includegraphics[width=0.98\columnwidth]{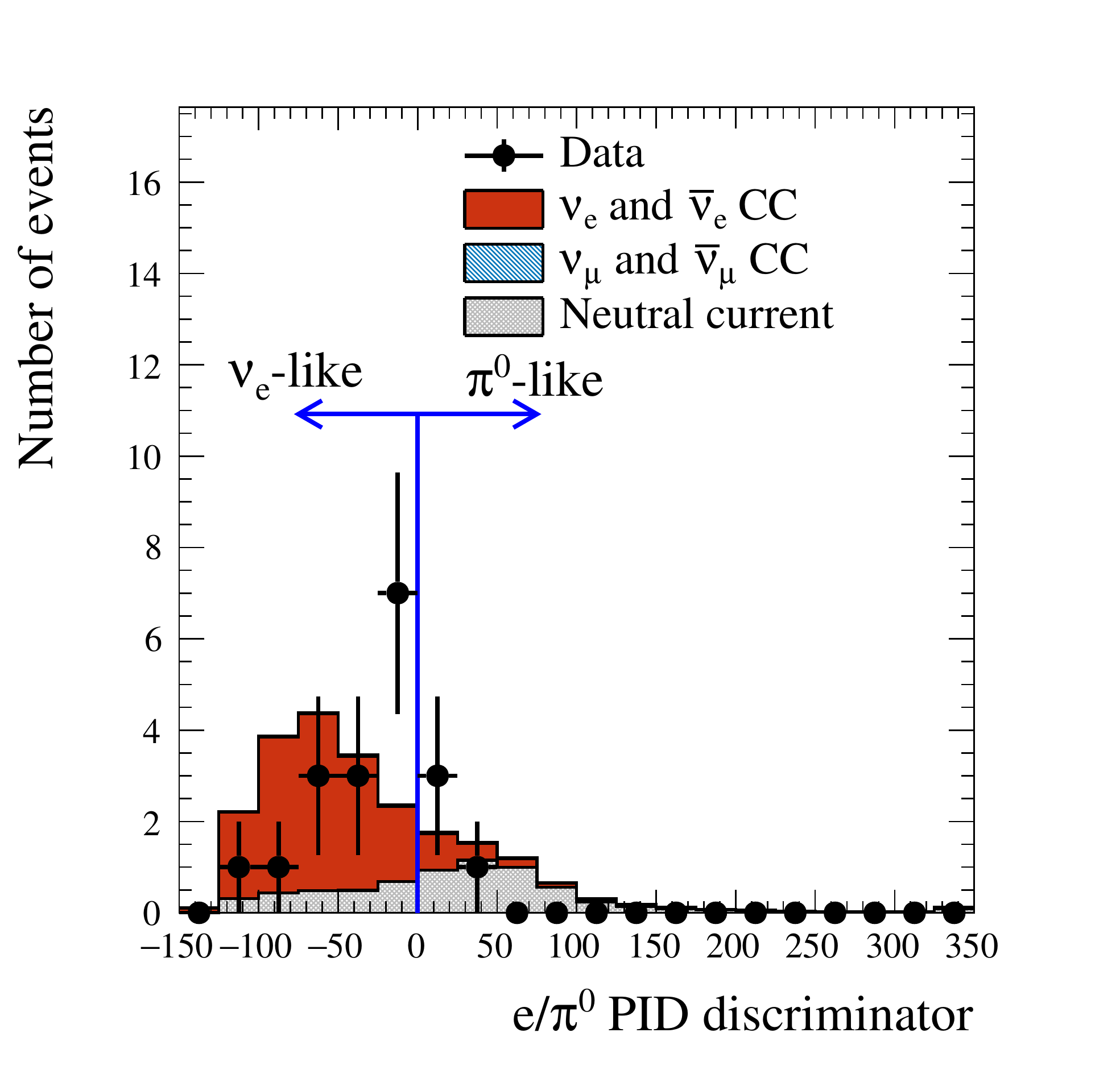}
  \end{subfigure}
  \caption {Distribution of the $e$/$\pi^{0}$ PID discriminator for 1R$_{e}$ events in FHC (top) and RHC (bottom) data.}
  \label{fig:SK_epi0_PID}
\end{figure}

\begin{figure}[htbp]
  \centering
  \includegraphics[width=0.46\textwidth]{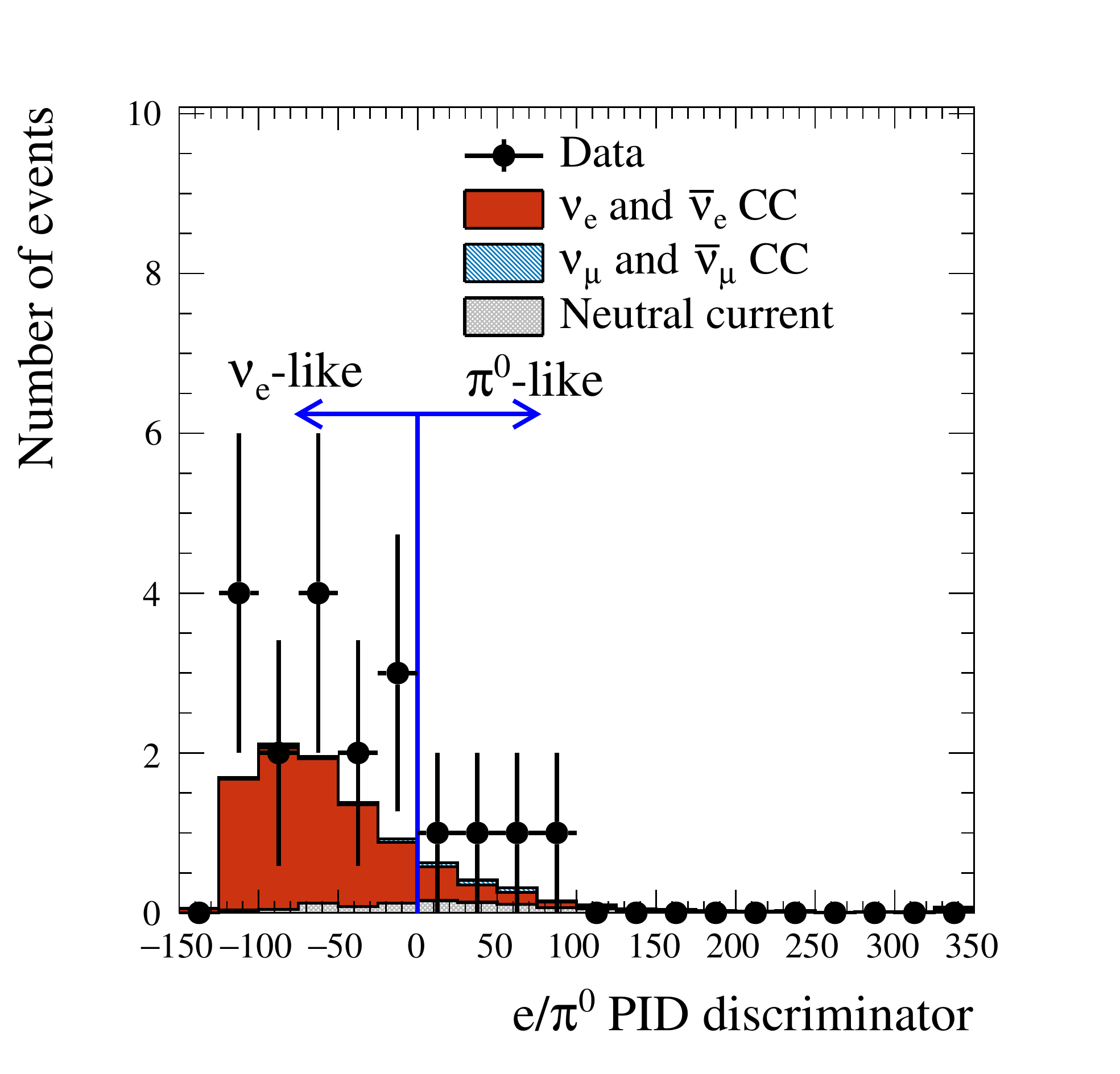}
  \caption {Distribution of the $e$/$\pi^{0}$ PID discriminator for 1R$_{e}$ $+$ 1 d.e. in FHC data.}
  \label{fig:SK_epi0_PID_nue1pi}
\end{figure}

\begin{figure}[htbp]
  \centering
  \begin{subfigure}[b]{0.47\textwidth}
    \includegraphics[width=0.98\columnwidth]{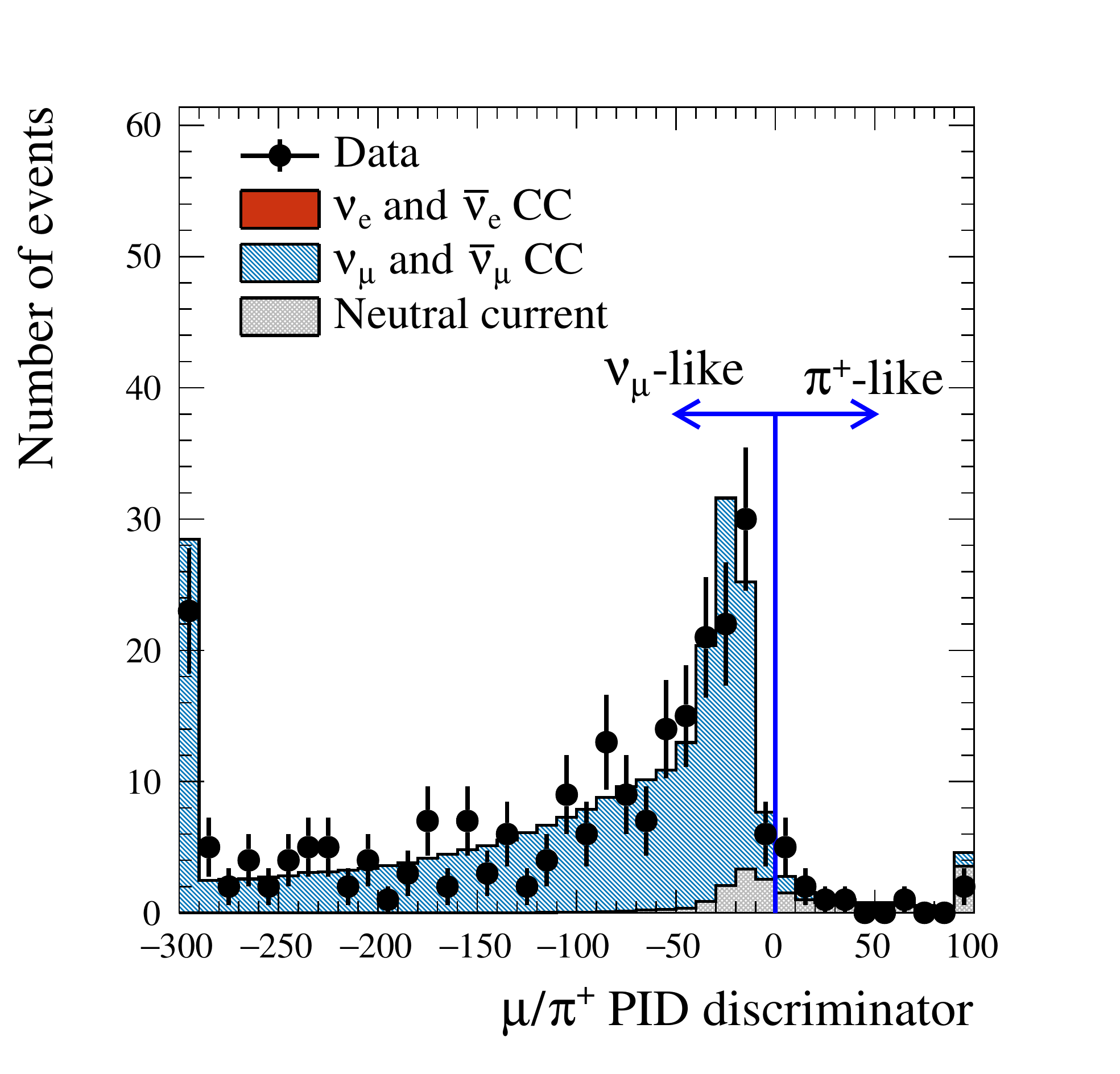}
  \end{subfigure}
  \begin{subfigure}[b]{0.47\textwidth}
    \includegraphics[width=0.98\columnwidth]{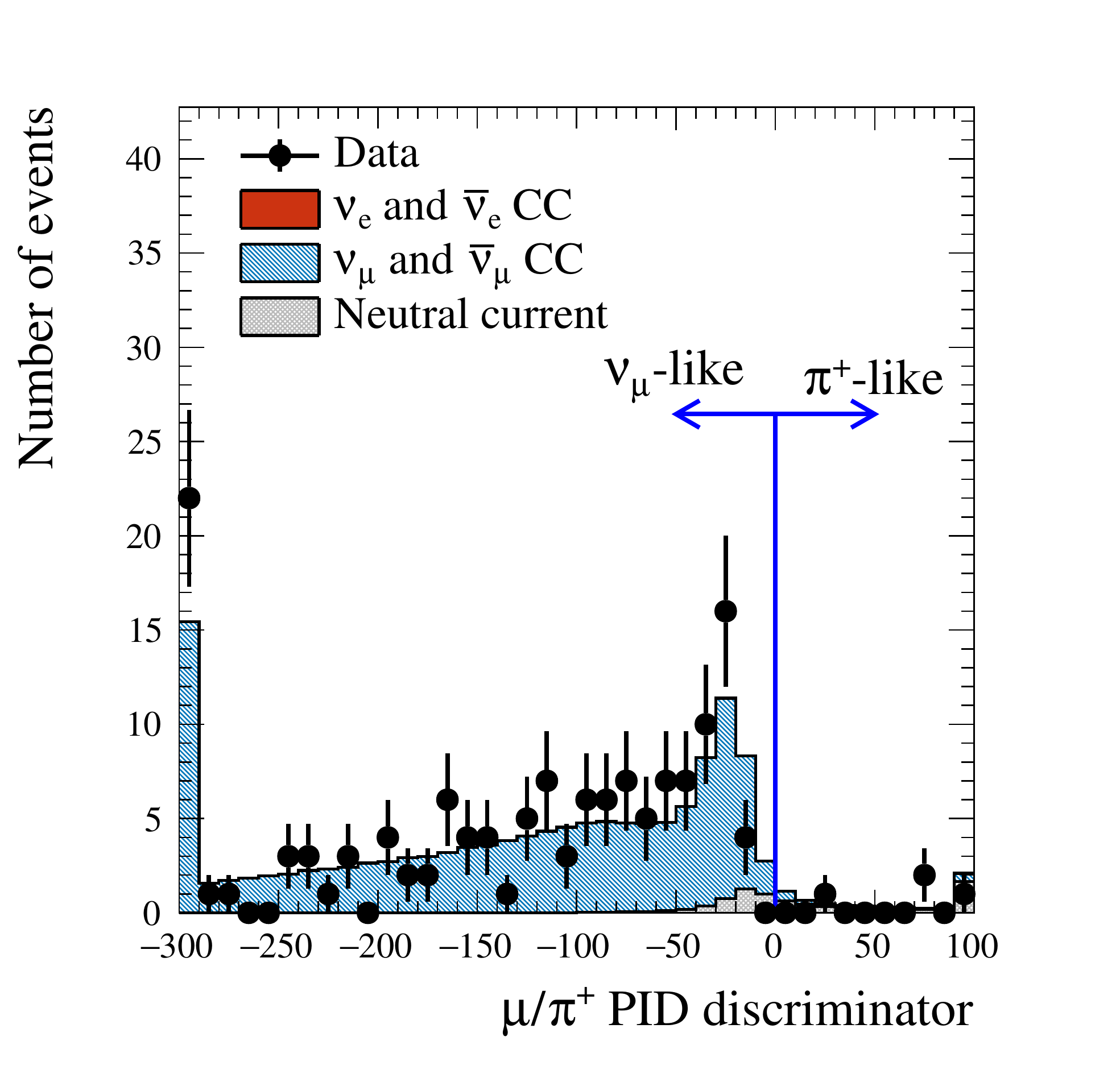}
  \end{subfigure}
  \caption {Distribution of the $\mu$/$\pi^{+}$ PID discriminator for 1R$_{\mu}$ events in FHC (top) and RHC (bottom) data.}
  \label{fig:SK_mupip_PID}
\end{figure}

The NC$\pi^{0}$ rejection criterion for the $e$-like samples, based on the likelihood ratio of the best-fit $\pi^{0}$ hypothesis to the $e$-like hypothesis ($\Lambda^{\pi^{0}}_{e}$) and the reconstructed $\pi^{0}$ mass ($m_{\gamma\gamma}$), is unchanged from previous analyses, where it was the only use of FiTQun reconstruction. Distributions of the neutral current $\pi^{0}$ rejection discriminator for the 1R$_{e}$ and 1R$_{e}$ $+$ 1 d.e. samples are shown in Figs.~\ref{fig:SK_epi0_PID} and \ref{fig:SK_epi0_PID_nue1pi} respectively. For 1R$_{\mu}$ events, the distribution of the neutral current $\pi^{+}$ discriminator is shown in Fig.~\ref{fig:SK_mupip_PID}.

Finally, $e$-like samples are required to have a reconstructed neutrino energy ($E_{\textrm{rec}}$) lower than 1250~MeV, as events beyond this energy are insensitive to neutrino oscillations and are more susceptible to systematic uncertainties associated with the neutral current rejection cut.

The selection criteria described above are summarized in Tab.~\ref{tab:SK_selection} and the breakdown of data and MC events passing each cut is given in Tabs.~\ref{table:expected_1-9_nue},~\ref{tab:tbl_run1_9_nue1pi} and~\ref{table:expected_1-9_numu}, for the samples targeting $\nu_{e}$ CCQE, $\nu_{e}$ CC 1$\pi^{+}$ and $\nu_{\mu}$ CCQE interactions, respectively.

\begin{figure}[htbp]
  \centering
  \begin{subfigure}[b]{0.47\textwidth}
    \includegraphics[width=0.98\columnwidth]{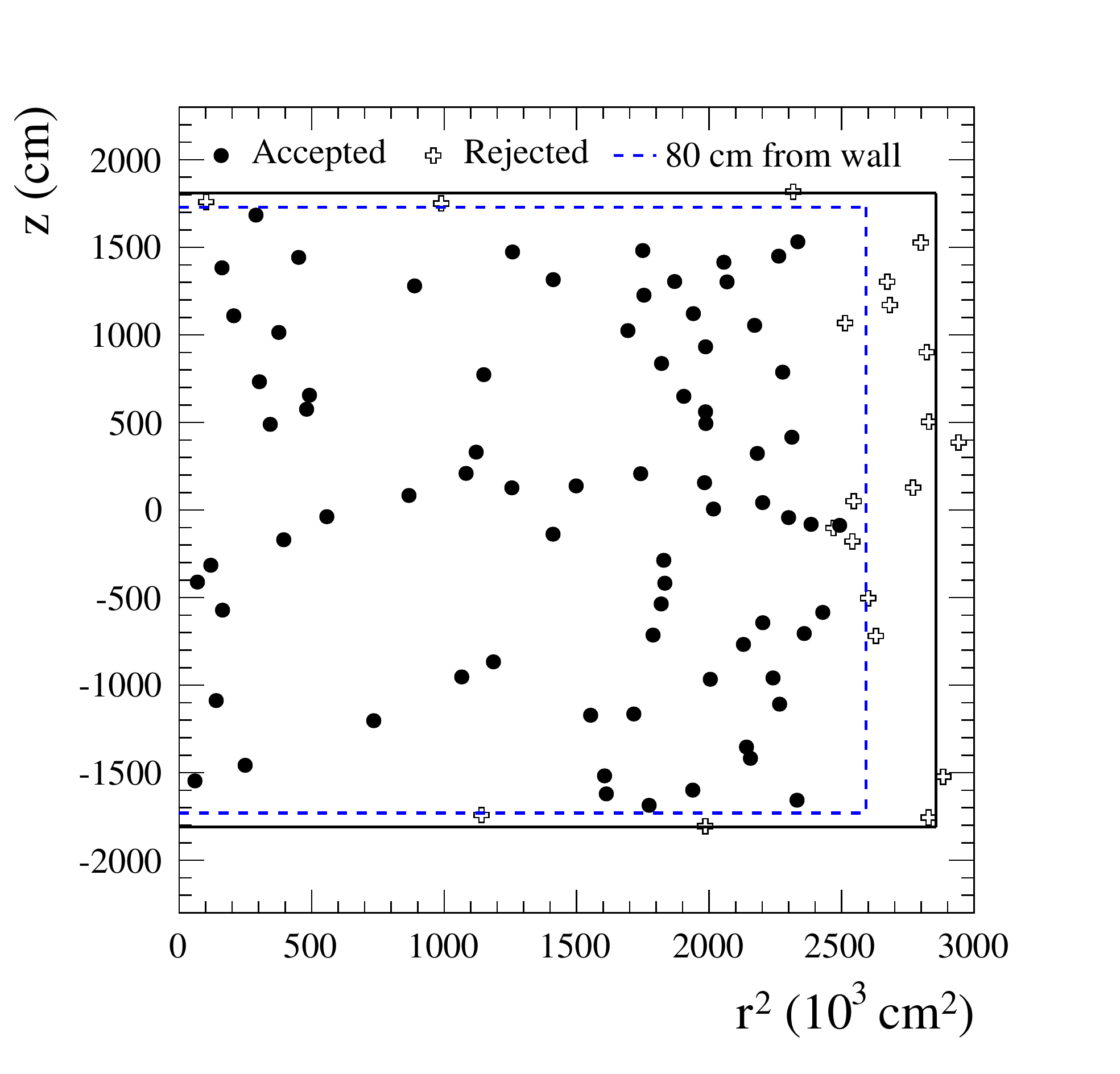}
  \end{subfigure}
  \begin{subfigure}[b]{0.47\textwidth}
    \includegraphics[width=0.98\columnwidth]{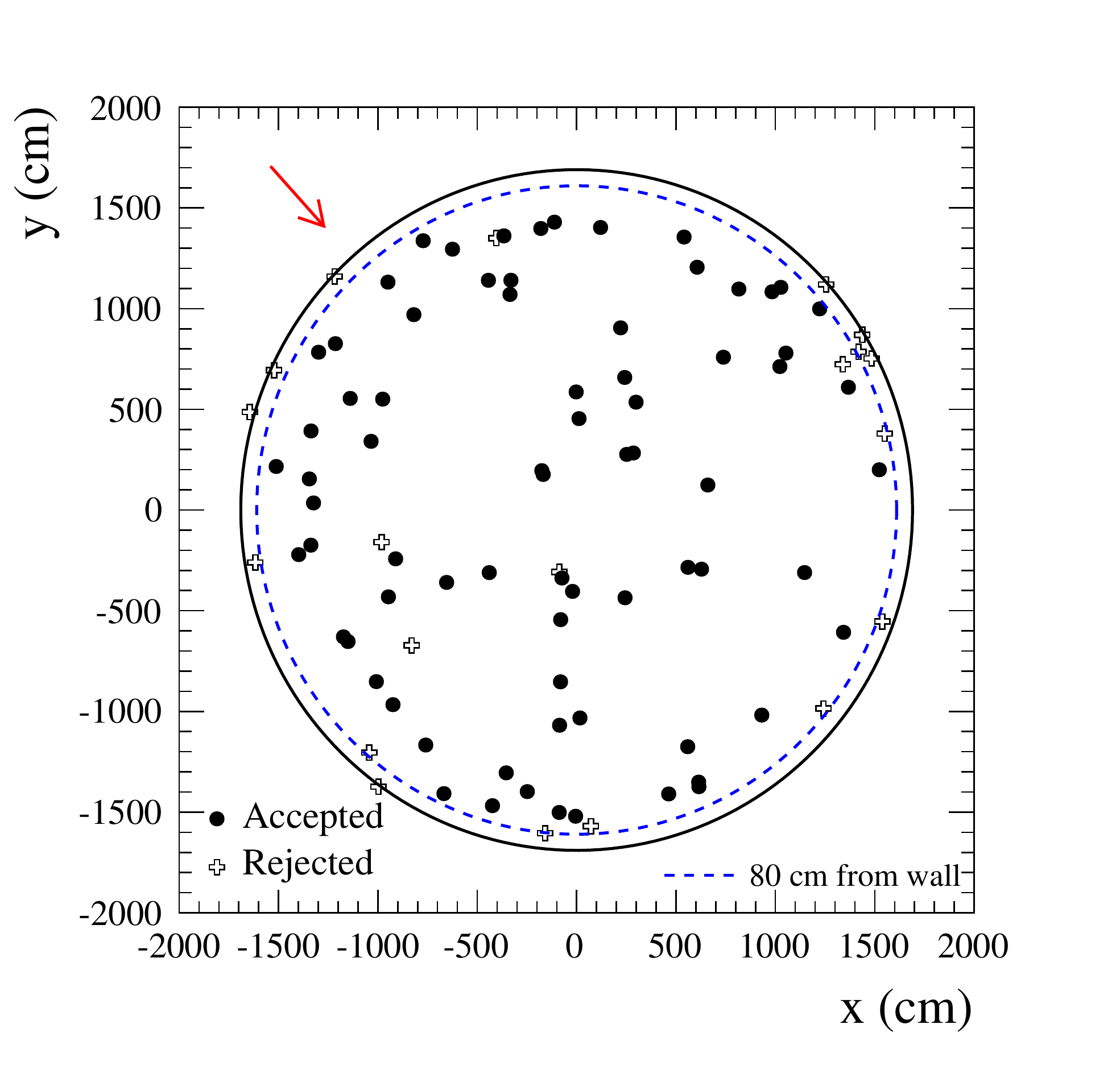}
  \end{subfigure}
  \caption{Reconstructed vertices of selected FHC data 1R$_{e}$ events projected in $z$ vs $r^2$ (top) and $y$ vs $x$ (bottom). The neutrino beam direction is shown as a red arrow and events which do not pass the fiducial volume criteria are shown as hollow crosses.}
  \label{fig:SK_vtx_fhc_nue}
\end{figure}

\begin{figure}[htbp]
  \centering
  \begin{subfigure}[b]{0.47\textwidth}
    \includegraphics[width=0.98\columnwidth]{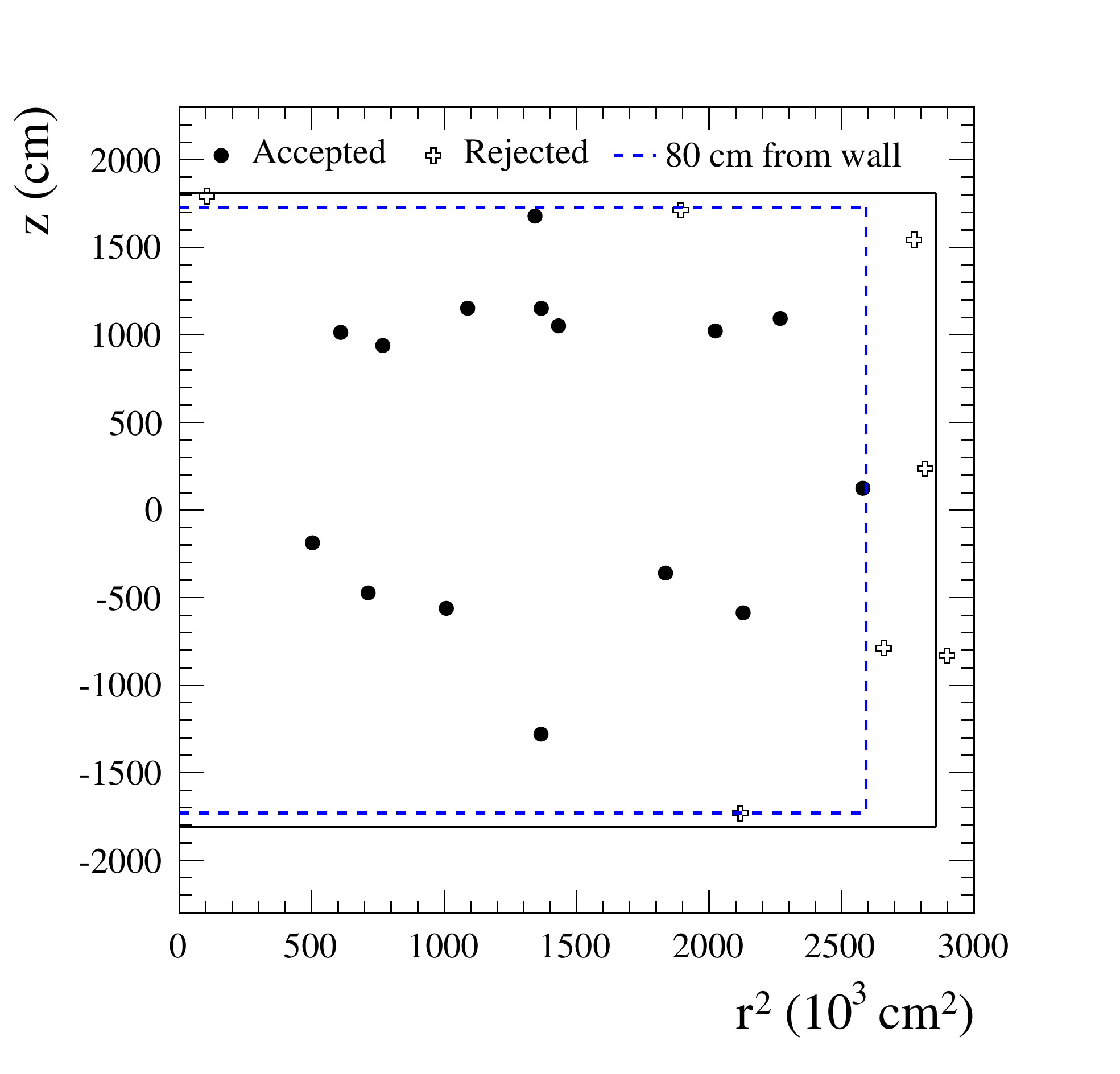}
  \end{subfigure}
  \begin{subfigure}[b]{0.47\textwidth}
    \includegraphics[width=0.98\columnwidth]{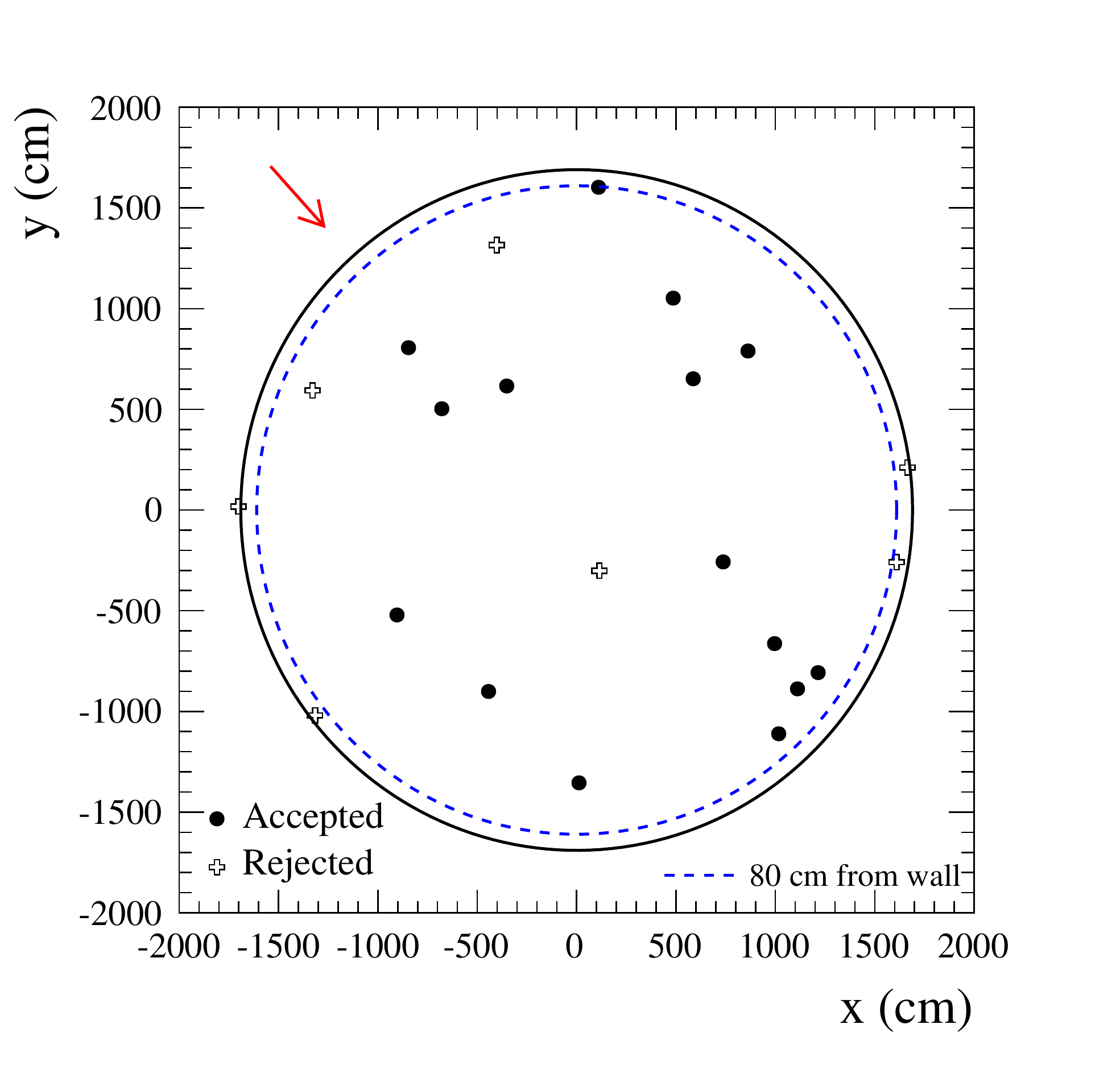}
  \end{subfigure}
  \caption{Reconstructed vertices of selected RHC data 1R$_{e}$ events projected in $z$ vs $r^2$ (top) and $y$ vs $x$ (bottom). The neutrino beam direction is shown as a red arrow and events which do not pass the fiducial volume criteria are shown as hollow crosses.}
  \label{fig:SK_vtx_rhc_nue}
\end{figure}

\begin{figure}[htbp]
  \centering
  \begin{subfigure}[b]{0.47\textwidth}
    \includegraphics[width=0.98\columnwidth]{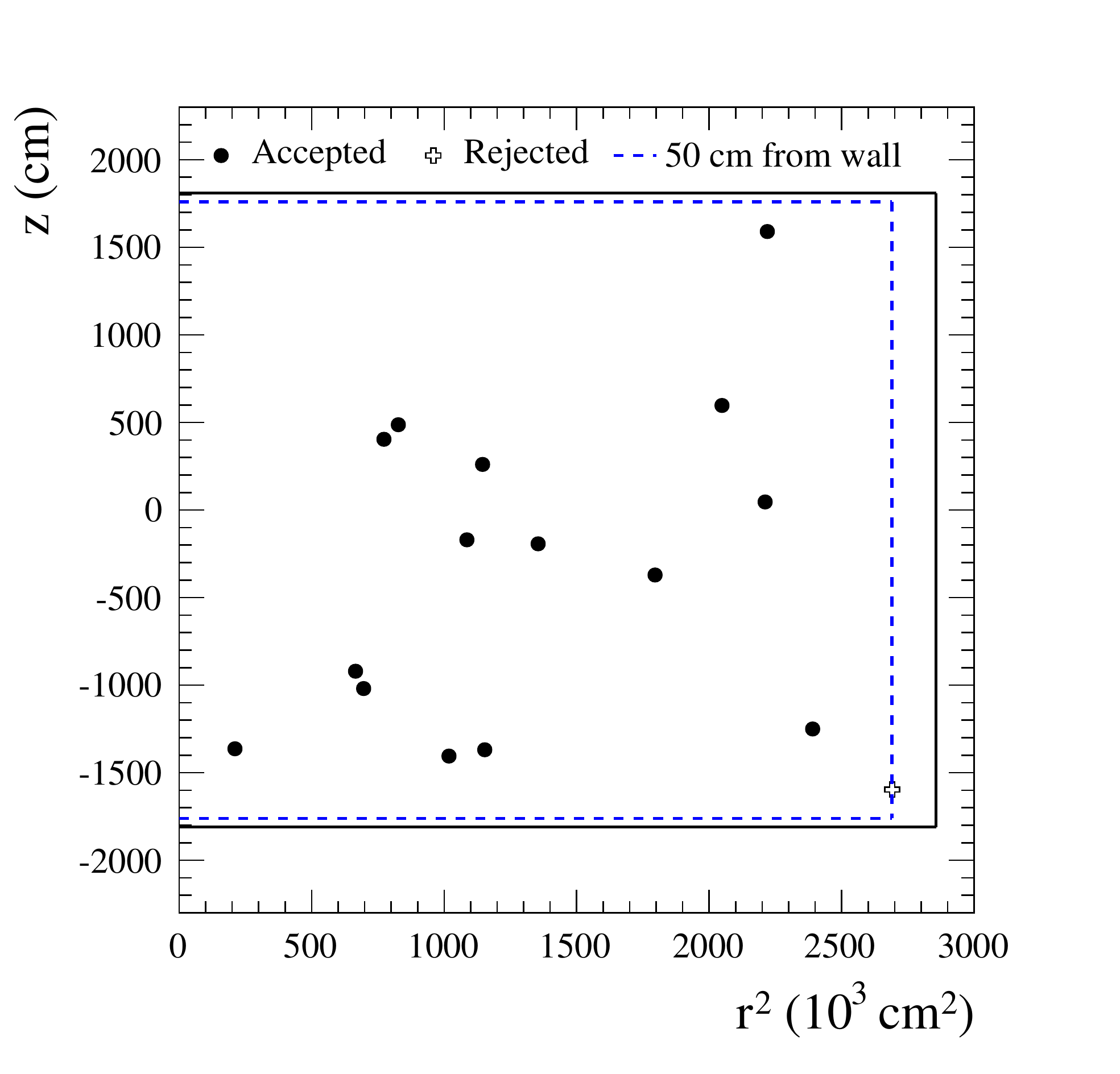}
  \end{subfigure}
  \begin{subfigure}[b]{0.47\textwidth}
    \includegraphics[width=0.98\columnwidth]{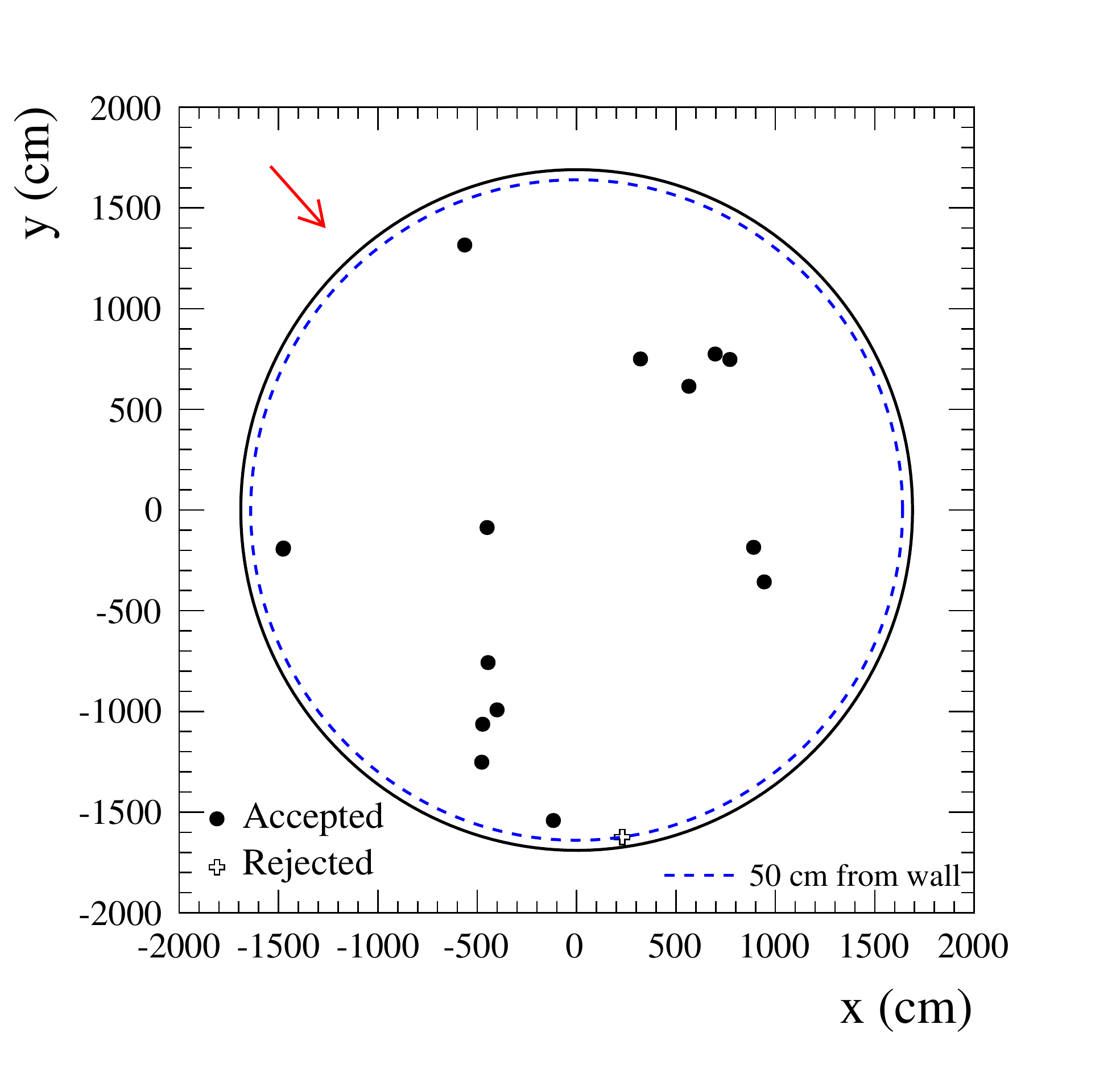}
  \end{subfigure}
  \caption{Reconstructed vertices of selected FHC data 1R$_{e}$ $+$ 1 d.e. events projected in $z$ vs $r^2$ (top) and $y$ vs $x$ (bottom). The neutrino beam direction is shown as a red arrow and events which do not pass the fiducial volume criteria are shown as hollow crosses.}
  \label{fig:SK_vtx_fhc_nue1pi}
\end{figure}

\begin{figure}[htbp]
  \center
  \begin{subfigure}[b]{0.47\textwidth}
    \includegraphics[width=0.98\columnwidth]{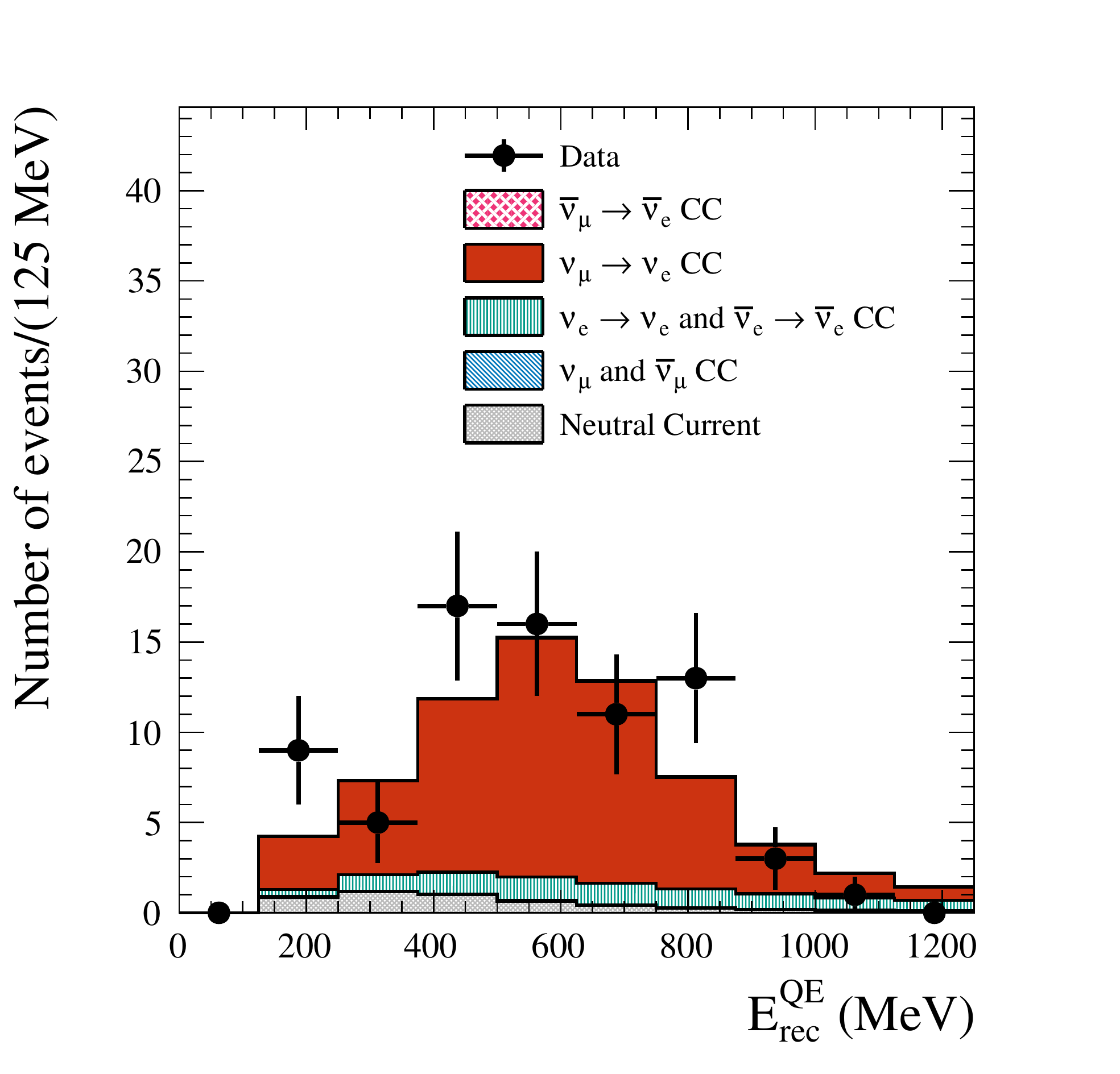}
  \end{subfigure}
  \begin{subfigure}[b]{0.47\textwidth}
    \includegraphics[width=0.98\columnwidth]{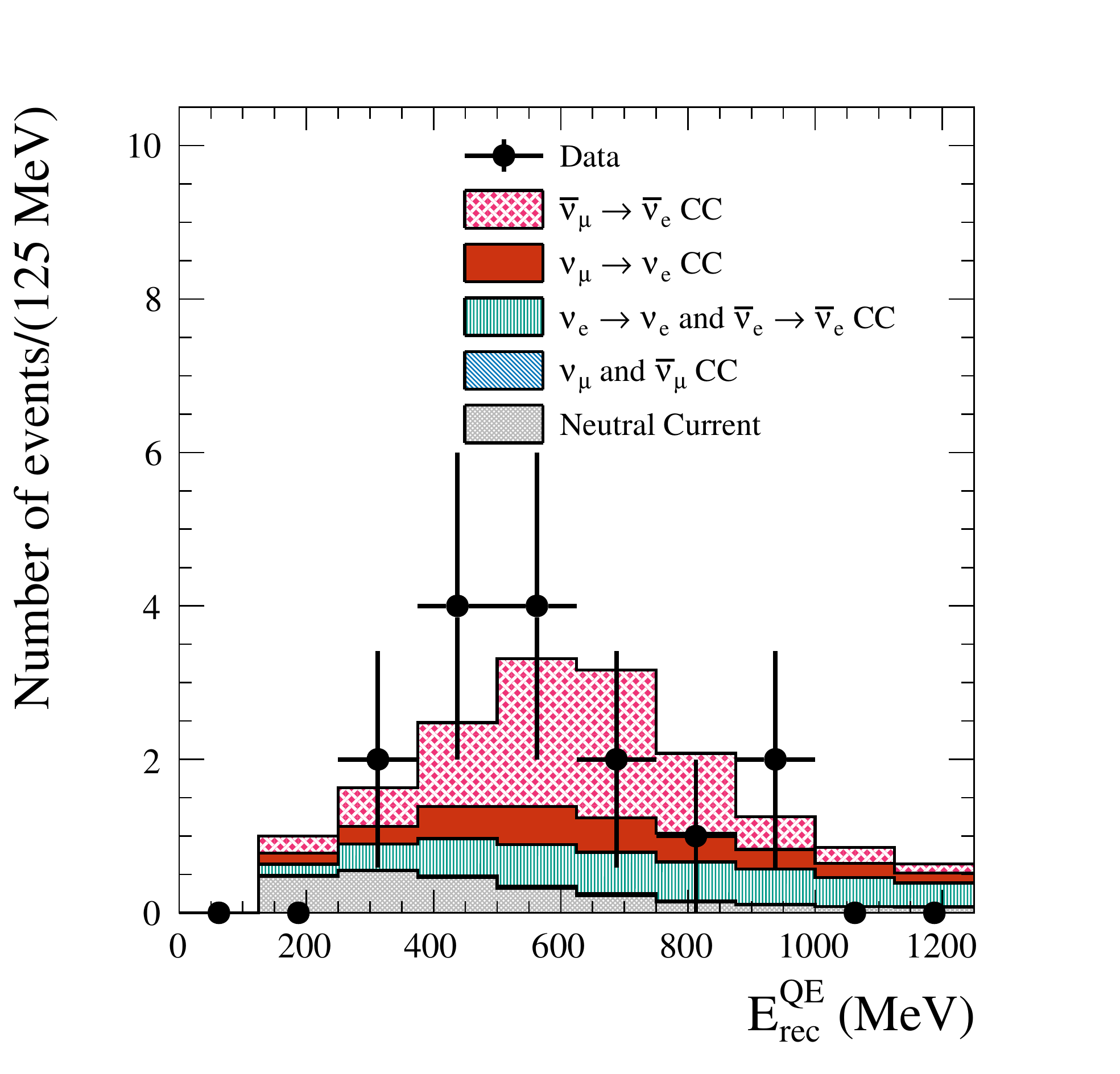}
  \end{subfigure}
  \caption {Reconstructed energy distribution for 1R$_{e}$ events in FHC (top) and RHC (bottom) data.}
  \label{fig:SK_Erec_nue}
\end{figure}

\begin{figure}[htbp]
  \centering
  \includegraphics[width=0.46\textwidth]{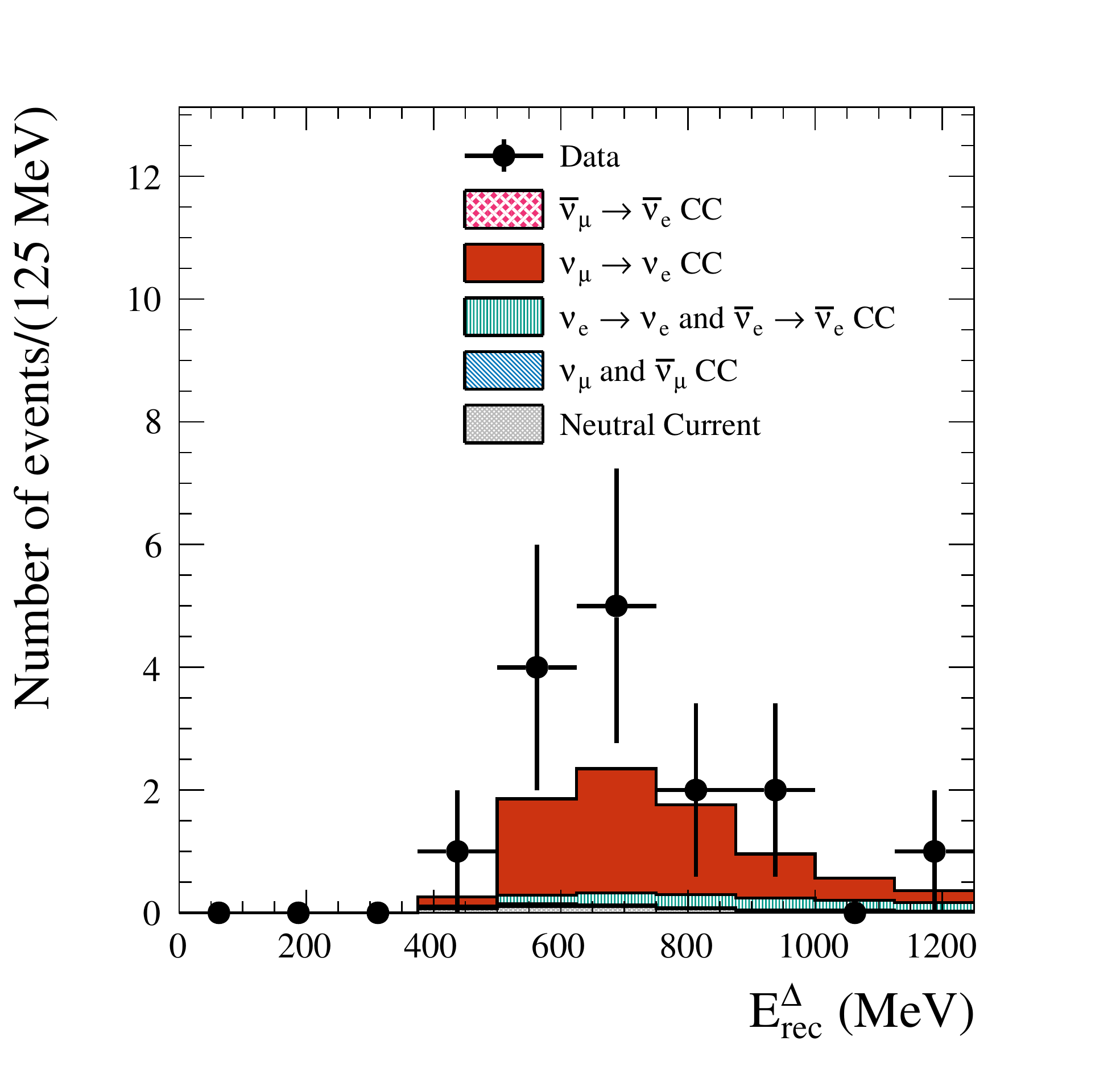}
  \caption {Reconstructed energy distribution for 1R$_{e}$ $+$ 1 d.e. events in FHC data. The $\Delta(1232)$ mass is used for the final-state nucleon in the reconstructed neutrino energy calculation.}
  \label{fig:SK_Erec_nue1pi}
\end{figure}

\begin{figure}[htbp]
  \centering
  \begin{subfigure}[b]{0.47\textwidth}
    \includegraphics[width=0.98\columnwidth]{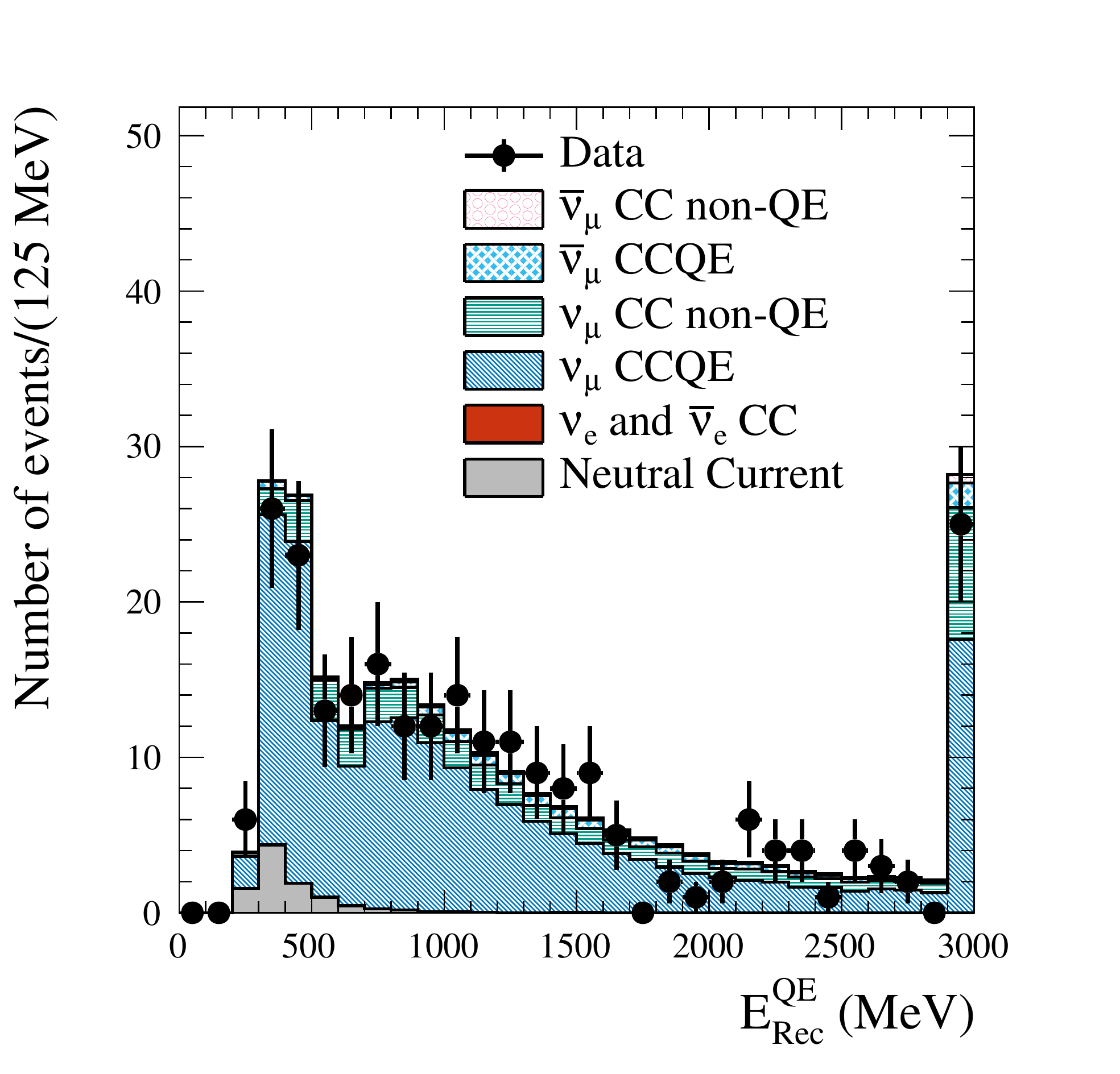}
  \end{subfigure}
  \begin{subfigure}[b]{0.47\textwidth}
    \includegraphics[width=0.98\columnwidth]{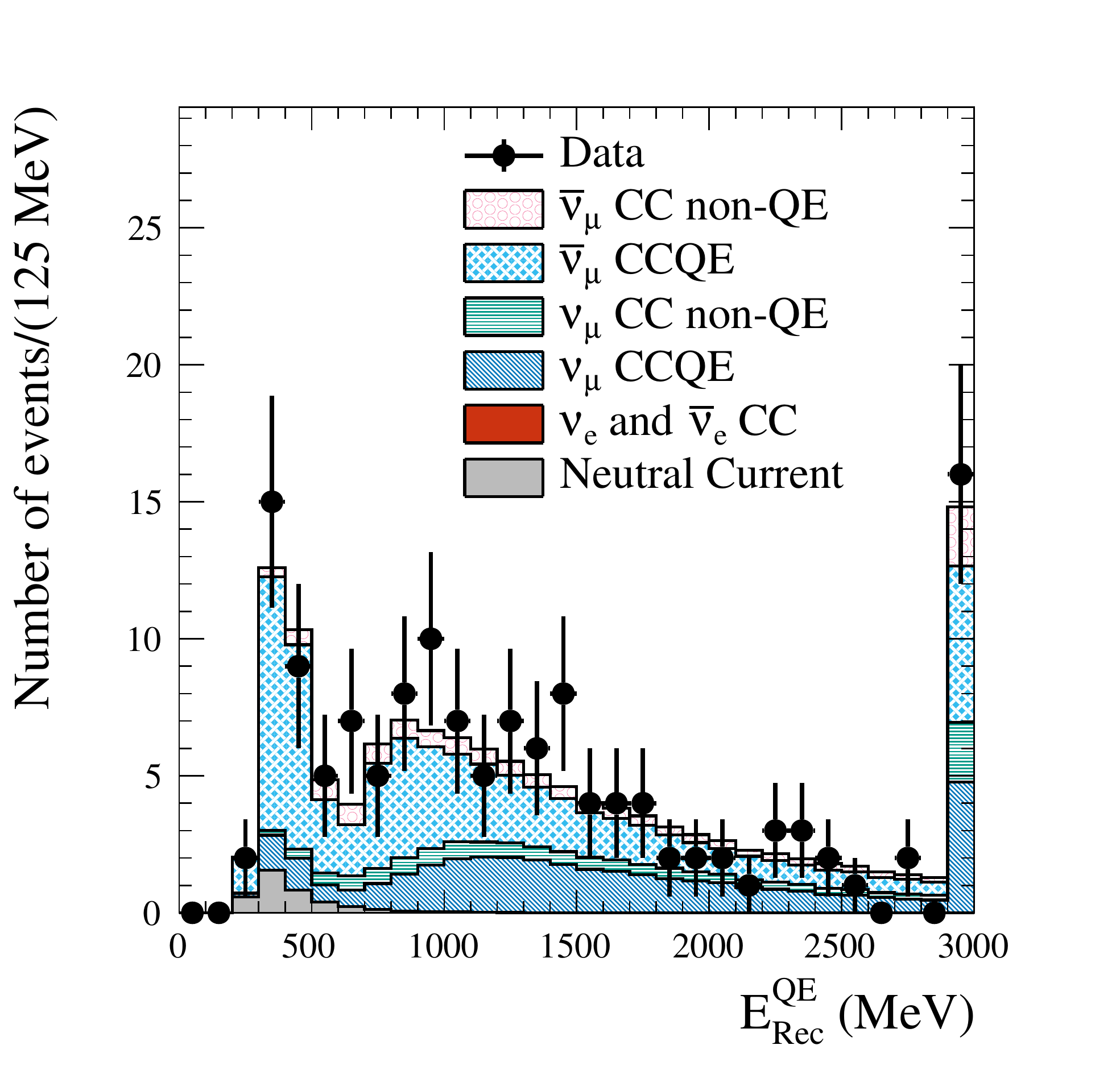}
  \end{subfigure}
  \caption {Reconstructed energy distribution for 1R$_{\mu}$ events in FHC (top) and RHC (bottom) data.}
  \label{fig:SK_Erec_numu}
\end{figure}

Distributions of reconstructed vertices of 1R$_{e}$ events in FHC and RHC data are shown in Figs.~\ref{fig:SK_vtx_fhc_nue} and~\ref{fig:SK_vtx_rhc_nue} respectively, and for 1R$_{e}$ $+$ 1 d.e. events in Fig.~\ref{fig:SK_vtx_fhc_nue1pi}. Reconstructed neutrino energy distributions are shown in Figs.~\ref{fig:SK_Erec_nue} and~\ref{fig:SK_Erec_nue1pi} for 1R$_{e}$ and 1R$_{e}$ $+$ 1 d.e. events, respectively, and in Fig.~\ref{fig:SK_Erec_numu} for 1R$_{\mu}$ events.

\begin{table*}[htbp]
  \centering
  \caption{SK selection criteria for the five fully-contained, single prompt particle, analysis samples.}
  \label{tab:SK_selection}
  \resizebox{\textwidth}{!}{
  \begin{ruledtabular}
  \begin{tabular}{lccc}
    & 1R$_{\mu}$ & 1R$_{e}$ & 1R$_{e}$ $+$ 1 d.e.\\ \hline
    \\\\[-16pt]
    \multicolumn{1}{l}{$wall$}  & $>\SI{50}{cm}$ & $>\SI{80}{cm}$ & $>\SI{50}{cm}$ \\[4pt]
    \multicolumn{1}{l}{$towall$}  & $>\SI{250}{cm}$ & $>\SI{170}{cm}$ & $>\SI{270}{cm}$ \\[4pt]
    $e$/$\mu$ identification & 
    $\Lambda^{e}_{\mu} < \frac{p_{e}}{\SI{5}{\MeV/\clight}}$ &
    $\Lambda^{e}_{\mu} > \frac{p_{e}}{\SI{5}{\MeV/\clight}}$ &
    $\Lambda^{e}_{\mu} > \frac{p_{e}}{\SI{5}{\MeV/\clight}}$ \\[4pt]
    $e$/$\mu$ momentum & $> \SI{200}{\MeV/\clight}$ 
    & $>\SI{100}{\MeV/\clight}$ & $> \SI{100}{\MeV/\clight}$ \\[4pt]
    Number of decay-$e$ & $\leq$ 1 & 0 & 1 \\[4pt]
    NC$\pi$ rejection & 
    $\Lambda^{\pi^{+}}_{\mu} < \frac{6\,p_{\mu}}{\SI{40}{\MeV/\clight}}$ &
    $\Lambda^{\pi^{0}}_{e} < 175 - \frac{35\,m_{\gamma\gamma}} {\SI{40}{\MeV/\clight^2}}$ &
    $\Lambda^{\pi^{0}}_{e} < 175 - \frac{35\,m_{\gamma\gamma}} {\SI{40}{\MeV/\clight^2}}$ \\[4pt]
    \multicolumn{1}{l}{\erecqe} & --- & $<\SI{1250}{\MeV}$ & $<\SI{1250}{\MeV}$ \\[4pt]
  \end{tabular}
  \end{ruledtabular}
}
\end{table*}

\begin{table*}[htbp]
\centering
\caption{Expected number of 1R$_{e}$ signal and background events passing each selection stage, compared to the data.}
\label{table:expected_1-9_nue}
\resizebox{\textwidth}{!}{
\begin{ruledtabular}
\begin{tabular}{lccccccc}
& $\numu+\numub$ & $\nue+\nueb$ & $\nu+\nubar$ & $\numu\rightarrow\nue$ & $\numub\rightarrow\nueb$ & & \\
FHC & CC & CC & NC & CC & CC & MC total & Data \\
\hline
 FC and FV             & 692.28 & 43.10 & 241.98 & 87.18 & 0.80 & 1065.34 & 1077 \\
 Single particle        & 307.47 & 22.18 &  44.31 & 73.09 & 0.61 &  447.65 &  451  \\
 Electron-like         &   8.72 & 22.16 &  26.38 & 72.99 & 0.61 &  130.85 &  151   \\
 p$_{e}$ $>$ 100 MeV/c &   3.22 & 22.00 &  18.45 & 71.56 & 0.61 &  115.84 &  131   \\
 No decay-e            &   0.88 & 18.73 &  15.57 & 64.60 & 0.59 &  100.37 &  108   \\
 Erec $<$ 1250 MeV     &   0.56 &  9.89 &  11.64 & 62.40 & 0.43 &   84.91 &  86   \\
 Not $\pi^0$           &   0.27 &  8.79 &   4.21 & 58.53 & 0.38 &   72.17 &  75   \\
\hline
\hline
RHC & & & & & & & \\
\hline
 FC and FV             & 311.20 & 21.48 & 122.86 &  5.81 & 10.31 & 471.67 & 497   \\
 Singe particle        & 144.49 & 10.88 &  22.61 &  4.13 &  8.81 & 190.93 & 215   \\
 Electron-like         &   2.81 & 10.88 &  13.83 &  4.13 &  8.80 &  40.44 &  42   \\
 p$_{e}$ $>$ 100 MeV/c &   1.41 & 10.83 &   9.92 &  4.06 &  8.75 &  34.97 &  32   \\
 No decay-e            &   0.41 &  9.48 &   8.60 &  3.47 &  8.58 &  30.53 &  28   \\
 Erec $<$ 1250 MeV     &   0.28 &  4.27 &   6.77 &  2.91 &  8.13 &  22.37 &  19   \\
 Not $\pi^0$           &   0.13 &  3.70 &   2.40 &  2.65 &  7.37 &  16.26 &  15   \\
\end{tabular}
\end{ruledtabular}
}
\end{table*}

\begin{table*}[htbp]
\centering
\caption{Expected number of 1R$_{e}$ $+$ 1 d.e. signal and background events passing each selection stage, compared to the data.} 
\label{tab:tbl_run1_9_nue1pi}
\resizebox{\textwidth}{!}{
\begin{ruledtabular}
\begin{tabular}{lccccccc}
& $\numu+\numub$ & $\nue+\nueb$ & $\nu+ \nubar$ & $\numu\rightarrow\nue$ & $\numub\rightarrow\nueb$ & & \\
FHC & CC & CC & NC & CC & CC & MC total & Data \\
\hline
 FC and FV             & 697.81 & 43.87 & 247.50 & 87.30 & 0.81 & 1077.27 & 1085   \\
 Single particle       & 303.60 & 22.24 &  44.39 & 72.83 & 0.61 &  443.67 &  443   \\
 Electron like         &   8.46 & 22.22 &  26.98 & 72.74 & 0.61 &  131.01 &  148   \\
 p$_{e}$ $>$ 100 MeV/c  &   2.81 & 22.05 &  18.73& 71.24 & 0.61 &  115.44 &  129   \\
 One decay-e           &   1.35 &  3.04 &   2.18 &  6.81 & 0.02 &   13.40 &   23   \\
 E$_\textrm{rec}$ $<$ 1250 MeV     &   0.46 &  1.10 &   1.01 &  6.27 & 0.01 &    8.85 &   19   \\
 Not $\pi^0$           &   0.16 &  0.93 &   0.38 &   5.64 & 0.01 &    7.13 &   15   \\ 
\end{tabular}
\end{ruledtabular}
}
\end{table*}

\begin{table*}[htp]
\centering
\caption{Expected number of 1R$_{\mu}$ signal and background events passing each selection stage, compared to the data.}
\label{table:expected_1-9_numu}
\resizebox{\textwidth}{!}{
\begin{ruledtabular}
\begin{tabular}{lccccccc}
& $\nue+\nueb$ & $\nu+\nubar$ & $\numu+\numub$ & $\numu$ & $\numub$ & & \\
FHC & CC & NC & CC non-QE & CCQE & CCQE & MC total & Data \\
\hline
 FC and FV             & 125.04 & 234.01 & 373.95 & 251.21 & 14.20 &  998.41 & 1002   \\ 
 Single particle       &  92.94 &  43.30 &  63.00 & 220.14 & 12.53 &  431.91 & 429   \\ 
 Muon like             &   0.10 &  17.78 &  58.93 & 215.84 & 12.44 &  305.08 & 285   \\ 
 p$_{\mu}$ $>$ 200 MeV/c &   0.10 &  17.66 &  58.89 & 215.63 & 12.44 &  304.71 & 284   \\ 
 0 or 1 decay-e        &   0.10 &  17.07 &  37.99 & 213.41 & 12.31 &  280.88 & 255   \\ 
 Not $\pi^+$           &   0.08 &   8.31 &  36.75 & 210.64 & 12.18 &  267.96 & 243   \\ 
\hline
\hline
RHC & & & & & & & \\
\hline
 FC and FV           & 34.68 & 118.94 & 169.35 & 44.95 & 72.95 & 440.87 & 454 \\
 Single particle            & 22.74 &  22.12 &  34.43 & 36.55 & 67.44 & 183.29 & 197 \\
 Muon like         &  0.02 &   8.72 &  33.08 & 36.19 & 66.45 & 144.45 & 159 \\
 p$_{\mu}$ $>$ 200 MeV/c              &  0.02 &   8.63 &  33.07 & 36.18 & 66.40 & 144.30 & 159 \\
 0 or 1 decay-e        &  0.02 &   8.37 &  25.22 & 35.76 & 65.71 & 135.07 & 144 \\
 Not $\pi^+$  &  0.02 &   3.89 &  24.71 & 35.36 & 65.00 & 128.97 & 140 \\
\end{tabular}
\end{ruledtabular}
}
\end{table*}

\subsection{Optimization of selection criteria}
The likelihood ratio of best-fit $e$ and $\mu$ hypotheses gives very good separation between these classes of events, with the separation improving at higher momentum. The cut line chosen to select $e$-like and $\mu$-like events accounts for the momentum dependence of the likelihood ratio and achieves mis-identification rates smaller than 1\% for true CCQE events across the T2K energy range.

Pion production in neutral current events forms one of the main backgrounds to both $\mu$-like and $e$-like selections. Furthermore, since the cross sections for these processes are not known precisely, these contributions carry significant systematic uncertainties into the signal samples. A simplified neutrino oscillation analysis framework is used to optimize the neutral current rejection criteria taking into account systematic uncertainties and with statistics corresponding to an exposure of $7.8\times10^{21}$~POT, evenly split between neutrino and antineutrino modes. In this simplified analysis framework, the systematic uncertainties (taking into account the near detector constraints) are propagated to the SK prediction as a covariance matrix in reconstructed neutrino energy. As a result, the SK samples do not constrain the nuisance parameters, and no correlations between these and the neutrino mixing parameters are taken into account. The four samples targeting $e$-like and $\mu$-like CCQE events in both FHC and RHC neutrino beam mode are fit simultaneously to an Asimov data set~\cite{asimov} to determine the experiment's sensitivity under different neutral current rejection cut points.

The criterion to reject NC$\pi^+$ events in the $\mu$-like samples, a line cut on the $\Lambda^{\pi^{+}}_{\mu}$ vs $p_{\mu}$ plane, is chosen to minimize the width of the $\sin^2\theta_{23}$ $1\sigma$ confidence interval. This cut, which was not available with the reconstruction techniques used in previous T2K neutrino oscillation analyses, reduces the NC contribution to the $\mu$-like samples by a factor of two, while selecting CCQE events with 99\% efficiency.

The NC$\pi^0$ rejection line cut in the $\Lambda^{\pi^{0}}_{e}$ vs $m_{\gamma\gamma}$ plane, applied to the $e$-like samples, is optimized based on the significance to exclude $\deltacp=0$. As the optimal cut line is very close to the one used in previous T2K neutrino oscillation analyses, this criterion is not updated for the analysis described here. It should be noted that the relative impact of this cut on the selected sample is significantly smaller in the analysis described here, where it reduces the NC contribution in the $e$-like samples by a factor of three, compared to previous analyses, where the reduction is of a factor of 6. This is due to the excellent performance of the multi-particle search algorithm, which reduces the NC background in $e$-like samples by a factor of five by correctly identifying the two $\gamma$s from $\pi^{0}$ decays with higher efficiency than the previously used algorithms.

\begin{figure}[htbp]
  \centering
  \begin{subfigure}[b]{0.47\textwidth}
    \includegraphics[width=0.98\columnwidth]{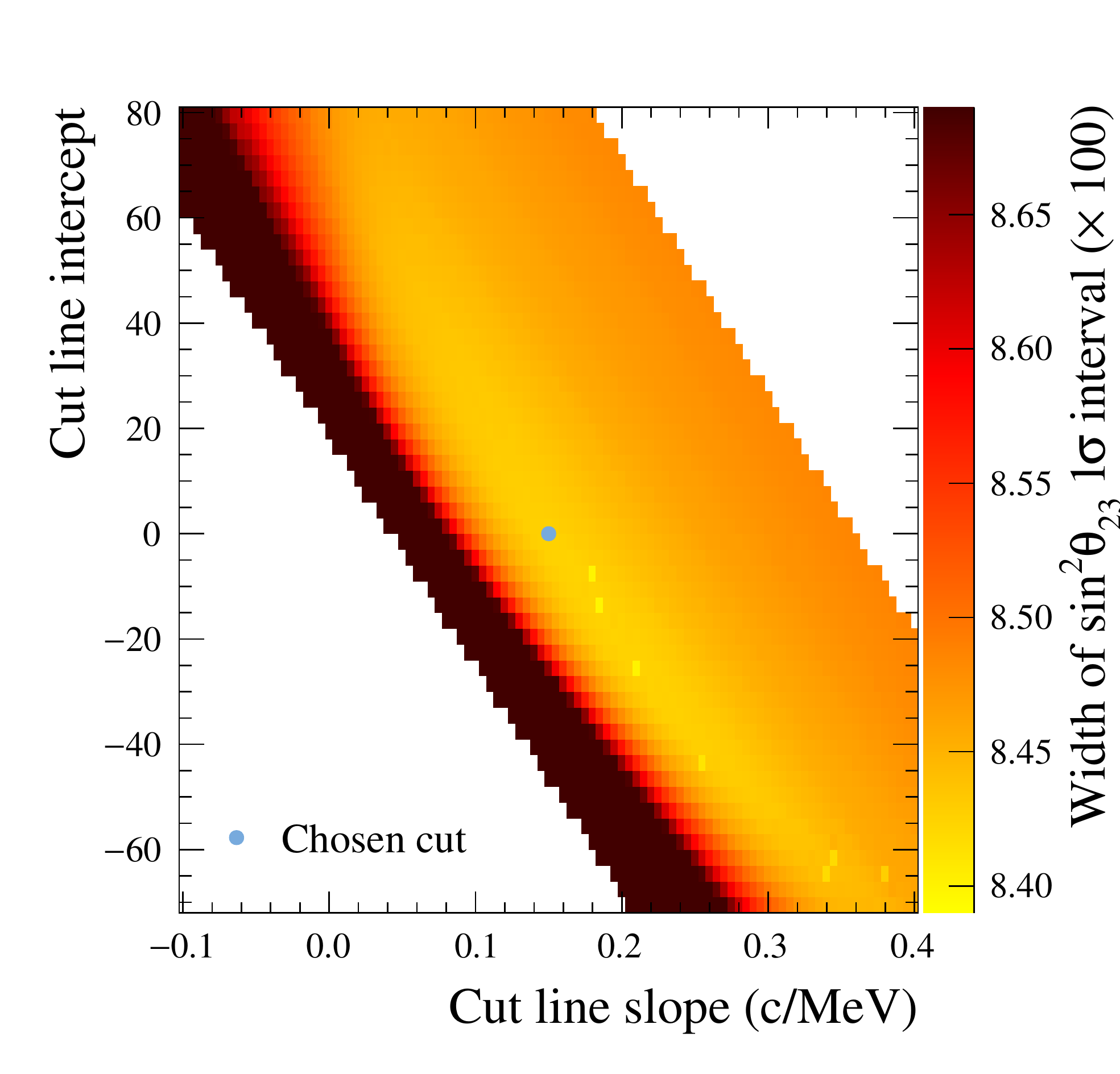}
  \end{subfigure}
  \begin{subfigure}[b]{0.47\textwidth}
    \includegraphics[width=0.98\columnwidth]{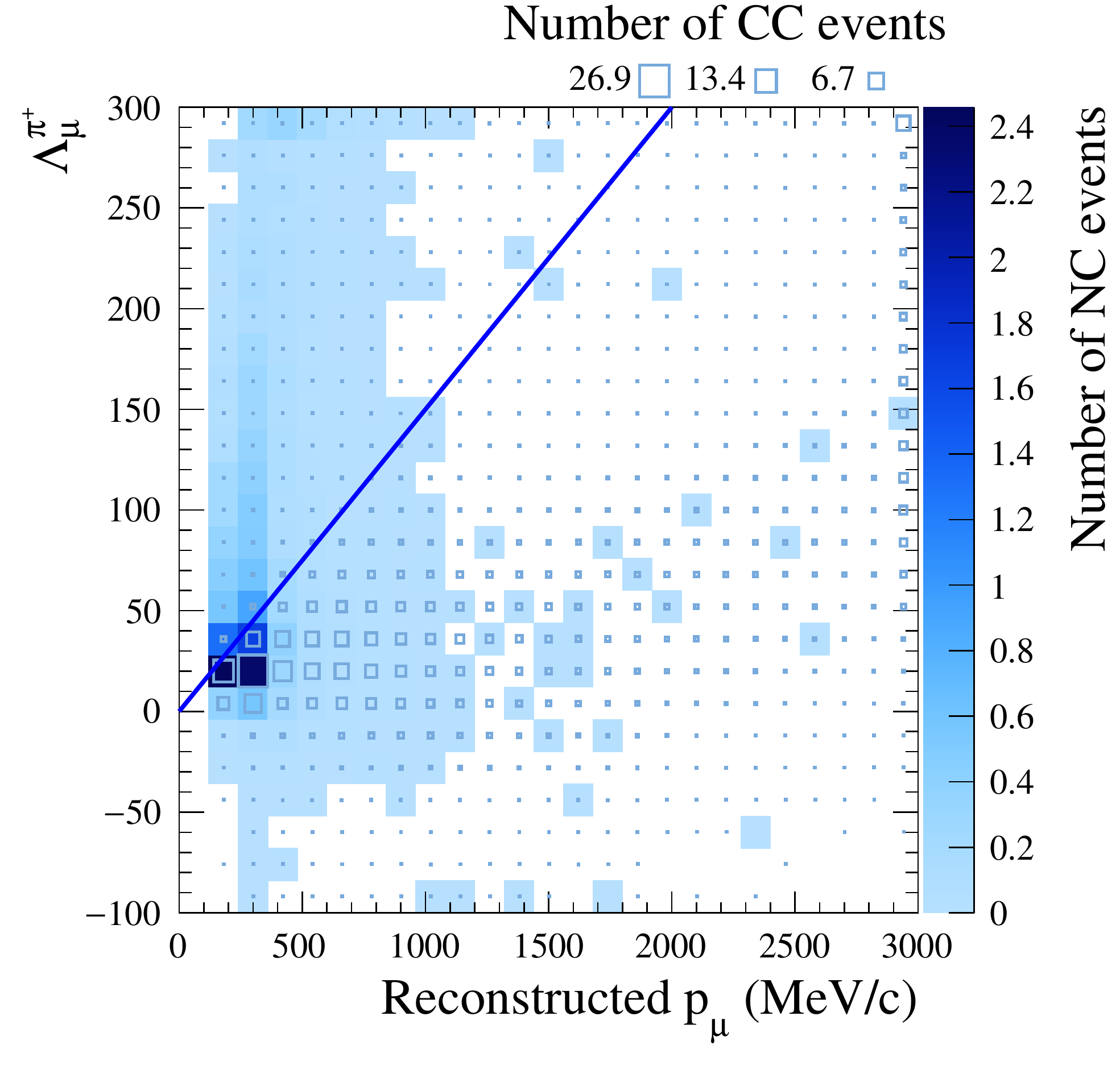}
  \end{subfigure}
  \caption {The width of $\sin^2\theta_{23}$ $1\sigma$ confidence interval as a function of the slope and intercept of the NC$\pi^{+}$ rejection cut line is shown on the top and the distribution of signal and background events is shown on the bottom, along with the line below which $\mu$-like events are selected.}
  \label{fig:SK_NCpip_opt}
\end{figure}

\begin{figure}[htbp]
  \centering
  \begin{subfigure}[b]{0.47\textwidth}
    \includegraphics[width=0.98\columnwidth]{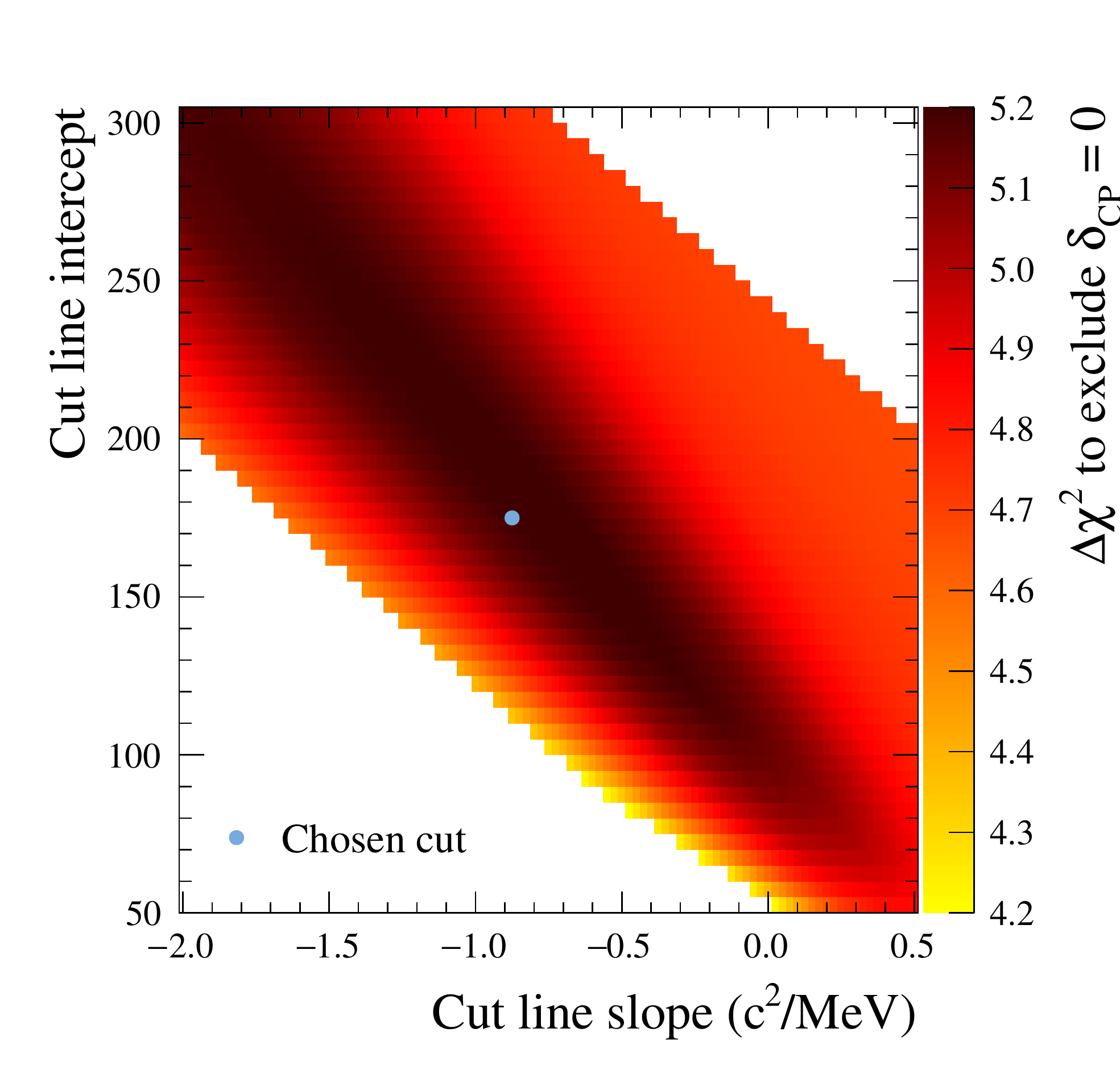}
  \end{subfigure}
  \begin{subfigure}[b]{0.47\textwidth}
    \includegraphics[width=0.98\columnwidth]{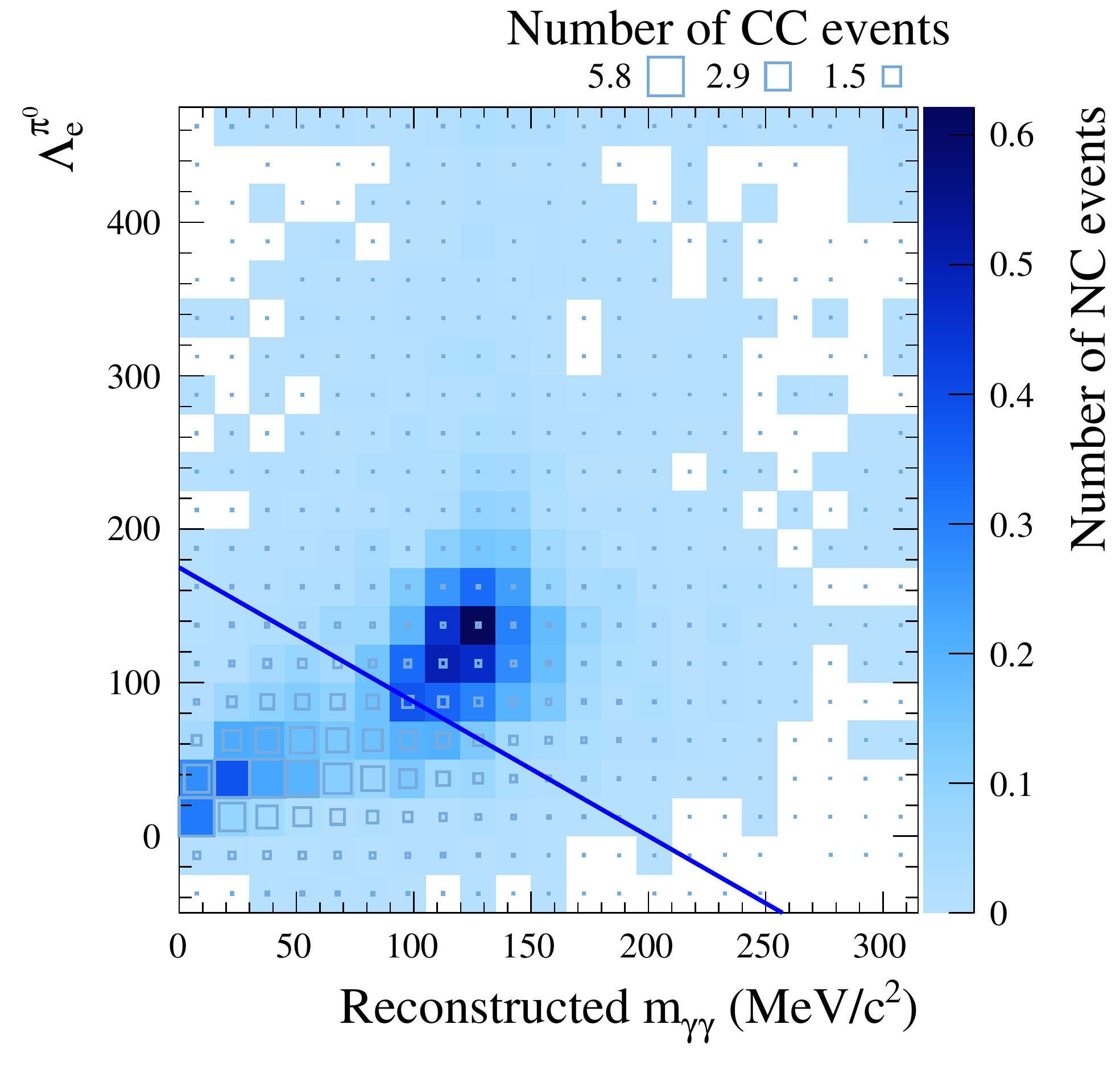}
  \end{subfigure}
  \caption{The sensitivity to exclude $\deltacp=0$ as a function of the slope and intercept of the NC$\pi^0$ rejection cut line is shown on the top and the distribution of signal and background events is shown on the bottom, along with the line below which $e$-like events are selected.}
  \label{fig:SK_NCpi0_opt}
\end{figure}

Optimization metrics for both neutral current rejection criteria are shown as a function of the cut parameters in Figs.~\ref{fig:SK_NCpip_opt}~and~\ref{fig:SK_NCpi0_opt}, along with the chosen cut points and distributions of the signal and background events in the cut variable planes. While this optimization is performed assuming the neutrino mixing parameters preferred by previous T2K results as given in Tab.~\ref{tab:asimova_params}, it should be noted that in both cases the optimization metrics show large regions around the optimal points where they are consistently good. Therefore, the sensitivity of this analysis does not depend strongly on the exact value of the cut points and the experimental sensitivity is not expected to depend strongly on the choice of neutrino mixing parameters used in the optimization procedure.

\subsection{Fiducial volume expansion}
\label{sec:sk_fv_opt}
In previous T2K neutrino oscillation analyses, events at SK were required to have $wall \ge 200$~cm in order to remove entering backgrounds and ensure the quality of reconstructed quantities. The new event selection presented here, benefiting from improvements in reconstruction, provides an opportunity to re-optimize the FV criterion for the T2K analysis samples, with the objective of increasing the event yield while maintaining a high purity of signal events in the selected samples and a low impact of detector systematic uncertainty. 

A two-dimensional parameterization of the FV criteria is chosen to allow for balancing two classes of effects. On one hand, the reduction of entering backgrounds and mitigation of the impact of known shortcomings of the simulated detector geometry are achieved with a $wall$ threshold, as in previous analyses of T2K data. On the other hand, important aspects of event reconstruction, such as particle identification, improve with the number of PMTs illuminated by the Cherenkov ring patterns. As this number grows with the distance to the detector walls along the particle direction of travel, $towall$, a threshold on this distance is used to select events with a high reconstruction performance.

The FV criteria are optimized in a fit to SK atmospheric neutrino data, from which the systematic uncertainty associated to the particle counting and identification is also extracted. 

\begin{figure}[htbp]
  \centering
  \includegraphics[width=0.46\textwidth]{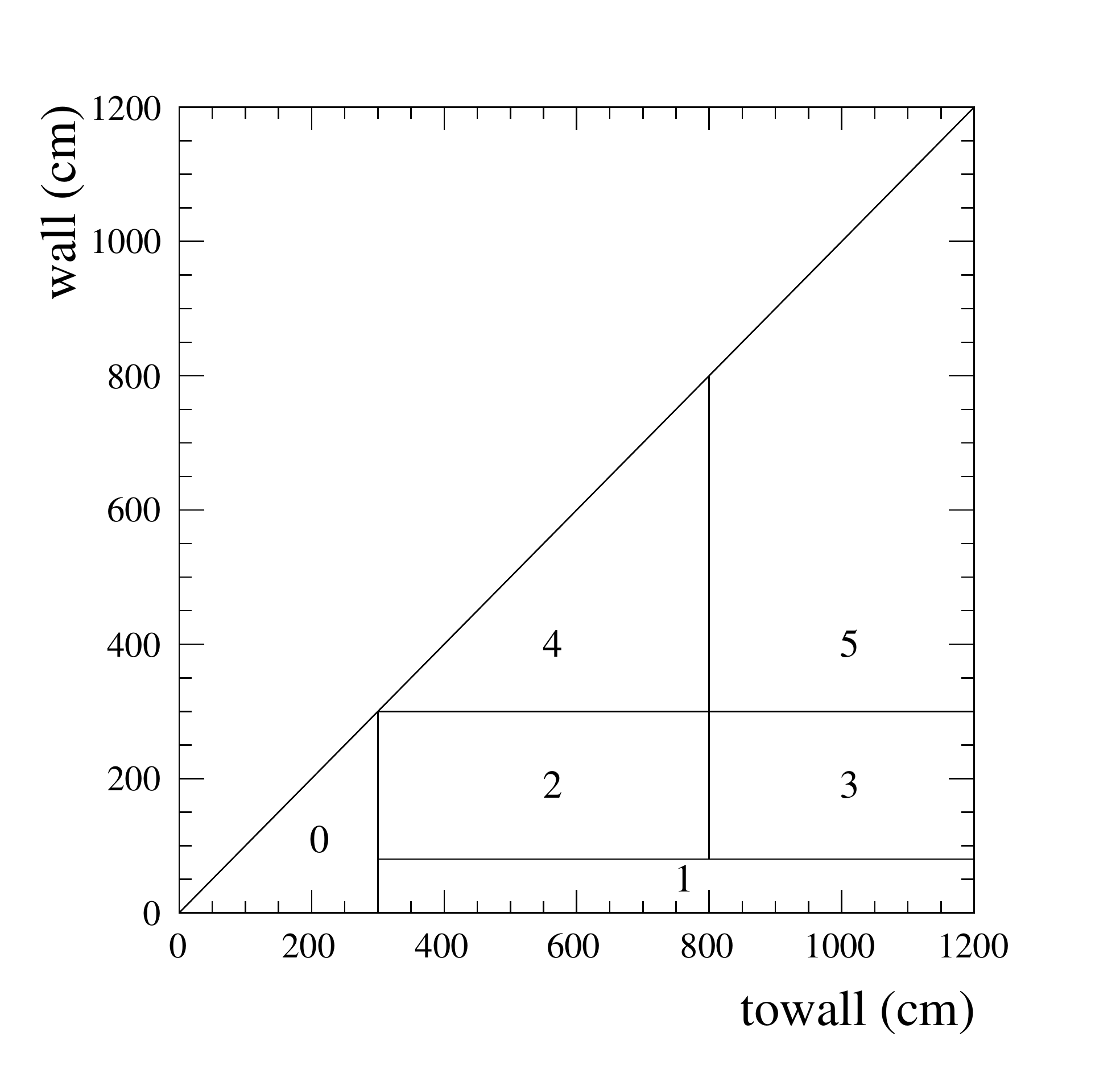}
  \caption{Detector regions used in detector systematic uncertainty estimation.}
  \label{fig:SK_FV_detregions}
\end{figure}

The SK atmospheric neutrino FC data is divided into 18 samples consisting of combinations of six detector regions, defined with $wall$ and $towall$ as shown in Fig.~\ref{fig:SK_FV_detregions}, and three classes of events discriminated by the number (0, 1 or 2+) of detected Michel electrons. The six detector regions were chosen to isolate areas where the systematic uncertainty associated to the detector model is expected to differ, while maintaining an adequate level of statistics in the SK atmospheric neutrino data sample.

These samples are projected into three particle identification variables ($\Lambda^{e}_{\mu}$, $\Lambda^{\pi^{0}}_{e}$ and $\Lambda^{\pi^{+}}_{\mu}$) and a continuous variable that discriminates single-particle from multi-particle events ($\Lambda^{2-particles}_{1-particle}$). 

In each of the samples the MC is split into six true event topologies consisting of: a single visible $e$, a single visible $\mu$, a visible $e$ with other visible particles, a visible $\mu$ with other visible particles, a single $\pi^{0}$, and finally events with a single visible $p$ or $\pi^{+}$. For each topology, the particle identification and counting variables are linearly transformed with two nuisance parameters each:
\begin{equation}
  \Lambda'^{\alpha}_{\beta} = a \Lambda^{\alpha}_{\beta} + b
\end{equation}

\noindent where $\Lambda'^{\alpha}_{\beta}$ is the transformed variable and $a$ and $b$ are ``scale'' and ``shift'' nuisance parameters, respectively. These ``scale'' and ``shift'' parameters are estimated with a Markov Chain Monte Carlo (MCMC) method~\cite{biometMCMC} that samples the Poisson likelihood for the observed data given the model that includes, in addition, parameters to capture the uncertainty on the atmospheric neutrino flux and cross sections. The SK atmospheric neutrino data used in this analysis were collected between October 2010 and May 2015. To ensure the validity of the atmospheric neutrino fit for the T2K oscillation analysis, high-statistics control sample data collected over the entire beam data period are used to monitor the detector stability.

The flux and cross-section parameterizations used in this procedure are simpler than those used in oscillation analyses and are based on Refs.~\cite{Abe:2011sj} and~\cite{Wendell:2010md}. The atmospheric neutrino flux is scaled by two parameters, one affecting events with neutrino energy lower than 1 GeV, with a prior uncertainty of 25\% and the other for events with energies higher than 1 GeV, with a prior uncertainty of 15\%. The ratio of $\nue + \nueb$ to $\numu + \numub$ events is controlled by a parameter with 5\% prior uncertainty. A prior uncertainty of 20\% is assigned to both CC non-QE and NC cross sections. The CCQE cross section has a more detailed parameterization, with events below 190 MeV having a prior uncertainty of 100\% and events in the 190 MeV to 1 GeV range being characterized by 11 parameters which scale the cross section in unevenly spaced energy bins with gradually decreasing prior uncertainties from 41\% to 2.2\%, with 0.6 GeV events being assigned 5.4\% prior uncertainty. CCQE events with energy in the 1 to 2 GeV and 2 to 3 GeV ranges are assigned 1.7\% and 0.9\% prior uncertainty, respectively. Finally, each sample used in the fit is assigned an unconstrained overall scale parameter.

The choice of fitted variables and conservative flux and cross-section prior uncertainties reduces the sensitivity of this fit to the neutrino oscillation parameters, which are fixed at the reference values given in Tab.~\ref{tab:asimova_params}.

\begin{figure*}[htbp]
  \centering
  \begin{subfigure}[b]{0.32\textwidth}
    \includegraphics[width=0.98\columnwidth]{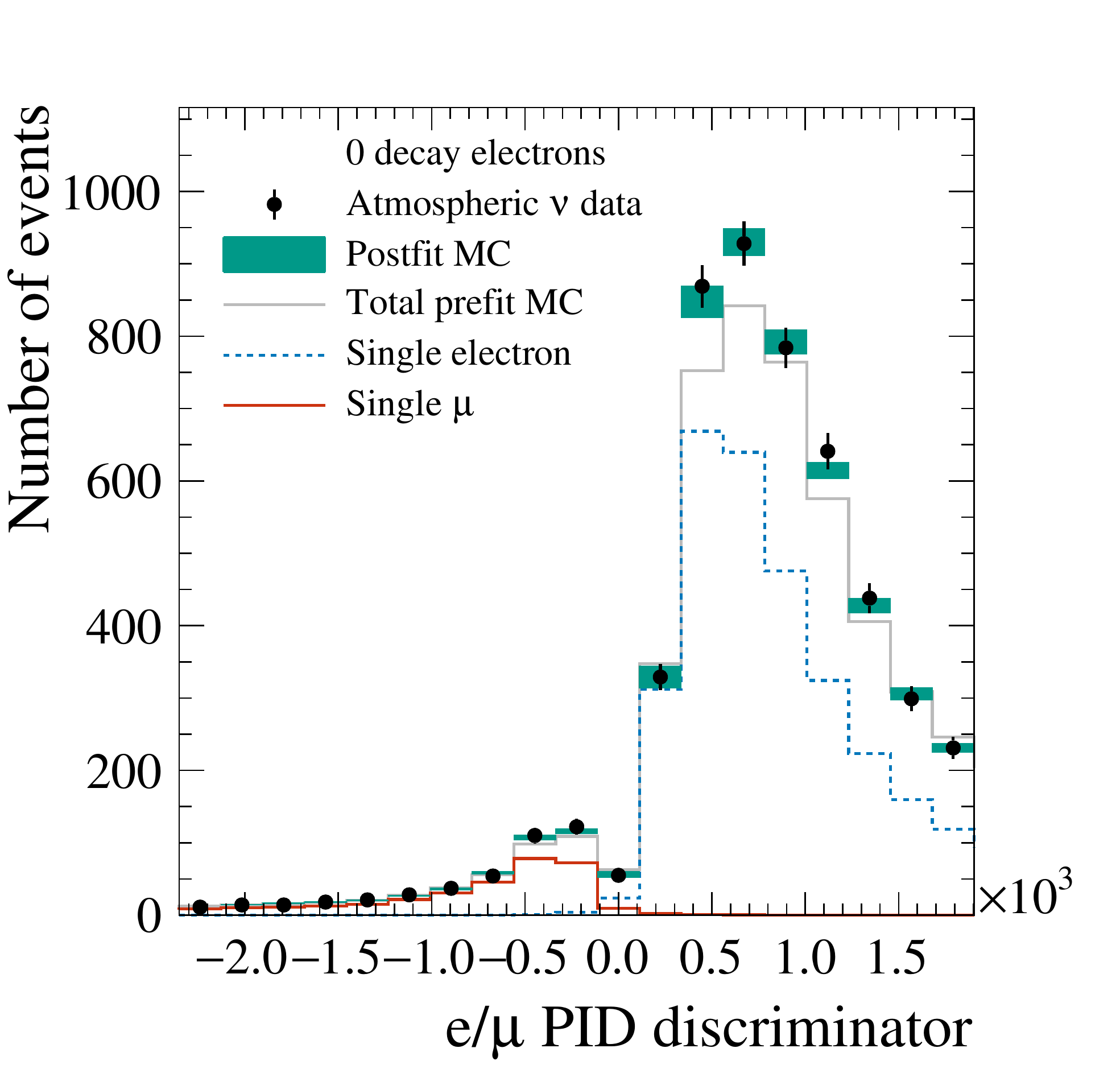}
    \caption{}
  \end{subfigure}
  \begin{subfigure}[b]{0.32\textwidth}
    \includegraphics[width=0.98\columnwidth]{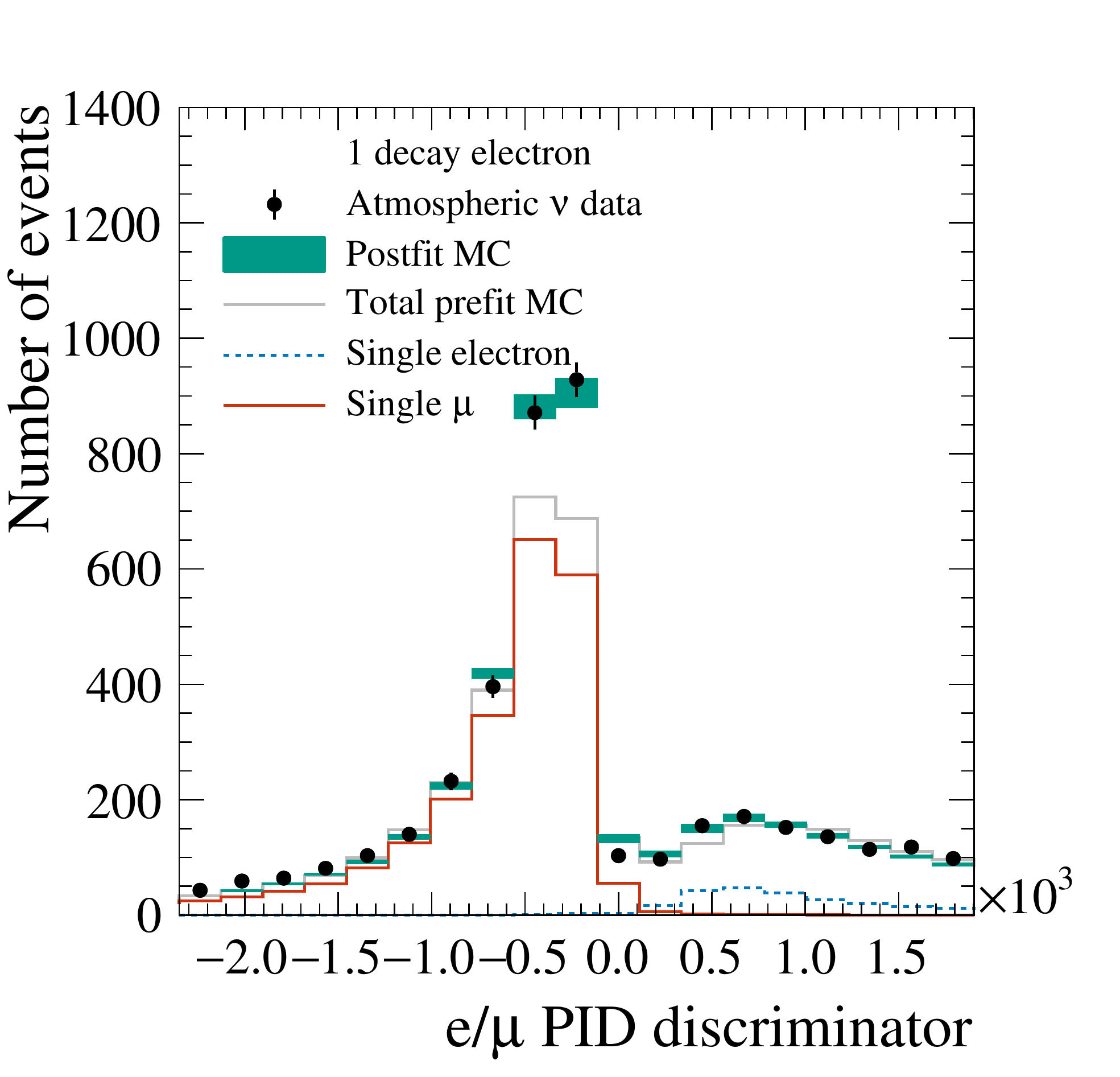}
    \caption{}
  \end{subfigure}
  \begin{subfigure}[b]{0.32\textwidth}
    \includegraphics[width=0.98\columnwidth]{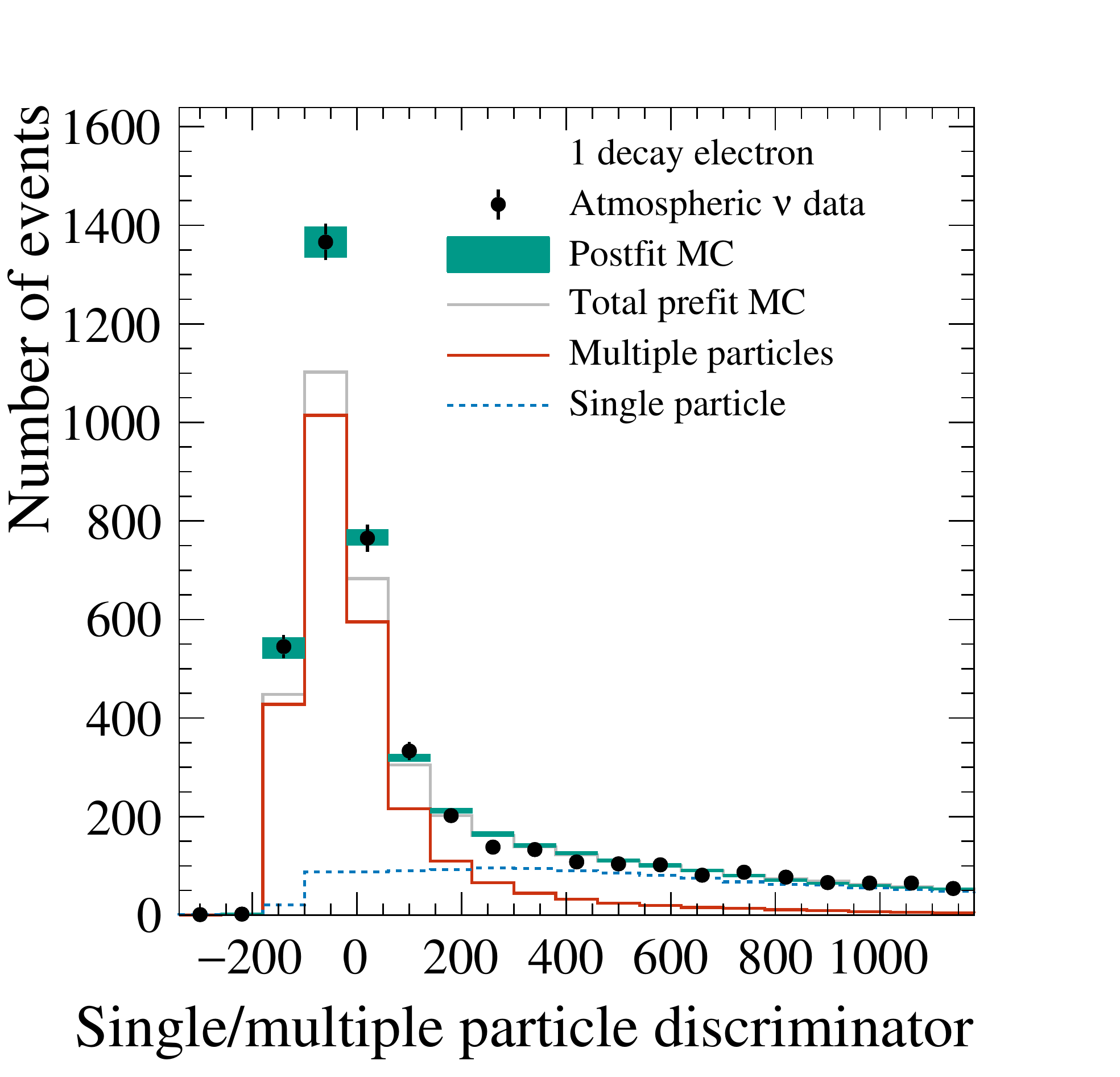}
    \caption{}
  \end{subfigure}
  \begin{subfigure}[b]{0.32\textwidth}
    \includegraphics[width=0.98\columnwidth]{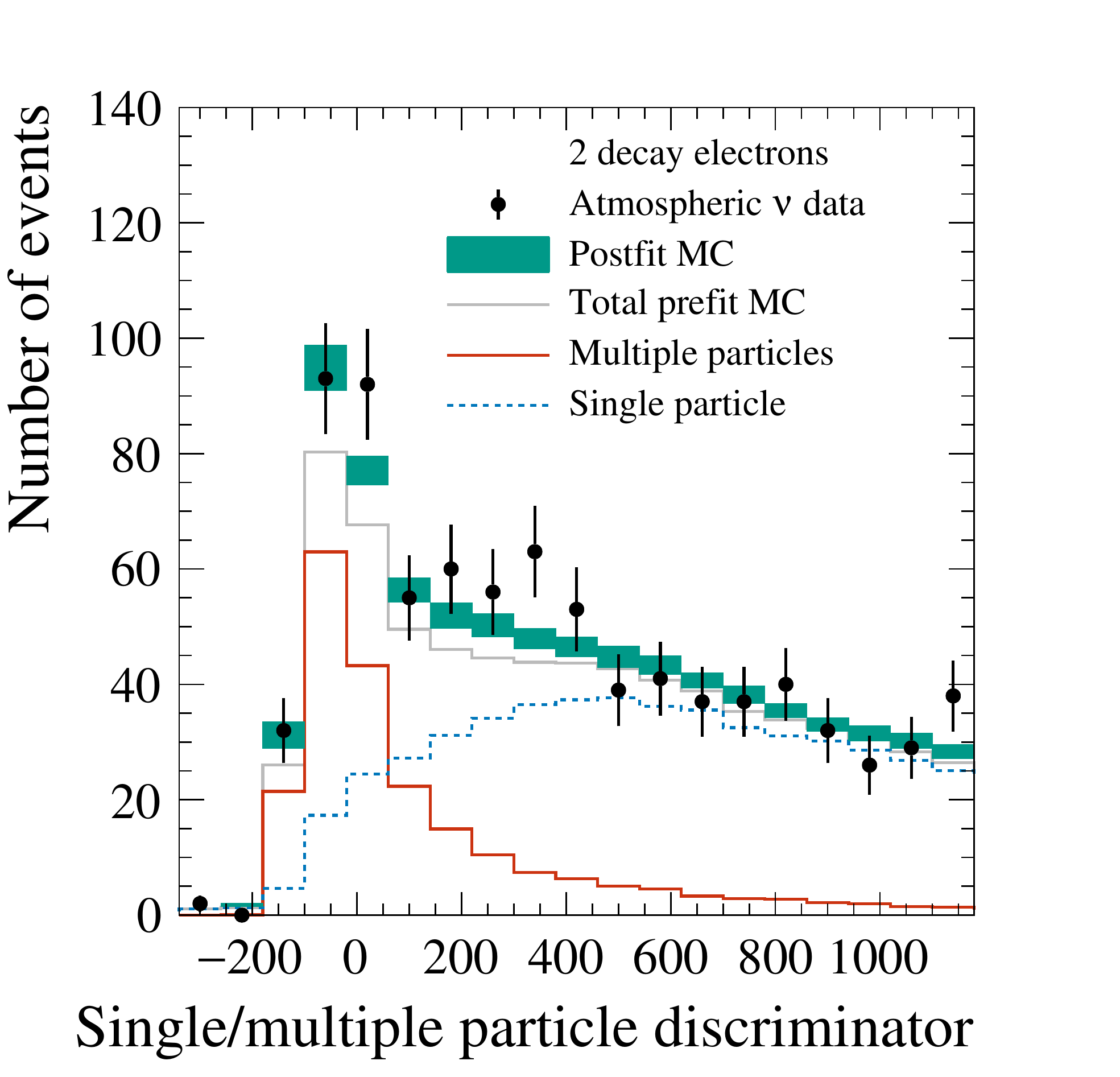}
    \caption{}
  \end{subfigure}
  \begin{subfigure}[b]{0.32\textwidth}
    \includegraphics[width=0.98\columnwidth]{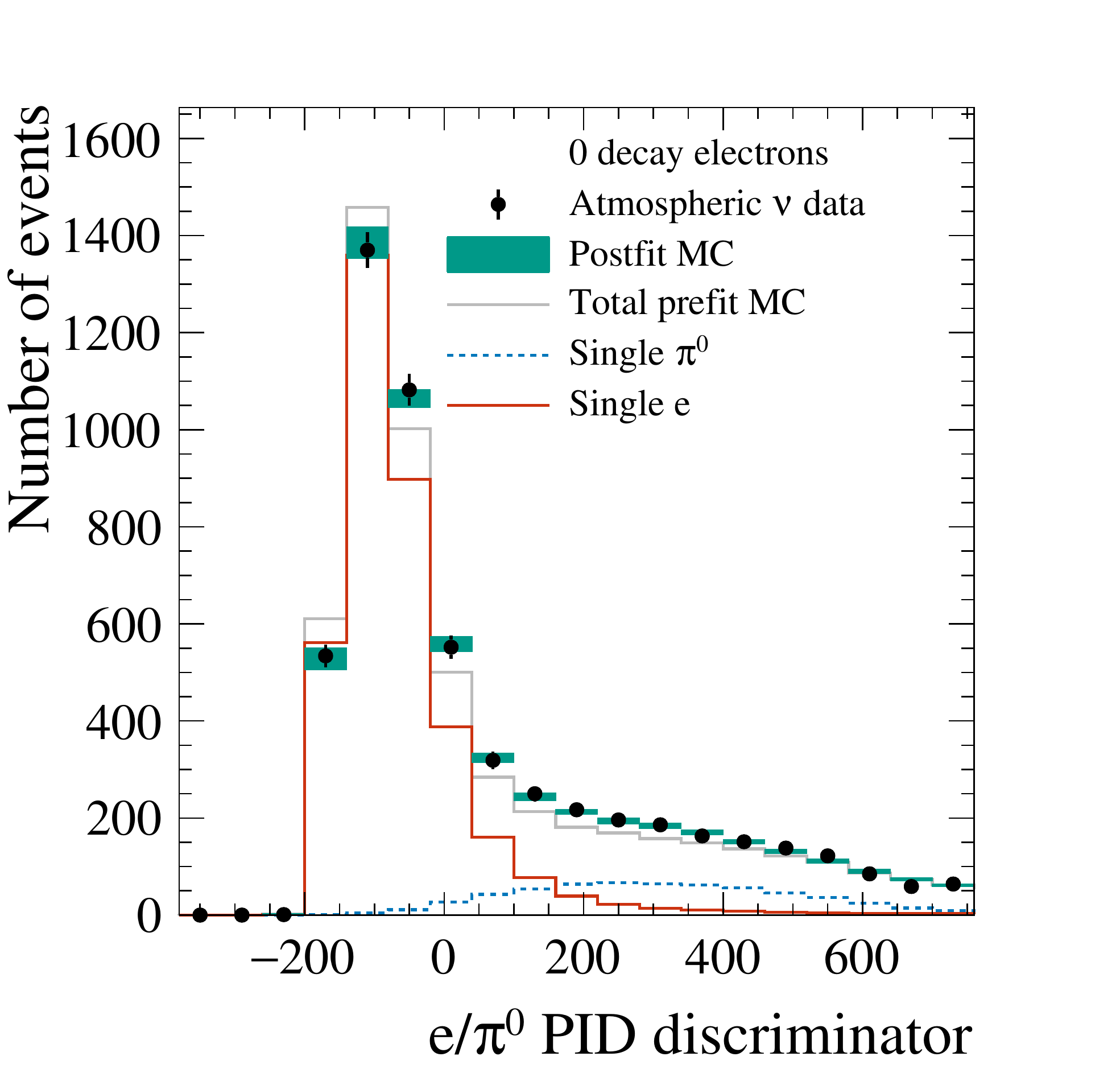}
    \caption{}
  \end{subfigure}
  \begin{subfigure}[b]{0.32\textwidth}
    \includegraphics[width=0.98\columnwidth]{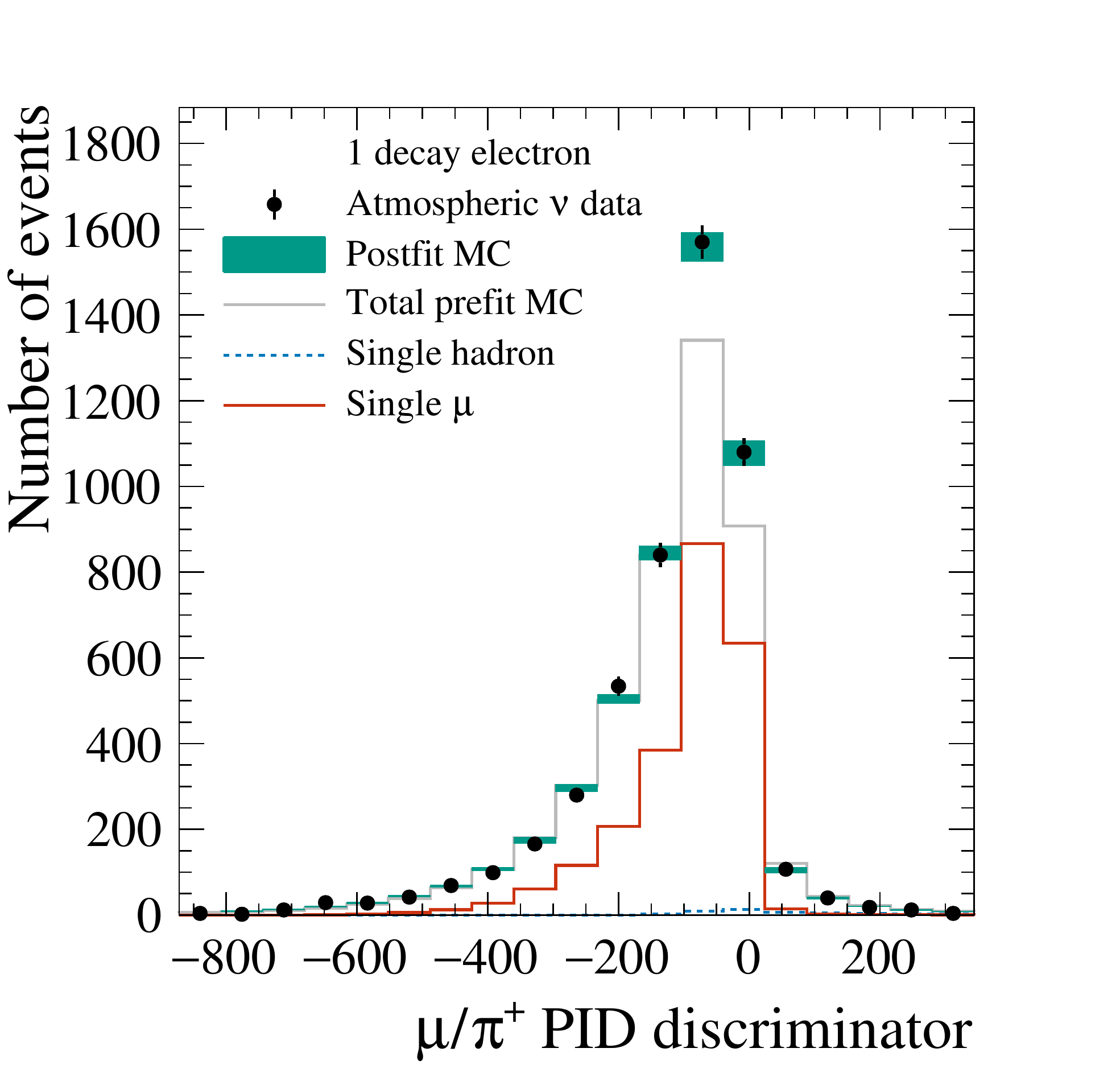}
    \caption{}
  \end{subfigure}
  \caption{Posterior predictive distributions of SK atmospheric MCMC fit to determine detector systematic uncertainties. The total nominal MC is shown in gray, with components targeted by each of the distributions shown in red and blue. The 68.27\% intervals of the posterior predictive distribution is shown in green. Observed data are shown as black circles. All distributions shown are for the largest detector region, 5. Distributions of the $e$/$\mu$ PID discriminator for 0 and 1 decay-$e$ events are shown in (a) and (b). In (c) and (d), the particle-counting parameter is shown for events with 1 and 2 decay-electrons, respectively. The $e$/$\pi^0$ discriminator distribution is shown in (e) for $e$-like events and and the $\mu$/$\pi^+$ distribution is shown in (f) for $\mu$-like events.}
  \label{fig:SK_FV_posteriorpredictive}
\end{figure*}

Examples of the posterior predictive distributions resulting from the MCMC sampling are shown in Fig.~\ref{fig:SK_FV_posteriorpredictive}.

To quantify the impact of the disagreement between data and MC on the T2K samples, the ``shift'' and ``scale'' parameters are sampled from the MCMC posterior and applied to the T2K beam MC. The T2K selection criteria other than FV are applied, and a fractional uncertainty is calculated for each analysis sample in each of the detector regions, with the MC separated in five true categories: CCQE, CC non-QE, CC events with mis-identified lepton flavor, neutral current, and entering backgrounds. In this procedure, the atmospheric neutrino flux and cross-section parameters used in the MCMC fit are marginalized over. The resulting fractional uncertainty on the expected number of events is taken to be the systematic uncertainty associated to the SK detector model.

With the detector systematic uncertainty estimated for each detector region, the optimal FV criteria are found by maximizing the following figure of merit, which quantifies the sensitivity of the sample with respect to changes in $\theta_{23}$ and $\deltacp$ for $\mu$-like and $e$-like samples, respectively:
\begin{equation}
F.O.M = \frac{  \left(\frac{ \partial \hat{N}}{\partial\theta}\right) ^2 }{\hat{N} + \sigma_{syst}^{2}}
\label{eq:SK_FV_FOM}
\end{equation}
\noindent where $\hat{N}$ is the expected number of events in a given sample, $\theta$ is the parameter of interest ($\theta_{23}$ for the $\mu$-like samples and $\deltacp$ for the $e$-like samples) and $\sigma_{syst}$ is the systematic uncertainty, including uncertainties associated to the detector and cross-section models as well as the uncertainty on the number of entering backgrounds.

\begin{figure*}[htbp]
  \centering
  \begin{subfigure}[b]{0.47\textwidth}
    \includegraphics[width=0.98\columnwidth]{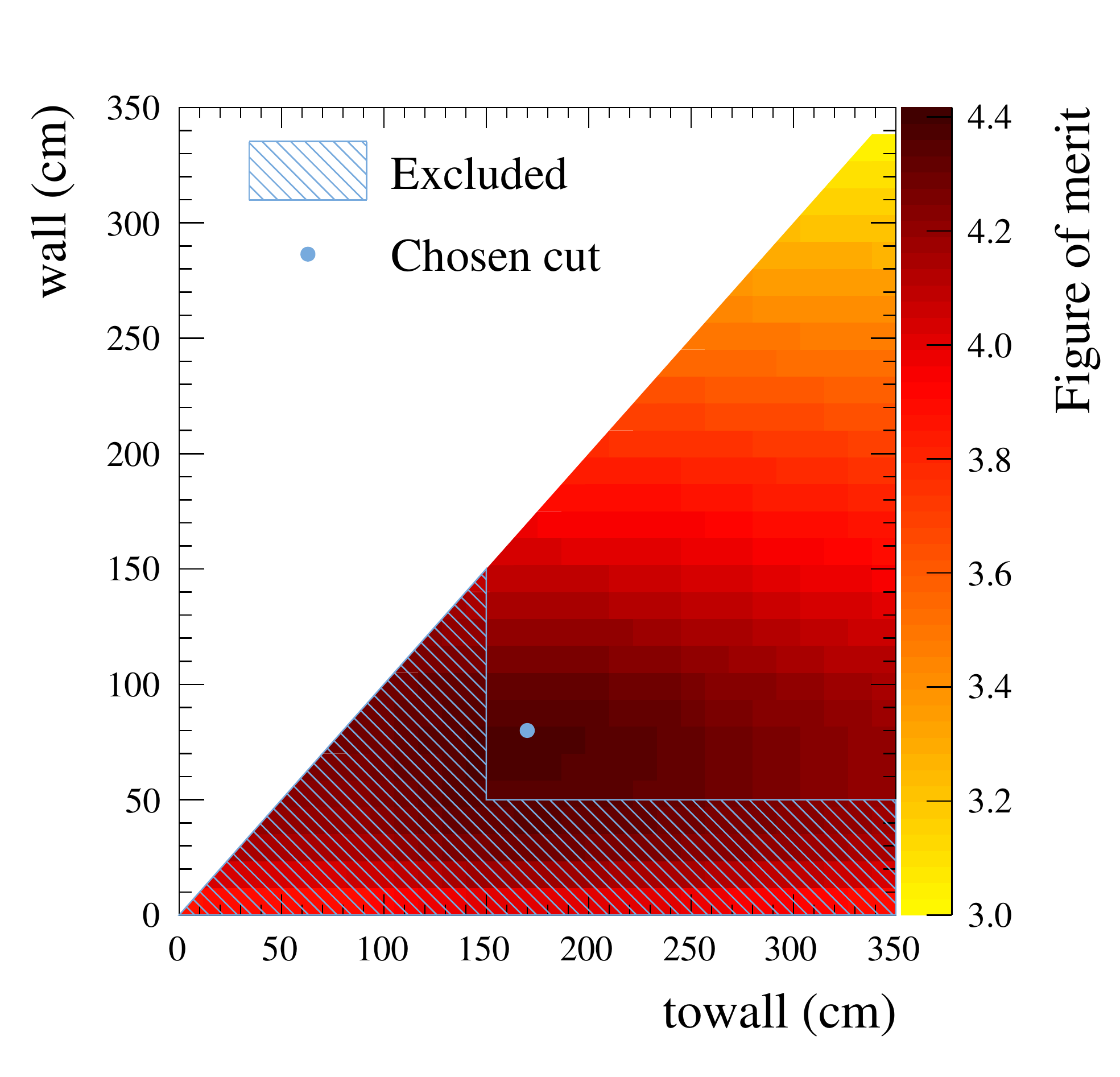}
  \end{subfigure}
  \begin{subfigure}[b]{0.47\textwidth}
    \includegraphics[width=0.98\columnwidth]{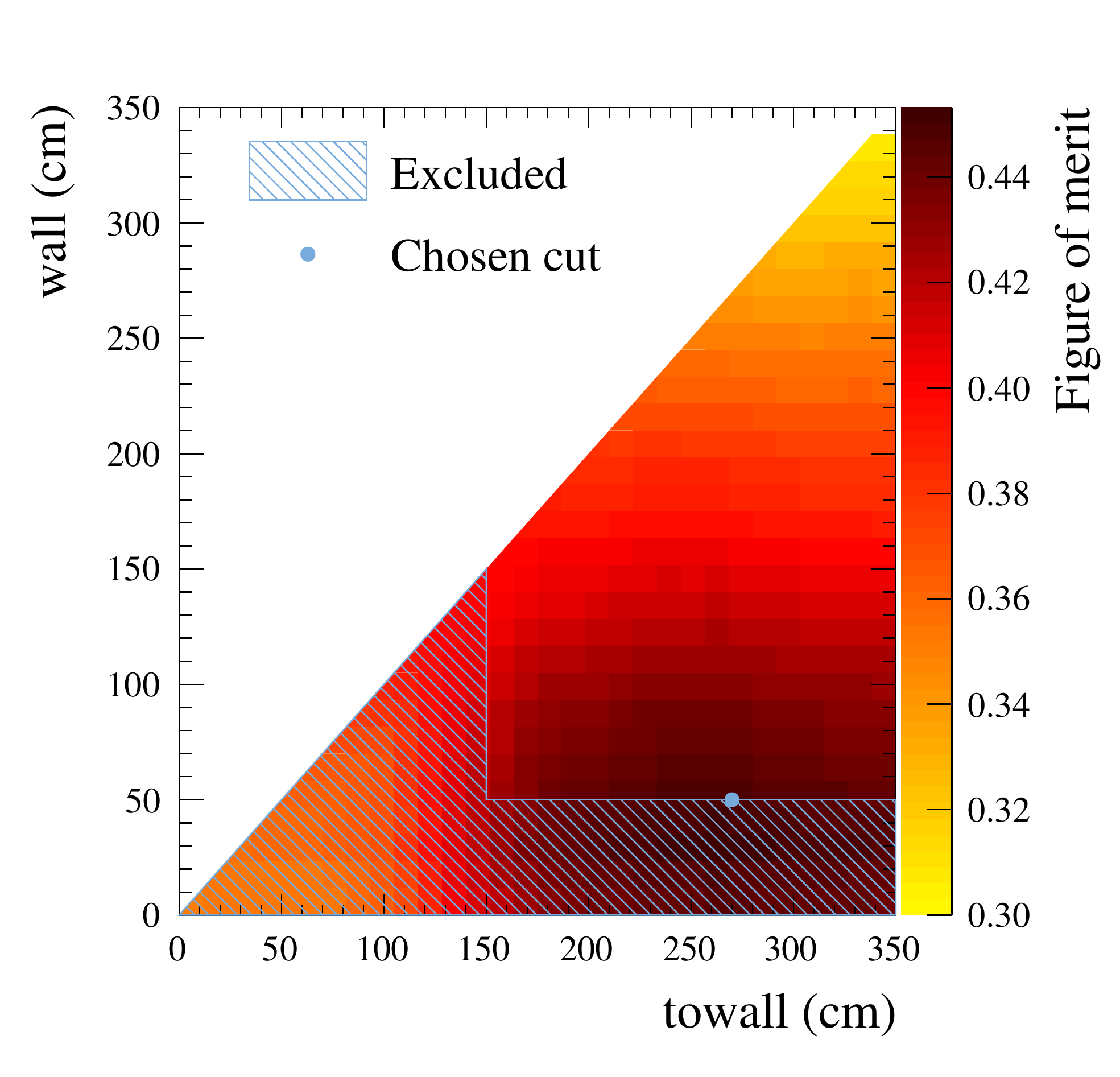}
  \end{subfigure}
  \begin{subfigure}[b]{0.47\textwidth}
    \includegraphics[width=0.98\columnwidth]{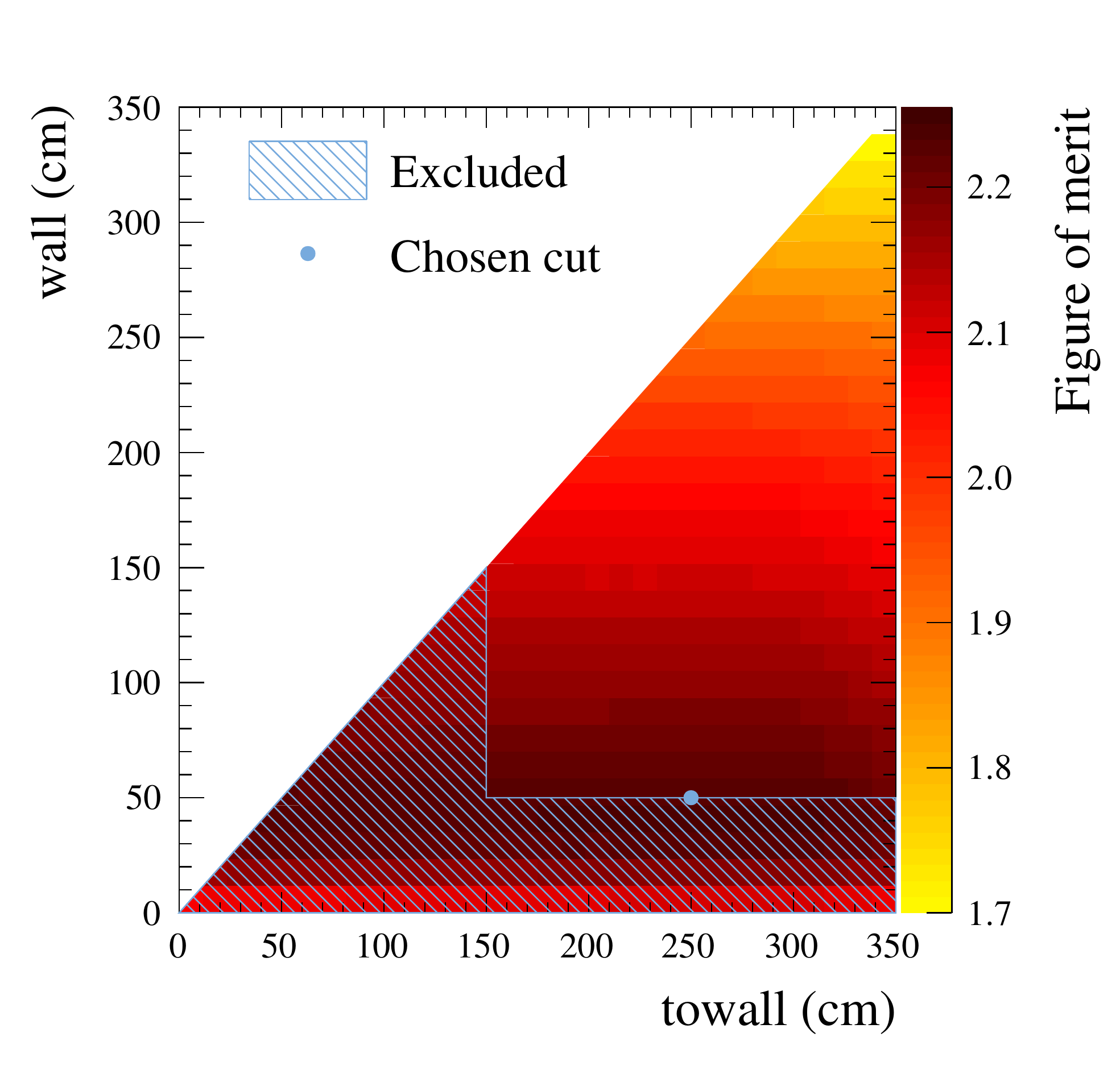}
  \end{subfigure}
    \caption{Scans of the figures of merit used to optimize the FV criteria in the $wall$ vs $towall$ space that defines the FV for the analysis described here. Figures of merit corresponding to the 1R$_{e}$ (top left), 1R$_{e}$ $+$ 1 d.e. (top right) and 1R$_{\mu}$ (bottom) samples are shown here. The chosen cut point is indicated in the figures as is the region which is ruled out by reconstruction performance considerations.}
  \label{fig:SK_FV_opt}
\end{figure*}

A minimum requirement of at least 50~cm in $wall$ and 150~cm in $towall$ is chosen based on MC studies which show deterioration in momentum reconstruction beyond those regions. Different FV criteria are chosen for the 1R$_{\mu}$, 1R$_{e}$ and 1R$_{e}$ $+$ 1 d.e. samples, with the same cuts applied in equivalent selections in neutrino and antineutrino modes. The figure of merit is shown as a function of $wall$ and $towall$ for the three samples in Fig.~\ref{fig:SK_FV_opt}. The optimal criteria resulting from this procedure are described in Tab.~\ref{tab:SK_selection}.

The combined effect of the new reconstruction algorithm, optimized neutral current rejection cuts and expanded FV volume has a significant impact across all analysis samples. The signal acceptance in the 1R$_{e}$ sample increases by 20\%, with the same purity as in previous analyses, while the signal events in the 1R$_{e}$ $+$ 1 d.e. sample increase by 30\%, with a reduction in the mis-identified muon background of 70\%. The 1R$_{\mu}$ samples have an increase in signal efficiency of 15\% and a reduction in backgrounds of 40\%.

\subsection{Systematic uncertainty}
The systematic uncertainty associated with the SK event selection is propagated to the oscillation analysis fitting frameworks as a covariance matrix in bins of either reconstructed neutrino energy or reconstructed lepton momentum, and broken down in true event topologies.

Systematic uncertainties on the particle count and identification are extracted from the atmospheric neutrino data with the MCMC method described above. 

The uncertainties on decay-$e$ tagging and mis-identification of muons as electrons are extracted from differences between the data and the MC in a control sample consisting of cosmic ray muons that stop within the ID. The cosmic ray muon events are weighted in momentum and $towall$ to match the expected distribution of beam-induced muons. The uncertainty on the decay-$e$ identification efficiency is 1\% and the uncertainty on the rate of spurious decay-$e$ tags is 0.2\%. The relative uncertainty on the mis-identification of muons as electrons is 30\%, though the contamination of $\numu$ CC events is smaller than 1\% in the $\nue$ samples without decay-electrons and around 2\% in the 1R$_{e}$ $+$ 1 d.e. sample.

Uncertainties introduced by the FV criteria are also estimated with MC to data comparisons in the cosmic ray muon sample, with both vertex and direction uncertainties taken into account. The reconstructed vertices in the cosmic ray muon sample cluster at the top or side walls of the detector, allowing for shifts in the MC relative to the data to be identified. The uncertainty on the direction is estimated by comparing the reconstructed muon direction to the equivalent quantity estimated using the muon and subsequent decay-$e$ vertices. The uncertainties are 2.5~cm for the vertex position and 0.24~degrees for the direction, corresponding to a 0.3\% to 0.4\% systematic uncertainty on the FV, depending on the analysis sample. This uncertainty is dominated by the uncertainty on the vertex position, with the direction playing a negligible role.

The uncertainty on the $\pi^0$ rejection efficiency in 1R$_{e}$ samples is estimated using hybrid $\pi^0$ sample constructed by superimposing an $e$-like event from the atmospheric neutrino or decay-$e$ from cosmic ray muon data with a simulated $\gamma$ with kinematics taken from NC$\pi^0$ events in the MC. The procedure is performed using both the MC and the real data as the source of the event and the difference in $\pi^0$ rejection efficiency between the data--MC and MC--MC samples is taken as the systematic uncertainty, binned in reconstructed lepton momentum and angle with respect to the beam. The overall uncertainty on the $\pi^0$ rejection efficiency is 26\%.

A summary of the uncertainties associated with the SK detector model is given in Tab.~\ref{table:SK_syst_summary}.

\begin{table*}[htbp]
\centering
\caption{Super-Kamiokande detector systematic uncertainties}
\label{table:SK_syst_summary}
\begin{ruledtabular}
  \begin{tabular}{lcc}
    Source & 1$\sigma$ uncertainty & Sample  \\ \hline
    Decay-$e$ tagging efficiency & 1.0\% & Cosmic ray muon \\
    Spurious decay-$e$ tagging rate & 0.2 / event & Cosmic ray muon \\
    $\mu \rightarrow e$ mis-identification & 30\% & Cosmic ray muon \\
    Fiducial volume acceptance & 0.3 - 0.4\% & Cosmic ray muon \\
    NC$\pi^0$ rejection efficiency & 26 \% & Hybrid $\pi^0$ \\
    $e$/$\mu$, $\pi^0$/$e$, $\pi^+$/$\mu$, single/multi particle identification & --- & Atmospheric neutrino\\ 
  \end{tabular}
\end{ruledtabular}
\end{table*}

\begin{figure*}[htbp]
  \centering
  \includegraphics[width=0.98\textwidth]{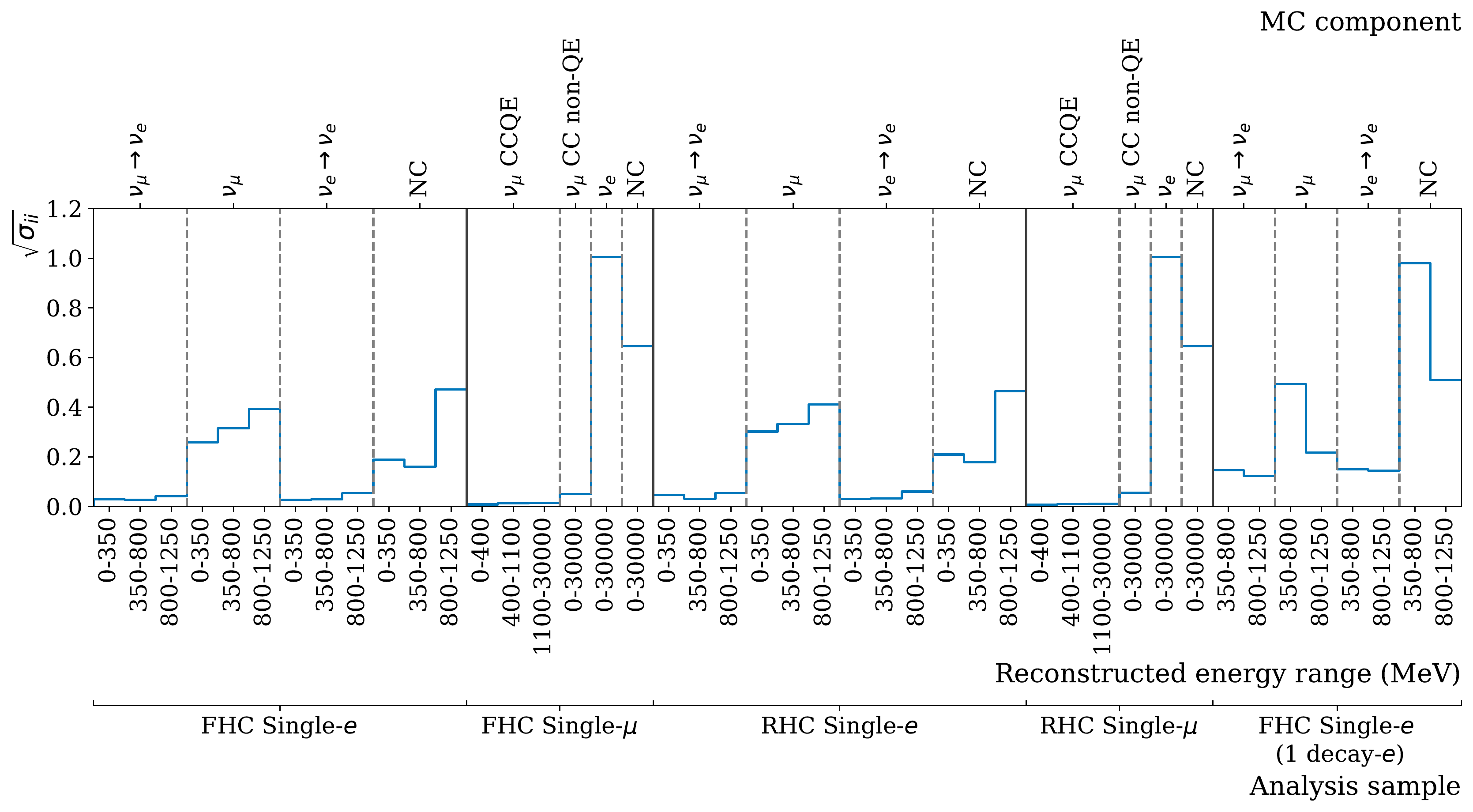}
  \caption{Square root of the diagonal elements of the covariance matrix describing the systematic uncertainty associated to the modeling of the SK detector.}
    \label{fig:SK_det_uncert_diag}
\end{figure*}

To propagate the systematic uncertainty on the SK event selection to the neutrino oscillation analysis frameworks, a covariance matrix is computed using the beam MC and the uncertainties quoted above. The MC is weighted with the flux and cross-section parameters at their central value from the fit to the near detector data and neutrino oscillation weights using the parameters in Tab.~\ref{tab:asimova_params} are applied. Variations of the MC are then produced with random throws of the systematic effects described above and projected into bins of true event topology and reconstructed neutrino energy or lepton momentum to produce the covariance matrix. The diagonal elements of the covariance matrix in reconstructed neutrino energy are shown in Fig.~\ref{fig:SK_det_uncert_diag}.

\section{Near/Far Extrapolation Fit}
\label{sec:ndfit}

Parameterized flux and cross-section models are used to calculate the predicted event rates at ND280 and SK. These models are fit to the high statistics, unoscillated near detector data to constrain the parameter uncertainties and tune their central values. An additional uncertainty is included in the flux covariance to account for the fact that the near detector fit results for Runs 2--6 are extrapolated to the far detector, which uses a larger data set from Runs 2--9. As in Ref.~\cite{Abe:2017vif_T2Krun7osc}, additional uncertainties which affect $\overset{\scriptscriptstyle(-)}{\nu_e}$ events have been introduced. These account for effects which may potentially affect $\overset{\scriptscriptstyle(-)}{\nu_e}$ but not $\overset{\scriptscriptstyle(-)}{\nu_{\mu}}$ cross sections.
    
The ND280 likelihood and fitting methods are unchanged since the analysis described in Ref.~\cite{Abe:2017vif_T2Krun7osc}. The 14 event samples are binned in $p_\mu$ and cos$\theta_\mu$, giving 1624 bins in total. The full likelihood includes a contribution from the binned $\chi^2$ data-model comparison, and a prior penalty contribution for each parameter. 

There are two near detector fitting frameworks used in this analysis. One fitter uses MINUIT to find the parameters which maximize the likelihood, while the other uses Markov Chain Monte Carlo (MCMC) methods to sample the parameter space. Both frameworks treat the systematics identically, and apply parameter variations on an event-by-event basis in the fit. The resulting parameter values from the MINUIT-based fit are used by two of the oscillation analyses (detailed in Sec.~\ref{sec:fitters}), with a covariance matrix describing their uncertainties. The MCMC analysis performs joint near and far detector data fits, but can also run near detector only fits for cross group validation.

The only change to the fitting frameworks since the last analysis was the treatment of the Fermi surface momentum systematics near their physical boundary. Previously, the covariance could not be calculated when the parameter was at its limit, causing the fit to not converge. For this analysis, the penalty contribution to the likelihood was `mirrored' around the physical boundaries for the $p_F$ parameters. This meant the parameters were allowed to pass beyond their physical boundaries. The mirroring was performed by setting the likelihood for values beyond the boundary to the value of the likelihood the same distance from the boundary on the other side. For example, for a physical boundary at $+1.0$, the likelihood at $+1.2$ would be the same as the likelihood at $+0.8$. This allowed the uncertainty to be calculated at the limit, and the fit to converge.

\begin{figure*}[htp]
\centering
\begin{subfigure}{0.47\textwidth}
  \includegraphics[width=0.98\columnwidth]{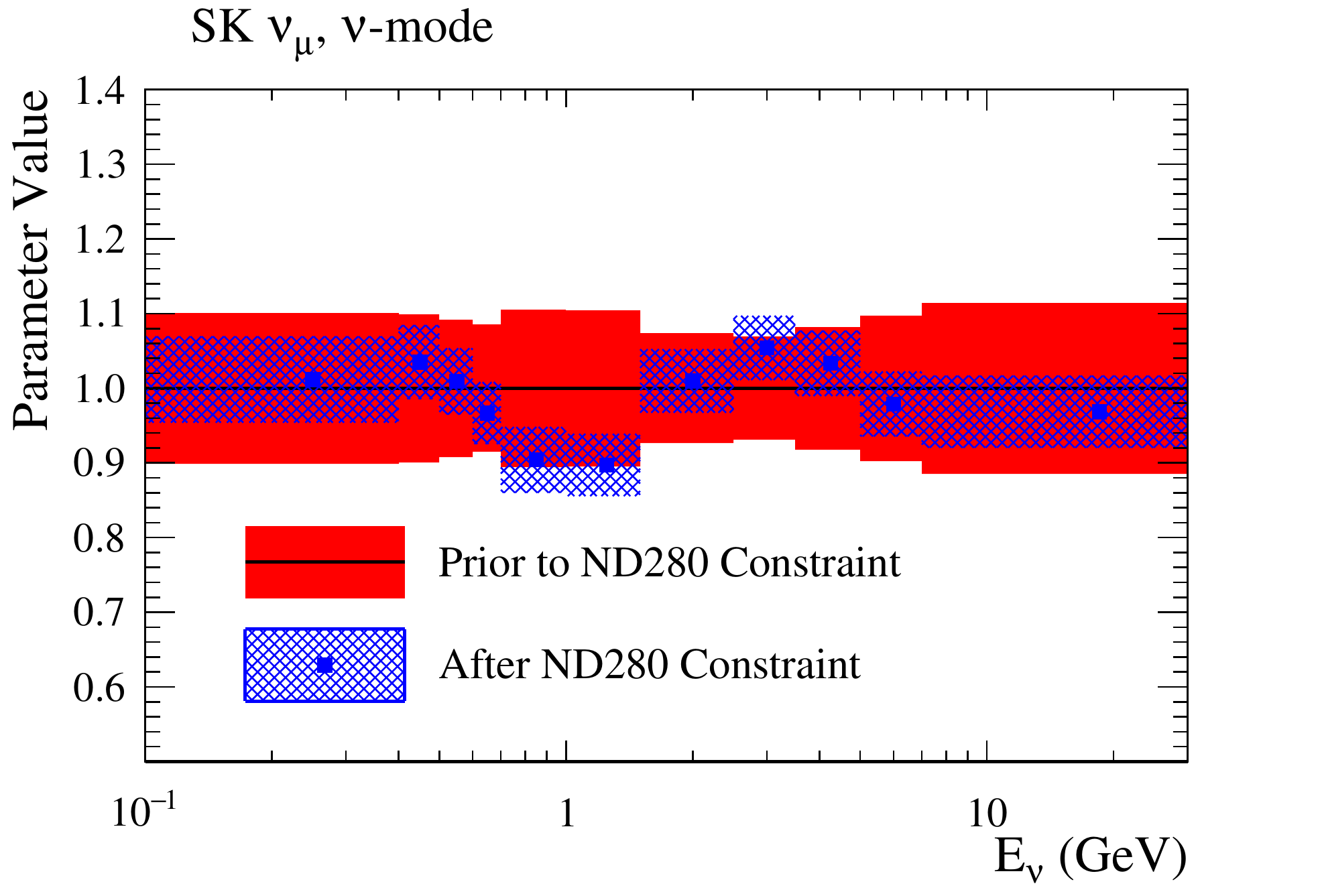} 
\end{subfigure}
\begin{subfigure}{0.47\textwidth}
  \includegraphics[width=0.98\columnwidth]{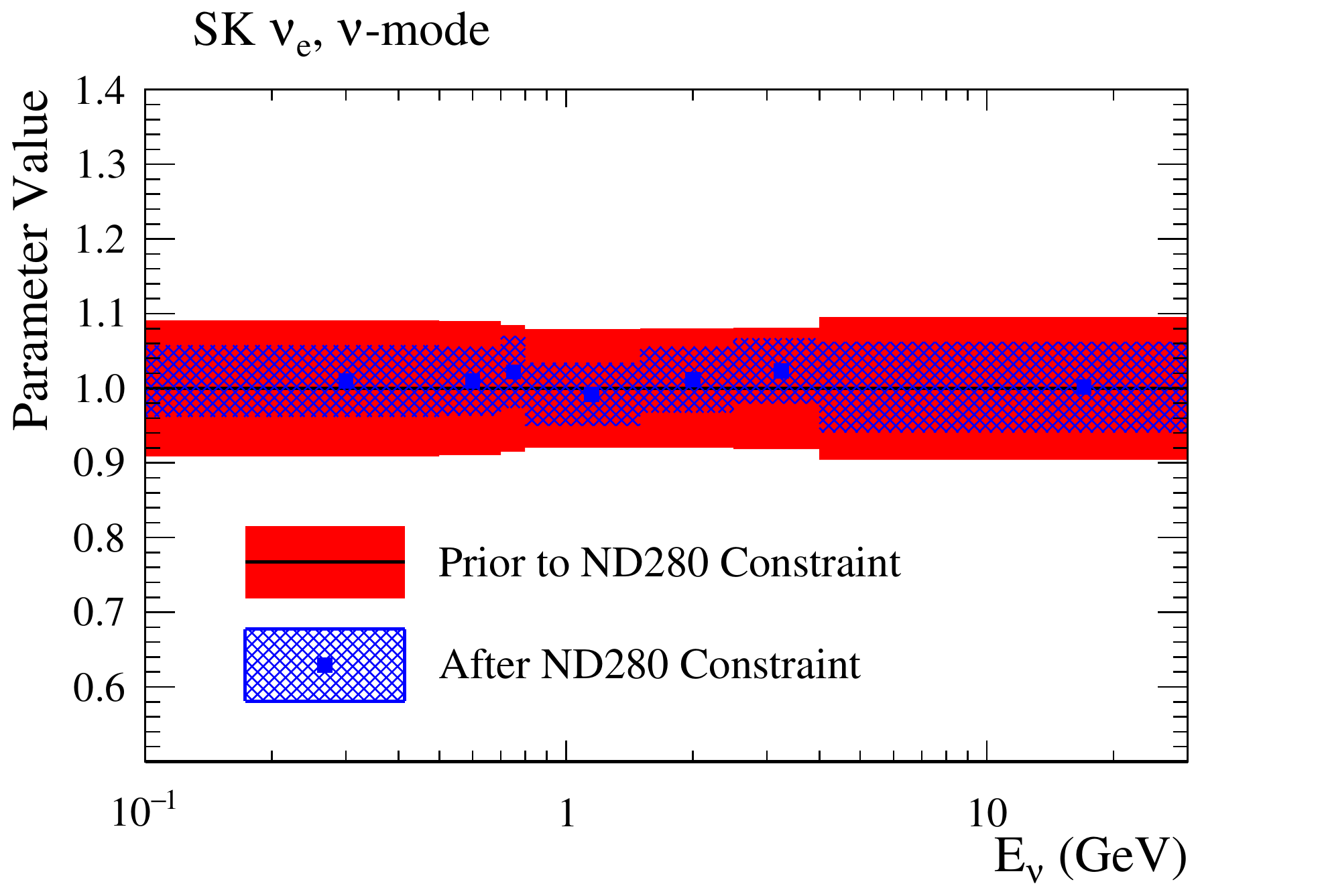}
\end{subfigure}
\begin{subfigure}{0.47\textwidth}
  \includegraphics[width=0.98\columnwidth]{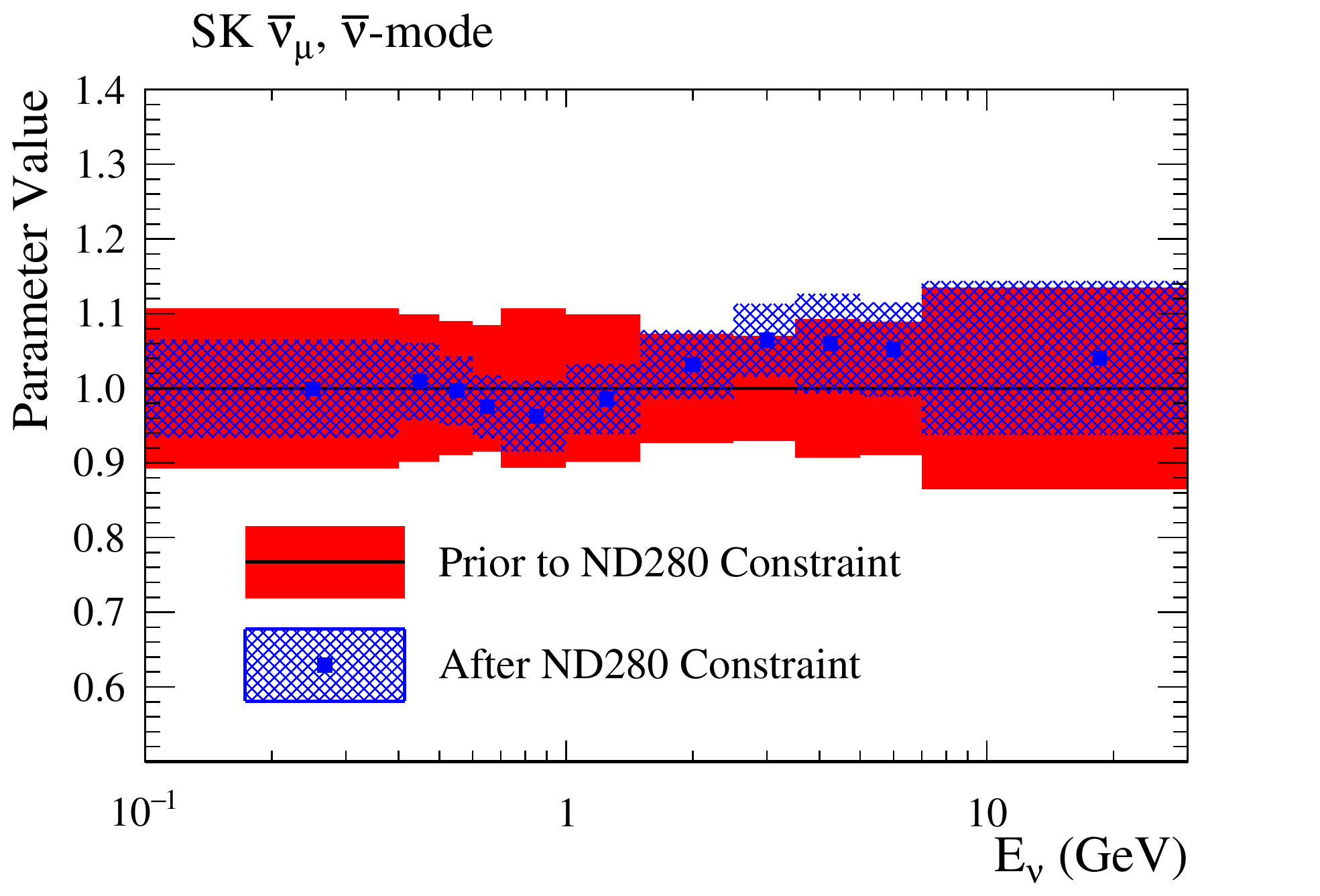} 
\end{subfigure}
\begin{subfigure}{0.47\textwidth}
  \includegraphics[width=0.98\columnwidth]{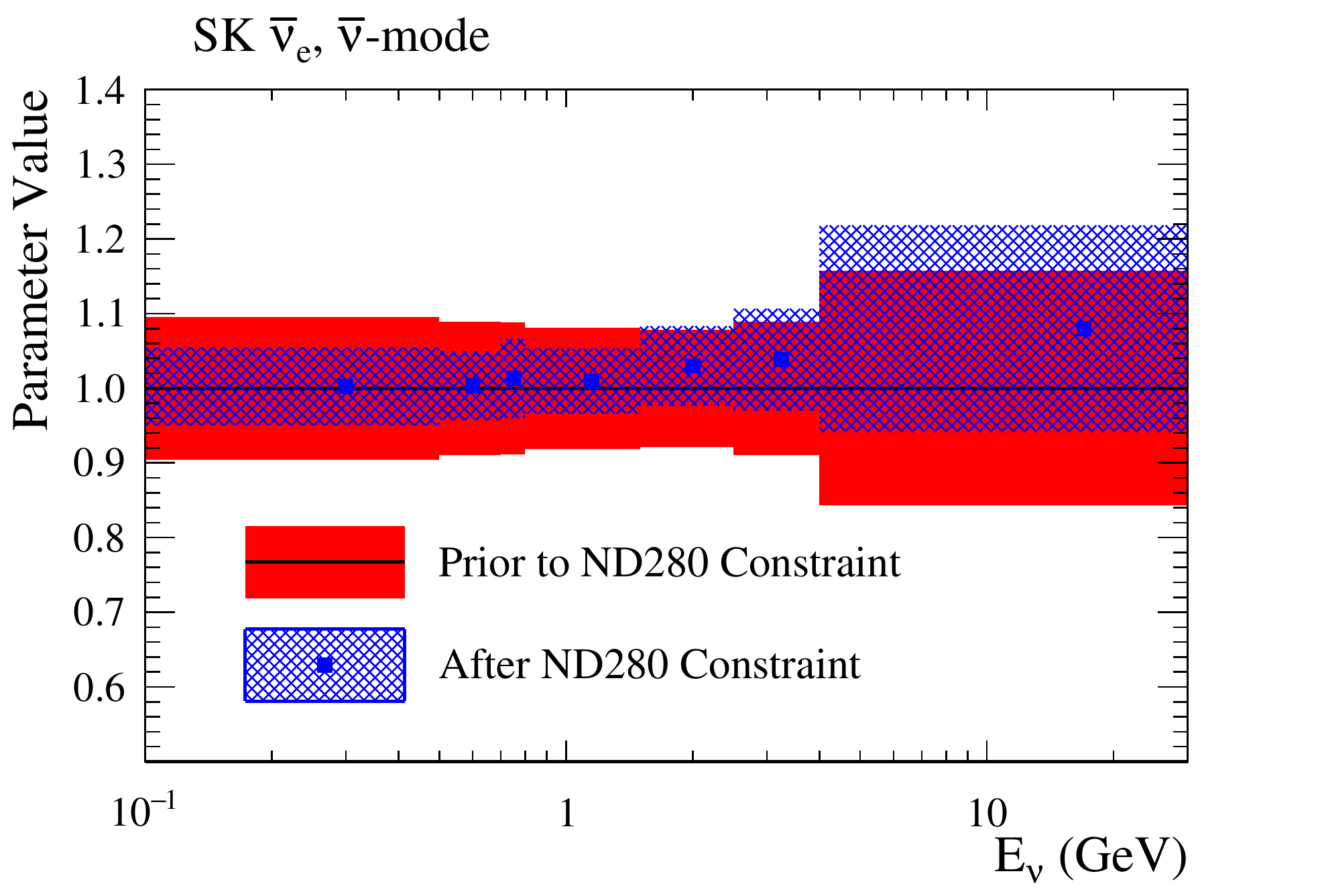}
\end{subfigure}
\caption{The SK flux parameters for the $\numu$ (top left) and $\nue$ (top right) neutrino species in FHC, and for the $\numub$ (bottom left) and $\nueb$ (bottom right) neutrino species in RHC, as a fraction of the nominal value. The bands indicate the 1$\sigma$ uncertainty on the parameters before (solid, red) and after (hatched, blue) the near detector fit.}
\label{fig:SKFluxPostfit}
\end{figure*}

\begin{figure}[htp]
\centering
  \includegraphics[width=0.47\textwidth]{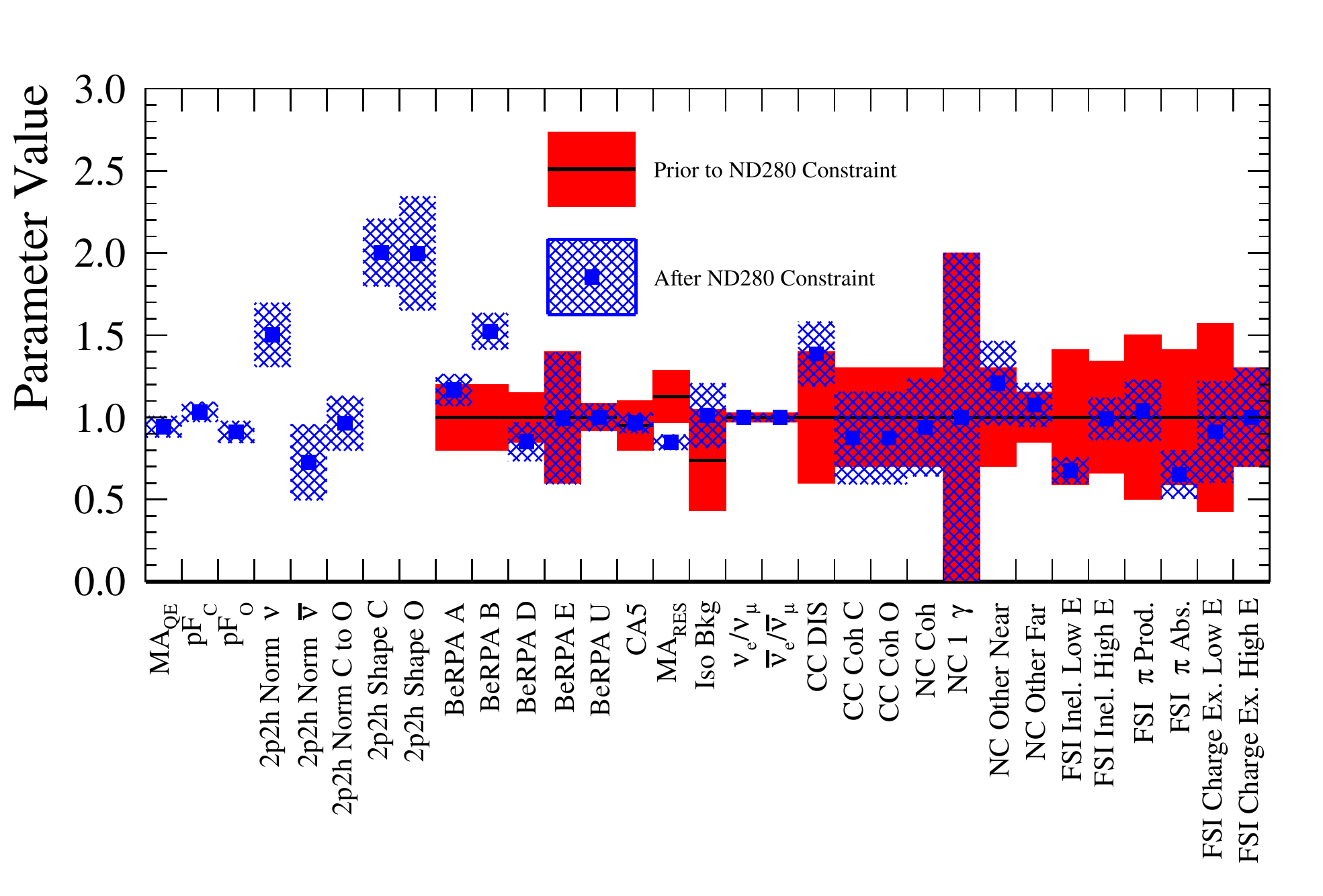}
\caption{The cross-section parameters as a fraction of the nominal value. The bands indicate the 1$\sigma$ uncertainty on the parameters before (solid, red) and after (hatched, blue) the near detector fit.}
 \label{fig:xsecPostfit}
\end{figure}

The prefit and postfit SK flux and cross-section parameter values and uncertainties are shown in Figs.~\ref{fig:SKFluxPostfit} and~\ref{fig:xsecPostfit}, as a fraction of the nominal values. The central values and uncertainties for all parameters are tabulated in the Appendix. There is a significant reduction in the postfit uncertainty for the majority of parameters. Those that are not constrained by the near detector fit are uncertainties that only apply to interactions with low statistics in the near detector. In the last analysis, Ref.~\cite{Abe:2017vif_T2Krun7osc}, the neutrino flux increased for all samples and species, but this effect is no longer present. The difference between the nominal simulation and data is now being absorbed by the movement of other parameters, in particular the BeRPA model.

\begin{figure}
\centering
\includegraphics[width=0.47\textwidth]{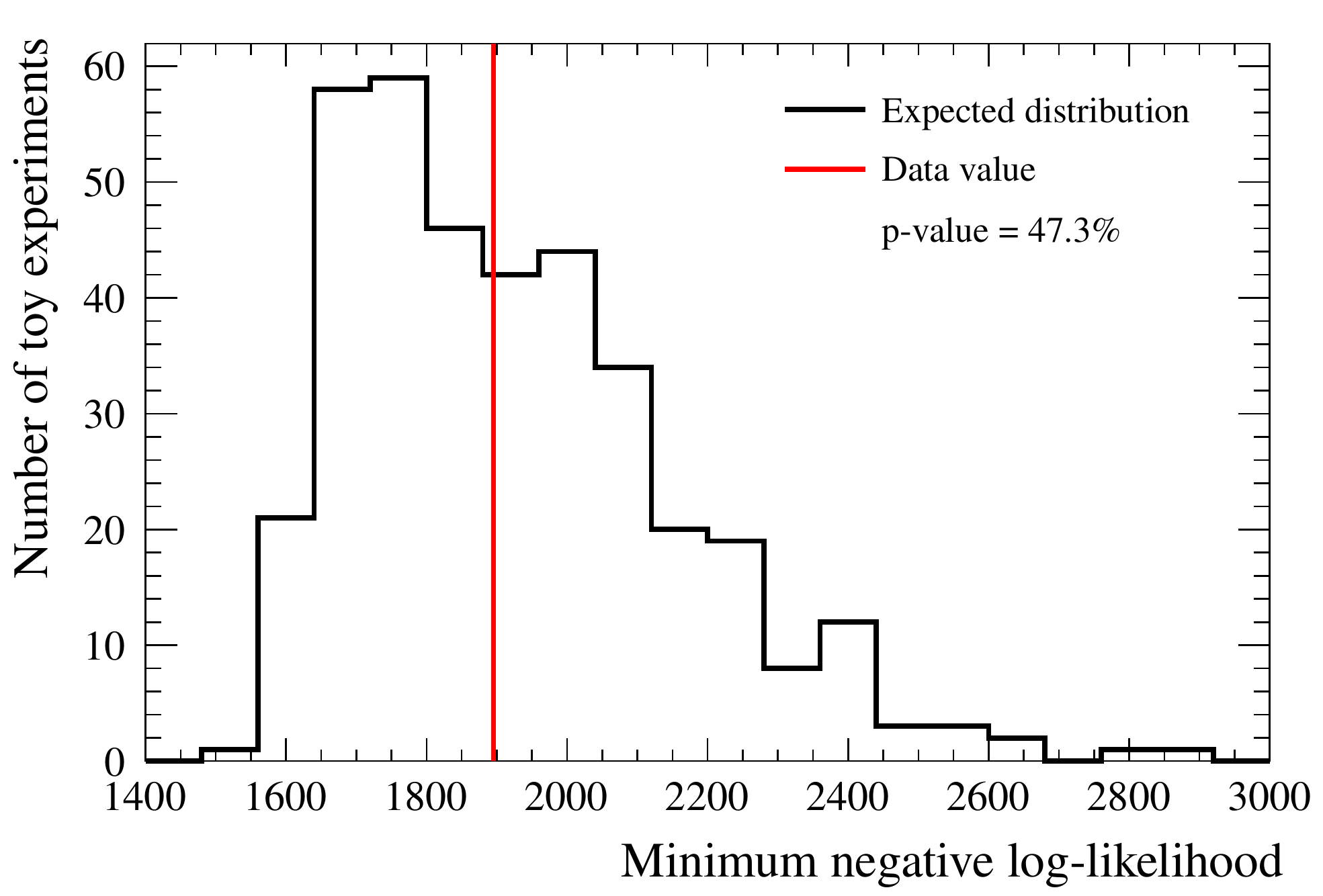} 
\caption{Distribution of the minimum negative log-likelihood values from fits to the mock data sets (black), with the value from the fit to the data superimposed in red. The $p$-value is $47.3\%$.}
\label{fig:banffpval}
\end{figure}

The goodness-of-fit for the near detector analysis was estimated by calculating the $p$-value in the MINUIT-based framework. Toy data sets were produced by throwing all systematics according to their prior covariance, and applying them to the nominal simulation prediction. The likelihood for each toy data set is shown in Fig.~\ref{fig:banffpval}, along with the likelihood from the data fit. The overall $p$-value for the fit is 47.3$\%$.
        
\begin{figure*}[htp]
\centering
\begin{subfigure}{0.47\textwidth}
  \includegraphics[width=0.98\columnwidth]{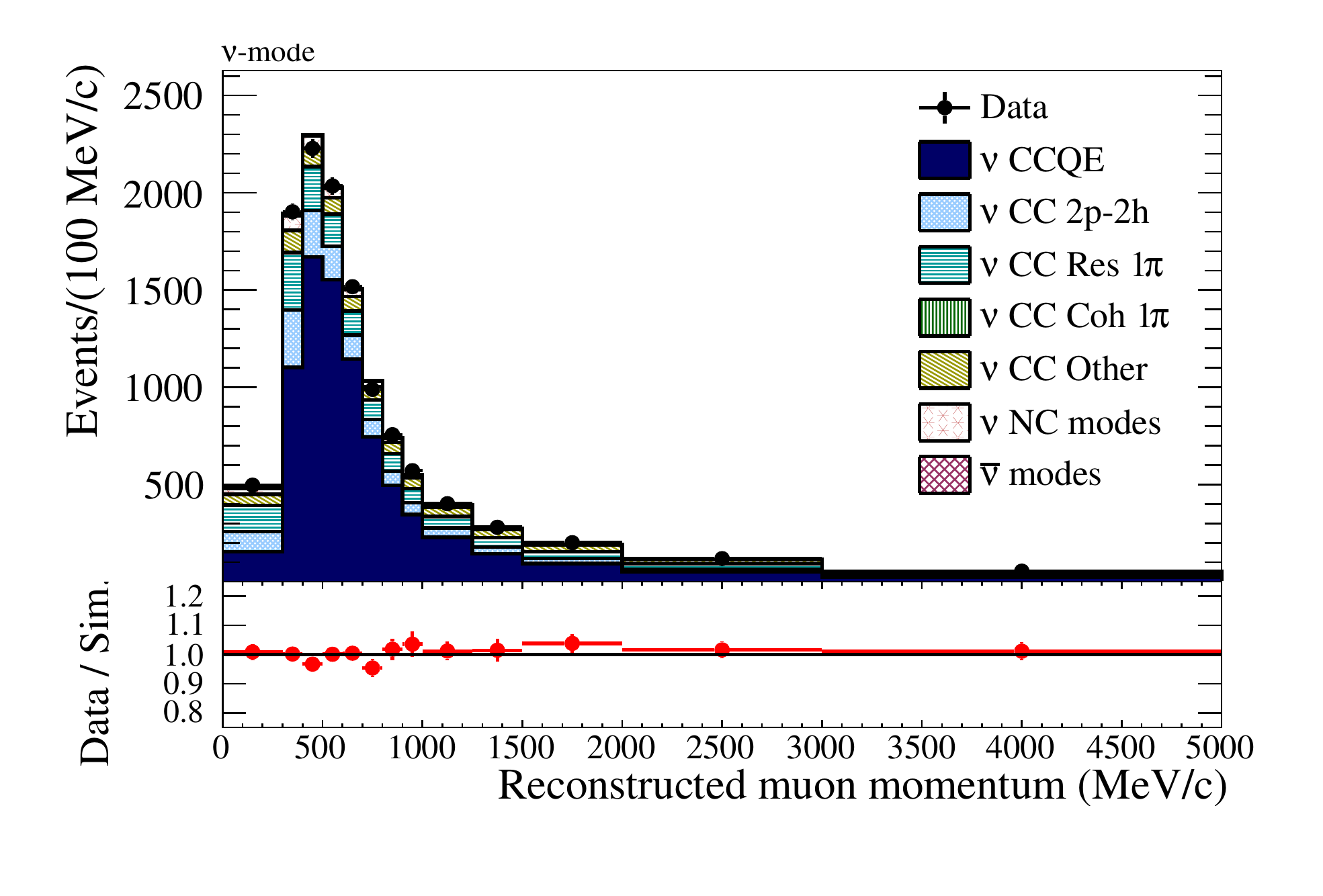} 
\end{subfigure}
\begin{subfigure}{0.47\textwidth}
  \includegraphics[width=0.98\columnwidth]{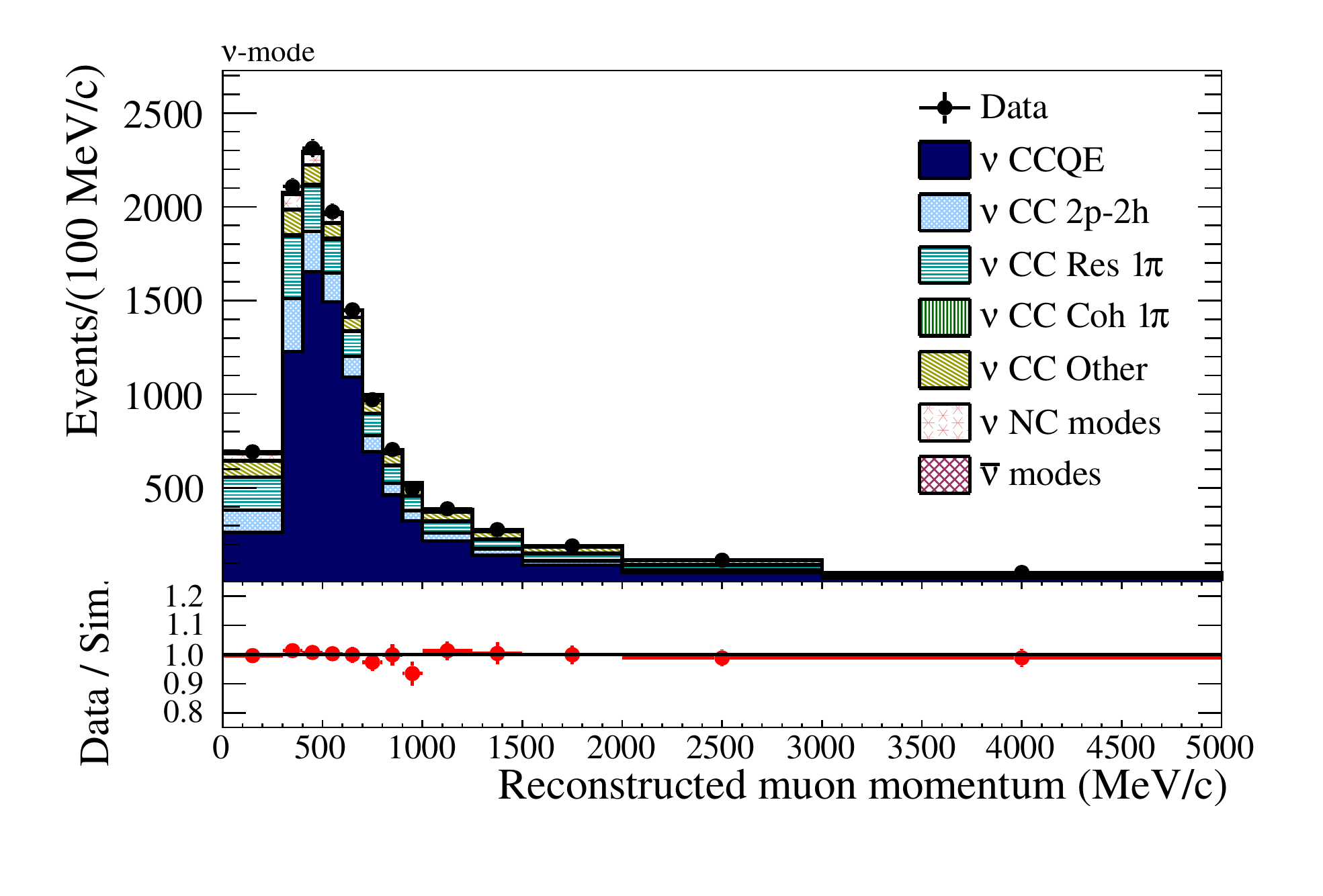}
\end{subfigure}
\begin{subfigure}{0.47\textwidth}
  \includegraphics[width=0.98\columnwidth]{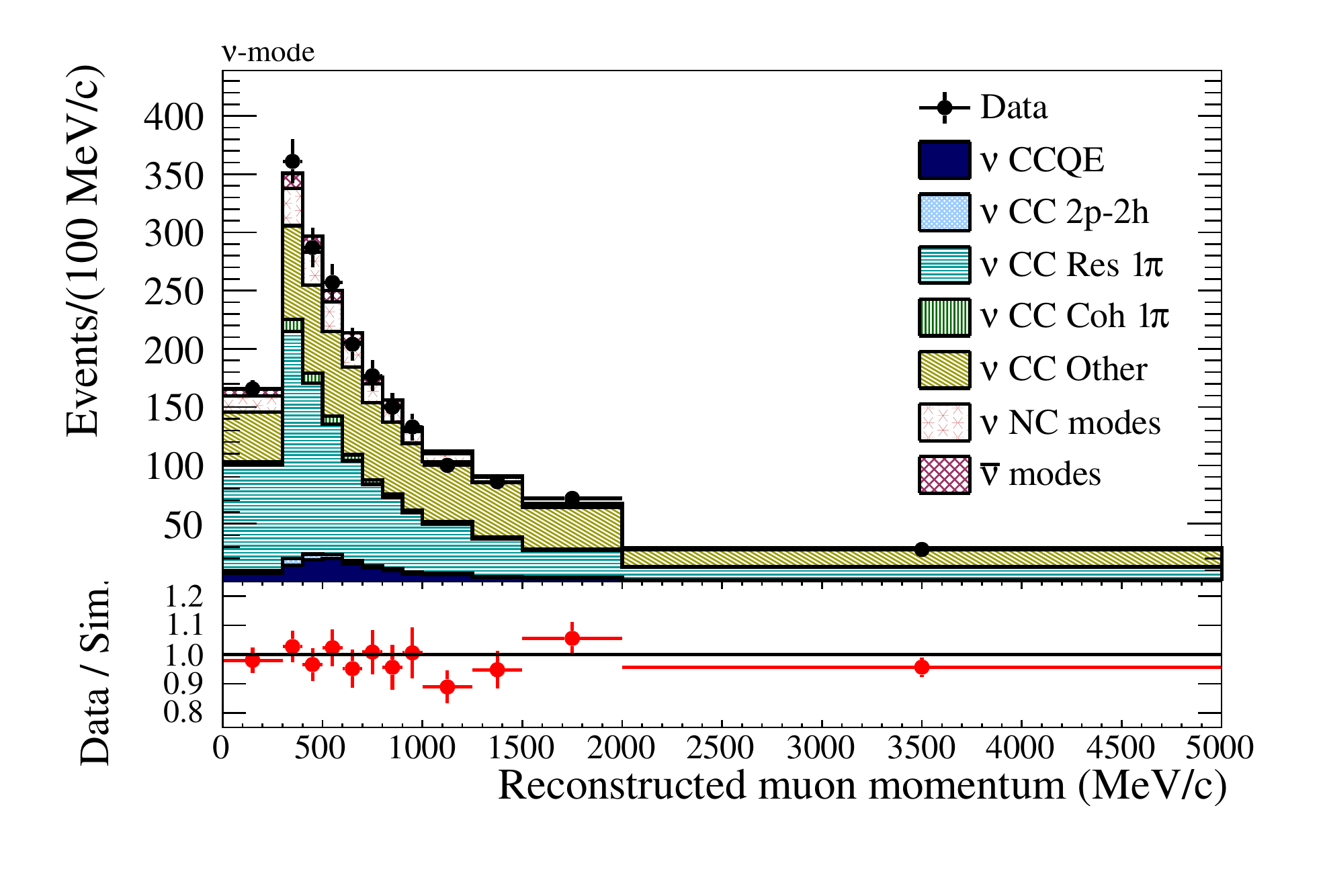} 
\end{subfigure}
\begin{subfigure}{0.47\textwidth}
  \includegraphics[width=0.98\columnwidth]{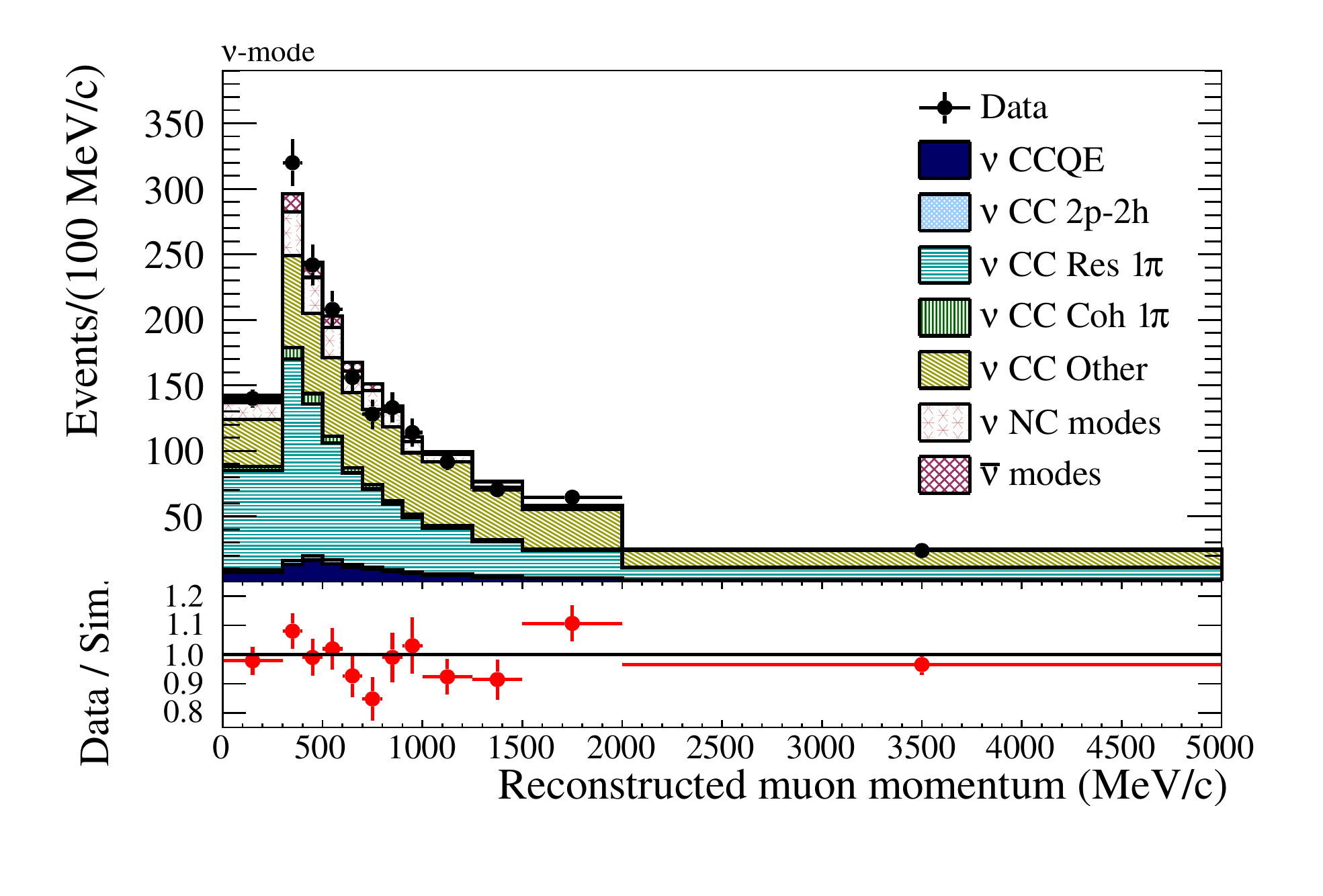}
\end{subfigure}
\caption{Post ND280 fit muon momentum distributions of the FHC $\nu_{\mu}$ CC 0$\pi$ (top), and FHC $\nu_{\mu}$ CC 1$\pi$ (bottom) samples in FGD1 (left) and FGD2 (right).}
\label{fig:FHCNumuPostfit}
\end{figure*}

\begin{figure*}[htp]
\centering
\begin{subfigure}{0.47\textwidth}
  \includegraphics[width=0.98\columnwidth]{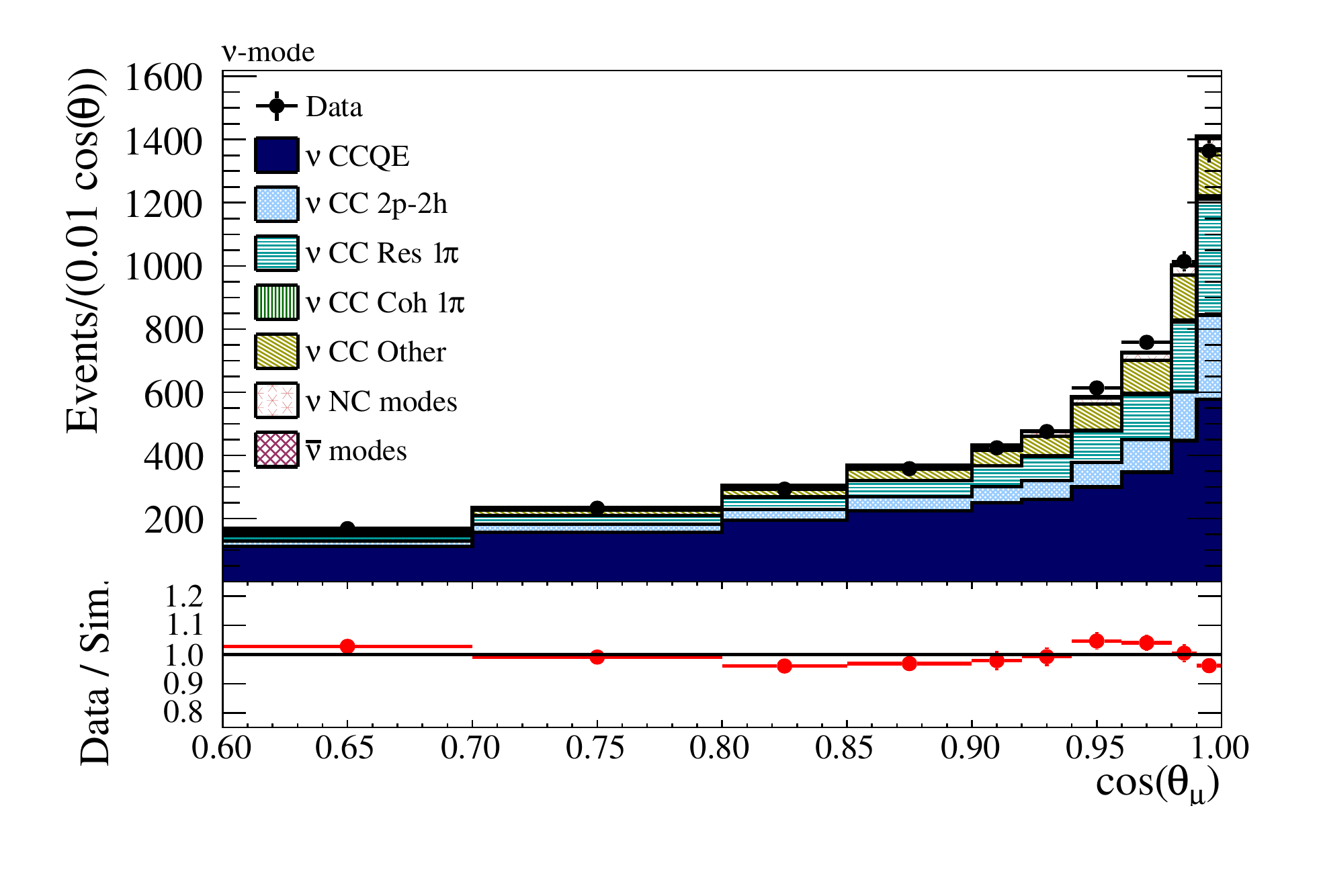} 
\end{subfigure}
\begin{subfigure}{0.47\textwidth}
  \includegraphics[width=0.98\columnwidth]{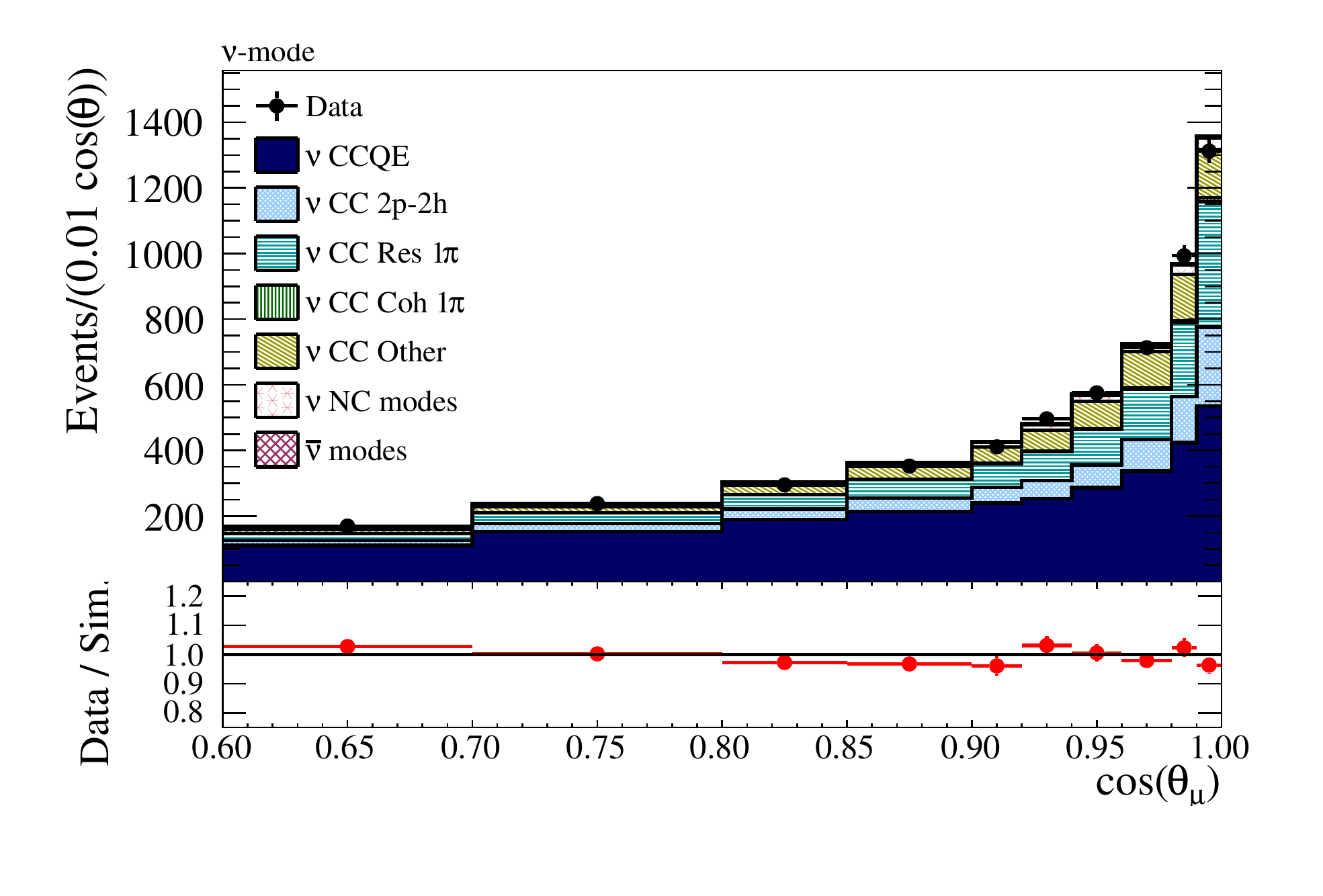}
\end{subfigure}
\begin{subfigure}{0.47\textwidth}
  \includegraphics[width=0.98\columnwidth]{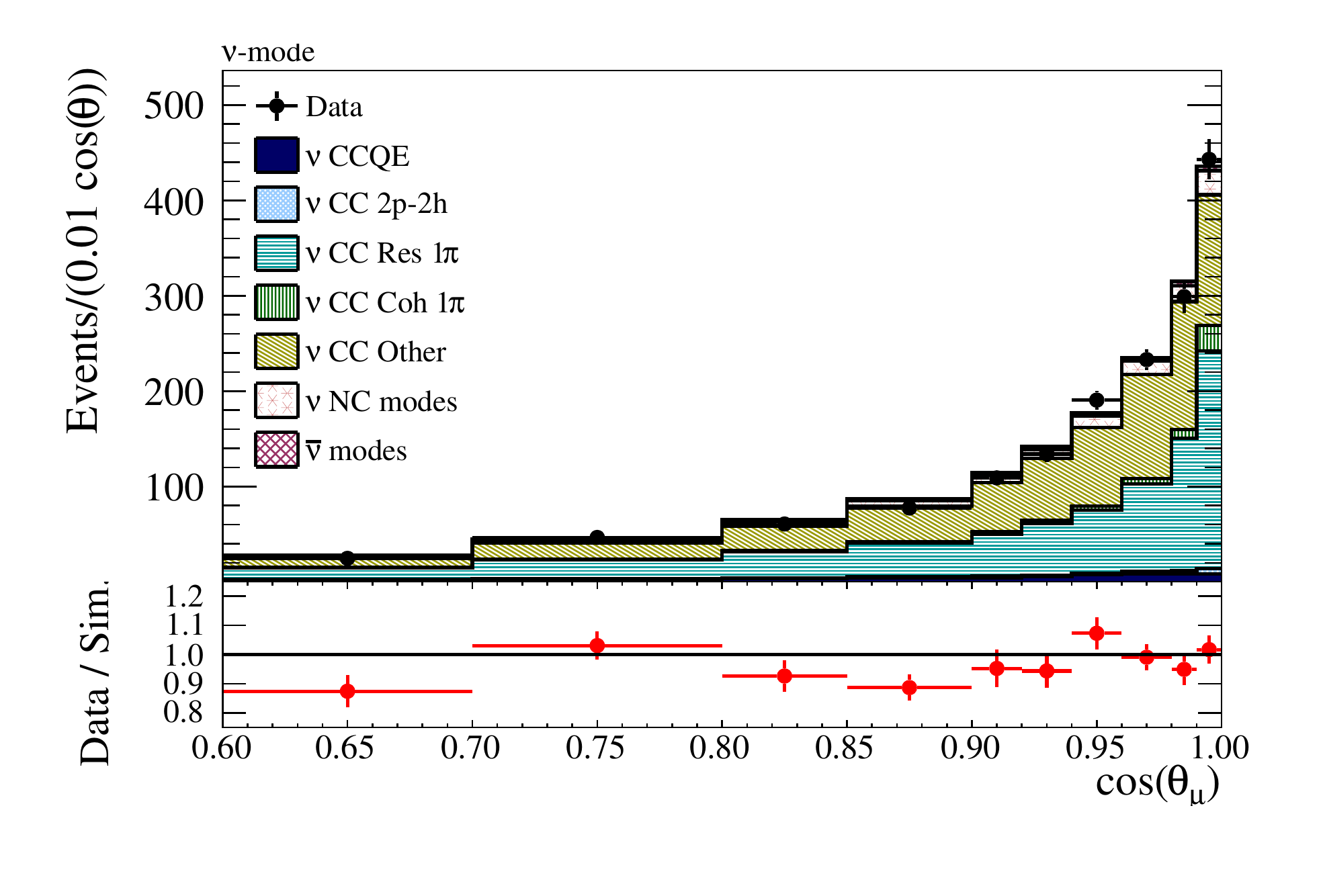} 
\end{subfigure}
\begin{subfigure}{0.47\textwidth}
  \includegraphics[width=0.98\columnwidth]{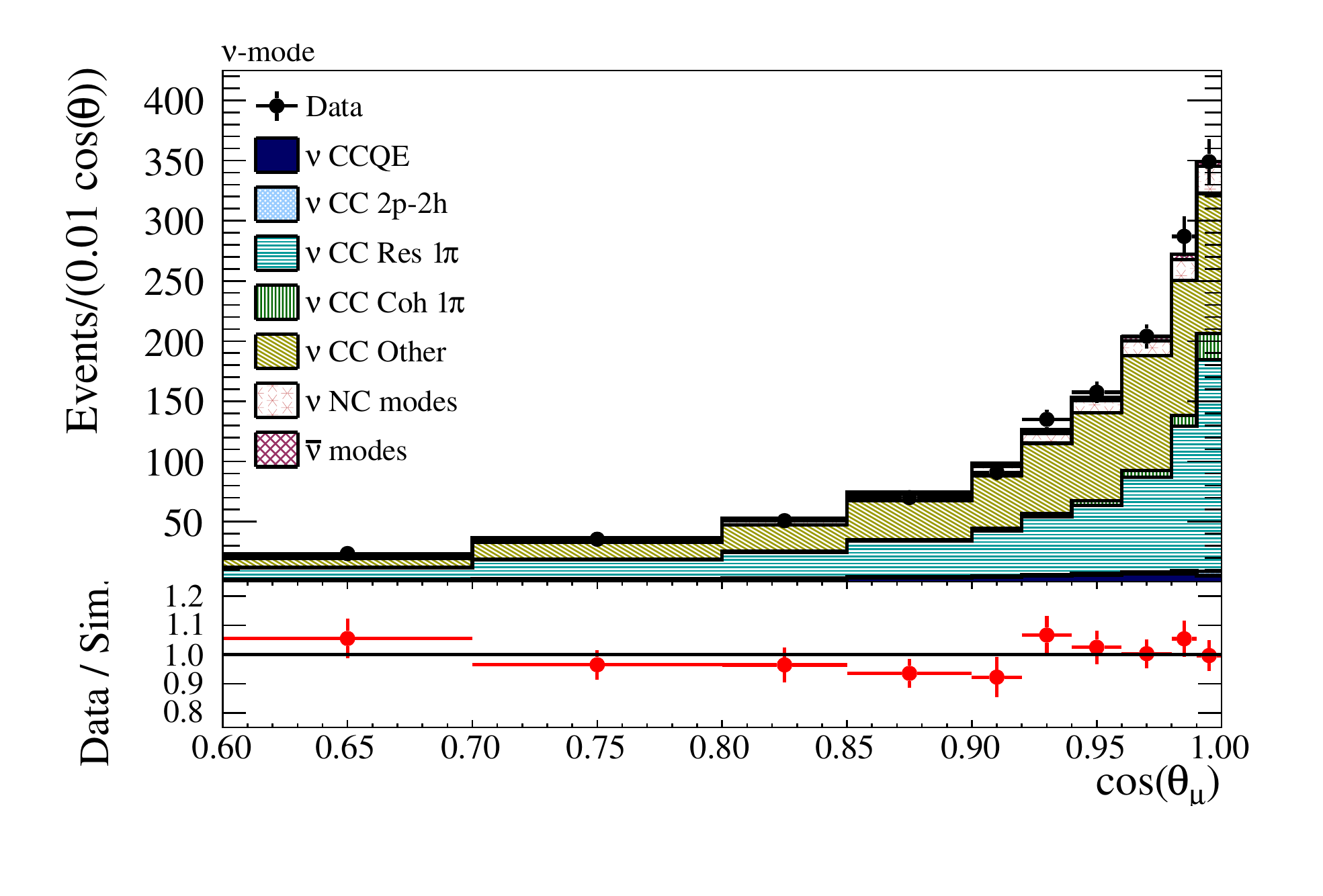}
\end{subfigure}
\caption{Post ND280 fit distributions of the final state muon angle of the FHC $\nu_{\mu}$ CC 0$\pi$ (top), and FHC $\nu_{\mu}$ CC 1$\pi$ (bottom) samples in FGD1 (left) and FGD2 (right).}
\label{fig:FHCNumuPostfit_theta}
\end{figure*}

\begin{figure*}[htp]
\centering
\begin{subfigure}{0.47\textwidth}
  \includegraphics[width=0.98\columnwidth]{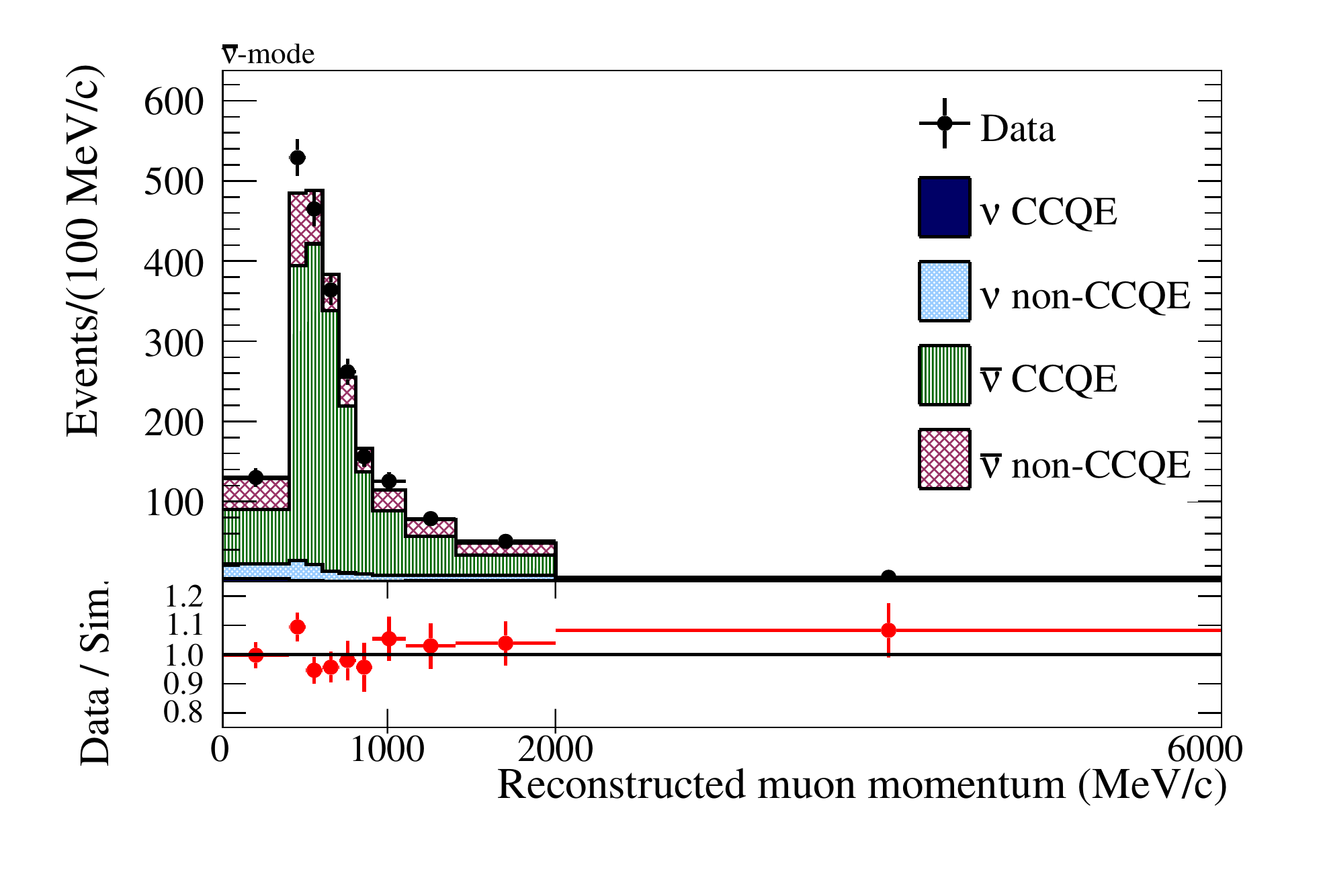} 
\end{subfigure}
\begin{subfigure}{0.47\textwidth}
  \includegraphics[width=0.98\columnwidth]{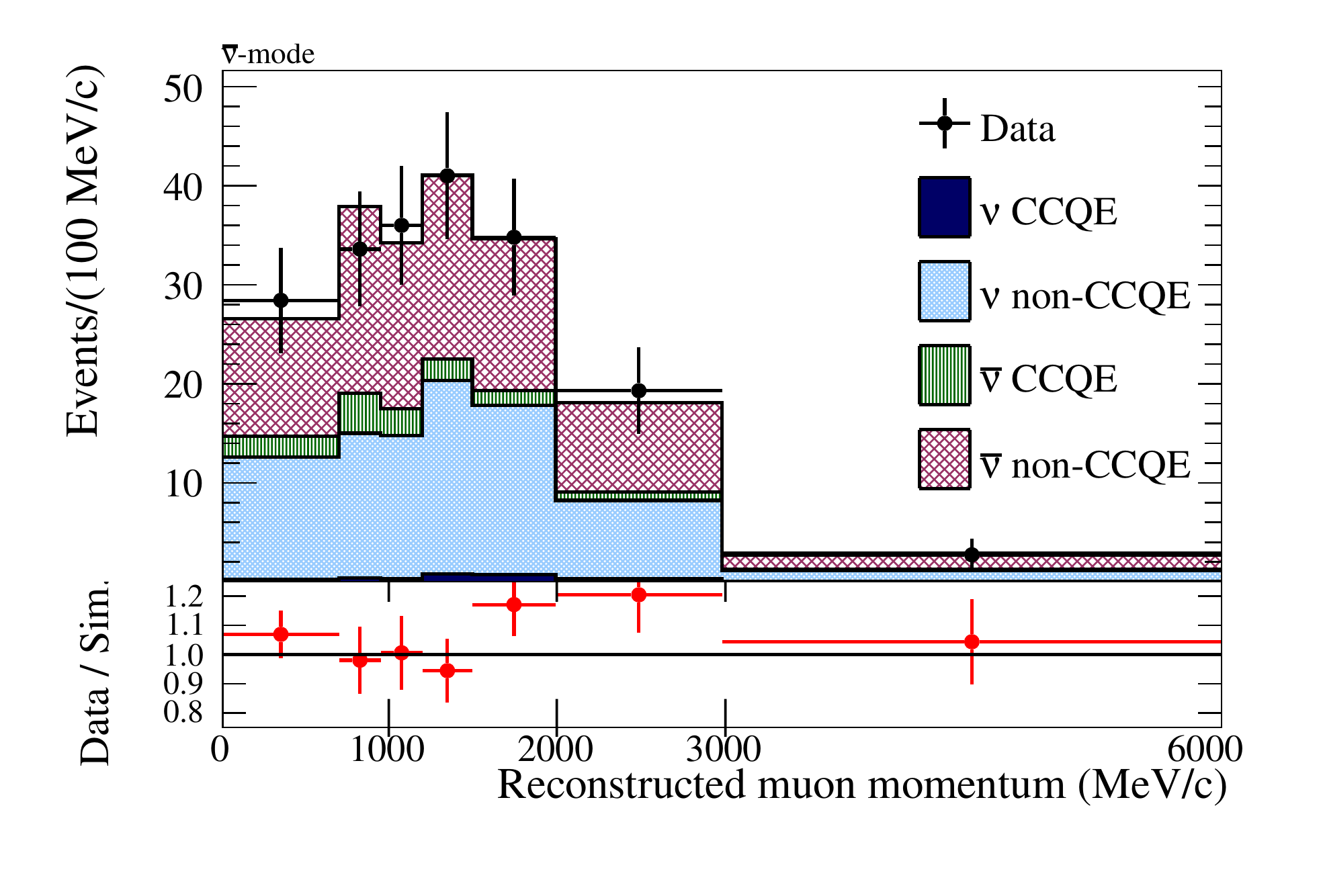}
\end{subfigure}
\begin{subfigure}{0.47\textwidth}
  \includegraphics[width=0.98\columnwidth]{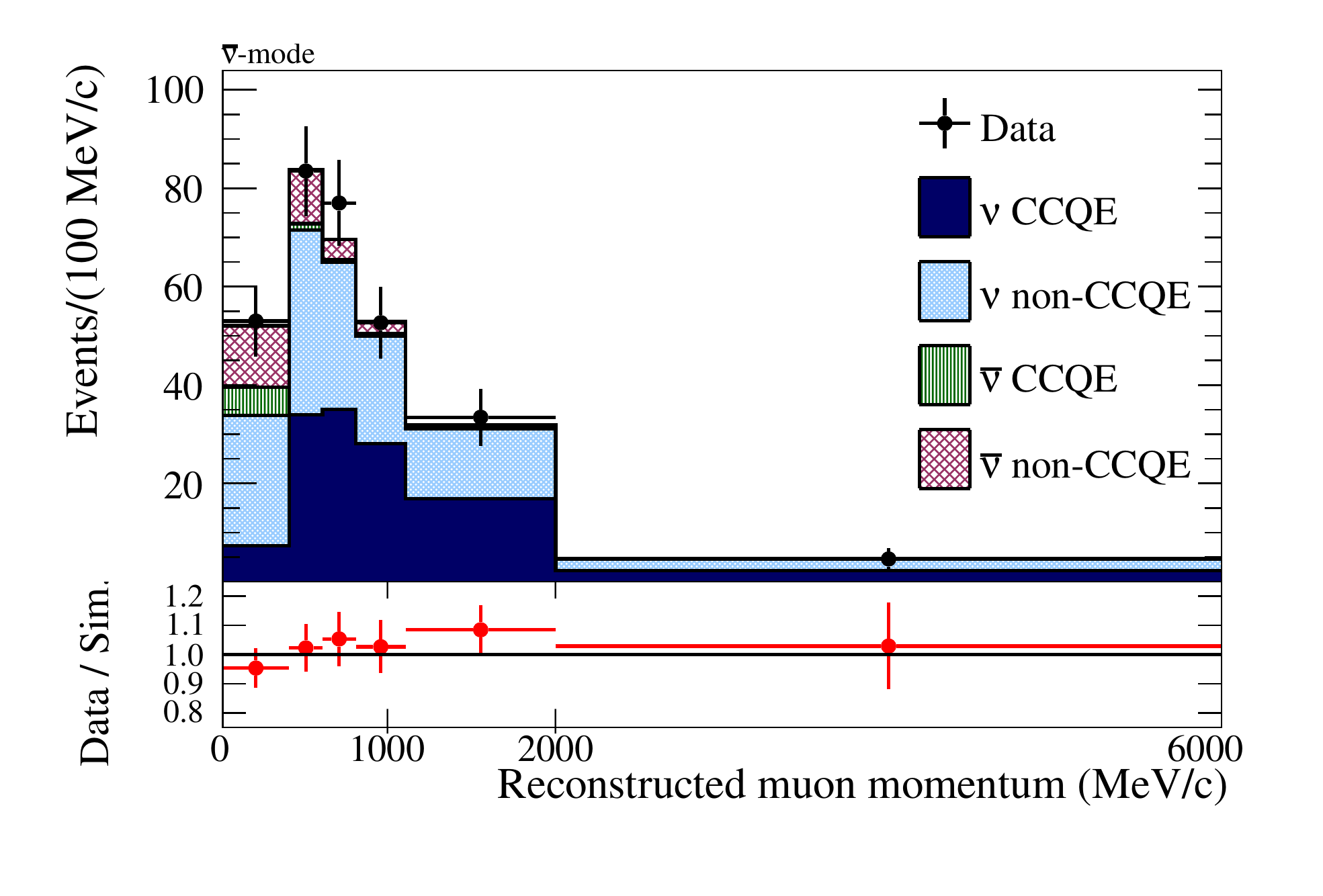} 
\end{subfigure}
\begin{subfigure}{0.47\textwidth}
  \includegraphics[width=0.98\columnwidth]{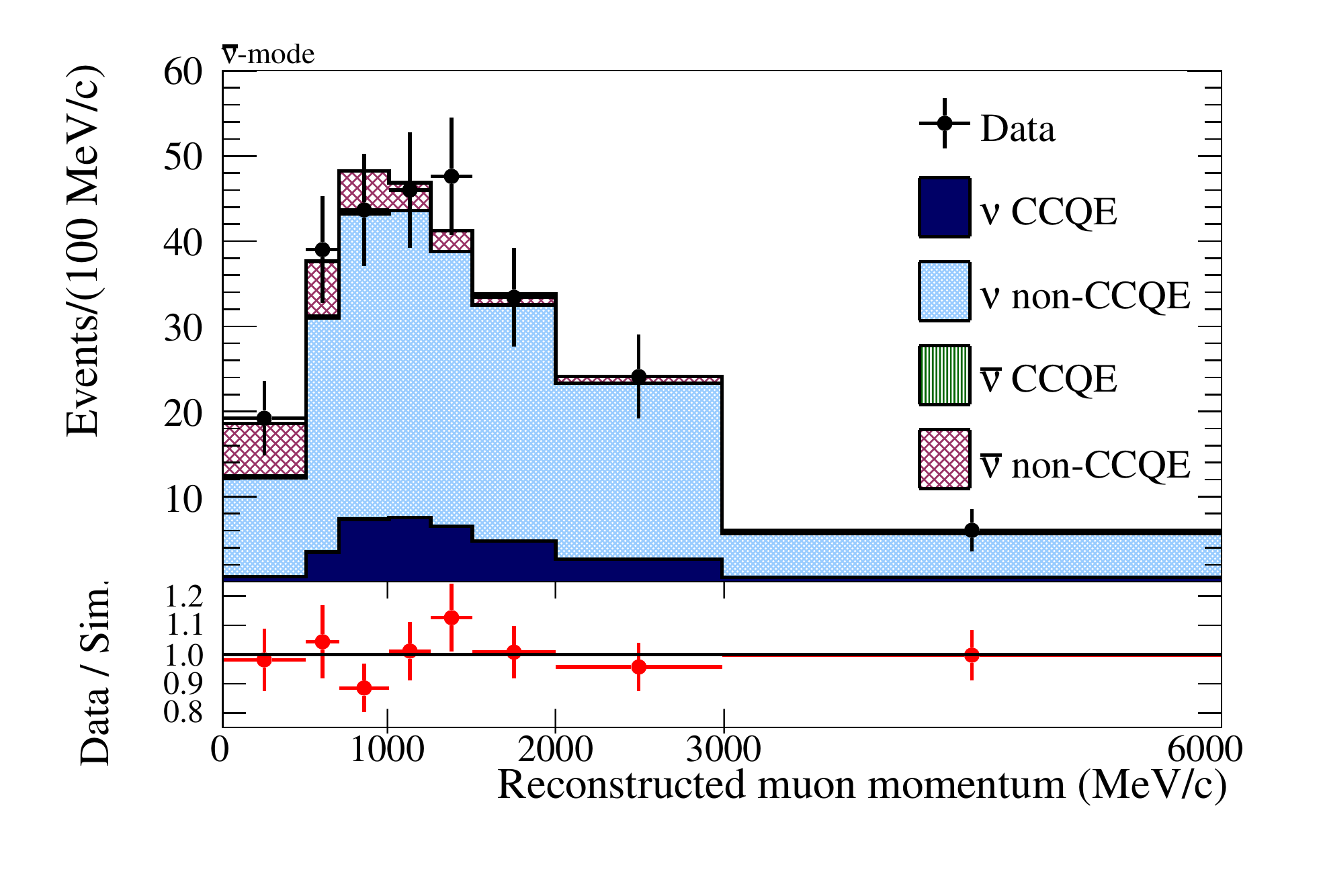}
\end{subfigure}
\caption{Post ND280 fit muon momentum distributions for the RHC $\numub$ (top) and $\numu$ (bottom) CC 1-track (left) and CC $N$-track (right) FGD1 samples.}
\label{fig:RHCNumuPostfit}
\end{figure*}

\begin{figure*}[htp]
\centering
\begin{subfigure}{0.47\textwidth}
  \includegraphics[width=0.98\columnwidth]{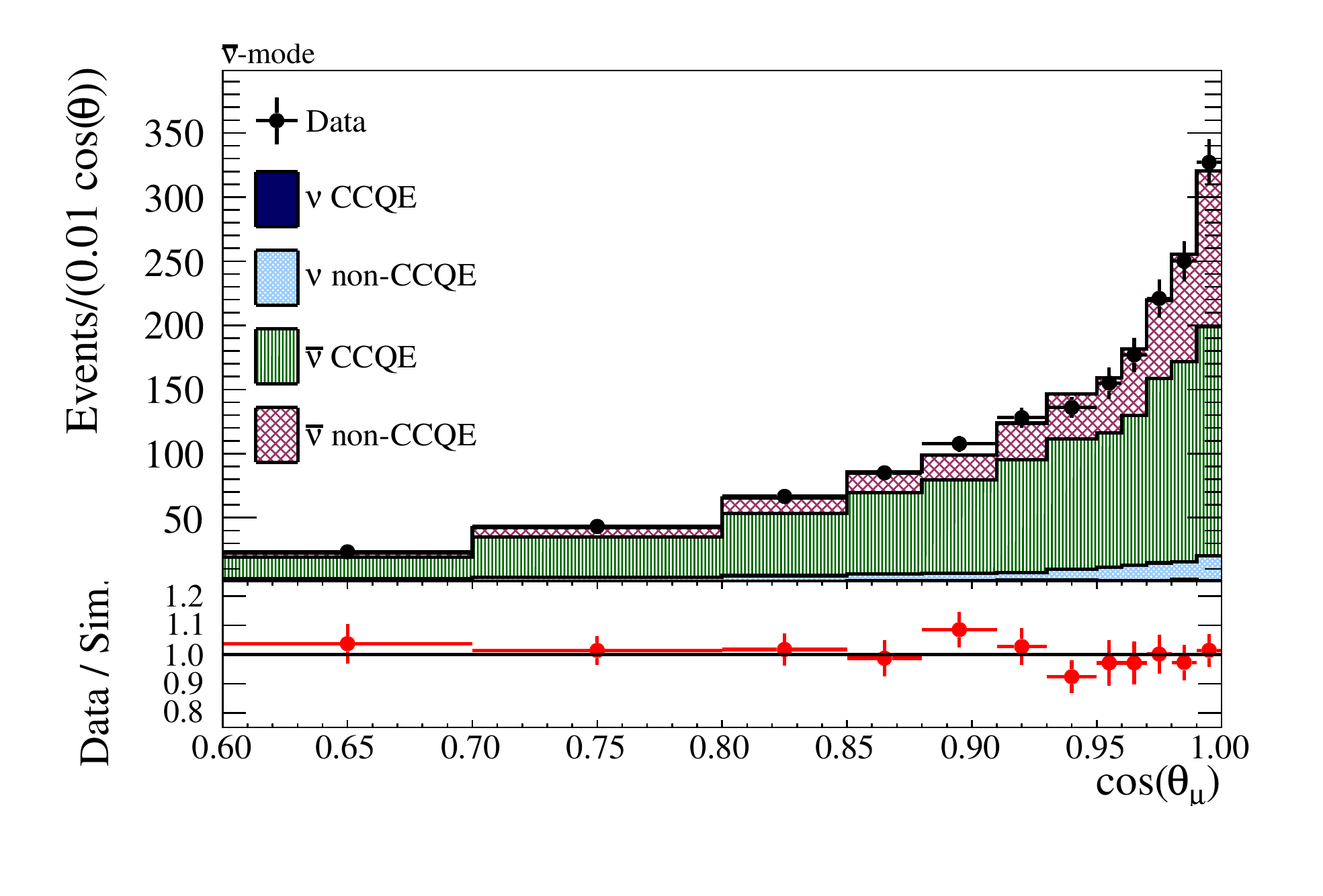} 
\end{subfigure}
\begin{subfigure}{0.47\textwidth}
  \includegraphics[width=0.98\columnwidth]{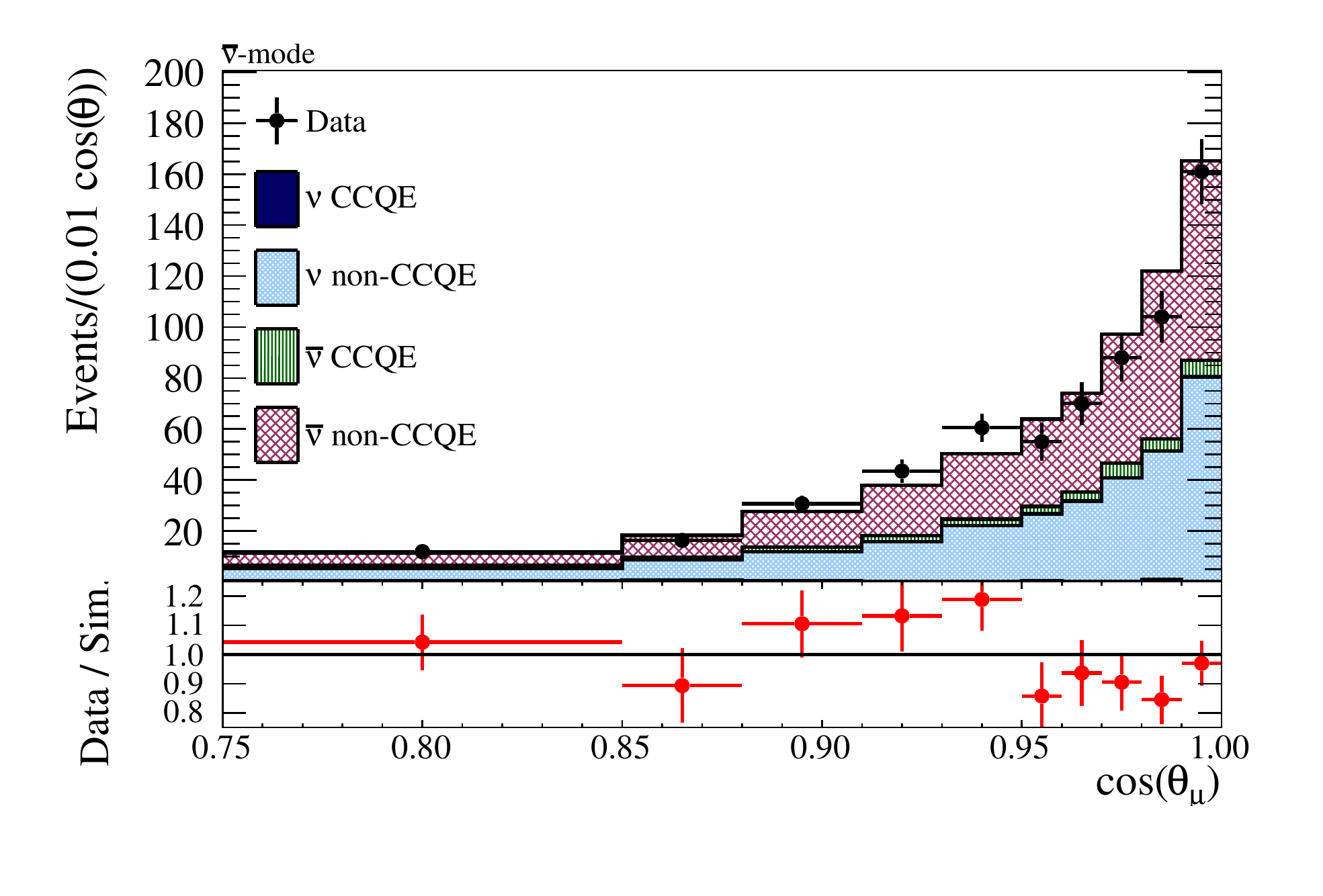}
\end{subfigure}
\begin{subfigure}{0.47\textwidth}
  \includegraphics[width=0.98\columnwidth]{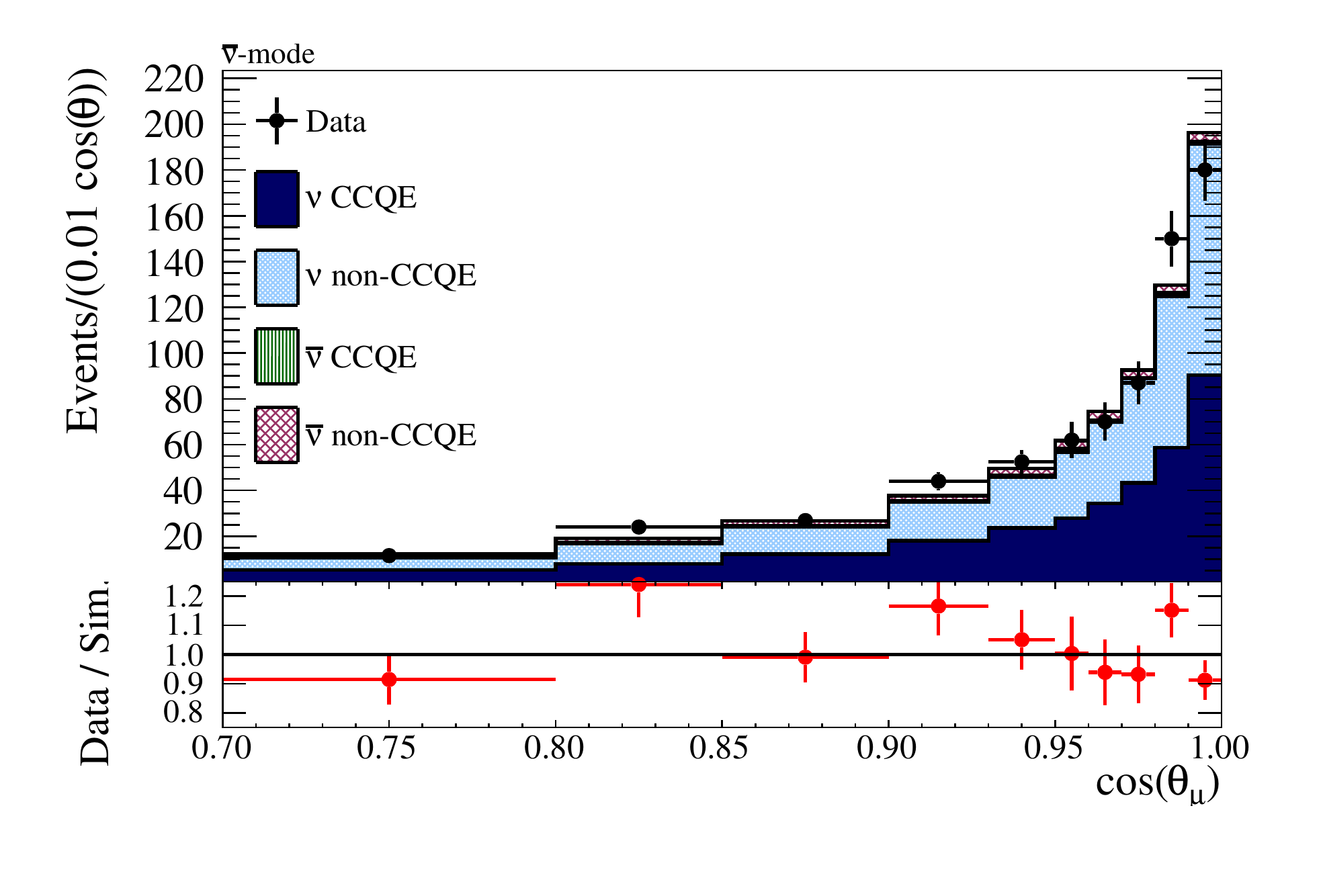} 
\end{subfigure}
\begin{subfigure}{0.47\textwidth}
  \includegraphics[width=0.98\columnwidth]{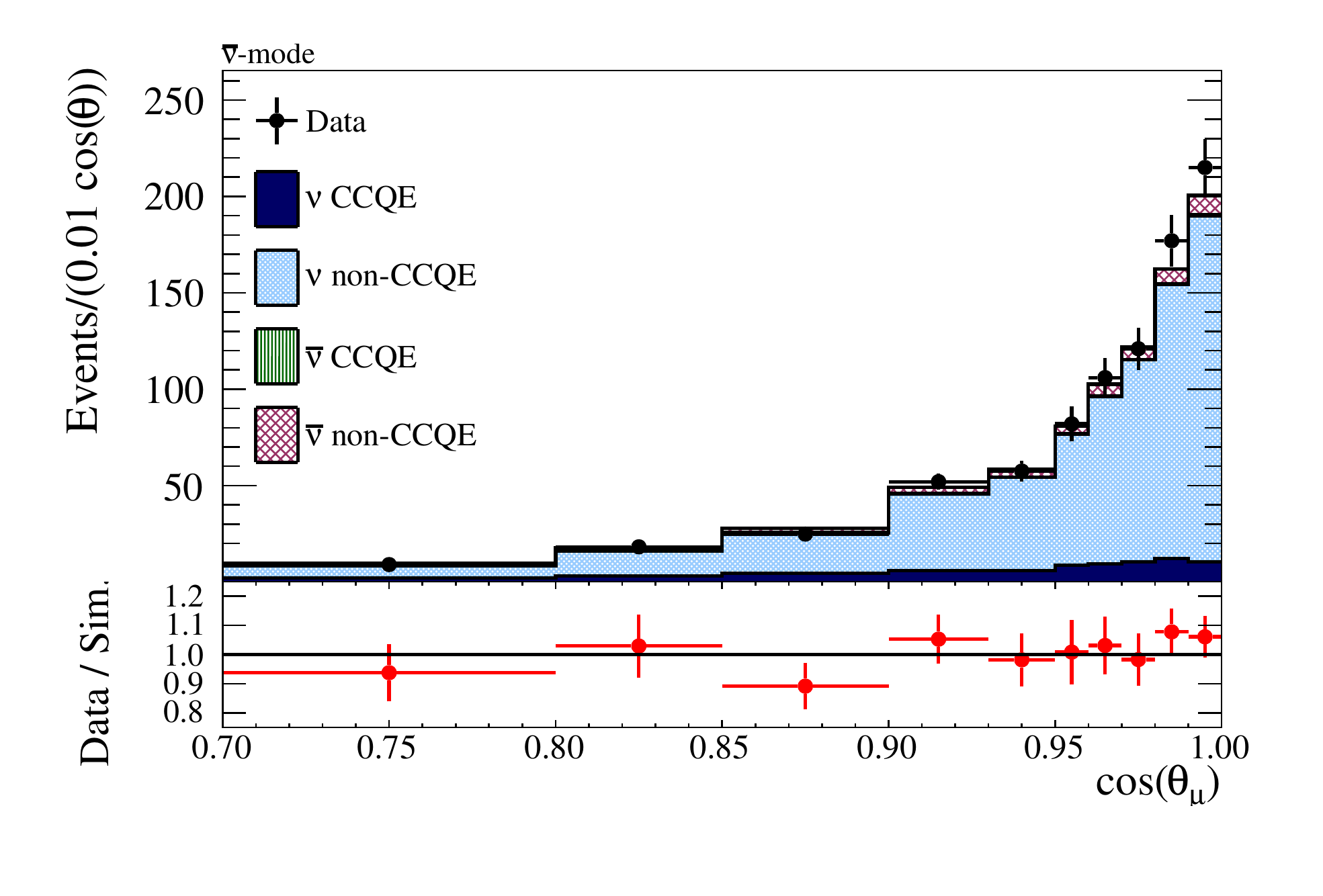}
\end{subfigure}
\caption{Post ND280 fit distributions of the final state muon angle for the RHC $\numub$ (top) and $\numu$ (bottom) CC 1-track (left) and CC $N$-track (right) FGD1 samples.}
\label{fig:RHCNumuPostfit_theta}
\end{figure*}

The postfit muon momentum and angle distributions of events are produced by applying the best-fit parameter values to the nominal simulation. These are shown broken down by interaction mode for each sample in Figures~\ref{fig:FHCNumuPostfit}, \ref{fig:FHCNumuPostfit_theta}, \ref{fig:RHCNumuPostfit} and~\ref{fig:RHCNumuPostfit_theta}, along with the observed distributions. There is much better agreement with the data than for the prefit distributions shown in Figures~\ref{fig:FHCNumuPrefit}, \ref{fig:FHCNumuPrefit_theta}, \ref{fig:RHCNumuPrefit} and~\ref{fig:RHCNumuPrefit_theta}. The numbers of postfit predicted events for all the 14 samples are shown in Tab.~\ref{tab:ndrates}.

\begin{figure}[htbp]
\centering
\begin{subfigure}{0.47\textwidth}
  \includegraphics[width=0.98\columnwidth]{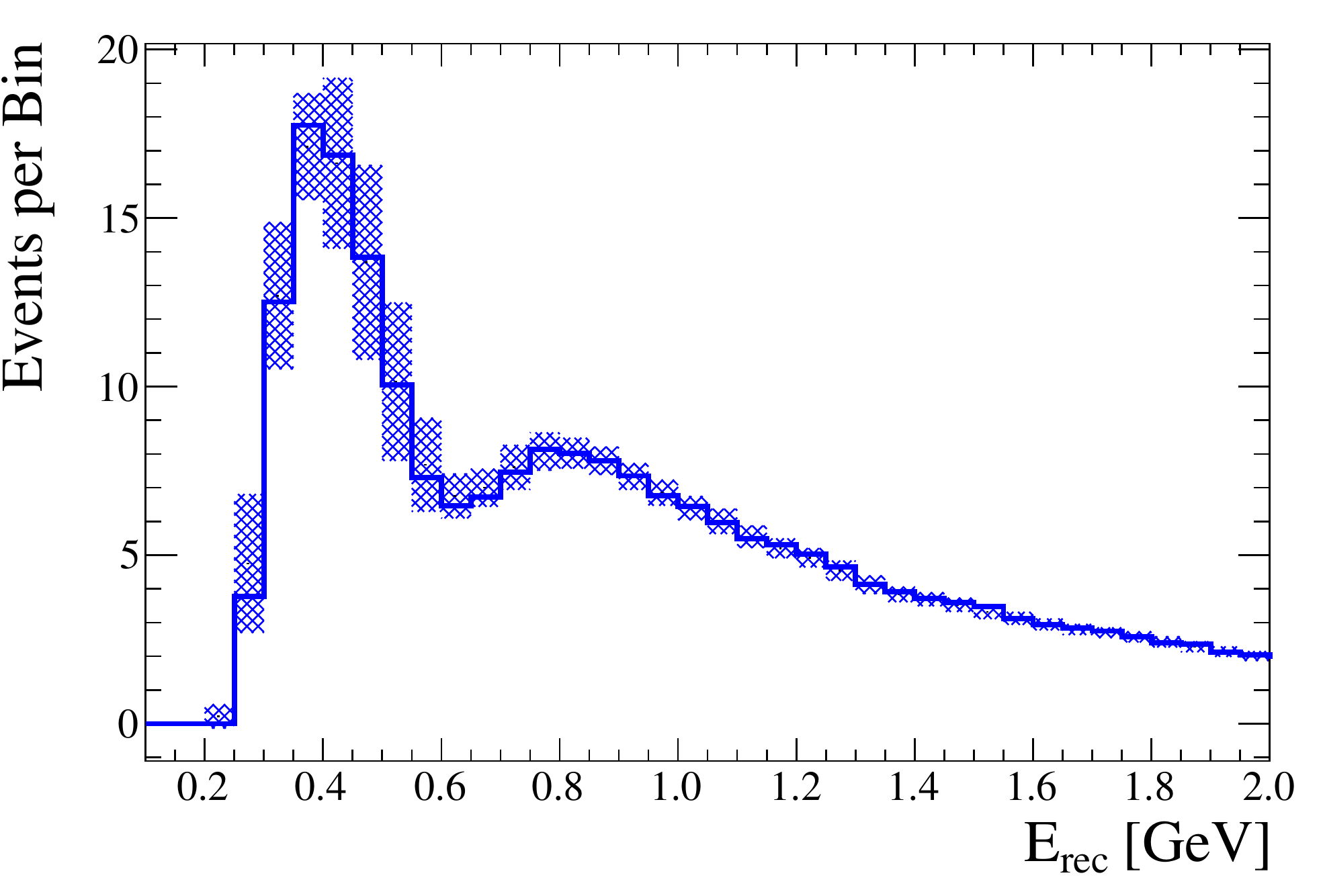}
\end{subfigure}
\begin{subfigure}{0.47\textwidth}
  \includegraphics[width=0.98\columnwidth]{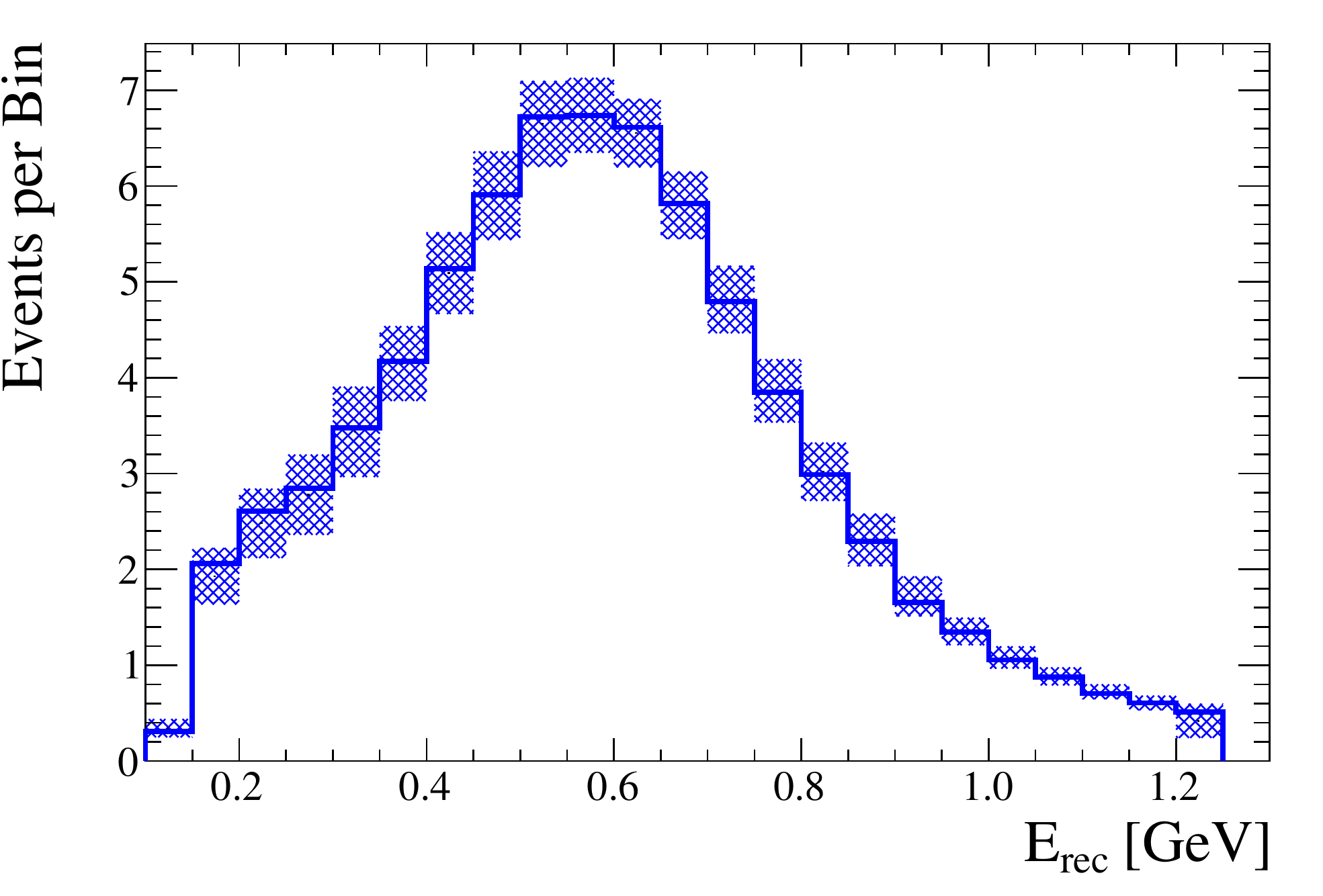}
\end{subfigure}
\begin{subfigure}{0.47\textwidth}
  \includegraphics[width=0.98\columnwidth]{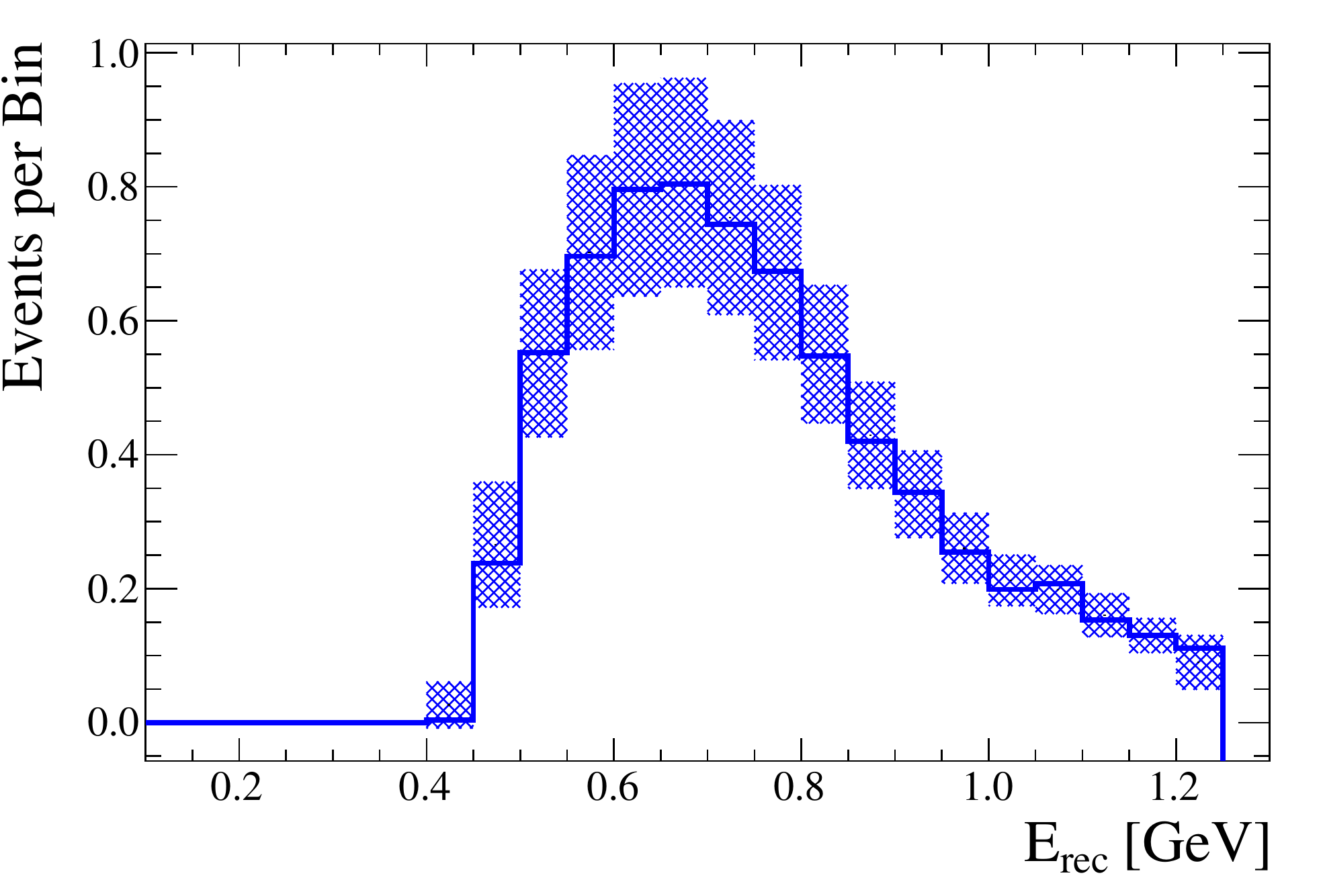}
\end{subfigure}
\caption{Post-ND280-fit predicted event spectrum for the FHC beam SK samples as a function of \erecqe.  The samples are: single-ring muon-like (top), single-ring electron-like without decay electrons (middle) and single-ring electron-like with a single decay electron (bottom).}
\label{fig:SKpredictionnu}
\end{figure}

\begin{figure}[htbp]
\centering
\begin{subfigure}{0.47\textwidth}
  \includegraphics[width=0.98\columnwidth]{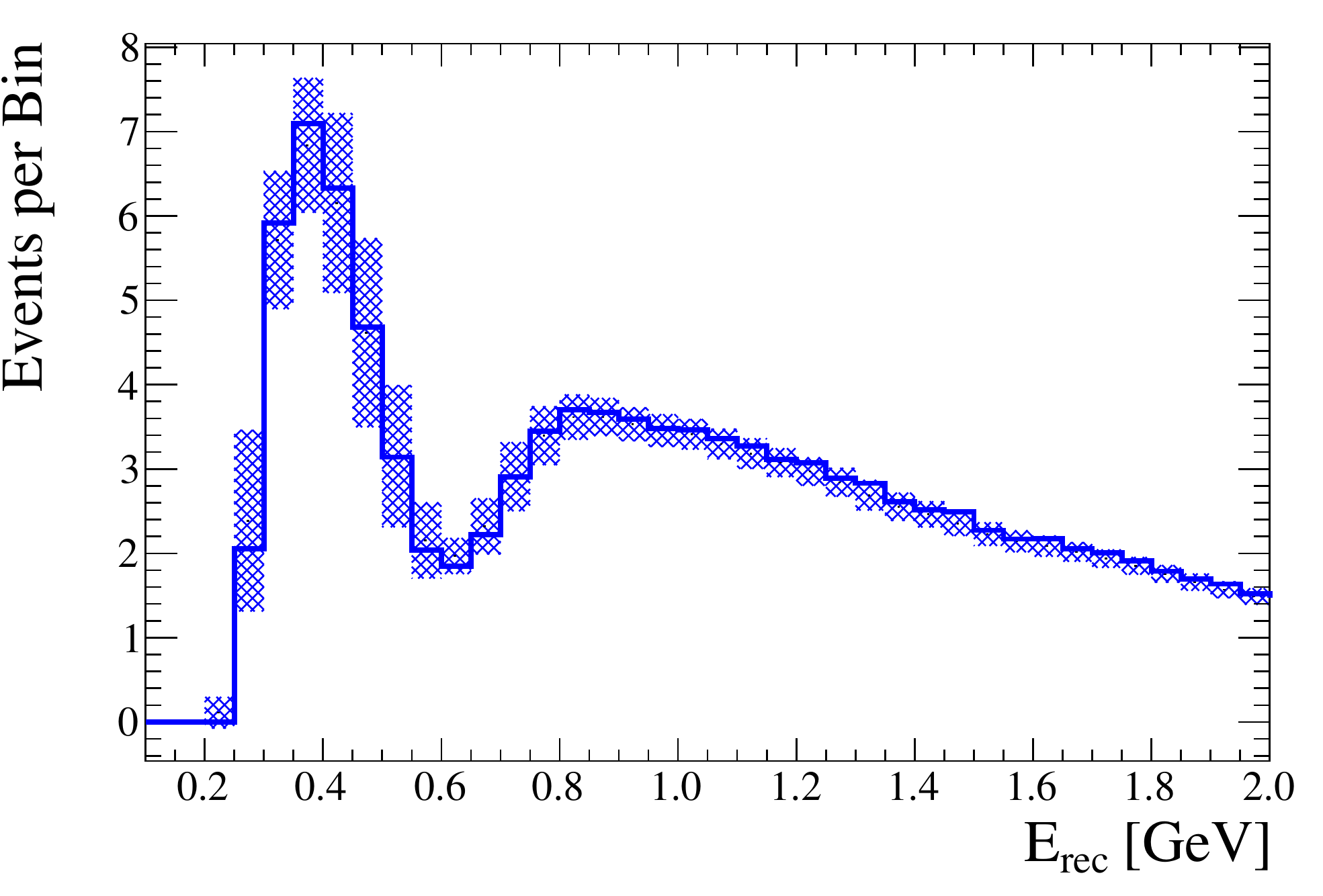}
\end{subfigure}
\begin{subfigure}{0.47\textwidth}
  \includegraphics[width=0.98\columnwidth]{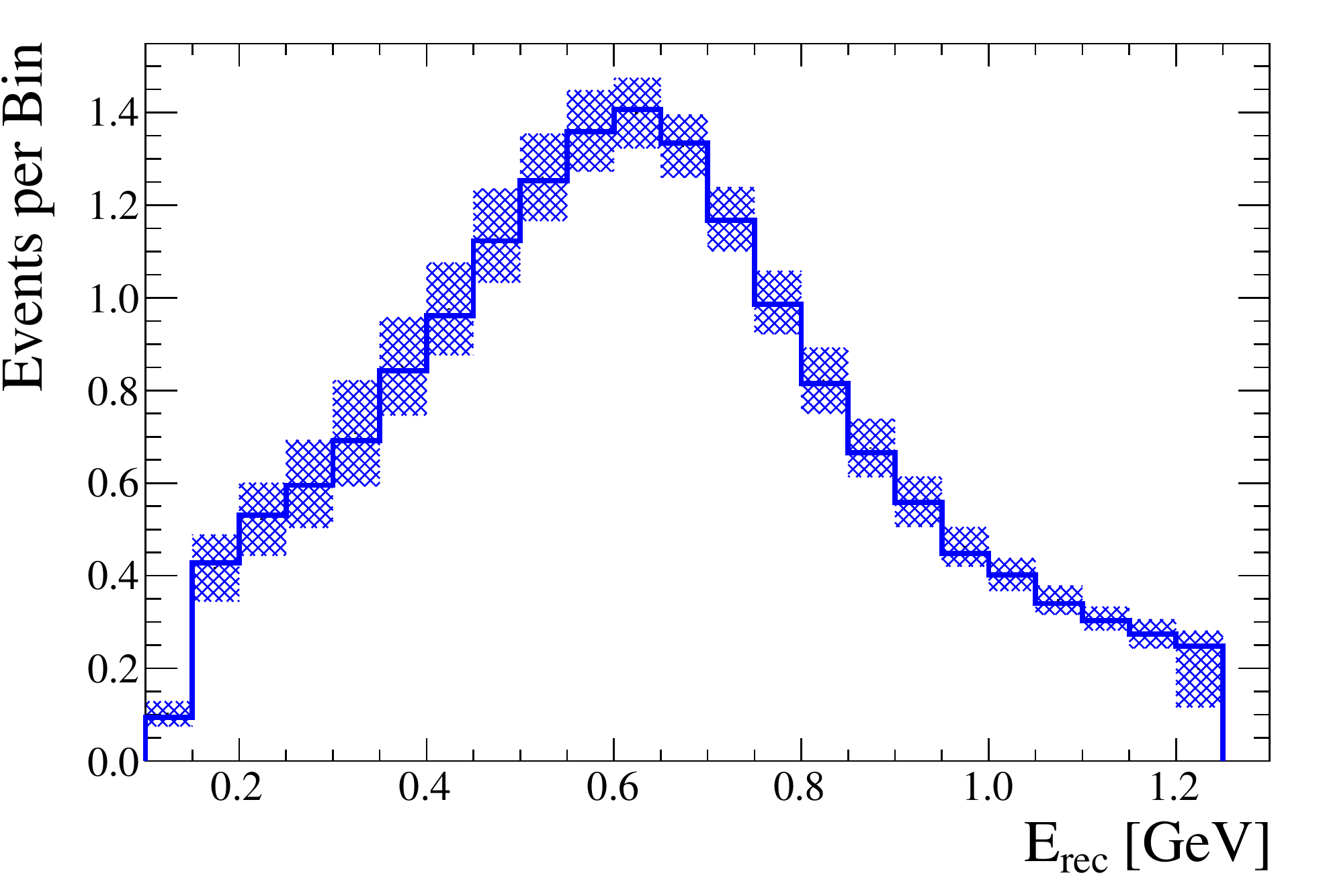}
\end{subfigure}
\caption{Post-ND280-fit predicted event spectrum for the RHC SK samples as a function of \erecqe.  The samples are: single-ring muon-like (top) and single-ring electron-like without decay electrons (bottom).}\label{fig:SKpredictionnub}
\end{figure}

The best-fit values are also used to produce SK event rate predictions, shown for the FHC 1R$_{\mu}$ and 1R${_e}$ SK samples in Figs.~\ref{fig:SKpredictionnu} and~\ref{fig:SKpredictionnub}. The total predicted and observed event rates are shown in Tab.~\ref{tab:postndfitratesSK}.

\begin{table}[htbp]
\sisetup{table-format=4.3}
\centering
\caption{Observed and predicted event rates for SK samples, pre- and post-ND280 fit.}
\label{tab:postndfitratesSK}
\begin{tabular}{l S[table-format=4.1] S r} 
\hline\hline
Sample & {Data} & {Prefit} & {ND280 Postfit}\\
\hline
FHC 1R$_{\mu}$ & 243 & 250.45 & 272.37 $\pm$ 6.05 \\
FHC 1R$_e$ & 75 & 64.17 & 72.79 $\pm$ 1.52 \\
FHC 1R$_{e}$ $+$ 1 d.e. & 15 & 7.80 & 6.87 $\pm$ 0.41\\
RHC 1R$_{\mu}$ & 140 & 130.53 & 139.42 $\pm$ 2.57\\
RHC 1R$_{e}$ & 15 & 15.73 & 16.77 $\pm$ 0.36\\
\hline\hline
\end{tabular}
\end{table}

\begin{figure}[htbp]
\centering
\begin{subfigure}{0.47\textwidth}
  \includegraphics[width=0.98\columnwidth]{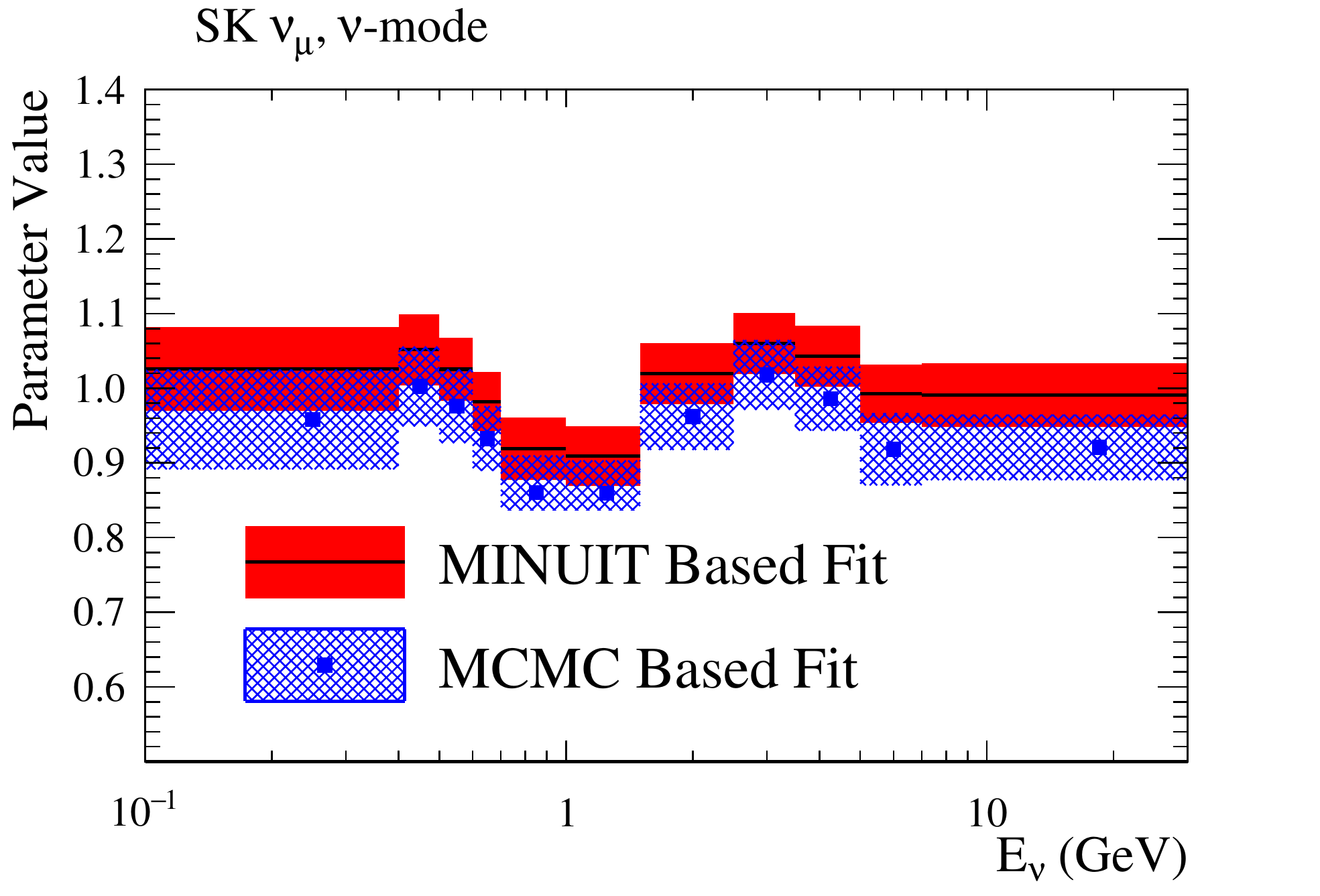} 
\end{subfigure}
\begin{subfigure}{0.47\textwidth}
  \includegraphics[width=0.98\columnwidth]{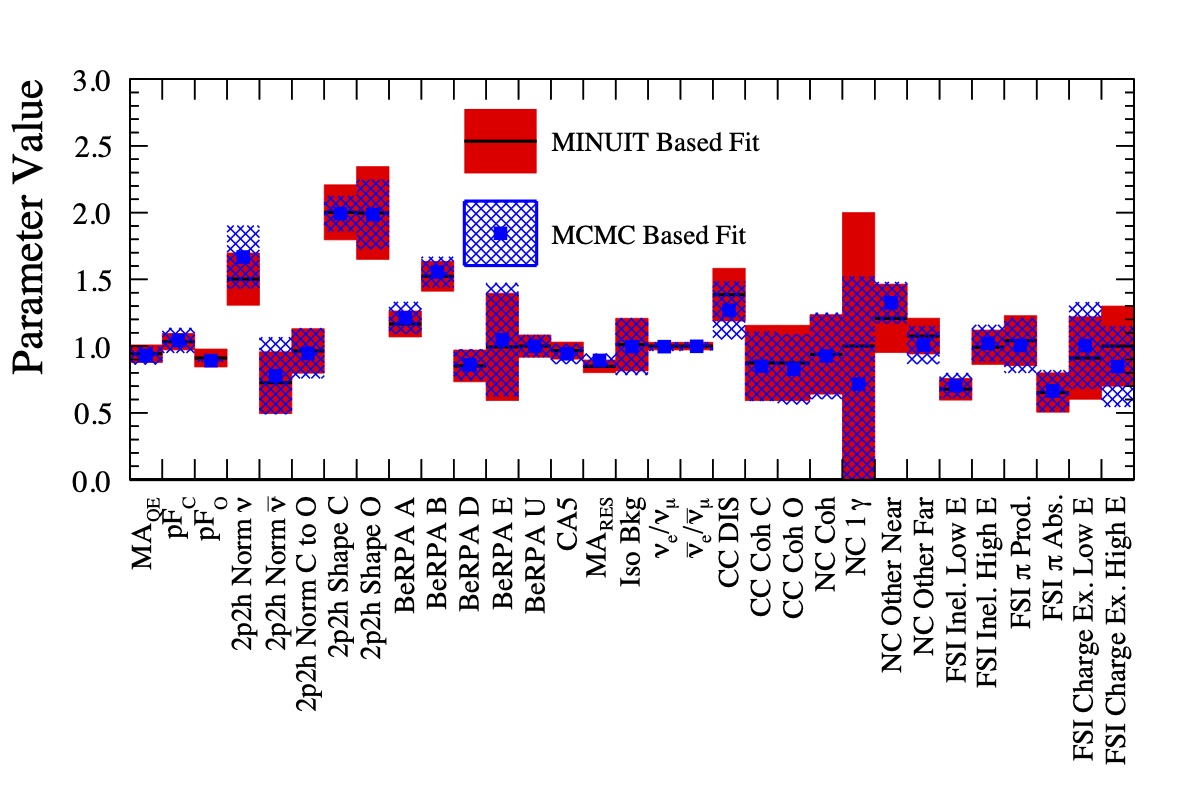}
\end{subfigure}
\caption{Comparisons of the fitted SK FHC flux parameters (top) and cross-section parameters (bottom) between the ND280 fit (red, solid) and the MCMC analysis (blue, hashed).}\label{fig:banffmach3postfit}
\end{figure}

Near detector only fits with the MCMC analysis framework were used to cross-check the two analyses. The postfit cross-section parameters for the two fits are compared in Fig.~\ref{fig:banffmach3postfit}, showing good agreement.

\section{Oscillation Analysis Fitters}
\label{sec:fitters}

To produce constraints on the three-flavor PMNS oscillation parameters, the rate and kinematic distributions of all five SK event samples are analyzed simultaneously. Systematic uncertainties in the flux, interaction and detector models are accounted for using systematic parameters applied as weights to the nominal prediction as described in~\cite{Abe:2017vif_T2Krun7osc}. Confidence regions and intervals are produced from marginal likelihood distributions as a function of parameters of interest.
    
The predicted kinematic distributions of each SK sample are generated from the nominal SK simulation to which weights are applied for each set of oscillation and systematic parameter values. The oscillation weights correspond to the three-flavor oscillation probability, calculated with matter effects and dependent on the true neutrino energy and flavor~\cite{Barger}, while the systematic weights are multiplicative factors. 
    
A likelihood, $L$, is calculated according to Eq.~\ref{eq:likelihood} as the product of the Poisson likelihood ratios for the number of events in each bin of the kinematic variables considered for each SK sample:

\begin{equation}
\label{eq:likelihood}
\begin{aligned}
    -2 \; \textrm{ln} \; L(\vec{o},\vec{f}) = 2 \sum_{s=0}^{5} \sum_{i=0}^{N_{s}-1} 
    &\Big( 
    n^{obs}_{s,i} \cdot \textrm{ln} \left( n^{obs}_{s,i} / n^{exp}_{s,i}  \right)\\
    &~+ \left(n^{exp}_{s,i}-n^{obs}_{s,i}\right)
    \Big).
\end{aligned}
\end{equation}

In Eq.~\ref{eq:likelihood}, $\vec{o}$ is a vector of the parameters of interest, $\vec{f}$ is a vector of nuisance parameters, $n^{obs}_{s,i}$ is the observed number of events in kinematic bin $i$ of SK sample $s$ which has a total of $N_s$ bins, $n^{exp}_{s,i} = n^{exp}_{s,i}(\vec{o},\vec{f})$ is the corresponding expected number of events.  The parameter(s) of interest correspond to one or more oscillation parameters among $\sin^{2} \theta_{13}$, $\Delta m^{2}_{32/31}$, $\sin^{2} \theta_{23}$, and \deltacp. The nuisance parameters correspond to the systematic parameters, and the oscillation parameters not chosen as parameters of interest in a given fit.   To obtain a likelihood which only depends on the parameter(s) of interest $\vec{o}$ and a data set $\vec{x}$ (the set of $n^{obs}_{s,i}$ in Eq.~\ref{eq:likelihood}), a marginal likelihood $L_{marg}$ is computed: the full likelihood, made of the product of the likelihood $L$ defined in Eq.~\ref{eq:likelihood} with the prior constraint on some of the parameters $\pi \left(\vec{o} ,\vec{f} \right)$, is numerically integrated over the nuisance parameters:
\begin{equation} \label{eq:marginal_likelihood}
    L_{\text{marg}} = L_{\text{marg}} \left( \vec{o}; \vec{x} \right)
    = \int L \left( \vec{o}, \vec{f}; \vec{x} \right) \pi \left(\vec{o} ,\vec{f} \right) \ud\mspace{-2mu} \vec{f}.
\end{equation}
    
External constraints are used for some of the oscillation parameters. The solar parameters, which have limited impact on the observed event distributions at T2K, are either kept fixed to their nominal values or have a Gaussian constraint applied, depending on the analysis considered. The nominal values and uncertainties used are $\sin^{2} 2 \theta_{12} = 0.846 \pm 0.021$ and $\Delta m^{2}_{21} = (7.53 \pm 0.18) \times 10^{-5}~\mbox{eV}^2 /\mbox{c}^4$~\cite{pdg_2014}. The three parameters $\sin^{2} 2 \theta_{23}$, $\Delta m^{2}_{23}$ and $\deltacp$ are unconstrained. T2K is sensitive to $\sin^{2} \theta_{13}$, but to date, the world's most accurate measurements of this parameter come from reactor neutrino experiments~\cite{Adey:2018zwh, Bak:2018ydk, Abe:2015rcp}. To obtain increased sensitivity to the other oscillation parameters, the reactor average of $\sin^{2} 2 \theta_{13} = 0.0830 \pm 0.0031$~\cite{pdg_2018} may be used as a Gaussian constraint, which will subsequently be referred to as the ``reactor constraint". It is also useful to evaluate the T2K experiment's constraint of the oscillation parameters without the contribution of the reactor experiments, and fits without this constraint on $\sin^{2} 2 \theta_{13}$ were also performed.
    
Three different analysis frameworks are used to perform the fit of the far detector data. They will be labeled as A, B and C, and follow the general procedure described above, but differ on a number of points summarized in Tab.~\ref{table:AnalysisDiff}:
\paragraph{Kinematic information used to fit the data from the electron-like samples} Two-dimensional distributions are used for those samples, either the combination of the momentum and angle with respect to the beam direction of the particle reconstructed as the lepton ($p_\mathrm{lep}$, $\theta_\mathrm{lep}$), or the reconstructed energy assuming CCQE kinematics (E$_\mathrm{rec}$) combined to this angle $\theta_\mathrm{lep}$.

\paragraph{Oscillation probability calculation} The events can either be binned in true neutrino energy with an oscillation probability corresponding to the mean true neutrino energy of the bin, or have individual oscillation probabilities computed for each event's true neutrino energy.

\paragraph{Use of the near detector data} The near detector data are either used in a simultaneous fit with the far detector data, or they can be fit separately to constrain the neutrino flux and interaction systematic uncertainties. In this second case, the constraint is propagated to the far detector analysis through a covariance matrix. The two methods are expected to lead to different results for the far detector fit if the constraint on the systematic parameters obtained with the near detector data cannot be properly described by a multi-variate normal distribution.

\paragraph{Fitting method} Two of the analyses use a grid search method, where the likelihood in terms of the parameters of interest is computed for different fixed values of those parameters while marginalizing over the other parameters. This marginalization over the nuisance parameters is done through numerical integration. The last analysis uses an MCMC method to sample the parameter space, where the density of the obtained samples follows the joint posterior probability density of the parameters. Those samples can then be binned to produce 1D or 2D posterior distributions for the parameter(s) of interest, effectively marginalizing over the nuisance parameters. A full description of the MCMC analysis can be found in~\cite{2020asztucthesis}. 
    
\begin{table}[htbp!]
\newcommand{\subcellw}[2]{\begin{tabular}{>{\centering\arraybackslash}p{#1\columnwidth}}#2\end{tabular}}
\centering
\caption{Differences between the three far detector analyses.} \label{table:AnalysisDiff}
\begin{tabular}{cccc}
\hline \hline
\multicolumn{1}{c}{}& A & B & C\\
\hline 
\subcellw{0.22}{$e$-like sample analysis bins}
    & ($E_\mathrm{rec}$, $\theta_\mathrm{lep}$) & ($E_\mathrm{rec}$, $\theta_\mathrm{lep}$) & ($p_\mathrm{lep}$, $\theta_\mathrm{lep}$)\\[2ex]
\subcellw{0.22}{ $P(\nu\rightarrow\nu)$ calculation} 
    & Binned & \subcellw{0.2}{Event-by-event} & Binned \\[2ex]
\subcellw{0.22}{ND280 constraint}
    & \subcellw{0.2}{Covariance matrix} & \subcellw{0.2}{Simultaneous fit}  & \subcellw{0.2}{Covariance matrix} \\[2ex]
\subcellw{0.22}{Fitting method}
    & Grid search & MCMC & Grid search \\
\hline \hline
\end{tabular}
\end{table}
    
\begin{figure*}[htbp]
\centering
\begin{subfigure}[b]{0.49\textwidth}
\includegraphics[width=0.9\textwidth]{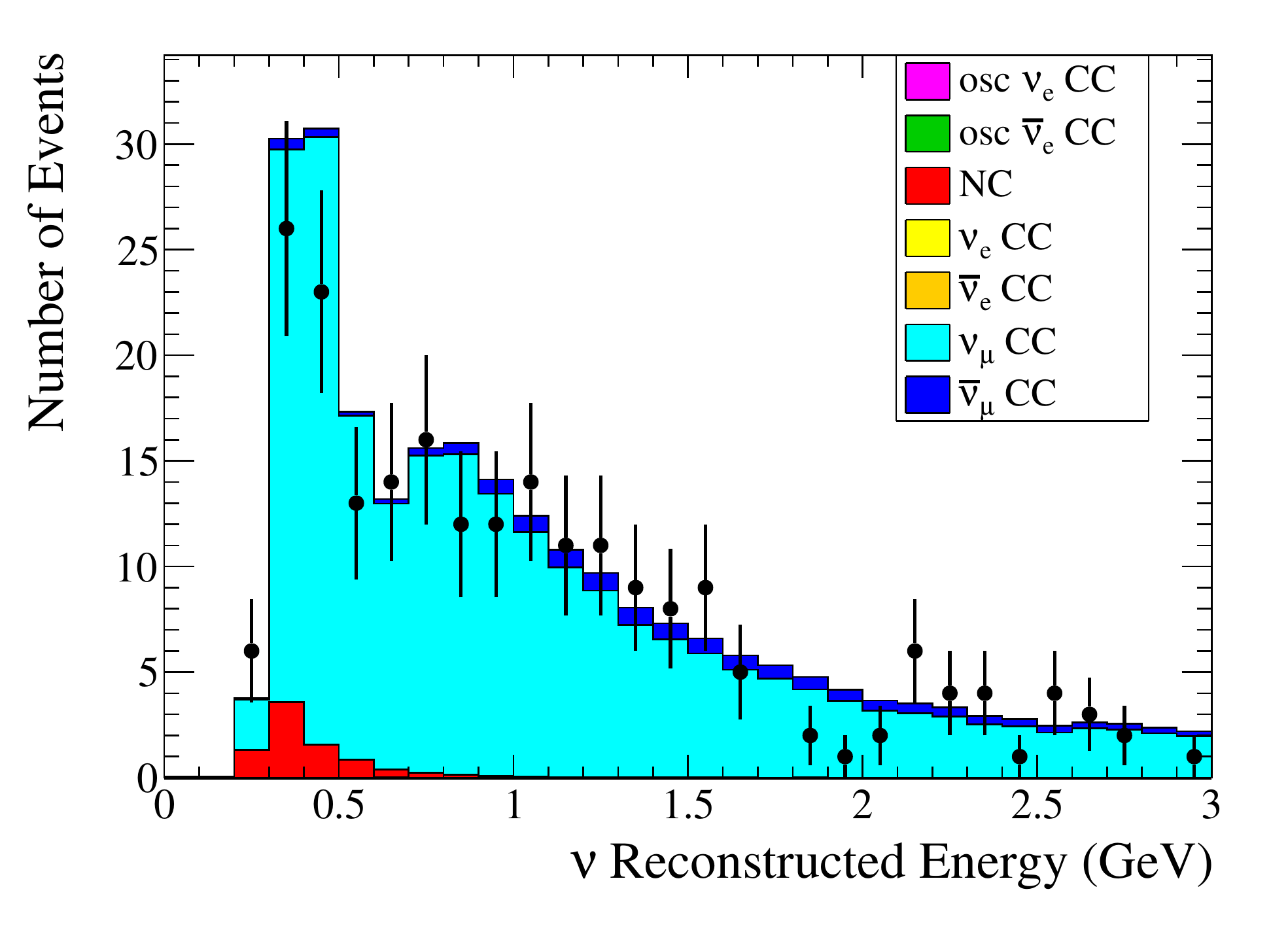}
\caption{FHC 1-Ring $\mu$-like}
\label{fig:spectra_FHC_1rmu}
\end{subfigure}
\begin{subfigure}[b]{0.49\textwidth}
\includegraphics[width=0.9\textwidth]{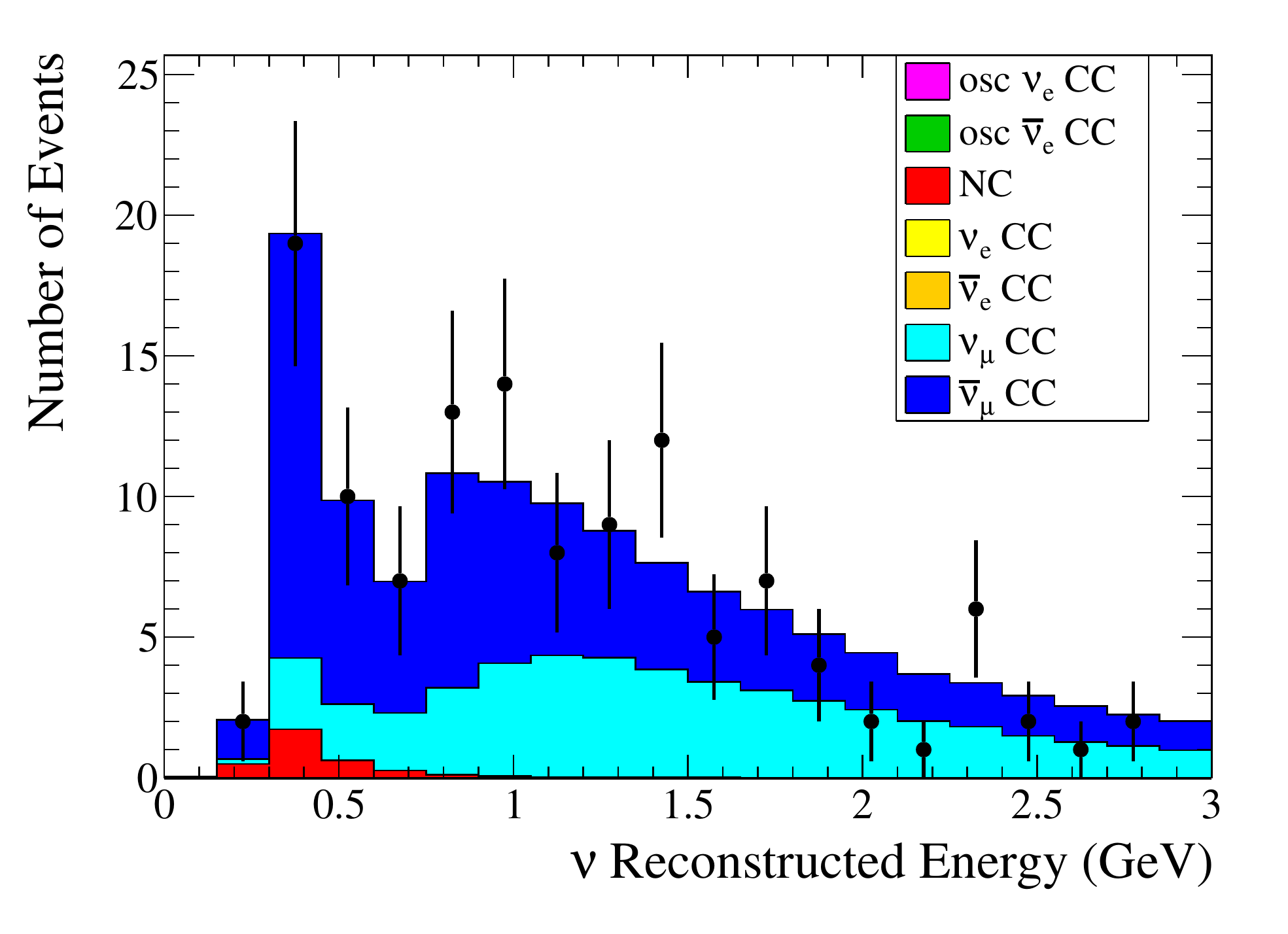}
\caption{RHC 1-Ring $\mu$-like}
\label{fig:spectra_RHC_1rmu}
\end{subfigure}
\\
\begin{subfigure}[b]{0.49\textwidth}
\includegraphics[width=0.9\textwidth]{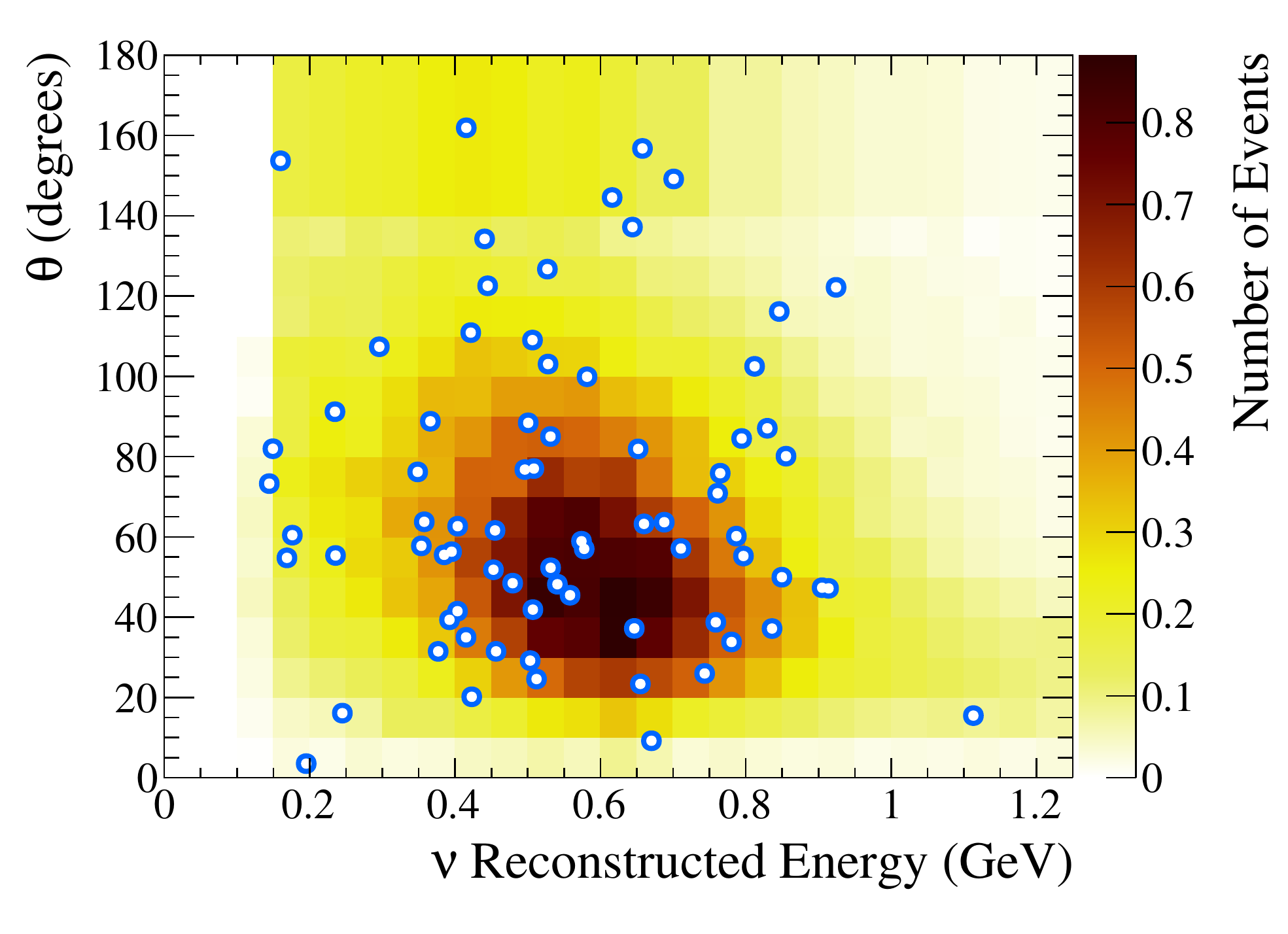}
\caption{FHC 1-Ring e-like}
\label{fig:spectra_FHC_1re}
\end{subfigure}
\begin{subfigure}[b]{0.49\textwidth}
\includegraphics[width=0.9\textwidth]{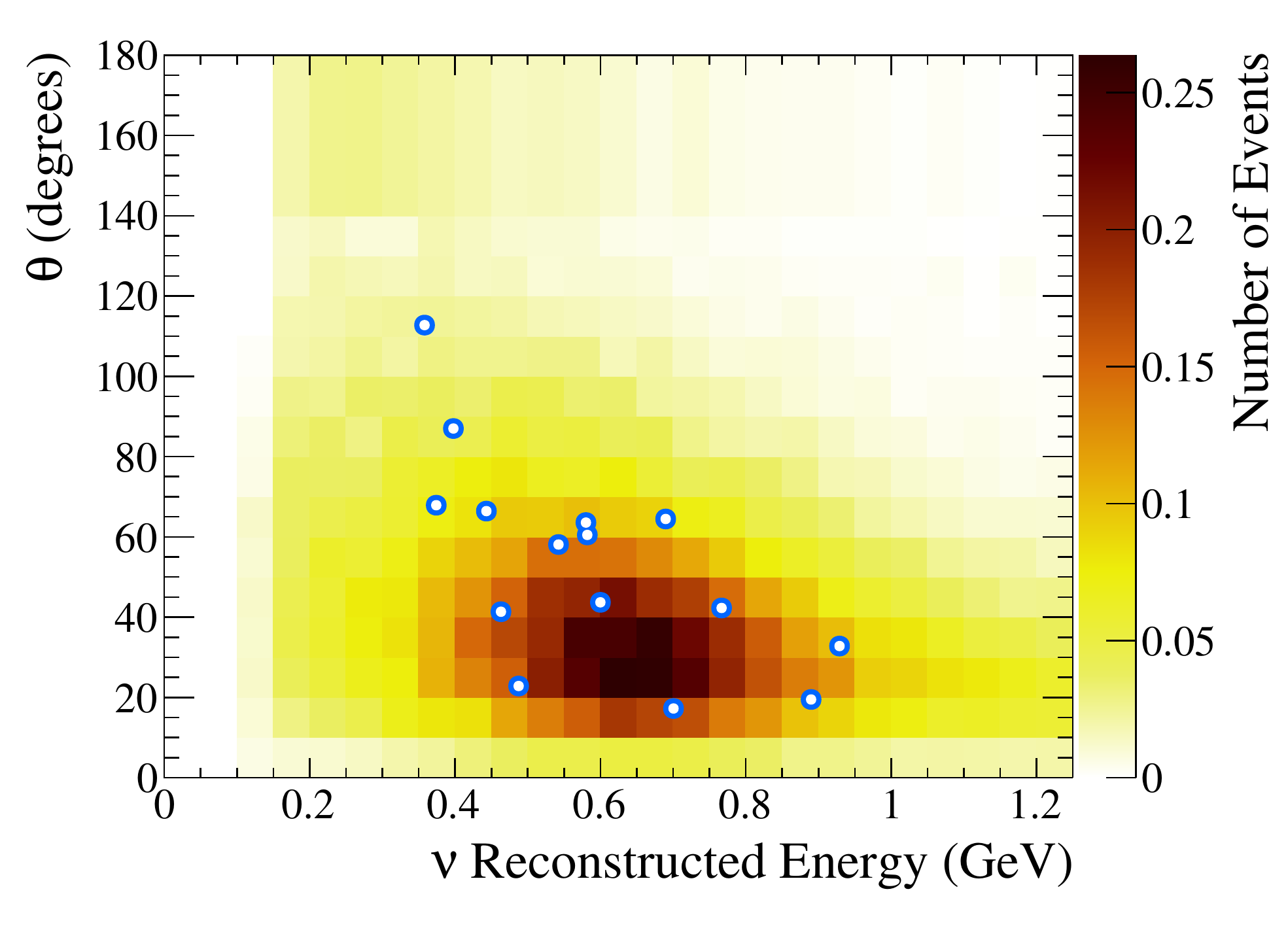}
\caption{RHC 1-Ring e-like}
\label{fig:spectra_RHC_1re}
\end{subfigure}
\\
\begin{subfigure}[b]{0.49\textwidth}
\includegraphics[width=0.9\textwidth]{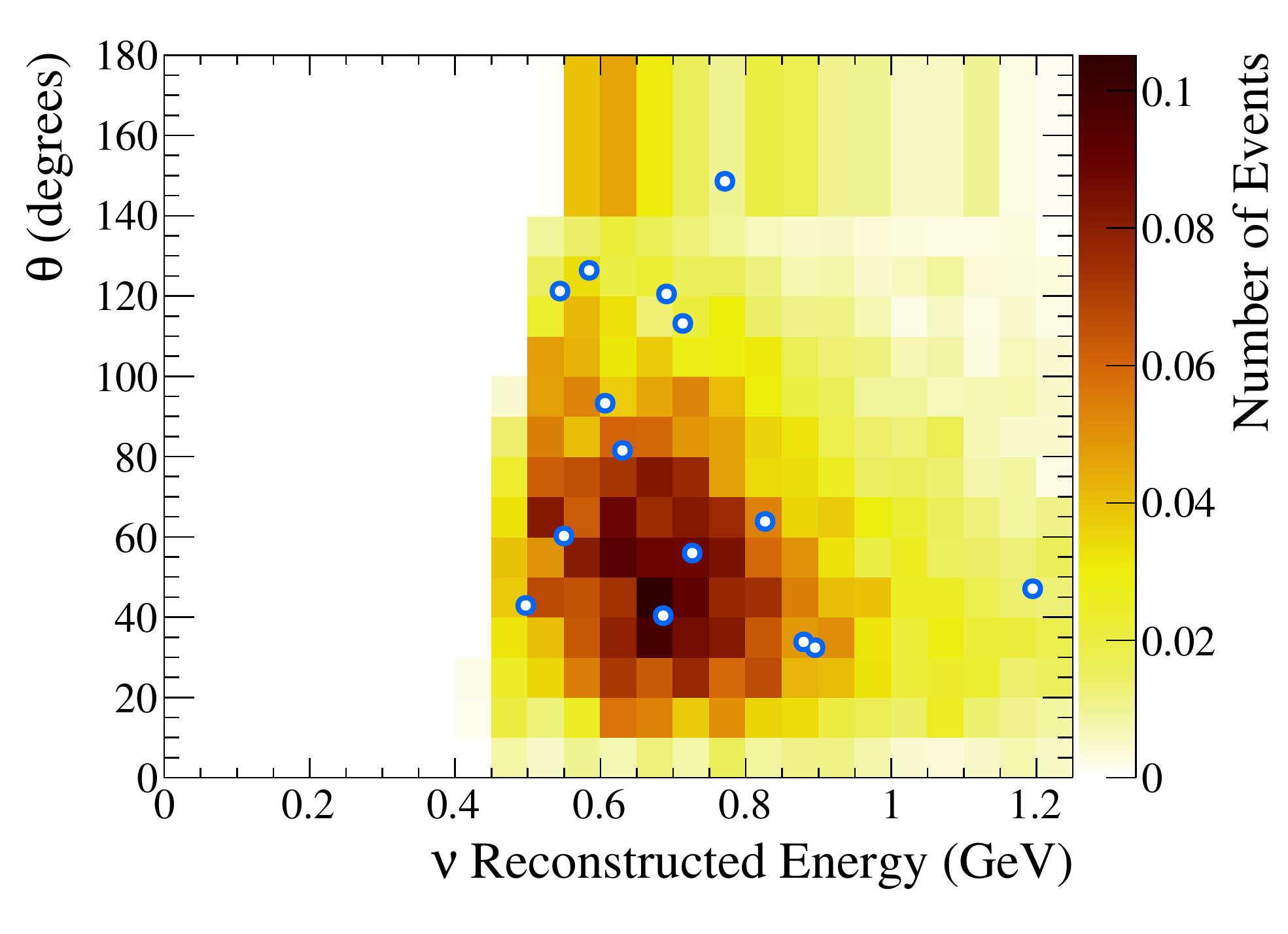}
\caption{FHC 1-Ring e-like + 1 d.e }
\label{fig:spectra_FHC_multire}
\end{subfigure}
\caption {
    Observed kinematic distributions compared to the expectations generated with oscillation parameters set to the values in Tab.~\ref{tab:asimova_params} for the different samples. The uncertainty shown around the data points in (a) and (b) accounts for statistical uncertainty only. The uncertainty range is chosen to include all points for which the measured number of data events is inside the 68\% confidence interval of a Poisson distribution centered at that point.
}
\label{fig:spectra}
\end{figure*}

\begin{figure}[htbp]
\centering
\includegraphics[width=0.47\textwidth]{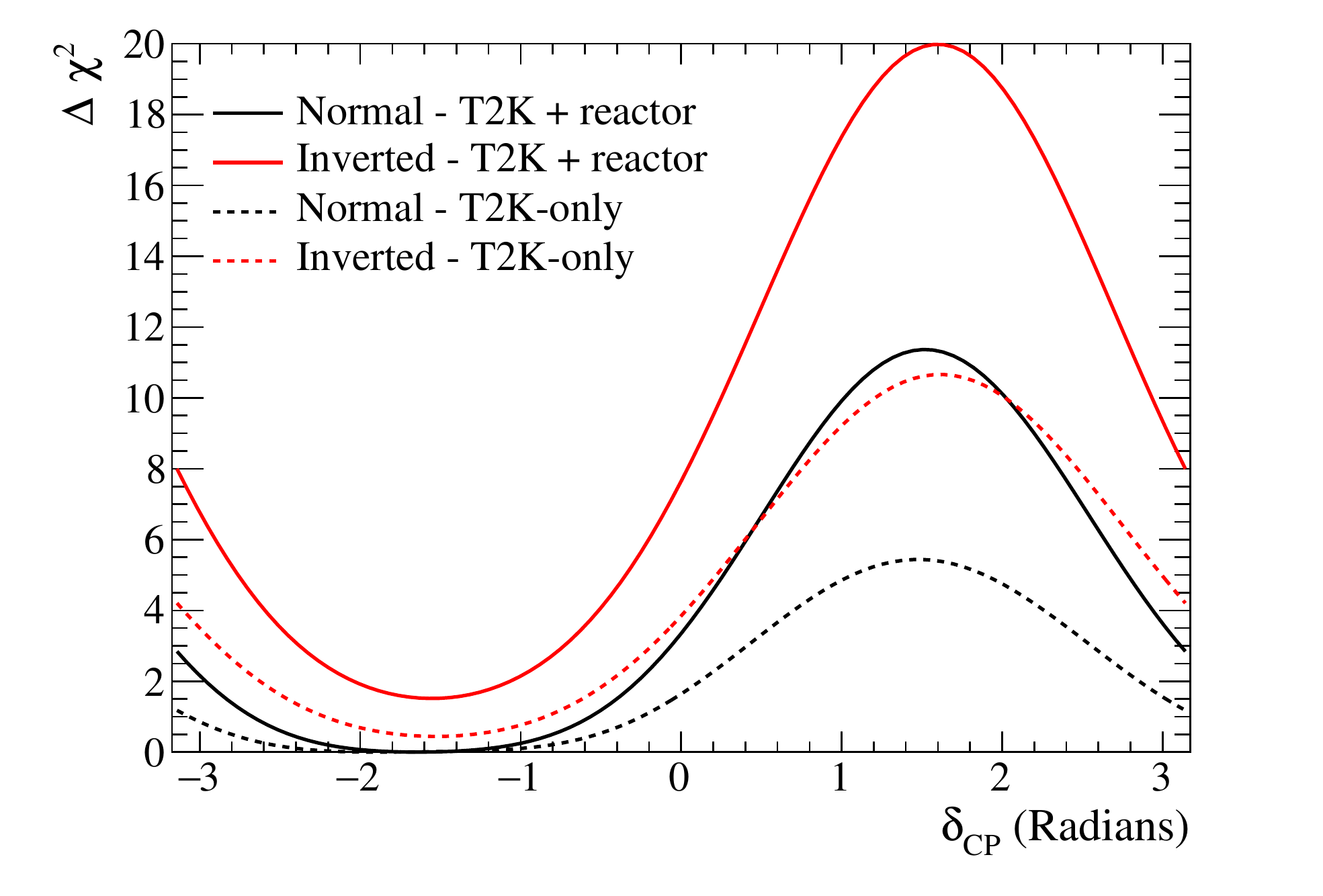}
\caption {
    The predicted $\Delta \chi^2=-2\ln \left[L/L_{max}\right]$ function as a function of $\deltacp$ with and without reactor constraint, for both mass orderings.
}
\label{fig:dcp_asimova}
\end{figure}

\begin{figure}[htbp]
\centering
\includegraphics[width=0.47\textwidth]{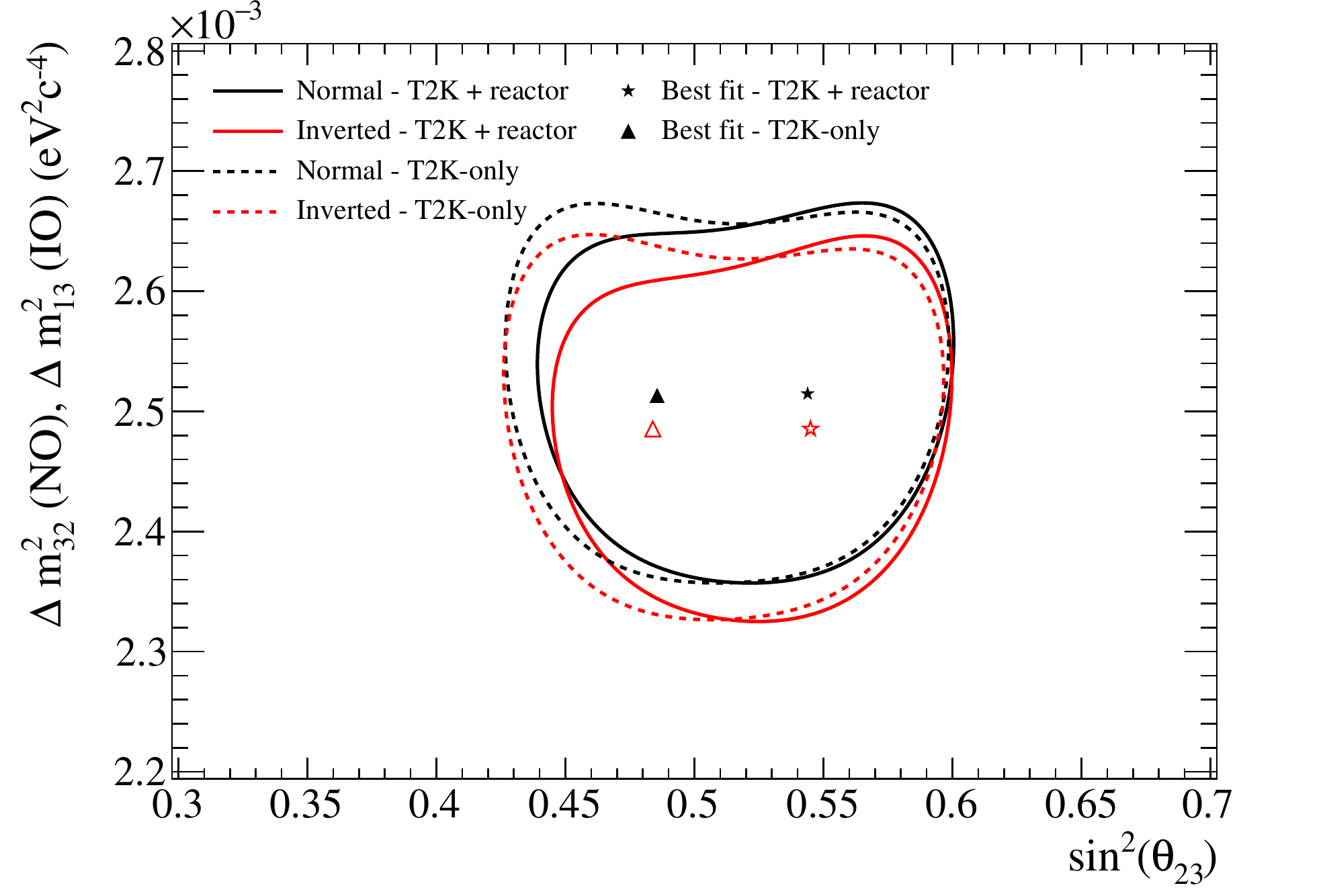}
\caption {
    Predicted 90\% CL regions for $\Delta m^{2}$ vs. $\sin^{2} \theta_{23}$ with and without the reactor constraint, for both mass orderings. Normal and inverted mass ordering contours are independent.
}
\label{fig:mixing23_dm23sq_asimova}
\end{figure}

The T2K experiment's expected median constraints on the oscillation parameters can be evaluated by fitting to an Asimov data set, generated for the true values of the oscillation parameters given by Tab.~\ref{tab:asimova_params}, and  nominal values of the systematic parameters. Using Analysis A, the expected sensitivity to $\deltacp$ is shown in Fig.~\ref{fig:dcp_asimova}, and for $\Delta m^2$ vs. $\sin^2 \theta_{23}$ in Fig.~\ref{fig:mixing23_dm23sq_asimova}. The predicted and observed event rates in each of the five SK samples are shown in Tab.~\ref{tab:events_asimov_a_dcp_var} at various  $\deltacp$ values, with other oscillation parameters fixed to the values in Tab.~\ref{tab:asimova_params} and with all systematic parameters set to their nominal values. Similarly, the predicted and observed kinematic distributions are shown in Fig.~\ref{fig:spectra}, with all oscillation parameters fixed to the values in Tab.~\ref{tab:asimova_params}.

\begin{table}[htbp]
\caption{
	Predicted and observed total number of events in each Super-K sample. The oscillation parameters other than $\deltacp$ were fixed to the values given in Table~\ref{tab:asimova_params} and all systematic parameters were set to their nominal values.
}
\label{tab:events_asimov_a_dcp_var}
\centering
\sisetup{table-format=3.1}
\begin{tabular} {  c@{\quad} S @{\quad} S @{\quad} S @{\quad} S @{\quad} S  }
	\hline
	\hline
	\multirow{2}{*}{$\deltacp$} &  {~FHC} & {~RHC} & {FHC} & {RHC} & {FHC \nue} \\
    & {$\mu$-like} & {$\mu$-like} & {$e$-like} & {$e$-like} & {CC $1\pi$}\\
    \hline
    $-\pi/2\phantom{-}$ & 272.4 & 139.5 &  74.4 &  17.1 &  7.0\\
    $0$      & 272.0 & 139.2 &  62.2 &  19.6 &  6.1\\ 
    $\pi/2$  & 272.4 & 139.5 &  50.6 &  21.7 &  4.9\\ 
    $\pi$    & 272.8 & 139.9 &  62.7 &  19.3 &  5.9\\
    \hline
    Observed & 243   & 140   &  75   &  15   & 15  \\
	\hline
	\hline
\end{tabular}
\end{table}

\begin{figure}[htbp]
\centering
\begin{subfigure}{0.47\textwidth}
\includegraphics[width=\columnwidth]{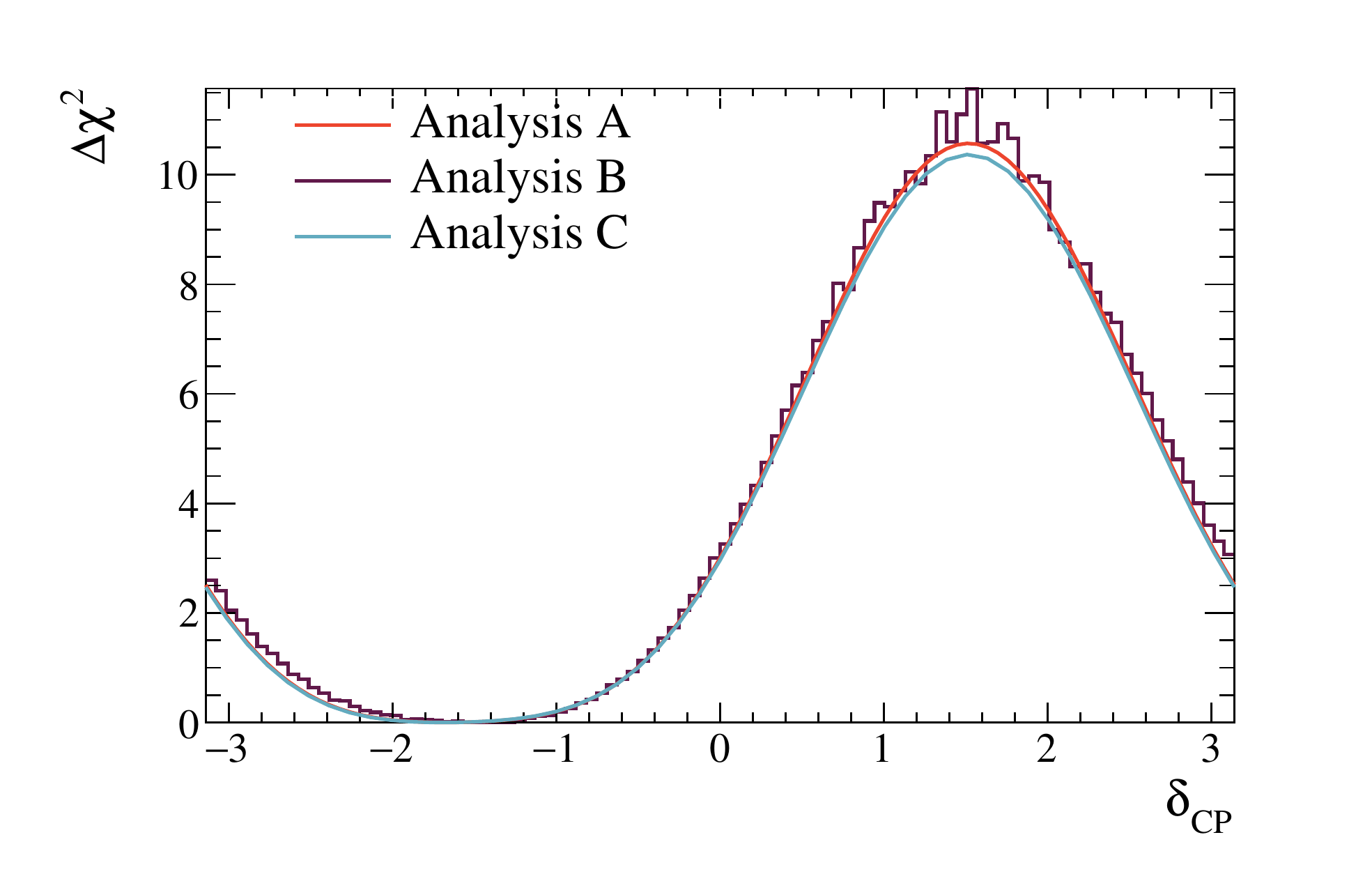}
\end{subfigure}
\begin{subfigure}{0.47\textwidth}
\includegraphics[width=\columnwidth]{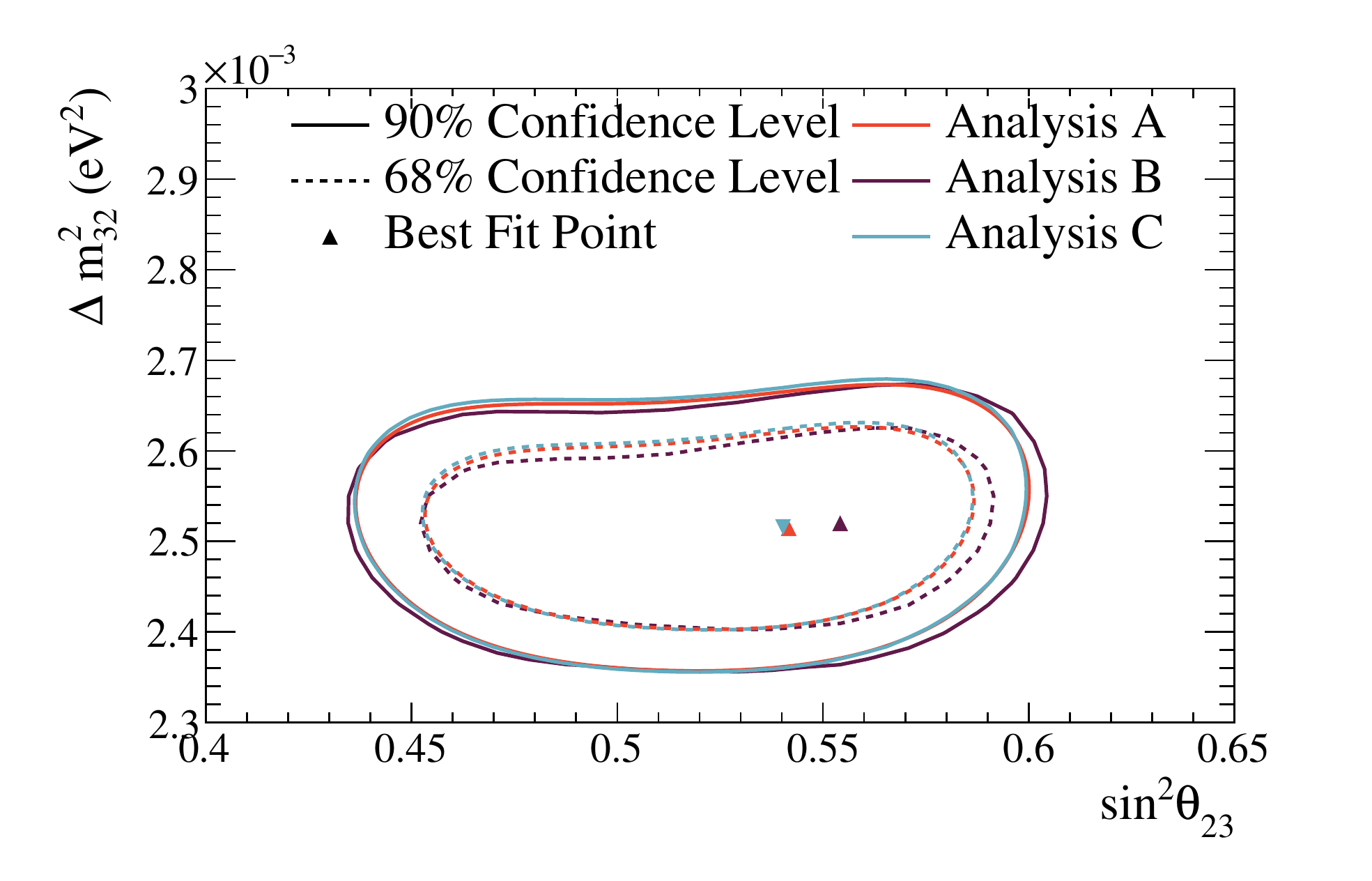}
\end{subfigure}
\caption {
    Comparison between analyses A, B and C. The normal mass ordering is assumed and the reactor constraint is applied. Top: the predicted $\Delta \chi^2$ functions for $\deltacp$. Bottom: the $1\sigma$ (dashed) and 90\% (solid) CL regions for $\Delta m^{2}$ vs. $\sin^{2} \theta_{23}$.
}
\label{fig:group_comparison_asimova}
\end{figure}

Using the differences between the three analyses, it can be checked that the results are not too sensitive  to  the   details  of  how  the  analysis  was performed. The expected sensitivities obtained by each analysis for $\deltacp$ and $\Delta m^2$ vs. $\sin^2 \theta_{23}$ are shown in Fig.~\ref{fig:group_comparison_asimova}. They show good agreement between each analysis, thus the differences in the analysis methods do not significantly impact the results. Additional comparisons between the three analyses were performed for each combination of parameter(s) of interest, mass ordering, use of reactor constraint. In each of these, the results produced by the three analyses were found to be consistent.


\section{Simulated Data Studies}
\label{sec:fake_data_studies}

Equation~\ref{eq:osc_prob} shows that the neutrino oscillation probability depends upon the energy of the neutrino and its path length from creation to interaction.
In long-baseline accelerator-based neutrino experiments the neutrino path length is fixed.
Accurately measuring the neutrino oscillation parameters therefore requires a precise understanding of the neutrino energy spectrum.
All neutrino experiments use models to link the observed final states back to the initial neutrino energy.
There is no set of models that describe the world’s neutrino data but there are a number of models that are in comparable agreement to the world’s data, as described in Ref.~\cite{ALVAREZRUSO20181}.  However, these models map true neutrino energy to reconstructed neutrino energy in different ways.
This is shown in Fig.~\ref{fig:qel_erec_bias},  where the bottom plot compares the reconstructed energy bias between quasi-elastic-like and $\Delta$-like 2p2h interactions.
The choice of interaction model therefore affects the neutrino energy distribution that experiments infer from their observed neutrino events.
This in turn can affect their measurement of the neutrino oscillation parameters, as has been shown in Ref.~\cite{PhysRevD.89.073015}.

In this analysis a comprehensive set of neutrino interaction models have been tested using simulated data studies to quantify their effect on the T2K oscillation result.
The simulated data procedure is described in Ref.~\cite{Abe:2017vif_T2Krun7osc}, while the model changes that are tested here are described in Section~\ref{sec:interaction_model}.
Simulated data is created for both the near and far detectors, where a fit is performed in the same way as for the real data.
The resultant oscillation parameter contours are then compared to those extracted from a fit to the Asimov data set.
If the T2K oscillation analysis is insensitive to the model change, or has the freedom to account for it correctly, then the simulated data contours and the Asimov contours should be very similar.
The oscillation parameter values used for this study are shown in Tab.~\ref{tab:asimova_params}.
Other parameter sets with \deltacp$=0$ and \ssqthtwothree$= 0.45$ were also studied, but showed no significant difference to the results presented here.

The likelihood distributions for each oscillation parameter are created for both the simulated data fit and the Asimov fit.
Any change in the center of the $2\sigma$ confidence interval for each parameter is taken as a bias due to the change in the interaction model introduced to the simulated data.
This is compared to the uncertainty on the parameter coming from the systematic uncertainties included in the analysis.
If a simulated data bias is greater than 50\% of the systematic uncertainty on a parameter then an additional uncertainty is added to the analysis to account for this.

\subsection{Simulated data study of the nucleon removal energy}

As described in Section~\ref{sec:interaction_model} the T2K neutrino event generator, NEUT 5.3.3, implements a Relativistic Fermi Gas (RFG) nuclear model.
To remove a nucleon from the nucleus requires energy to overcome the nuclear potential. The nucleon removal energy (NRE), can be measured by electron scattering experiments, but is not known perfectly. Even the definition of this quantity is not simple~\cite{Bodek2019}.
In addition, the RFG is a very simple model of the nuclear structure.
More advanced models, such as spectral function models, provide a much more detailed description of the nucleon energies within the nucleus, each of which has its own NRE.

\begin{figure}[htbp]
	\centering
	\includegraphics[width=0.47\textwidth]{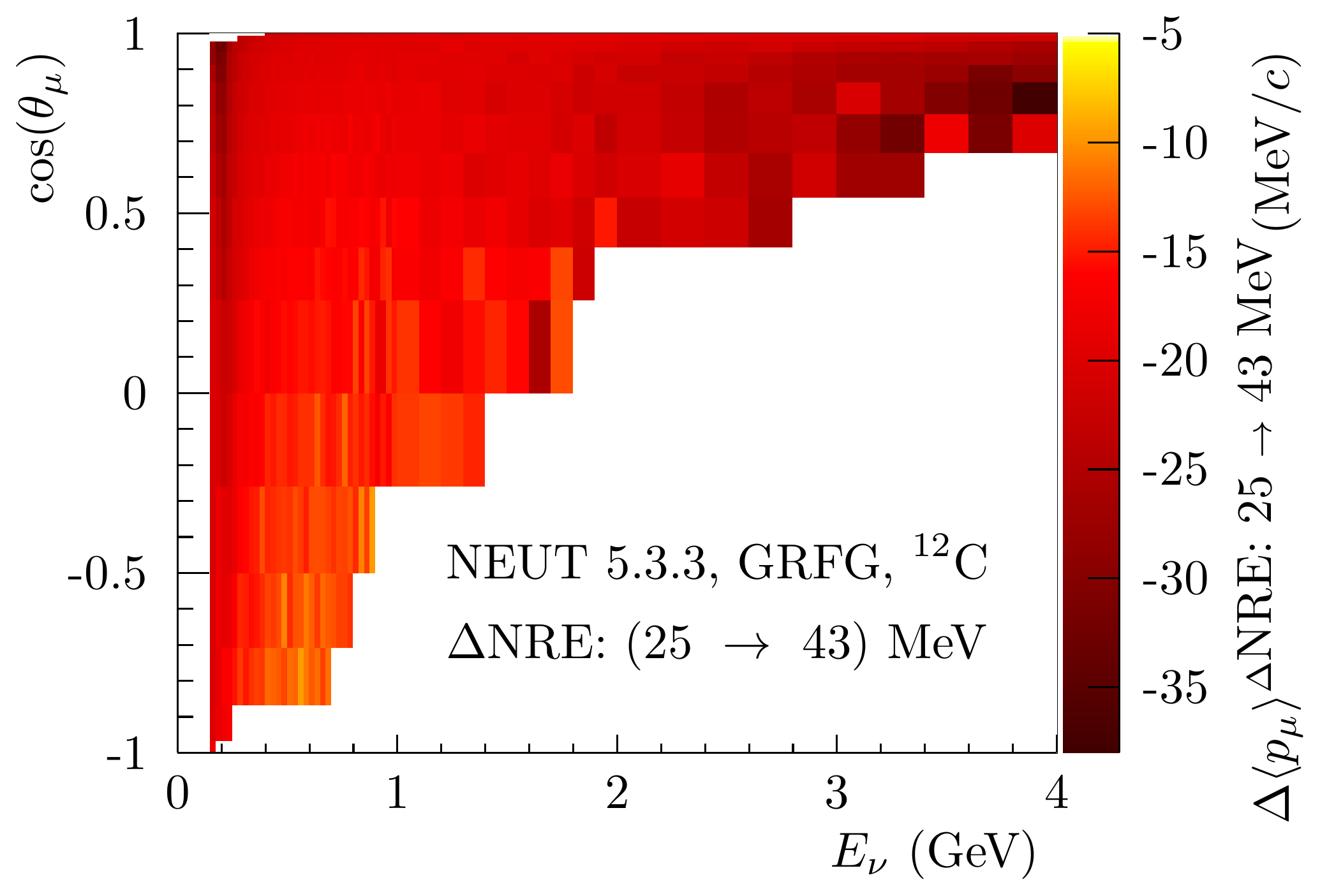}
	\caption {The momentum shift caused by increasing the nucleon removal energy in NEUT 5.3.3 by 18\,MeV.}
	\label{fig:nre}
\end{figure}

\begin{figure}[htbp]
	\center
	\includegraphics[width=0.47\textwidth]{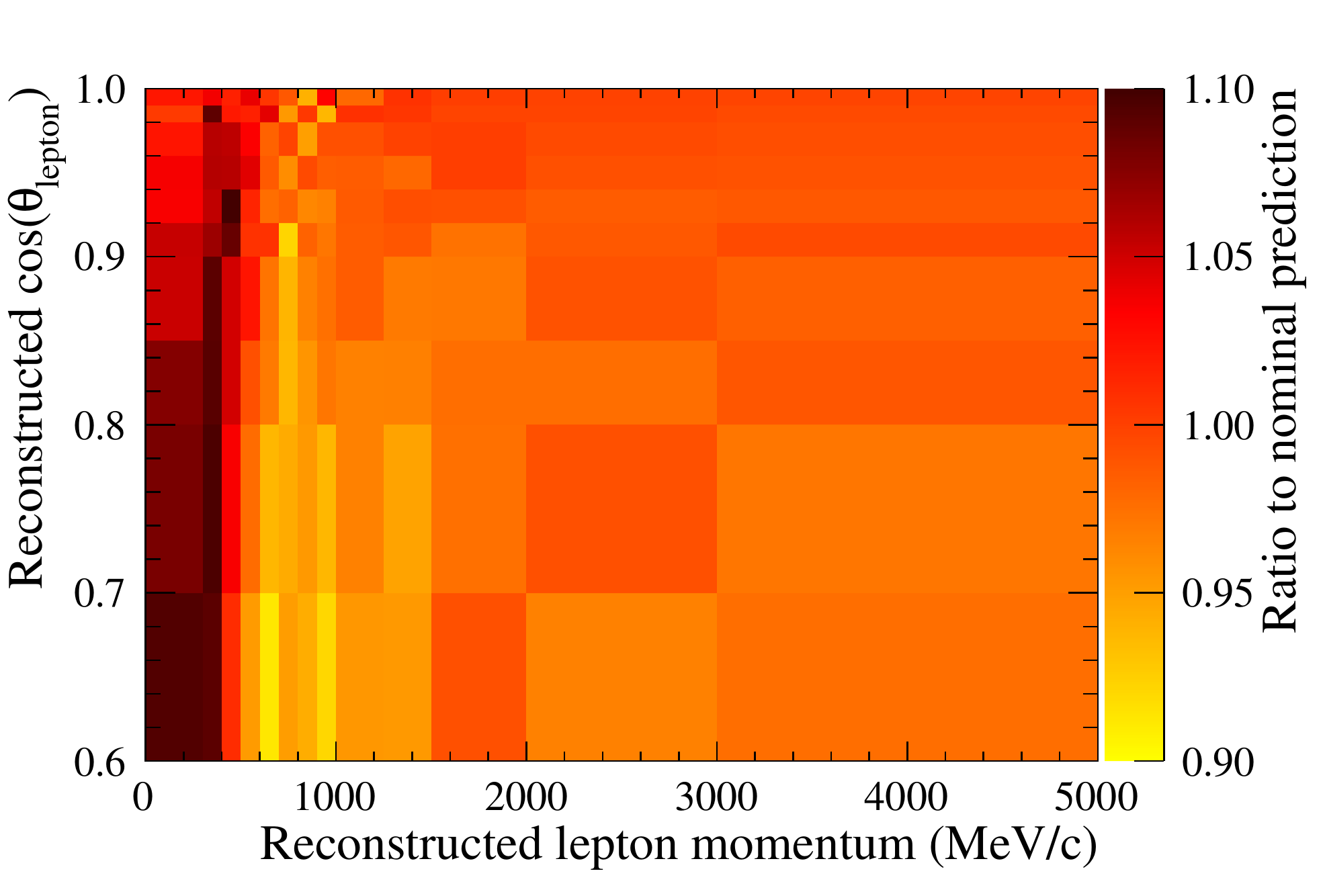}
	\caption {The ratio of the NRE simulated data to the nominal prediction for the ND280 CC $0\pi$ FGD1 sample.  The histogram axes have been truncated for clarity.}
	\label{fig:nd280_eb_fd}
\end{figure}

\begin{figure}[htbp]
\centering
\begin{subfigure}{0.47\textwidth}
\includegraphics[width=0.98\columnwidth]{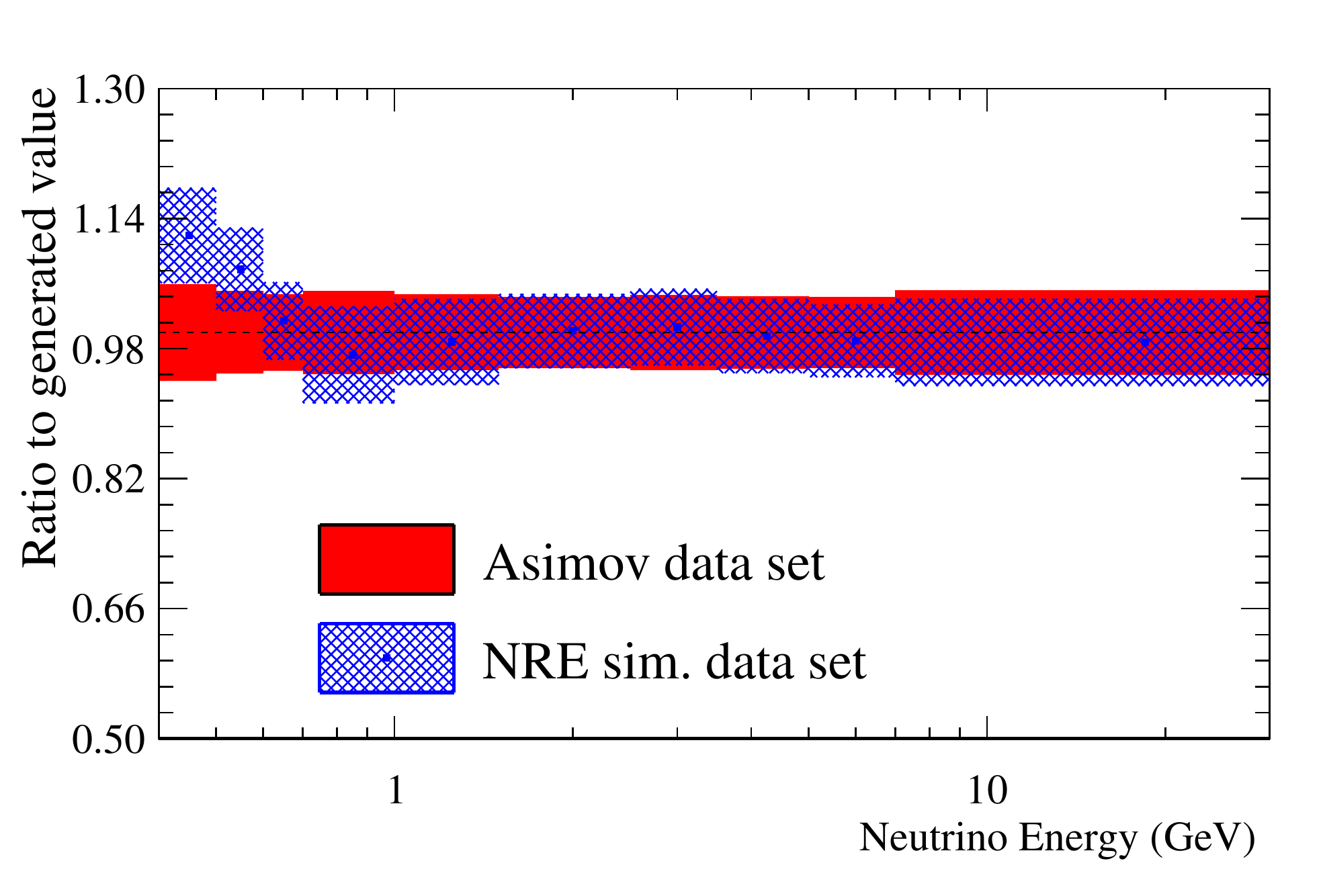}
\end{subfigure}
\begin{subfigure}{0.47\textwidth}
\includegraphics[width=0.98\columnwidth]{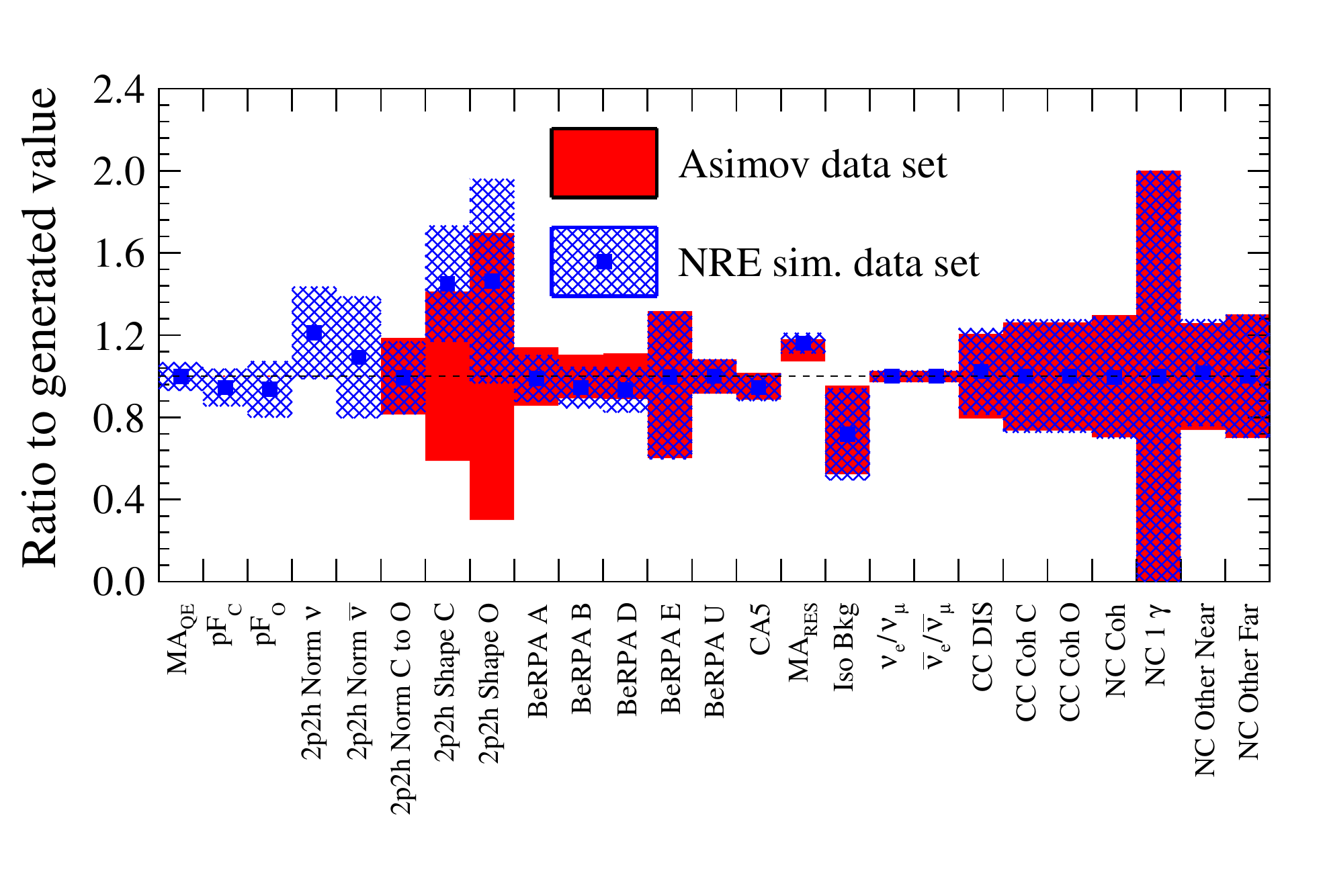}
\end{subfigure}
\caption {Results of the fit to the near detector simulated data with an increased nucleon removal energy.  The best fit values for the $\nu_{\mu}$ flux parameters at SK (top) and cross-section parameters (bottom) are shown for the NRE fit (blue) compared to the Asimov data set fit (red).}
\label{fig:nd280_eb_fit}
\end{figure}

The T2K oscillation analysis does not have a parameter to account for the NRE uncertainty directly in the fit.
Therefore the effect of the NRE uncertainty on the oscillation parameter measurements is evaluated using a simulated data study.
Neutrino events were generated with the RFG nucleon removal energy increased by 18~MeV from the nominal value (25\,MeV for interactions on Carbon and 27\,MeV for interactions on Oxygen).
Increasing the NRE results in a decrease in the energy available to the lepton produced in the neutrino interaction.
As a consequence, the events simulated with a larger NRE produced leptons with lower momenta than the nominal MC.
This momentum shift was calculated as a function of neutrino energy and lepton angle, shown in Fig.~\ref{fig:nre}.
The momentum shift in Fig.~\ref{fig:nre} was applied to the reconstructed momenta of the charged lepton for the full near and far detector MC.
The MC was then scaled to match the POT exposure of the data, creating a simulated data set.
The ratio of the simulated data to the nominal MC is shown in Fig.~\ref{fig:nd280_eb_fd} for the ND280 CC $0\pi$ sample, highlighting the shift in events from high momentum bins to lower momentum bins.
A fit is then performed on this simulated data using the analysis framework described earlier in this paper.
There are no parameters in the T2K cross-section model that change the lepton momentum directly, but combinations of parameters are able to mimic this effect.
The result of fitting the simulated near detector data is presented in Fig.~\ref{fig:nd280_eb_fit}, which shows the best fit values for the flux and cross-section model parameters.
Figure~\ref{fig:nd280_eb_fit} shows that the low energy flux is increased, to increase the rate of low momentum events.
The 2p2h shape parameters control whether the 2p2h events in the MC are produced more by $\Delta(1232)$ resonances (values above 1.0) or other modes (values below 1.0).
The $\Delta(1232)$ resonance produces leptons with a lower momentum than the other modes.
This means that increasing the shape parameter increases the rate of 2p2h events in the low momentum region.

\begin{figure}[htbp]
\centering
\begin{subfigure}{0.47\textwidth}
\includegraphics[width=0.98\columnwidth]{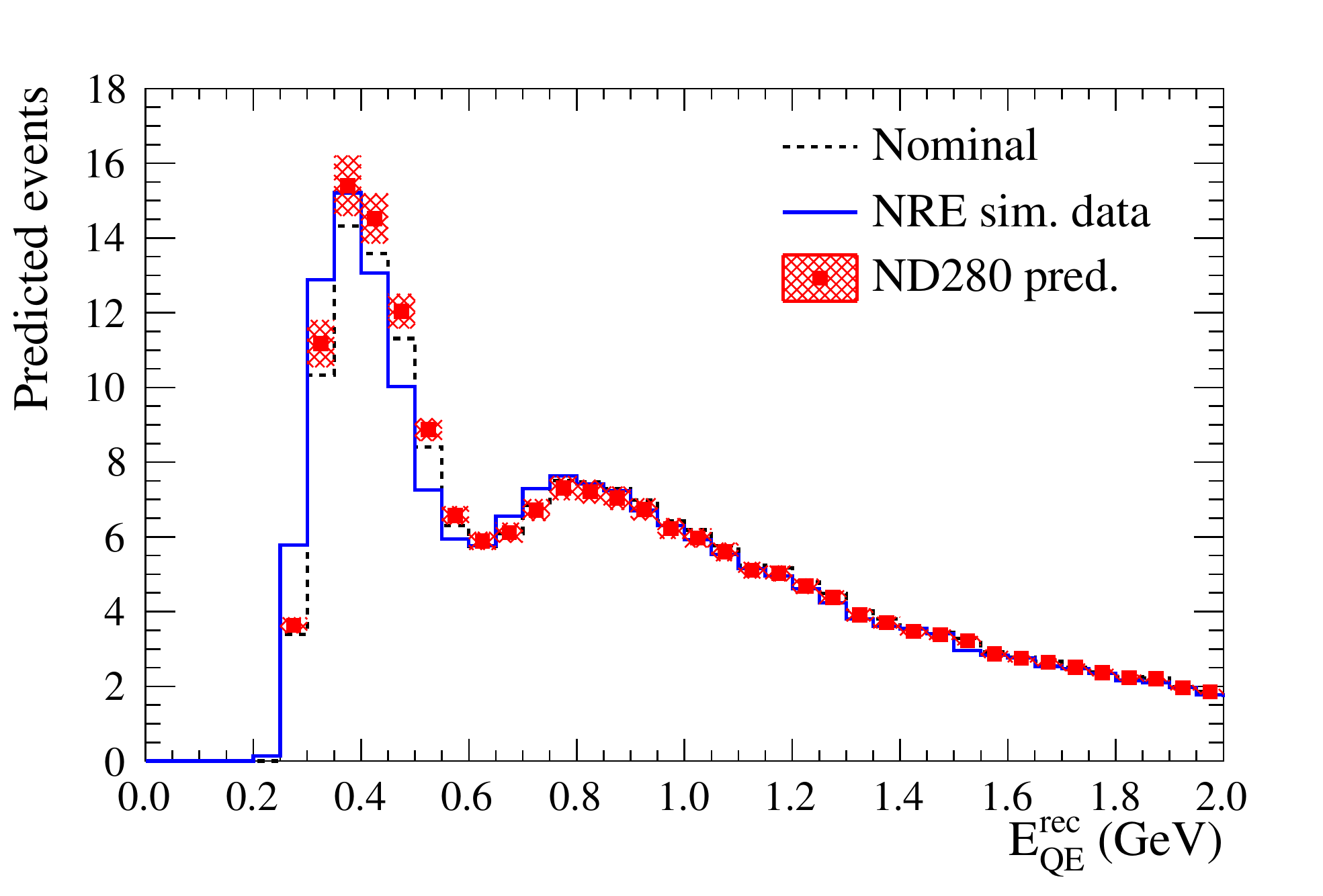}
\end{subfigure}
\begin{subfigure}{0.47\textwidth}
\includegraphics[width=0.98\columnwidth]{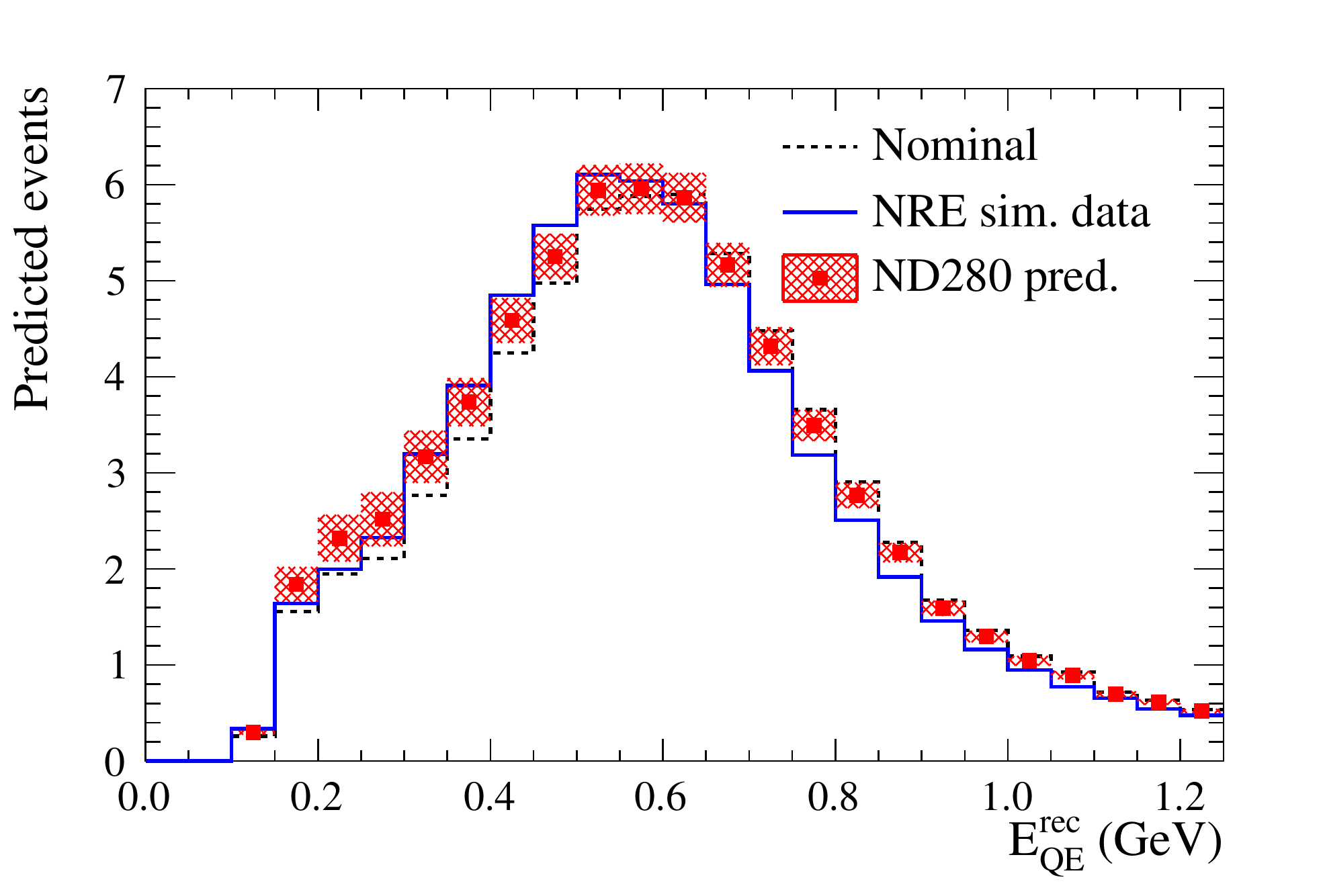}
\end{subfigure}
\caption {Comparison of the nominal MC prediction (dashed black), simulated data (solid blue) and prediction from the near detector fit (shaded red), for the far detector muon-like (top) and electron-like (bottom) neutrino beam event samples.}
\label{fig:sk_eb_fd}
\end{figure}

\begin{figure*}[htbp]
\centering
\begin{subfigure}{0.47\textwidth}
\includegraphics[width=0.98\columnwidth]{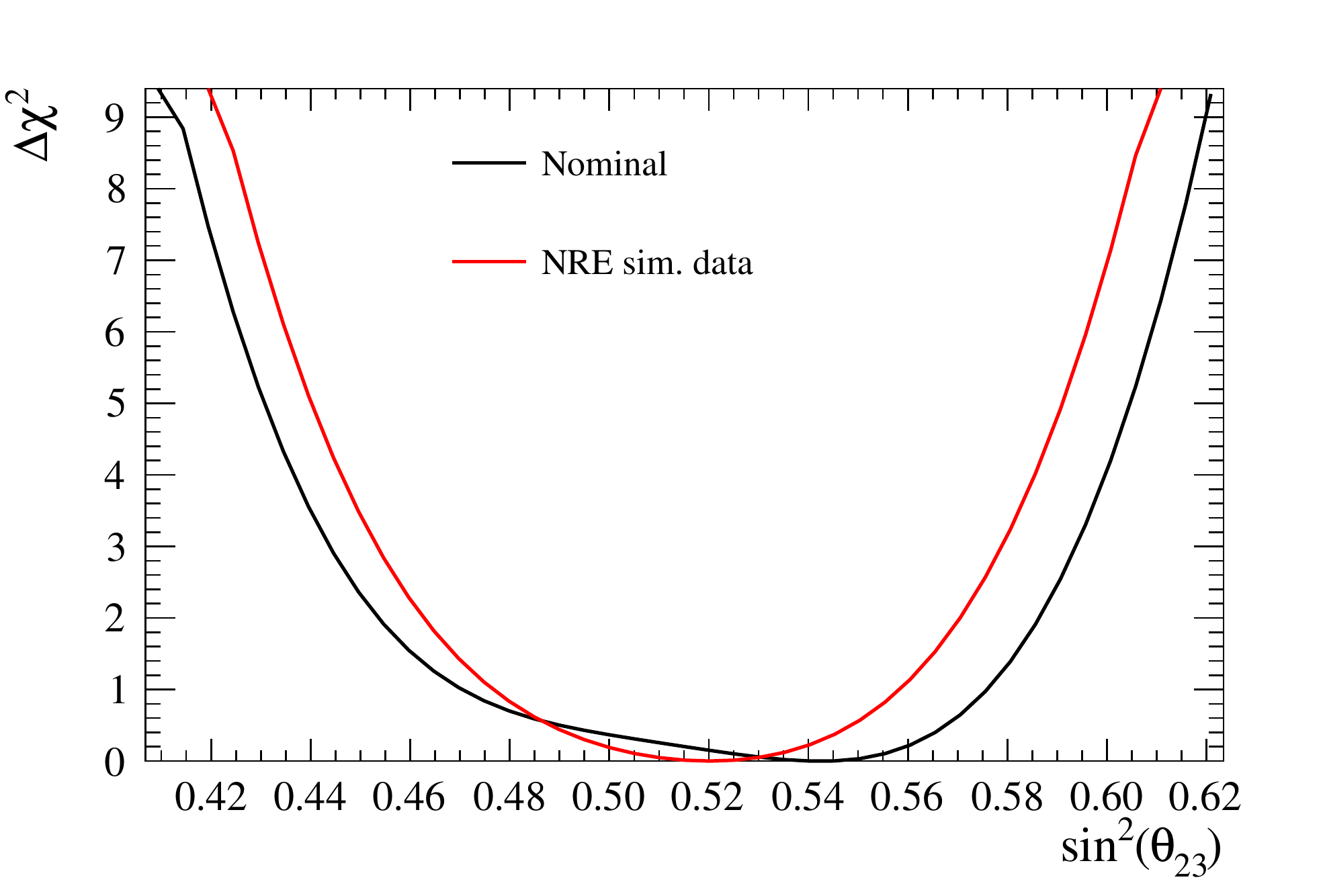}
\end{subfigure}
\begin{subfigure}{0.47\textwidth}
\includegraphics[width=0.98\columnwidth]{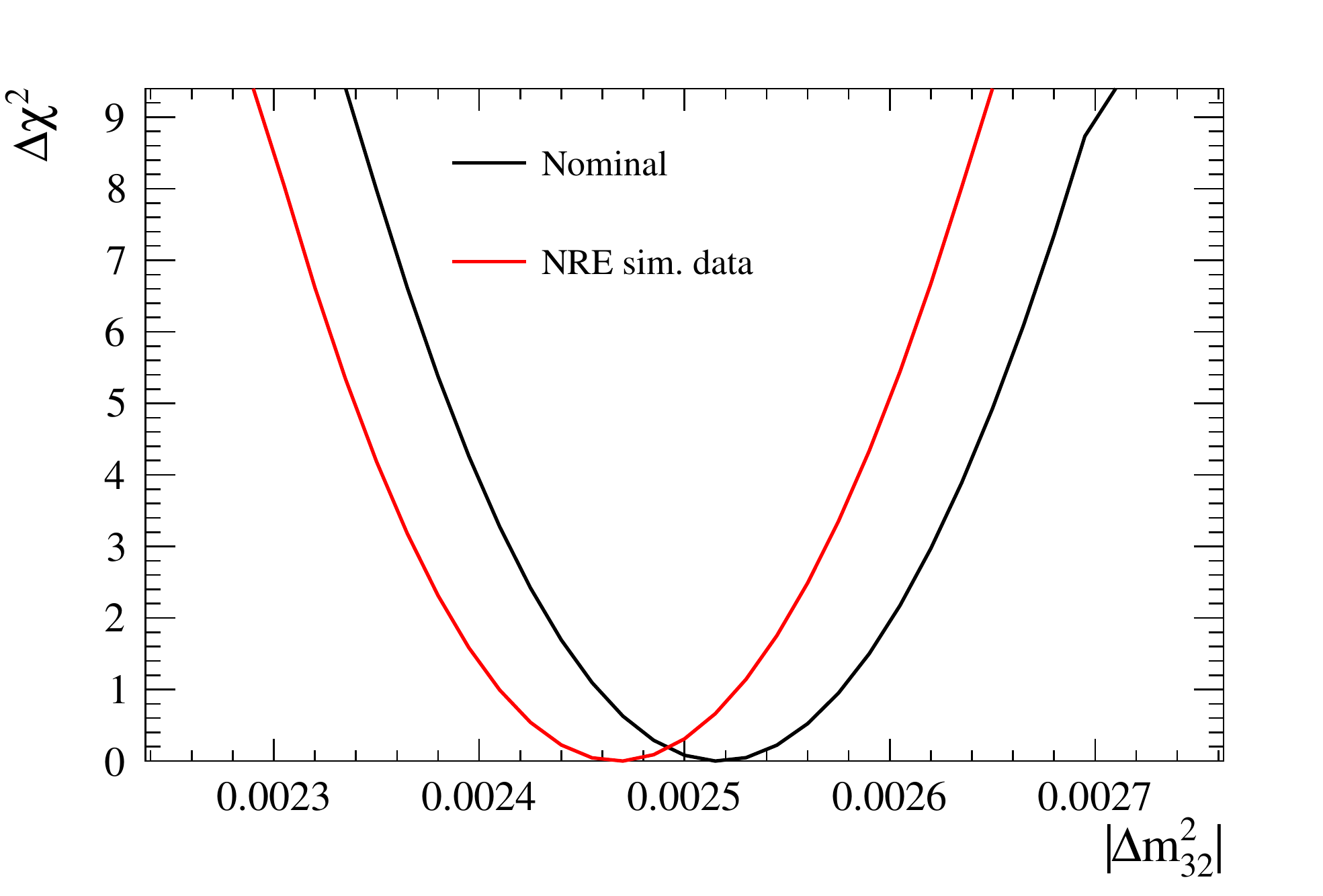}
\end{subfigure}
\begin{subfigure}{0.47\textwidth}
\includegraphics[width=0.98\columnwidth]{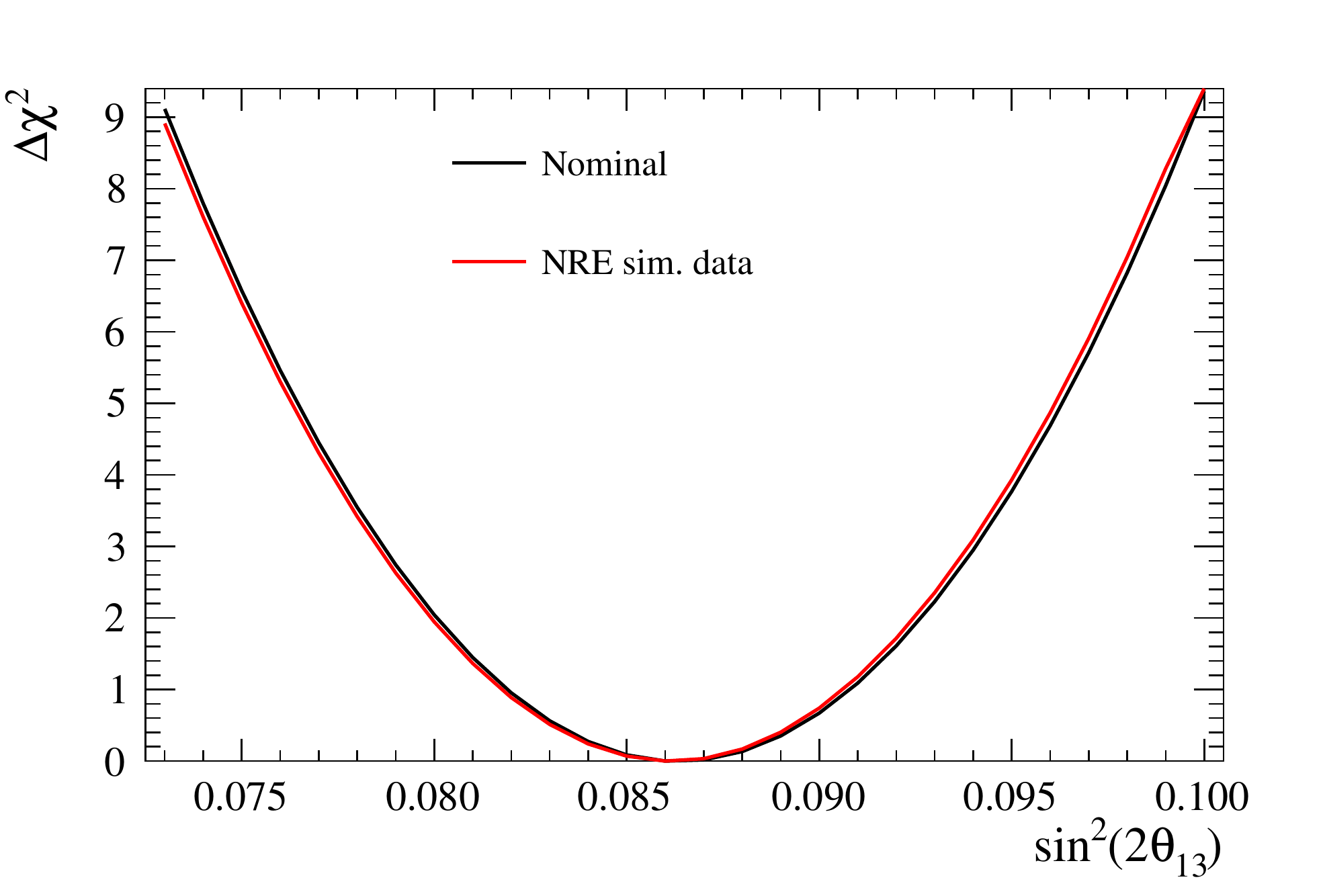}
\end{subfigure}
\begin{subfigure}{0.47\textwidth}
\includegraphics[width=0.98\columnwidth]{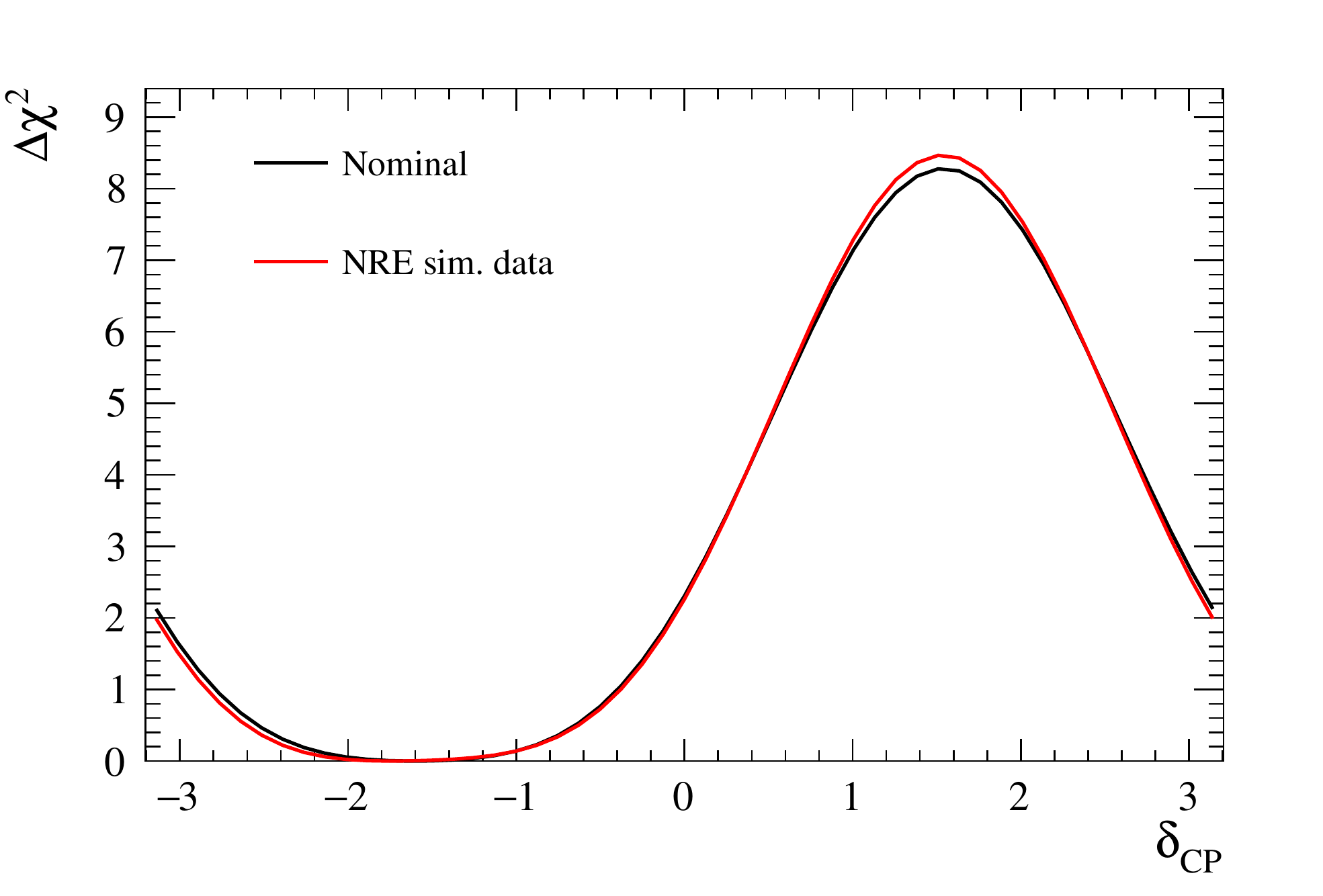}
\end{subfigure}
\caption {Comparison of the nominal MC oscillation parameter $\Delta \chi^{2}$ function (black) to those from the simulated data fits (red).}
\label{fig:eb_osc_comp}
\end{figure*}

The near detector fit result is used to predict the oscillated event distribution at the far detector.
This is shown in Fig.~\ref{fig:sk_eb_fd}, which compares the nominal MC to the simulated data and the near detector prediction.
The near detector prediction is closer to the simulated SK data than the nominal MC simulation, but does not match well below the oscillation dip.
The nominal prediction is fit to the simulated SK data to extract the best fit oscillation parameter values and their $2\sigma$ confidence intervals.
These are compared to oscillation parameter values and confidence intervals produced by fitting to the Asimov data set, as described in Sec.~\ref{sec:fitters}.
This comparison is shown in Fig.~\ref{fig:eb_osc_comp}.

Changing the NRE changes the shape of the simulated data, not the normalization.
Since the constraint on \ssqthonethree and \deltacp is largely driven by the normalization of the electron-like sample, it is unsurprising that these parameters are relatively unaffected.
In addition \ssqthonethree is strongly constrained by the reactor experiment measurements.
However, both \ssqthtwothree and \dmsqtwothree show a significant difference between the nominal and the simulated data results, with \ssqthtwothree shifting towards maximal disappearance and \dmsqtwothree decreasing.

The changes in the oscillation parameter contours indicate that the T2K cross-section model parameterization cannot account for changes to the NRE.
The near detector fit mis-attributes the NRE change to the flux and 2p2h model parameters.
The near detector post-fit prediction has a different neutrino energy distribution to the nominal MC, and so produces different far detector event distributions when the neutrino oscillation probability is applied.
As a result the oscillation parameters extracted from the simulated data fit no longer match those from the nominal MC analysis.

To account for this in the T2K analysis an additional uncertainty is introduced.
A spline is created for each bin of the five far detector sample histograms.
The value of the spline is the ratio between the simulated far detector data and the far detector prediction calculated using the near detector simulated data fit result.
Spline knots are created using NREs of 18~MeV, 27~MeV (nominal) and 45~MeV at both near and far detector.
The splines are 100\% correlated across all sample bins, and produce a multiplicative weight to scale the far detector prediction in each bin.
The spline takes into account the change in the far detector prediction due to both the changing NRE and the mis-fitting at the near detector.
This means that it does not provide a measurement of the true NRE, but an `effective NRE' taking into account both of these effects.
A prior constraint is placed on this effective NRE parameter, setting the central value to 27~MeV, with an uncertainty of $\pm$18~MeV.

\begin{figure}[ht!]
\centering
\begin{subfigure}{0.47\textwidth}
\includegraphics[width=0.98\columnwidth]{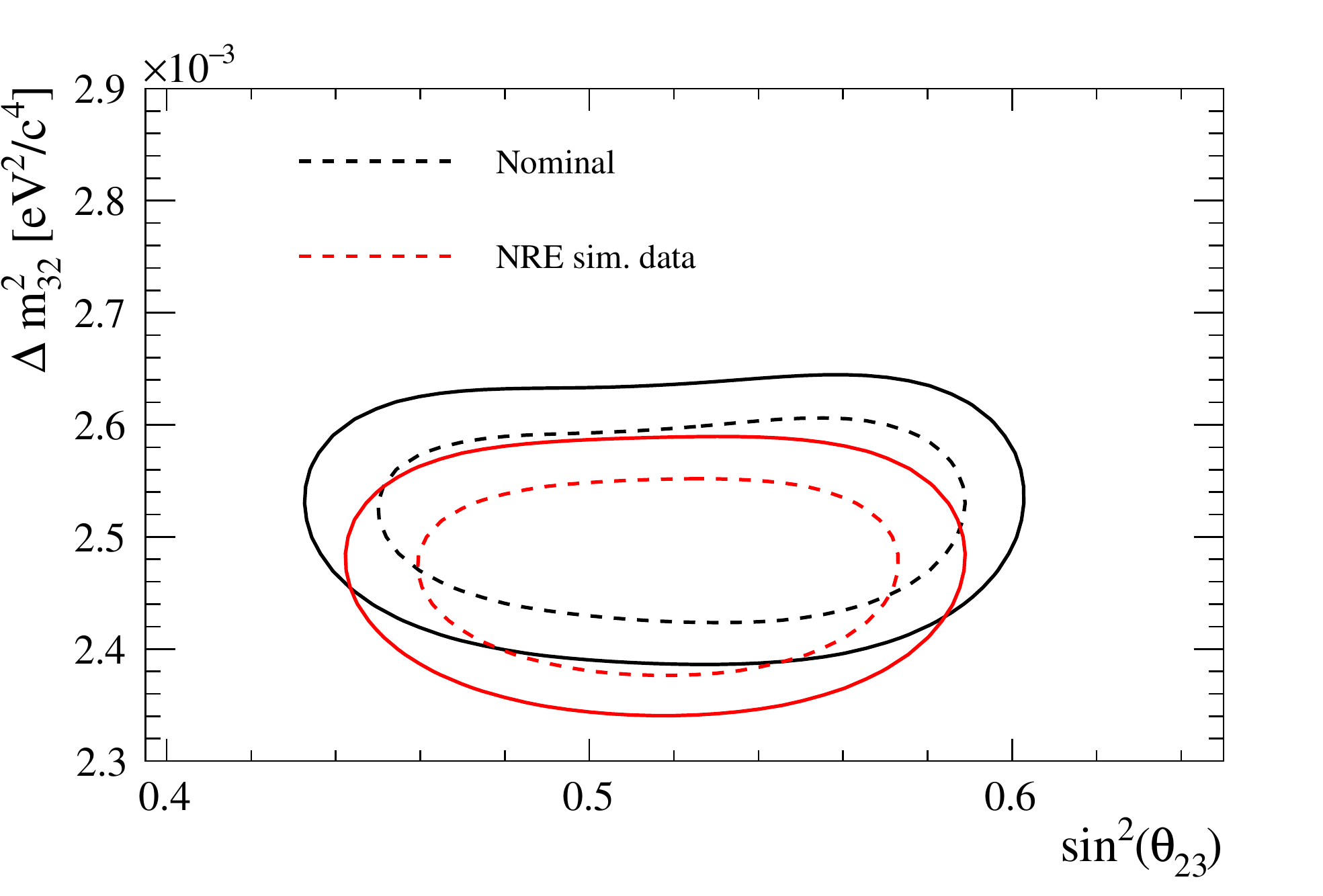}
\end{subfigure}
\begin{subfigure}{0.47\textwidth}
\includegraphics[width=0.98\columnwidth]{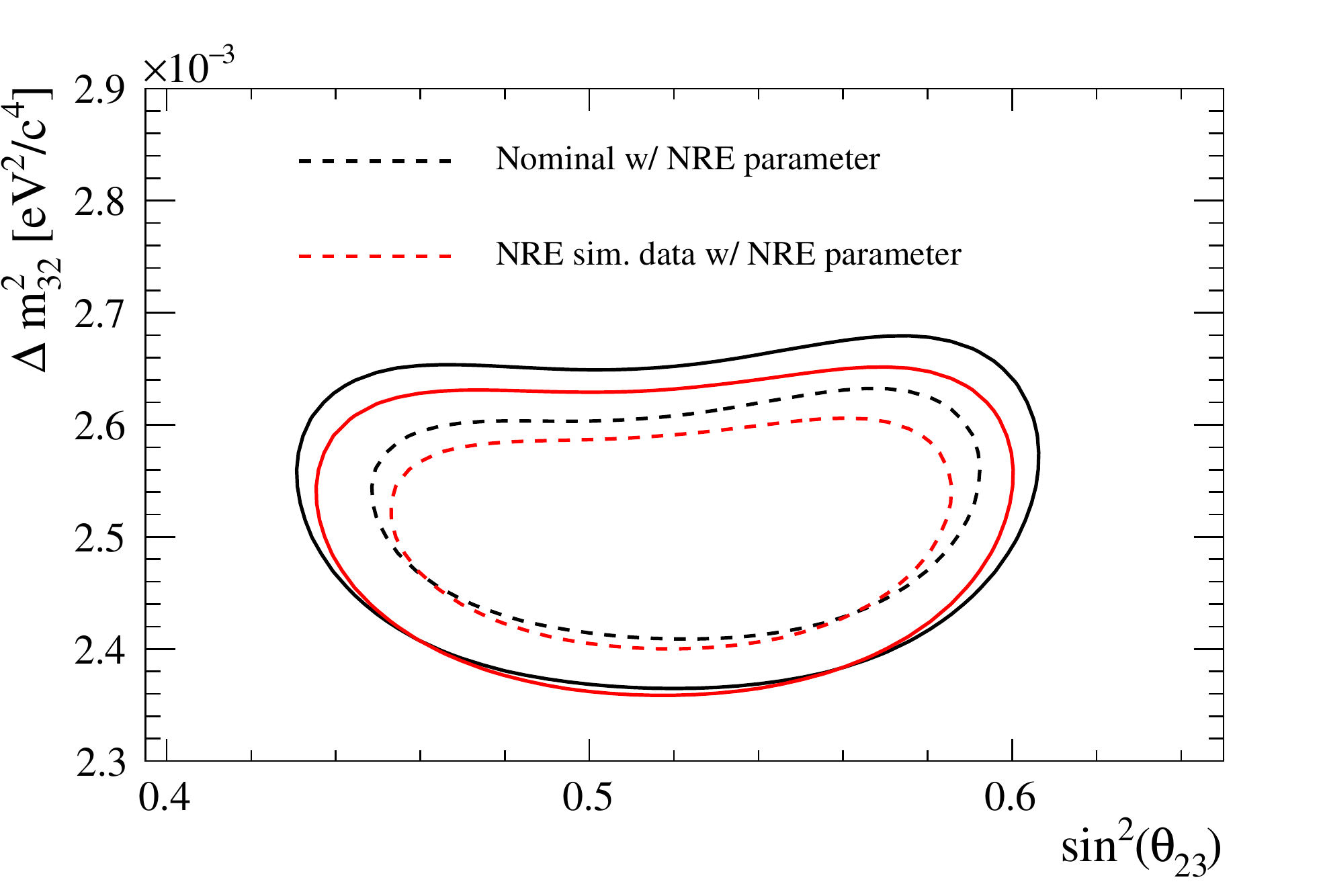}
\end{subfigure}
\begin{subfigure}{0.47\textwidth}
\includegraphics[width=0.98\columnwidth]{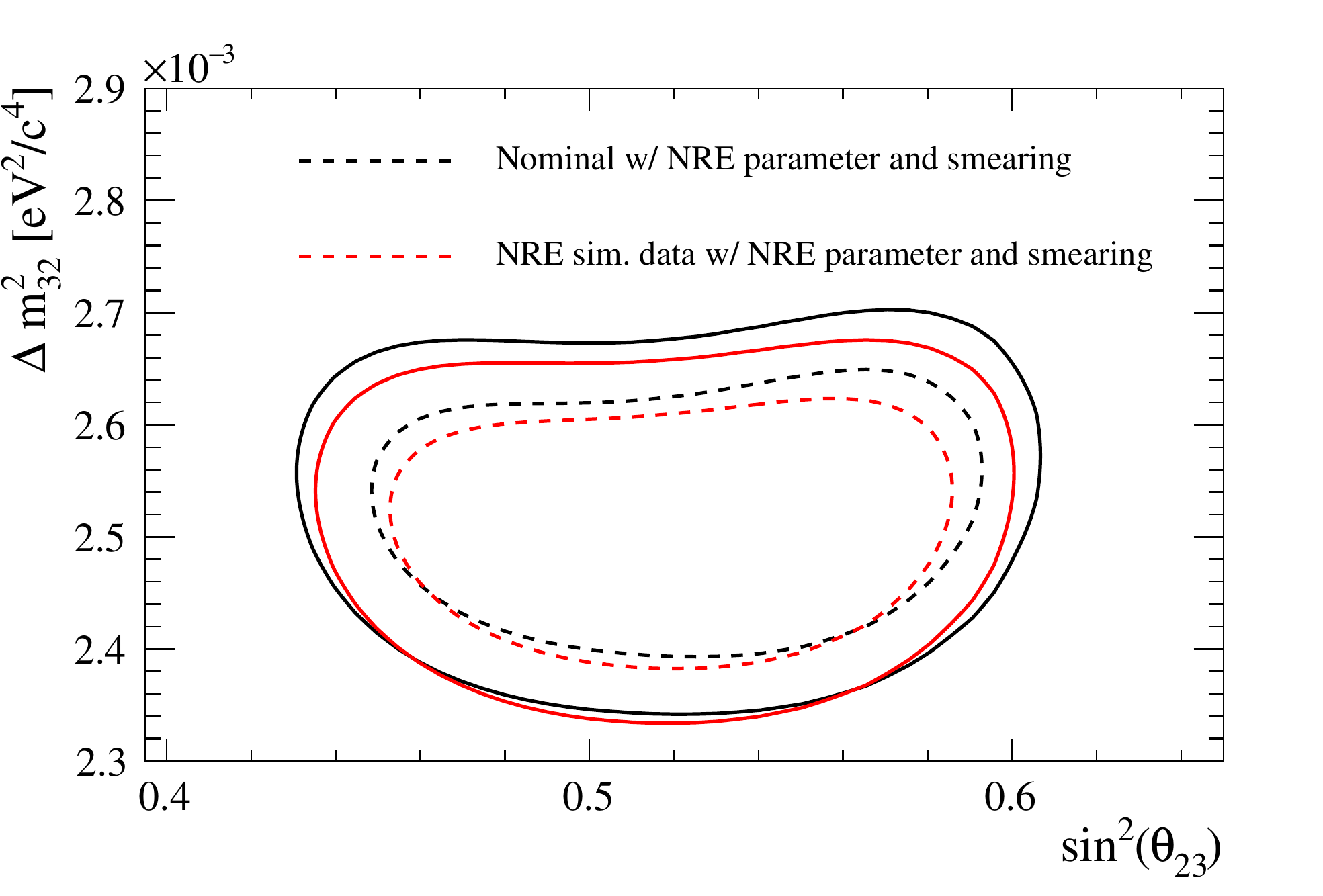}
\end{subfigure}
\caption {Comparison of the \ssqthtwothree-\dmsqtwothree parameter contours using the nominal cross-section model (top), after the addition of the `effective NRE' parameter (middle) and after the additional smearing is applied to \dmsqtwothree (bottom).  In all cases the expected result, as described in the text, is shown in black and the simulated data result is in red.  The solid lines represent the 90\% confidence region and the dashed lines represent the $1\sigma$ confidence region.}
\label{fig:eb_osc_spline}
\end{figure}

The result of including this parameter in the simulated data study is shown in Fig.~\ref{fig:eb_osc_spline}.
The new parameter increases the size of the oscillation contours whilst shifting the simulated data result to be in much better agreement with the expectation.
There is some residual difference between the contours, particularly in the \dmsqtwothree best fit point.
This difference is included as an additional uncertainty in the analysis by smearing the \dmsqtwothree likelihood surface.
The far detector oscillation fit likelihood as a function of \dmsqtwothree has a Gaussian distribution.
This distribution is convolved with a Gaussian of unit area, centered at 0, with a width given by the shift in the \dmsqtwothree best fit point between the simulated data fit and the expectation.
This is equivalent to adding this shift as an uncertainty on \dmsqtwothree in quadrature with the existing uncertainties in the analysis.
The result of this smearing is shown in Fig.~\ref{fig:eb_osc_spline}.
\subsection{Summary of simulated data studies}
Table~\ref{tab:final_bias} shows the final bias table for simulated data sets studied in the analysis, after the addition of the NRE uncertainty parameter.  For the data-driven $E_{\nu}-Q^2$ category the largest effect from the three simulated data studies is shown.
In all cases the observed bias on \ssqthtwothree and \deltacp was insignificant compared to existing systematic uncertainty on the parameter and so no additional uncertainty was introduced.
Non-negligible bias was observed for \dmsqtwothree.
The quadrature sum of the observed biases, $4.1 \times 10^{-5}$~eV$^{2}$c$^{-4}$, was added as an additional uncertainty on \dmsqtwothree using the method described above.
\begin{table}[htp]
\centering
\caption{Oscillation parameter biases (as percentages of the total and systematic uncertainties) observed in the simulated data studies including the additional uncertainties on the NRE.}
\label{tab:final_bias}
\begin{tabular}{p{0.32\columnwidth}@{\quad}c@{\quad}r@{\quad\,}r@{\quad\,}r}
\hline
\hline
Simulated data set  &  Relative to  &  \ssqthtwothree & \dmsqtwothree & \multicolumn{1}{c}{\deltacp} \\
\hline
  \multirow{2}{*}{Martini 2p2h}
   & ~Total & 9.0\,\% & 16\,\% & 0.1\,\%\\
   & ~Syst. & 20\,\% & 22\,\% & 0.3\,\% \\[0.5ex]
  Data-driven CC $0\pi$ 
  & ~Total & 15\,\% & 14\,\%  & 4.0\,\% \\
  $E_{\nu}-Q^{2}$ dependence 
  & ~Syst. & 34\,\% & 20\,\% & 17\,\%  \\[0.5ex]
  \multirow{2}{*}{BSF\,1p1h}
  & ~Total & 1.5\,\% & 22\,\% & 0.1\,\% \\
  & ~Syst. & 3.4\,\% & 31\,\% & 0.3\,\% \\[0.5ex]
  \multirow{2}{*}{Nieves LFG 1p1h}
  & ~Total & 4.0\,\% & 25\,\% & 7.0\,\% \\
  & ~Syst. & 8.3\,\% & 35\,\% & 20\,\% \\[0.5ex]
%
  Nucleon
  & ~Total & 5.0\,\% & 33\,\%  & 0.1\,\%   \\
  removal energy
  & ~Syst. & 10\,\%  & 46\,\% & 0.6\,\% \\[0.5ex]
 \multirow{2}{*}{Coulomb \rlap{correction}} 
 & ~Total & 1.0\,\%  & 0.1\,\%  & 0.1\,\% \\
 & ~Syst. & 2.3\,\% & 0.1\,\% & 0.3\,\%  \\[0.5ex]
  Kabirnezhad 
  & ~Total & 8.0\,\% & 34\,\% & 0.0\,\% \\
  single pion 
  & ~Syst. & 20\,\% & 50\,\% & 1.0\,\% \\
\hline
\hline
\end{tabular}
\end{table}
The effect of the systematic parameters on the predicted event rates on each SK event sample, including the additional NRE uncertainty, is shown in Tab.~\ref{tab:sk_errors_summary}.
The effect of the prior uncertainties on the (typically marginalized) oscillation parameters \ssqthonetwo , \dmsqonetwo and \ssqthonethree is also shown.
\begin{table*}[htp]
	\centering
	\caption{
	Fractional uncertainty (\%) on event rate by error source and sample, calculated with expected events rates generated according to the nominal oscillation parameter values from Table \ref{tab:asimova_params}. Final column is the fractional uncertainty (\%) on the ratio of FHC/RHC events in the one-ring $e$ sample. The final row, `All Systematics', does not include the effects of any oscillation parameters.
	}
	\label{tab:sk_errors_summary}
	\sisetup{table-format=3.3}
	\begin{tabular}{lSS@{\hspace{1.5em}}SSSS}
		\hline
		\hline
		\rule{0mm}{3mm}
		& \multicolumn{2}{c}{{1-Ring $\mu$}} & \multicolumn{4}{c}{{1-Ring $e$}} \\[1ex]
		Error source & {~\,FHC} & {~\,RHC} & {~\,FHC} & {~\,RHC} & 
		$\substack{\text{FHC}\\\text{1 d.e.}}$ & $\nicefrac{\text{FHC}}{\text{RHC}}$\\[1ex]
		\hline
SK Detector & 2.4 & 2.0 & 2.8 & 3.8 & 13.2 & 1.5 \\
SK FSI+SI+PN & 2.2 & 2.0 & 3.0 & 2.3 & 11.4 & 1.6 \\
		Flux + Xsec (ND unconstrained) & 14.3 & 11.8 & 15.1 & 12.2 & 12.0 & 1.2 \\
Flux + Xsec (ND constrained) & 3.3 & 2.9 & 3.2 & 3.1 & 4.1 & 2.7 \\
Nucleon Removal Energy & 2.4 & 1.7 & 7.1 & 3.7 & 3.0 & 3.6 \\
$\sigma(\nue)/\sigma(\nueb)$ & 0.0 & 0.0 & 2.6 & 1.5 & 2.6 & 3.0 \\
NC1$\gamma$ & 0.0 & 0.0 & 1.1 & 2.6 & 0.3 & 1.5 \\
NC Other & 0.3 & 0.3 & 0.2 & 0.3 & 1.0 & 0.2 \\
$\ssqthtwothree + \dmsqonetwo$ & 0.0 & 0.0 & 0.5 & 0.3 & 0.5 & 2.0 \\
$\sin^2 \theta_{13}$ PDG2018 & 0.0 & 0.0 & 2.6 & 2.4 & 2.6 & 1.1 \\
		\hline
All Systematics & 5.1 & 4.5 & 8.8 & 7.1 & 18.4 & 6.0 \\
		\hline
		\hline
	\end{tabular}
\end{table*}

\section{Oscillation Analysis Results}
\label{sec:results}

Following the recommendations in ~\cite{cousins2018lectures}, we produce results using different statistical approaches, both frequentist and Bayesian (with analysis of sensitivity to prior) and test the frequentist properties of our Bayesian methods when possible. Using the three far detector analyses described previously, point and interval estimations are made for the parameters \ssqthtwothree, \deltacp and \dmsqtwothree (normal ordering) or \dmsqonethree (inverted ordering). Two types of intervals are produced: confidence intervals (with approximate coverage based on the constant $\Delta \chi^2$ method in most cases, and with exact coverage using the Feldman--Cousins unified approach for \deltacp) and credible intervals. The mass ordering was studied using mainly Bayesian hypothesis testing, with additional frequentist checks.

\subsection{Measurements of the parameters of the 3 flavor oscillation model}

\subsubsection{$\Delta \chi^2$ and frequentist results} \label{sec:OA_frquentist_results}
\begin{figure}[htbp]
\centering
\includegraphics[width=0.47\textwidth]{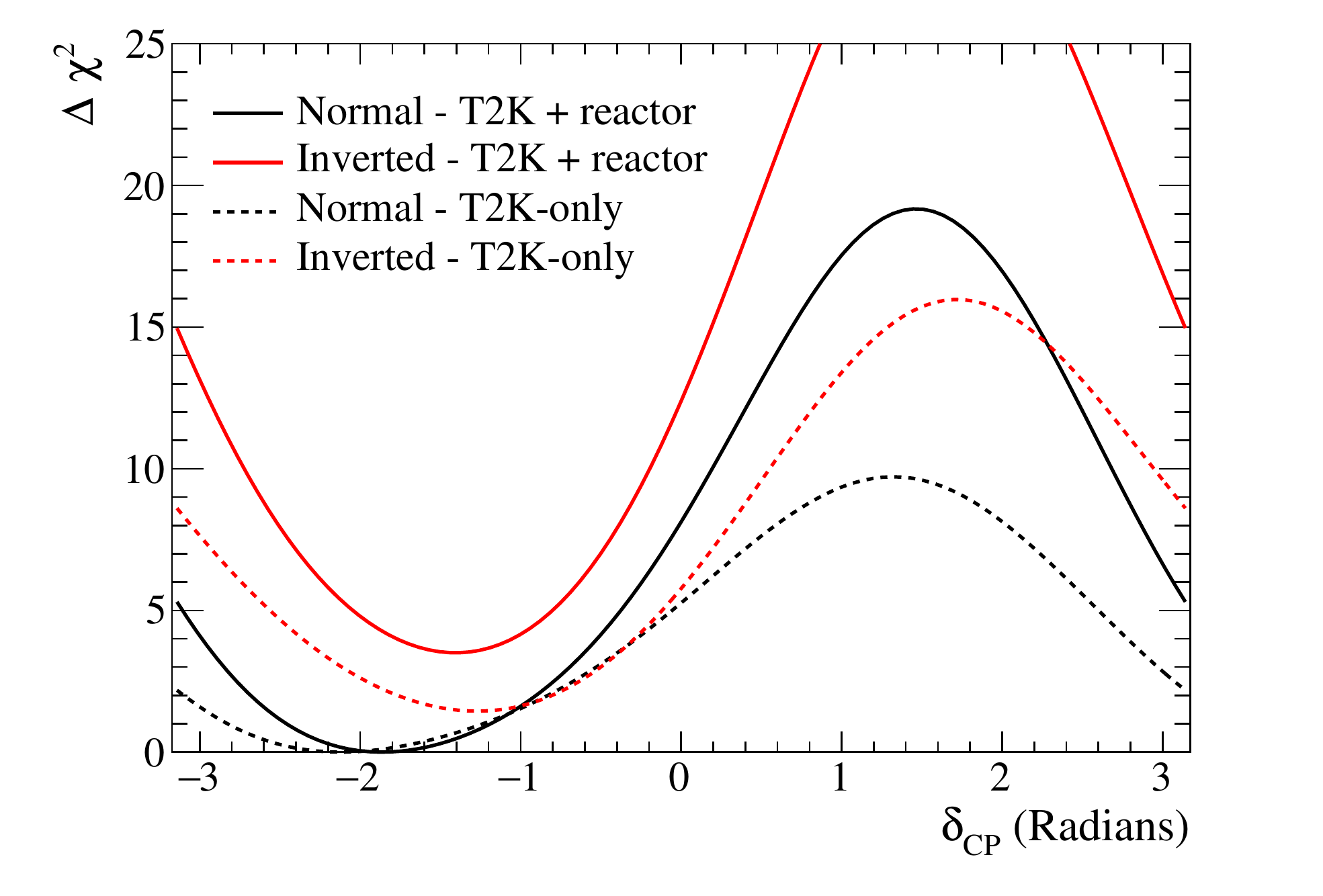}
\caption{
    The observed $\Delta \chi^2$ function of \deltacp, with and without the reactor constraint. The $\Delta \chi^2$ is computed with respect to the best fit over the two mass orderings, and separate best fit points are used for the T2K-only and the T2K+reactor cases.
}
\label{fig:dcp_data}
\end{figure}

\begin{figure}[htbp]
\centering
\includegraphics[width=0.47\textwidth]{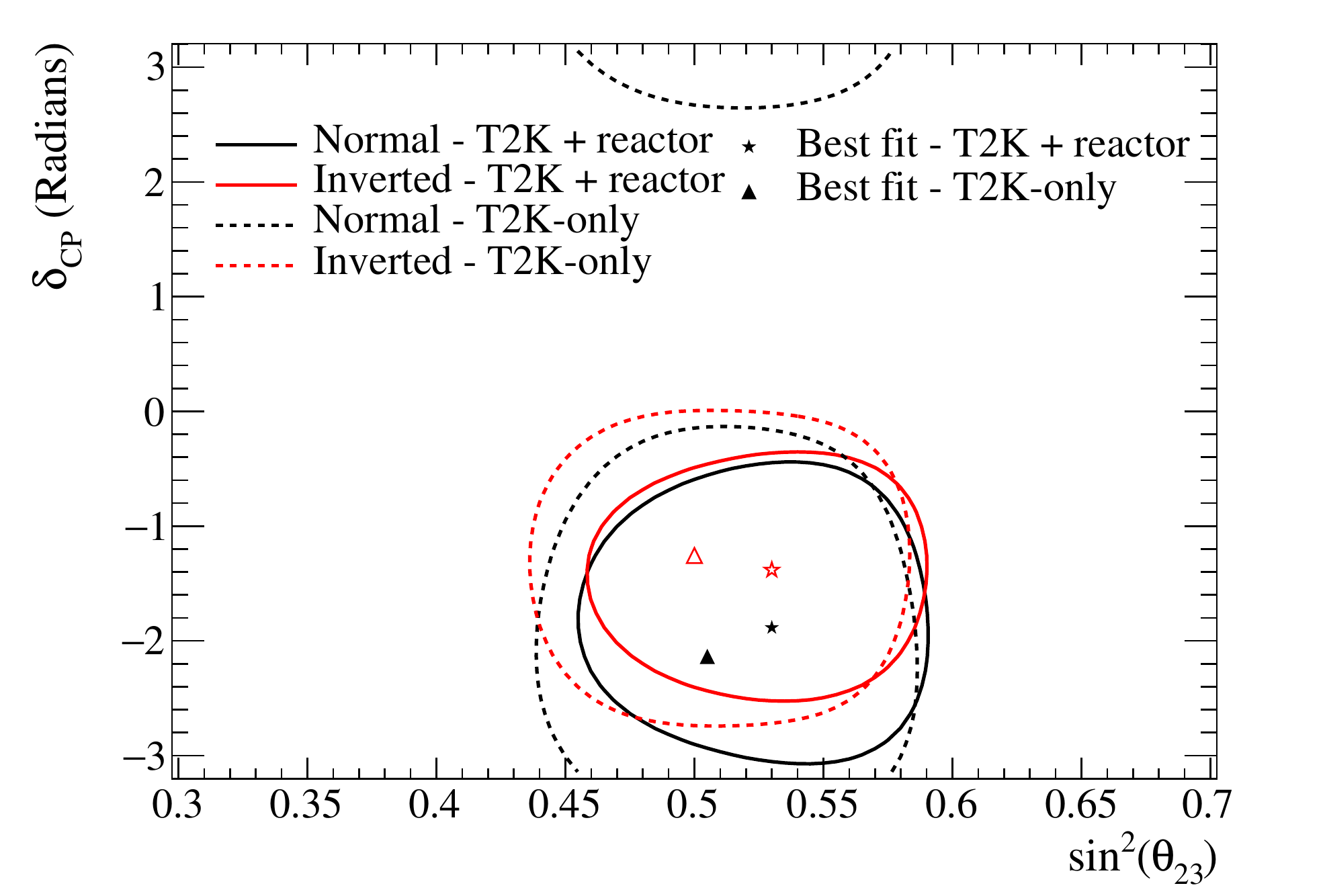}
\caption{
    The observed constant $\Delta \chi^2$ 90\% confidence regions of \ssqthtwothree and \deltacp with normal and inverted mass orderings and with and without the reactor constraint. Normal and inverted mass ordering contours are independent. $\Delta \chi^2$ values are calculated independently for the functions with and without the reactor constraint.
}
\label{fig:mixing23_deltacp_data}
\end{figure}

\begin{figure}[htbp]
\centering
\includegraphics[width=0.47\textwidth]{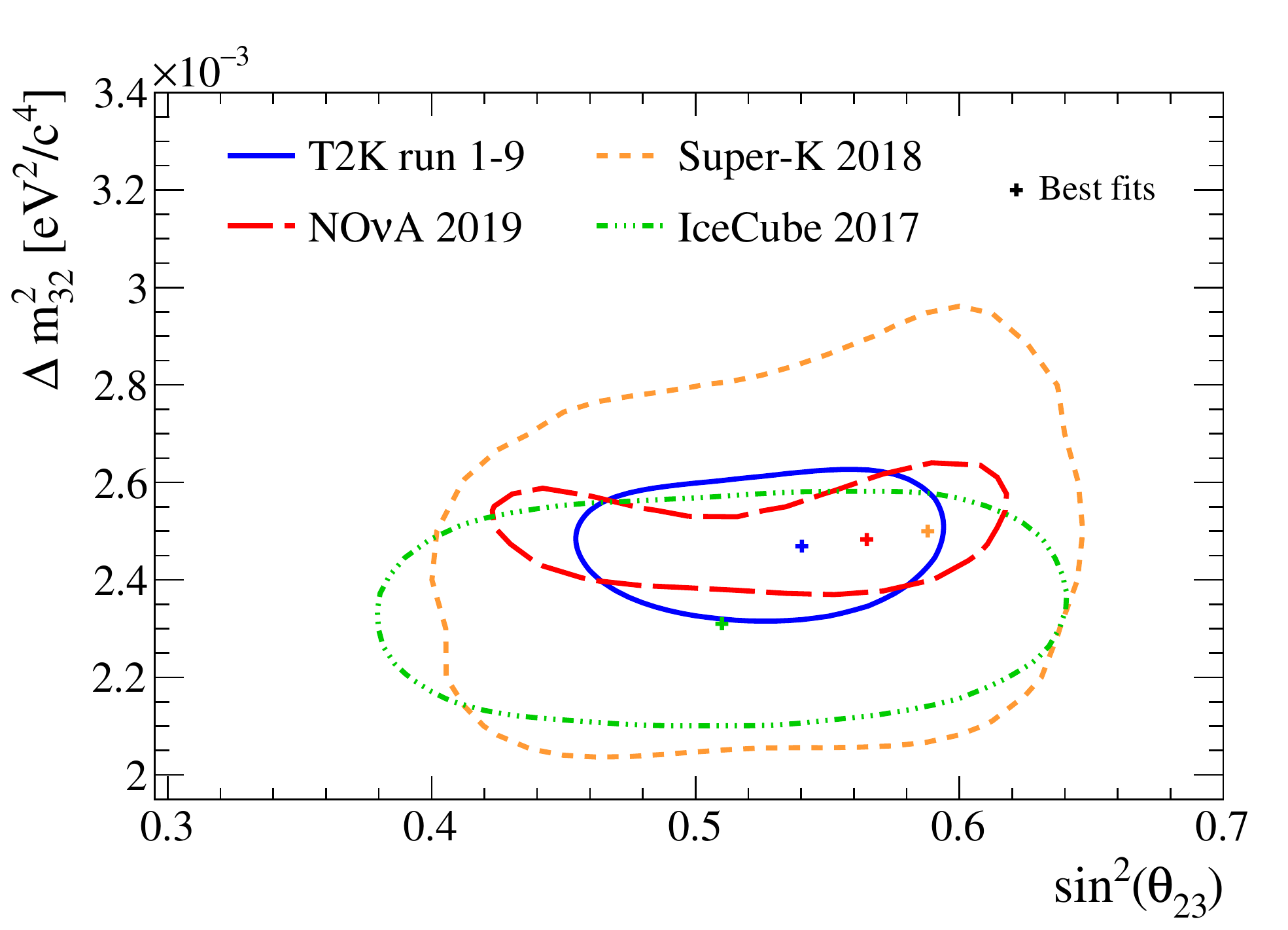}
\caption{
    The observed T2K constant $\Delta \chi^2$ 90\% confidence regions (from analysis C) of \ssqthtwothree and \dmsqtwothree with the reactor constraint and compared to the results from Super-K~\cite{Abe:2017aap_skatm}, \nova~\cite{Acero:2019ksn_nova} and IceCube~\cite{PhysRevLett.120.071801}. 
}
\label{fig:mixing23_dm23sq_data_comparison_other_expt}
\end{figure}

Intervals based on the constant $\Delta \chi^2$ method are produced for the different parameters using analyses A and C (analysis B can produce similar intervals for comparison purpose, although its main results are the credible intervals described in \ref{BayesianResults}). As their results are in good agreement, only the results obtained with analysis A are shown in this section, unless otherwise indicated. The best fit values and $1\sigma$ confidence intervals obtained for the different parameters in both mass ordering scenarios are summarized in Tab.~\ref{tab:data_bestfit_t2k_only} and in Tab.~\ref{tab:data_bestfit} with and without using the results of reactor experiments to constrain \ssqthonethree, respectively. The global best fit was found to be for the NO, and the data show a preference for the upper octant. These preferences will be quantified in part~\ref{section_MH}.  The $\Delta \chi^2=-2\ln \left(L/L_{max}\right)$ functions obtained for \deltacp with and without using the results of reactor experiments to constrain \ssqthonethree are displayed in Fig.~\ref{fig:dcp_data}. The favored and disfavored values of \deltacp are similar between the two cases, but the constraint on \deltacp becomes stronger when the reactor experiments results are used.
The obtained 90\% confidence regions for (\ssqthtwothree, \deltacp) are displayed in Fig.~\ref{fig:mixing23_deltacp_data}. The largest parts of the confidence regions are located in the upper octant, especially when the constraint from reactor experiments is used, but the results are still compatible with maximal mixing.
For the atmospheric parameters, the obtained normal ordering 90\% confidence region for (\ssqthtwothree, \dmsqtwothree) is shown in Fig.~\ref{fig:mixing23_dm23sq_data_comparison_other_expt}  together with the measurements from other neutrino oscillation experiments. Good agreement is seen between all of the experiments. 

\begin{table}[htp]
    \renewcommand{\dot}{$&$}
    \newcommand{\blank}[1]{\multicolumn{#1}{c}{~}}
	\centering
	\caption{ The measured oscillation parameter best-fit and the $\pm 1\sigma$ intervals, shown for the T2K-only (without reactor constraint) fit and for normal and inverted hierarchies with respect to the hierarchy best-fit. The $\pm 1\sigma$ interval corresponds to the values for which $\Delta \chi^2 \leq 1$.}
	\label{tab:data_bestfit_t2k_only}
	\begin{minipage}{\columnwidth}
	\centering
	\begin{tabular}{ c r@{\,}l@{\quad} r@{\,}l }
		\hline\hline  
        \multirow{2}{*}[0.5ex]{Parameter} & \multicolumn{4}{c}{Best-fit and $1\sigma$ interval}  \\
		 & \multicolumn{2}{c}{NO} & \multicolumn{2}{c}{IO}\\
		\hline\\[-1.5ex]
		\deltacp 
		& $-2.14$ & $_{-0.69}^{+0.90}$ &  $-1.26$ & $_{-0.69}^{+0.61}$ \\[1ex]
		\ssqthonethree \si{\per10^{-3}}  
		& $26.8$ & $_{-4.3}^{+5.5}$ & $30.0$ & $_{-5.0}^{+5.9}$ \\[1ex]
		\ssqthtwothree 
		& $0.512$ & $_{-0.042}^{+0.045}$ &  $0.500$ & $_{-0.036}^{+0.050}$ \\[1ex]
		\dmsqtwothree \si{\per10^{-3} \eV^2 \clight^{-4}}
		& $2.46$ & $_{-0.07}^{+0.07}$ & \blank{2}  \\[1ex]
		$|$\dmsqonethree$|$ \si{\per10^{-3} \eV^2 \clight^{-4}}
		& \blank{2} & $2.43$ & $_{-0.08}^{+0.07}$ \\[1ex]
		\hline
		\hline
	\end{tabular}
    \end{minipage}

\end{table}

\begin{table}[htp]
    \renewcommand{\dot}{$&$}
    \newcommand{\blank}[1]{\multicolumn{#1}{c}{~}}
	\centering
	\caption{ The measured oscillation parameter best-fit and the $\pm 1\sigma$ intervals, shown for the T2K + reactor fit and for normal and inverted hierarchies with respect to the hierarchy best-fit. The $\pm 1\sigma$ interval corresponds to the values for which $\Delta \chi^2 \leq 1$.}
	\label{tab:data_bestfit}
	\begin{minipage}{\columnwidth}
	\centering
	\begin{tabular}{ c r@{\,}l@{\quad} r@{\,}l }
		\hline\hline  
        \multirow{2}{*}[0.5ex]{Parameter} & \multicolumn{4}{c}{Best-fit and $1\sigma$ interval}  \\
		 & \multicolumn{2}{c}{NO} & \multicolumn{2}{c}{IO}\\
		\hline\\[-1.5ex]
		\deltacp 
		& $-1.89$ & $_{-0.58}^{+0.70}$ & $-1.38$ & $_{-0.55}^{+0.48}$ \\[1ex]
		\ssqthtwothree  
		& $0.532$ & $_{-0.037}^{+0.030}$ & $0.532$ & $_{-0.035}^{+0.029}$ \\[1ex]
		\dmsqtwothree \si{\per10^{-3} \eV^2 \clight^{-4}}  
		& $2.45$ & $_{-0.07}^{+0.07}$ &\blank{2} \\[1ex]
		$|$ \dmsqonethree $|$ \si{\per10^{-3} \eV^2 \clight^{-4}}  
		&\blank{2}& $2.43$ & $_{-0.07}^{+0.07}$ \\[1ex]
		\hline
		\hline
	\end{tabular}
    \end{minipage}

\end{table}

\begin{figure}[htbp]
\centering
\begin{subfigure}[b]{0.47\textwidth}
\includegraphics[width=\columnwidth]{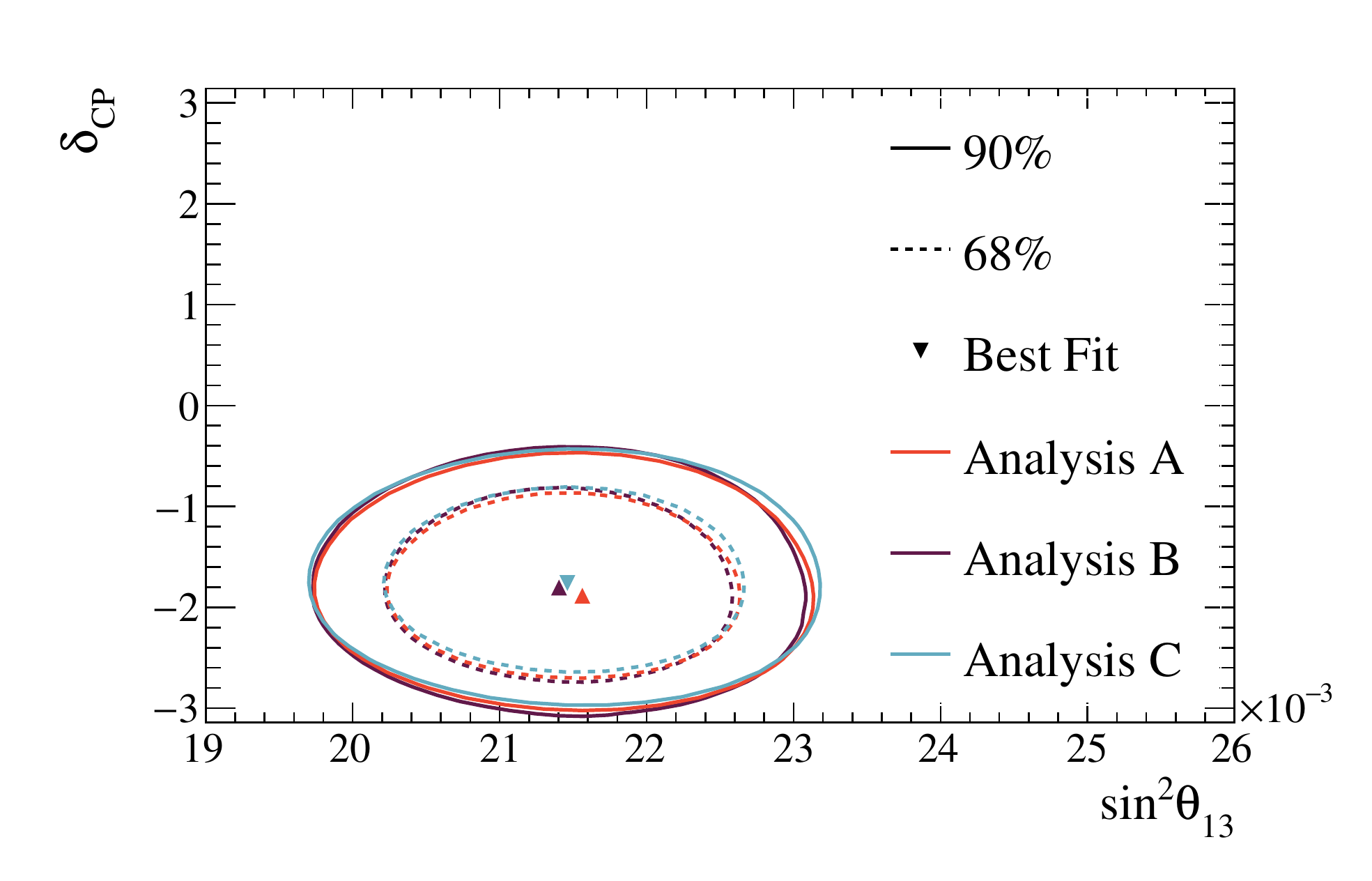}
\end{subfigure}
\begin{subfigure}[b]{0.47\textwidth}
\includegraphics[width=\columnwidth]{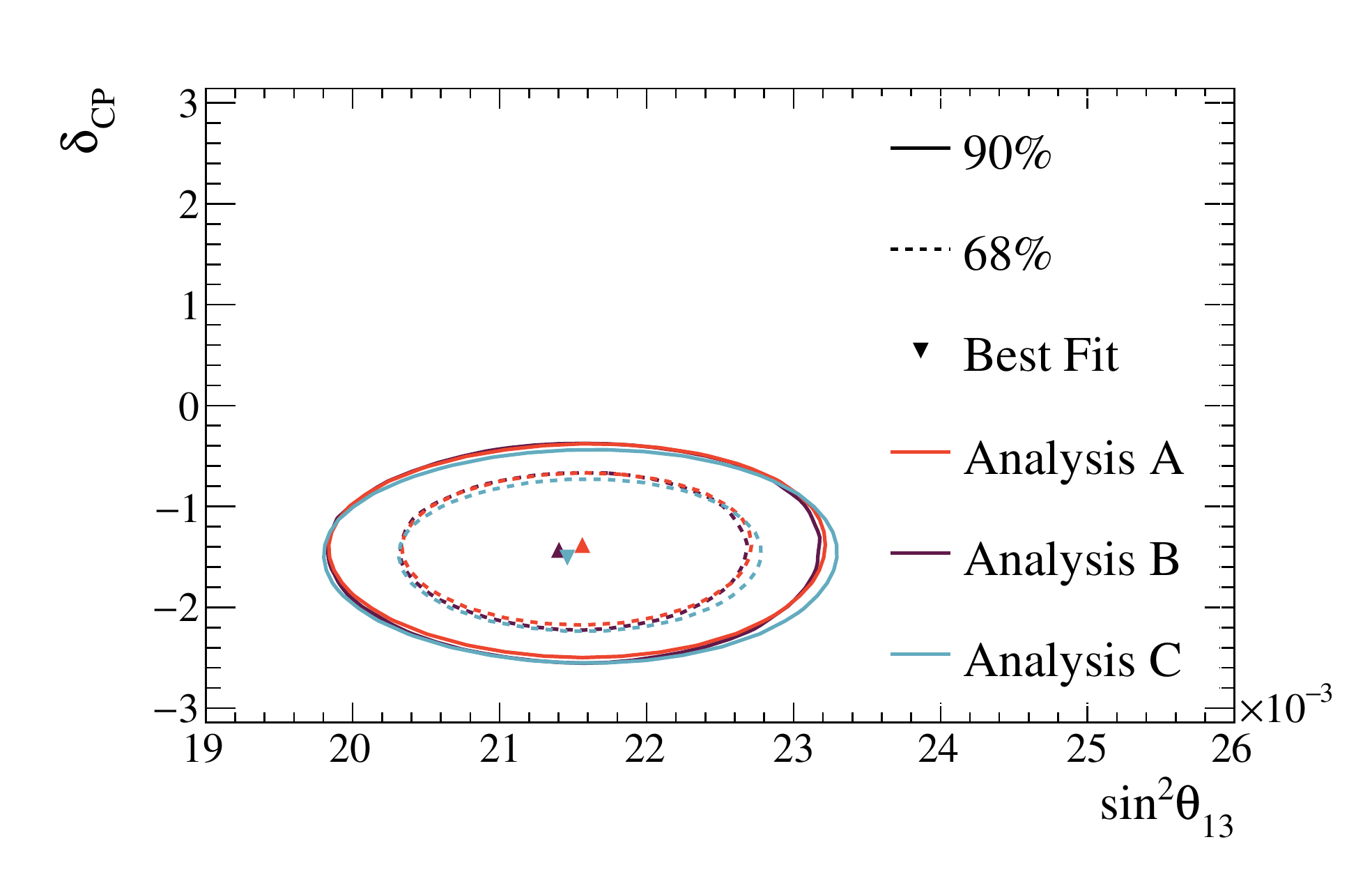}
\end{subfigure}
\caption {
    Comparison between the observed $1\sigma$ (dashed) and 90\% (solid) CL regions for \appearance produced using analyses A, B and C, assuming normal mass ordering (top) or inverted mass ordering (bottom). The reactor constraint is applied.
}
\label{fig:group_comparison_deltacp_mixing13_data}
\end{figure}

\begin{figure}[htbp]
\centering
\begin{subfigure}[b]{0.47\textwidth}
\includegraphics[width=\columnwidth]{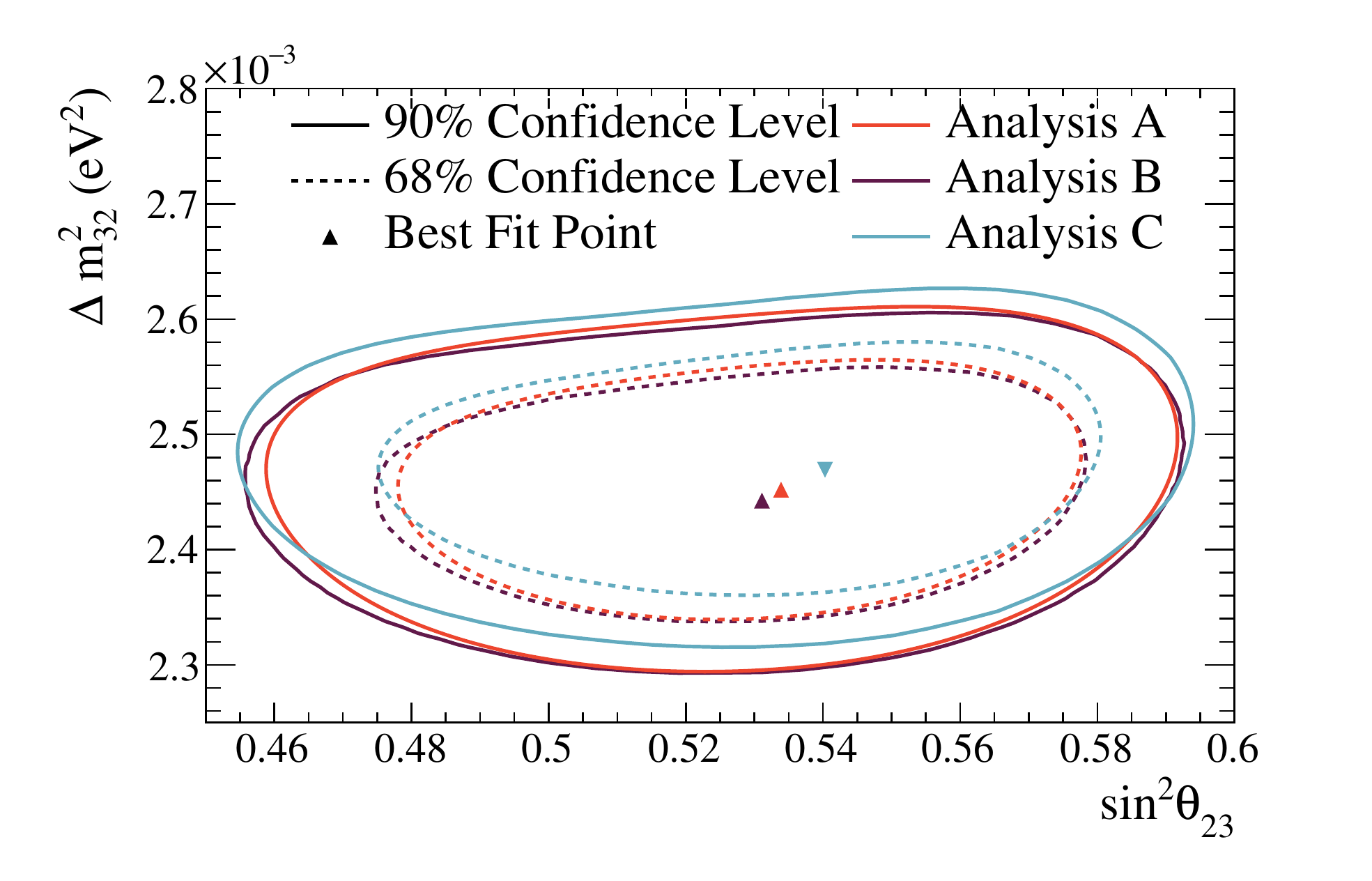}
\end{subfigure}
\begin{subfigure}[b]{0.47\textwidth}
\includegraphics[width=\columnwidth]{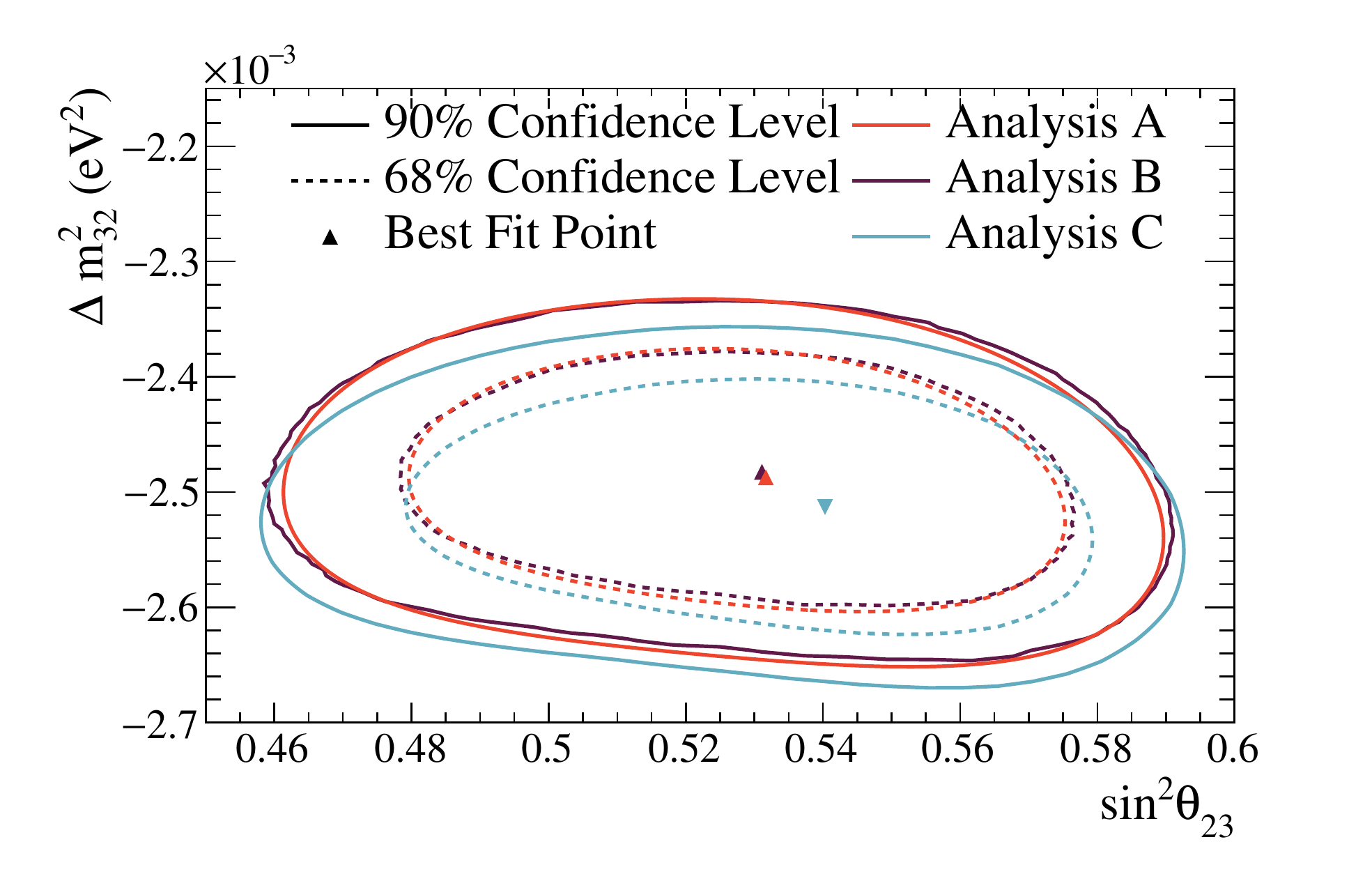}
\end{subfigure}
\caption {
    Comparison between the observed $1\sigma$ (dashed) and 90\% (solid) CL regions for \disappearance produced using analyses A, B and C, assuming normal mass ordering (top) or inverted mass ordering (bottom). The reactor constraint is applied.
}
\label{fig:group_comparison_dm23sq_mixing23_data}
\end{figure}

Section~\ref{sec:fitters} demonstrated that analyses A, B and C have similar sensitivities (Fig.~\ref{fig:group_comparison_asimova}), their data fit results are now compared for \appearance in Fig.~\ref{fig:group_comparison_deltacp_mixing13_data} and \disappearance in Fig.~\ref{fig:group_comparison_dm23sq_mixing23_data}. The largest differences are observed for analysis C, in particular for the atmospheric parameters. Those differences reduce to a negligible level if analysis C is repeated with the appearance samples binned in the same E$_\mathrm{rec}$  variable as analyses A and B, indicating this is primarily an effect coming from the choice of kinematic variables.
    
The intervals listed in Tabs.~\ref{tab:data_bestfit_t2k_only} and~\ref{tab:data_bestfit} were constructed using constant $\Delta \chi^2$ critical values. This treatment gives proper coverage when Wilks' theorem holds~\cite{Wilks:1938dza}, but can result in poor coverage when this is not the case. In particular, it is not expected to give proper coverage for \deltacp, due to the cyclic nature of the parameter and the presence of physical boundaries at $\pm \pi/2$. In these cases intervals with exact coverage can be formed directly from the likelihood ratio, by computing the appropriate $\Delta \chi^2$ critical value for each value of the parameters considered, as proposed by Feldman and Cousins~\cite{Feldman:1997qc}. This method is very CPU-intensive, so it is only used for the one-dimensional \deltacp interval.
    
For analyses A and C, critical values of $\Delta \chi^2$ (values of the parameter of interest for which the $\Delta \chi^2$ is lower than the critical value for a given confidence level are included in the confidence interval for this level) were calculated for multiple values of \deltacp as follows. Pseudo-experiments were generated at \NpointsFC\ evenly-spaced grid points of \deltacp in each mass ordering, and at two additional points near the intersection of the 3$\sigma$ critical value and  $\Delta \chi^2$ curves. The systematic and other oscillation parameters are randomly varied to generate the pseudo-experiments, with different procedures for the different parameters. The systematic parameters, $\sin^{2} 2 \theta_{12}$, \dmsqonetwo and $\sin^{2} 2 \theta_{13}$ are drawn from the prior probabilities described in Sec.~\ref{sec:fitters}, with $\sin^{2} 2 \theta_{13}$ constrained by reactor data. The two oscillation parameters which do not have a prior constraint in the analysis, \ssqthtwothree and \dmsqtwothree, are drawn from the 2D likelihood resulting from the fit of an Asimov data set generated at the best fit point obtained in the T2K+reactor data fit. Each set of parameter values is used to generate predicted kinematic distributions for each sample, which are then sampled assuming a Poisson probability in each reconstructed variable(s) bin to obtain a pseudo-experiment. The pseudo-experiments are fit in the same manner as the real data, and the $\Delta \chi^2$ between the true and best fit \deltacp and mass ordering is recorded. The N$^{\text{th}}$ percentile of this distribution then forms the N\% critical value for this combination of \deltacp and mass ordering.
    
\begin{figure}[htbp]
\centering
\includegraphics[width=0.47\textwidth]{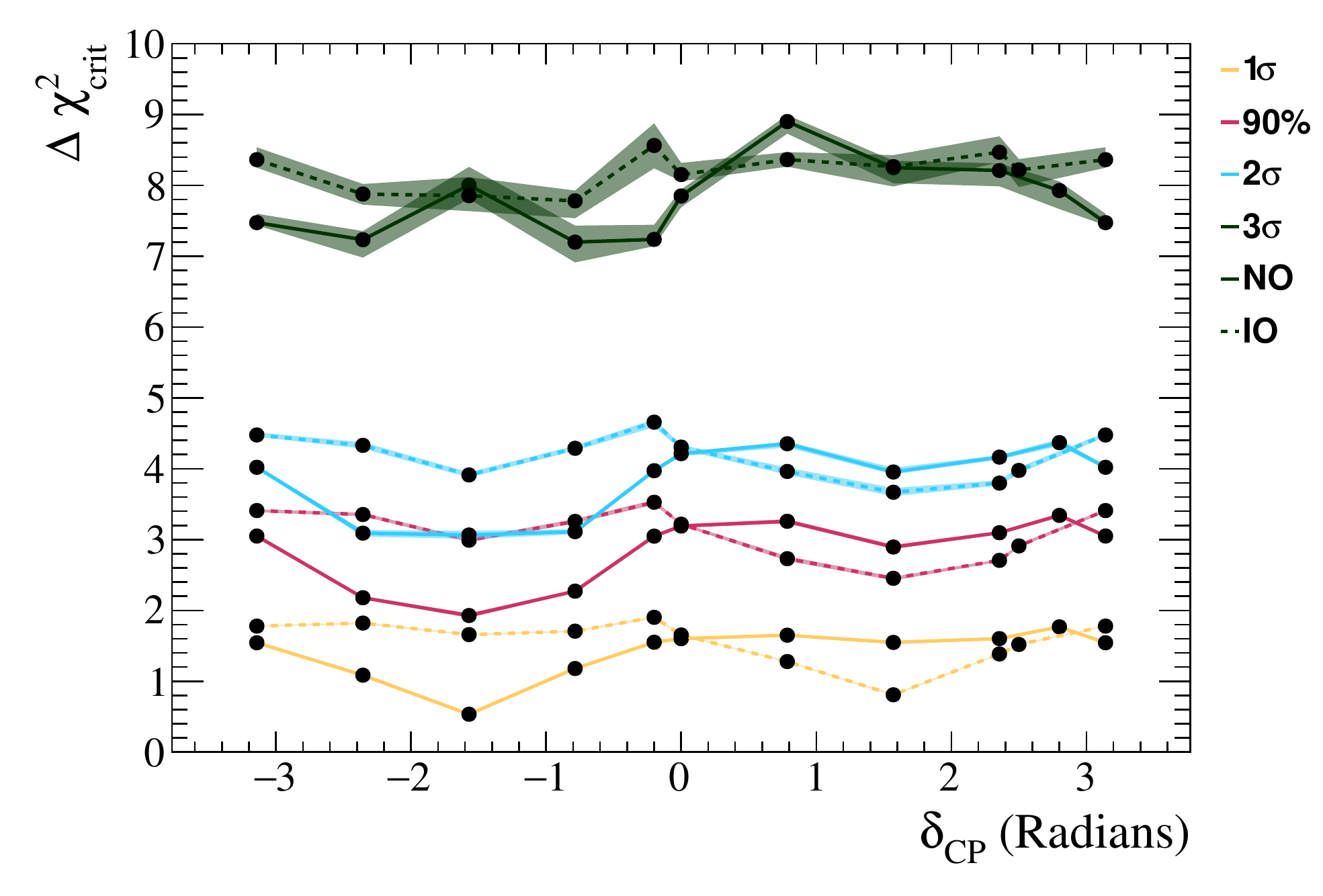}
\caption {
    Feldman--Cousins critical values for \deltacp. The shaded bands represent the $\pm 1\sigma$ Monte Carlo statistical uncertainty on the calculated critical values.
}
\label{fig:fc_critvals}
\end{figure}

The obtained critical values for \deltacp are displayed in Fig.~\ref{fig:fc_critvals}. For all confidence levels, significant deviations from the values expected for a parabolic log likelihood function (as assumed by the constant $\Delta \chi^2$ method) are observed, demonstrating the necessity of the method. The critical values obtained by Analysis A were compared to those of Analysis C and were generally found to be in good agreement, with minor differences mostly explained by the different kinematic variables used by the two analyses for appearance samples.

In order to better understand the structure of the critical values, the effects of the \deltacp-mass ordering and cos(\deltacp) degeneracies, the physical boundaries around $\deltacp = \pm \pi/2$ and of the beam exposure were studied and found to have a significant effect, due to the following factors:
	
\begin{enumerate}
	\item The amount by which the $\nu_{\mu}\rightarrow\nu_{e}$ and $\numub \rightarrow\nueb$ oscillation probabilities can be changed by varying \deltacp is limited. This creates an effect similar to physical boundaries at the values of \deltacp corresponding to the maximum and minimum number of events in the appearance samples. The critical values in those regions are therefore lower than those expected under the assumption of a parabolic log-likelihood.
		
	\item Mass ordering is a discrete parameter and works as an additional degree of freedom, raising critical values, though without sufficient freedom to make the critical values behave as if the problem had 2 degrees of freedom.
		
	\item The effect of physical boundaries is more visible for the $3\sigma$ critical values, since the critical values at this confidence level are usually determined by the pseudo-experiments corresponding to more extreme statistical fluctuations.
		
	\item The critical values increase with statistics, and the values obtained have increased relative to their values in previous T2K analyses~\cite{Abe2015k, Abe:2018wpn_T2Krun8osc, Abe:2017vif_T2Krun7osc}. 
\end{enumerate}
	
To understand this last point, critical values for the $1,2,3\sigma$ and $90\%$ confidence levels were computed assuming different exposures, and were found to non-linearly increase with exposure in all cases. The leading cause was found to be the approximate degeneracy between \deltacp and $\pi - \deltacp$ (T2K observables are mainly sensitive to $\sin\deltacp$, with $\cos\deltacp$ having a much smaller effect), which acts as an additional pseudo-degree-of-freedom that is negligible at low exposures but becomes more important as more data is taken. The above physical boundary effects also contribute at all exposures; however, for true \deltacp values away from the boundaries, the boundary effect decreases with increased exposure.	
	
The finite number of pseudo-experiments used to compute the critical values introduces a Monte Carlo statistical uncertainty, and the number of pseudo-experiments is chosen to make the size of this uncertainty negligible for the $2\sigma$ critical values; however, the number required to reduce the $3\sigma$ critical value statistical uncertainties to this level would be computationally intractable, so additional care is taken to ensure the validity of the $3\sigma$ confidence intervals. The confidence intervals are calculated for the different combinations of the nominal and $\pm 1\sigma$ binomial error values of the critical $\Delta \chi^2$, and the largest interval is kept. A second source of error comes from the interpolation of the critical $\Delta \chi^2$ between the points where it was computed. To minimize this effect, critical values have been calculated at two additional \deltacp values, each close to the boundaries of the $3\sigma$ confidence interval. 

\begin{figure}[htbp]
\centering
\includegraphics[width=0.47\textwidth]{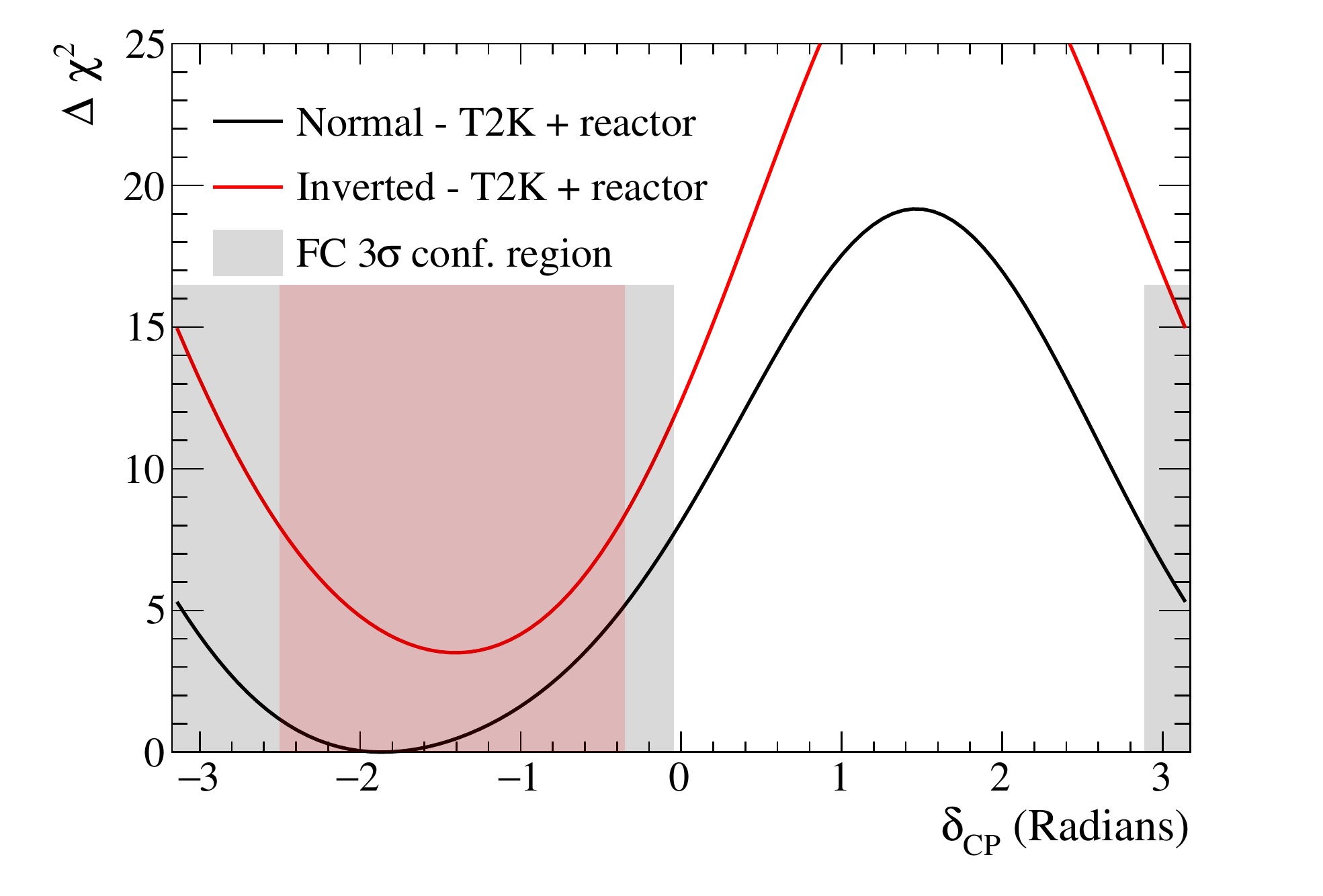}
\caption {
    The observed $3\sigma$ Feldman--Cousins (FC) confidence intervals for \deltacp. The $\Delta \chi^2$ is computed with respect to the best fit over the 2 mass orderings.
}
\label{fig:fc_deltacp_3sigma_data}
\end{figure}

The resulting confidence intervals obtained for \deltacp are listed in Tab. \ref{tab:data_deltacp_fc_intervals} and the obtained 3$\sigma$ intervals are displayed in Fig.~\ref{fig:fc_deltacp_3sigma_data}, superimposed on to the observed $\Delta \chi^2$ function. 
	
\begin{table}[htp]
	\caption{ Feldman--Cousins confidence intervals for \deltacp with the reactor constraint applied in both normal and inverted mass orderings. }
	\label{tab:data_deltacp_fc_intervals}
	\centering
	\begin{tabular}{ c  c  c }
		\hline \hline
		Confidence & \multicolumn{2}{c}{Interval (Radians)} \\
		Level & Normal Ordering & Inverted Ordering \\
		\hline
		$1\sigma$ & $\left[ -2.51, -1.26 \right]$ & --- \\
		$90\%$    & $\left[ -2.80, -0.84 \right]$ & --- \\
		$2\sigma$ & $\left[ -2.97, -0.63 \right]$ & $\left[ -1.80, -0.98 \right]$ \\
		$3\sigma$ & $\left[ -\pi,  -0.03 \right] \cup \left[ 2.87, \pi \right]$ & $\left[ -2.54,  -0.32  \right]$ \\
		\hline \hline
	\end{tabular}
\end{table}

In both mass orderings, the two CP conserving values of 0 and $\pi$ are outside of the 2$\sigma$ confidence intervals. Our data therefore exclude the conservation of CP symmetry in neutrino oscillation at the 2$\sigma$ level. For the inverted mass ordering, both \CP-conserving values are outside of the $3\sigma$ confidence intervals. For the normal mass ordering, $\deltacp = 0$ is just outside of the $3\sigma$ confidence interval, while $\deltacp = \pm \pi$ is inside. The robustness of the exclusion of $\deltacp = 0$ given the model uncertainties is studied in more details in Sec.~\ref{additonal_model_checks}, where it was found that the boundary of the $3\sigma$ interval is so close to $\deltacp = 0$ that this point can move in and out of the $3\sigma$ interval due to changes in the model. On the contrary, the exclusion of the CP-conserving values at the $2\sigma$ level was found to be robust.
~\\

\subsubsection{Bayesian results}\label{BayesianResults}
\begin{figure}[htbp]
  \centering
  \includegraphics[width=0.47\textwidth]{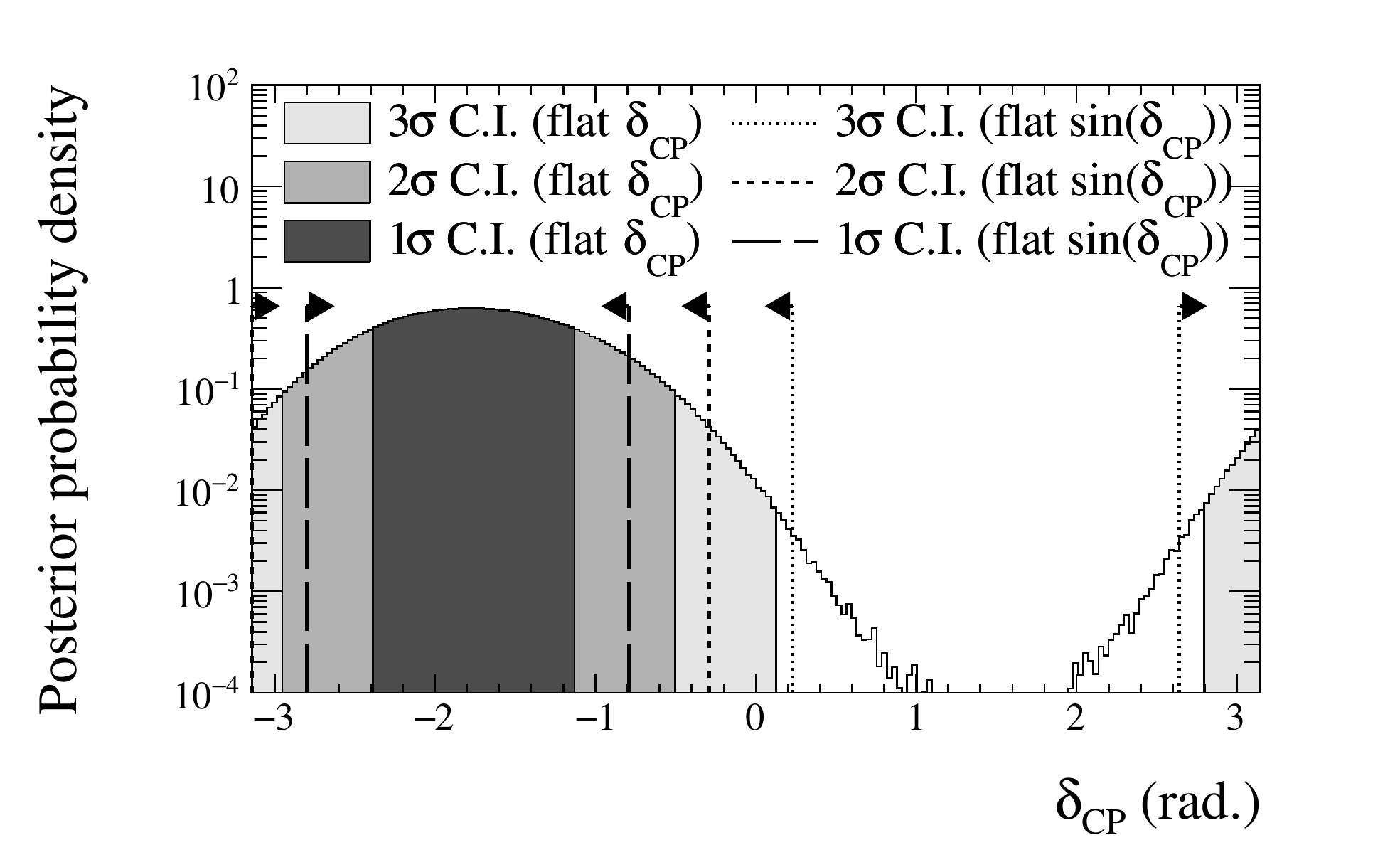}
  \caption{Posterior probability density for $\deltacp$ and credible
  intervals obtained using a prior uniform in $\deltacp$ compared against
  credible intervals obtained with a prior uniform in $\sin(\deltacp)$. The \CP conserving
  values are outside of the $2\sigma$ credible intervals in both cases.}
  \label{fig:bayesian:1Ddcp_compare_prior}
\end{figure}

Bayesian results for the oscillation parameters were produced using Analysis B. The posterior probability of $\deltacp$ marginalized over both mass orderings obtained in a fit of T2K data with $\ssqthonethree$ constrained by the reactor experiments is shown in Fig.~\ref{fig:bayesian:1Ddcp_compare_prior} with two different prior probabilities for the parameter of interest. The \CP conserving values of $\deltacp$ are found to be outside of the 2\,$\sigma$ credible interval in both cases, with the 1\,$\sigma$ range still covering the maximal \CP violation value of $\deltacp = -\pi/2$. Most of the results in this section use a prior probability uniform in $\deltacp$, the alternative prior probability was tested to see the effect of the choice of prior. The alternative was chosen uniform in $\sin(\deltacp)$ as this is both the variable involving $\deltacp$ to which our observables are most sensitive to, and the one relevant for CP violation. The detailed comparison between the intervals obtained with the two prior probabilities can be found in Tab.~\ref{tab:bayesian:dcpprior}, where it can be seen that this choice affects the size of the intervals obtained, but does not change the main conclusions of the analysis.

\begin{table}[htbp]
\centering
\caption{
Credible intervals for $\deltacp$ with uniform prior probabilities on either $\deltacp$ or on $\sin(\deltacp)$.
}
\label{tab:bayesian:dcpprior}
\begin{tabular}{l c@{\quad} c }
    \hline\hline    
    Prior uniform \rlap{in} & \deltacp & $\sin\deltacp$\\
    \hline
    68.26\,\% ($1\sigma$) & $[-2.39,\,-1.13]$ & $[-2.80,\,-0.79]$\\
    95.45\,\% ($2\sigma$) & $[-2.76,\,-0.72]$ & $[-3.14,\,-0.29]$\\
    99.73\,\% ($3\sigma$) & $[-\pi,\,0.13] \cup [2.80,\,\pi]$ & $[-\pi,\,0.23] \cup [2.65,\,\pi]$\\ 
    \hline \hline
\end{tabular}
\end{table}

\begin{figure}[htbp]
  \centering
  \includegraphics[width=0.47\textwidth]{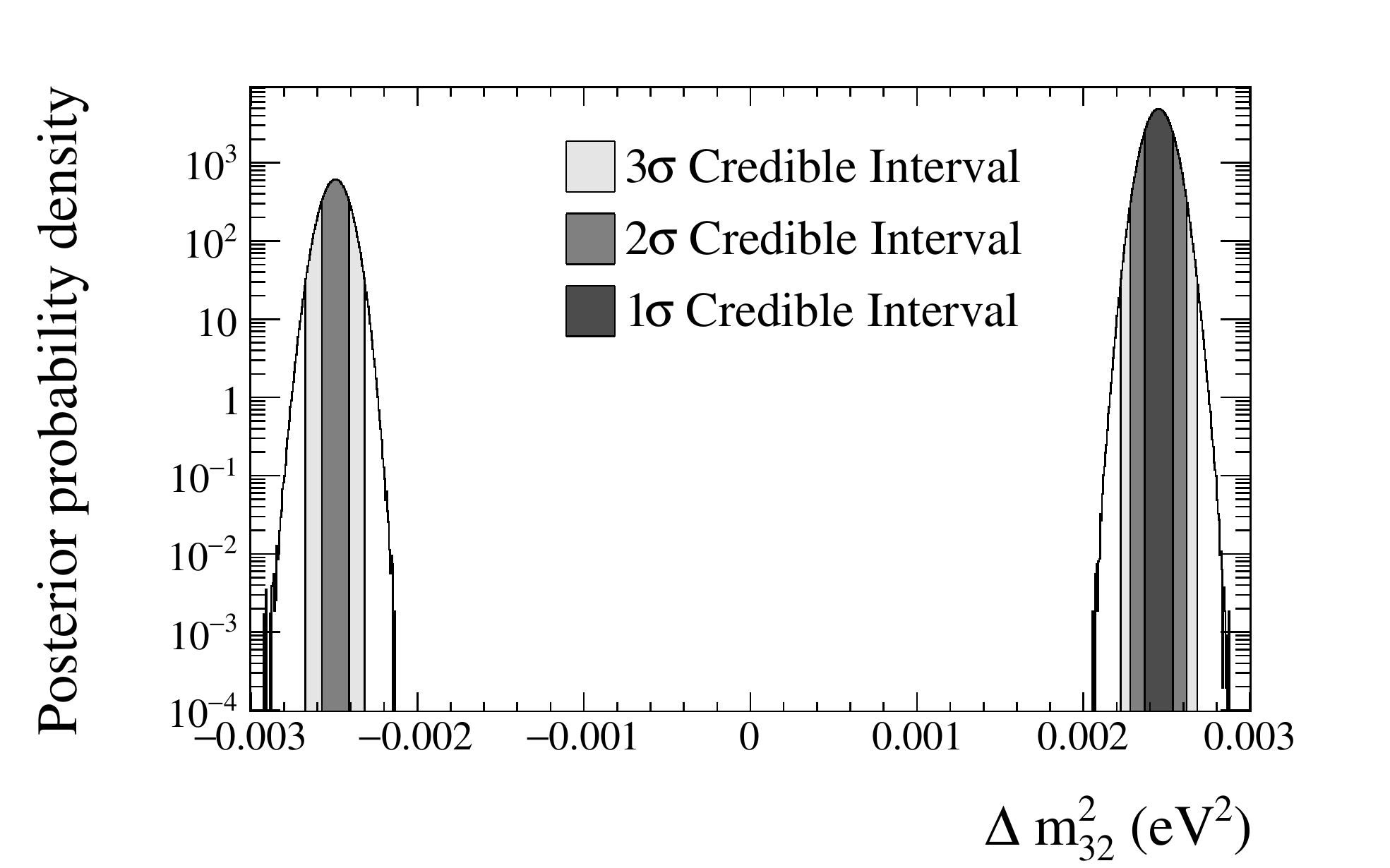}
  \caption{Posterior probability distribution for $\Delta
  m^2_{32}$ covering both mass orderings with 1, 2 and $3\sigma$
  Bayesian Credible Intervals marked. The normal ordering contains 89\% of the posterior
  probability.}
  \label{fig:bayesian:1Ddm23both}
\end{figure}

The proposal function for choosing the random steps used in Analysis B's MCMC can jump between mass orderings, with a 50\% chance at each step to propose a point with the opposite sign of $\Delta m^2_{32}$. It therefore produces a single posterior probability for $\Delta m^2_{32}$, covering both positive and negative values. The posterior probability obtained in a fit of T2K data with constraint from the reactor experiments, is shown in Fig.~\ref{fig:bayesian:1Ddm23both}. A clear preference towards the normal ordering, which contains 89\% of the total posterior density, can be seen, with in particular the lack of an observed 1\,$\sigma$ credible region in the $\Delta m^2_{32} < 0$ part of the figure. 
\begin{figure}[htbp]
  \centering
  \includegraphics[width=0.47\textwidth]{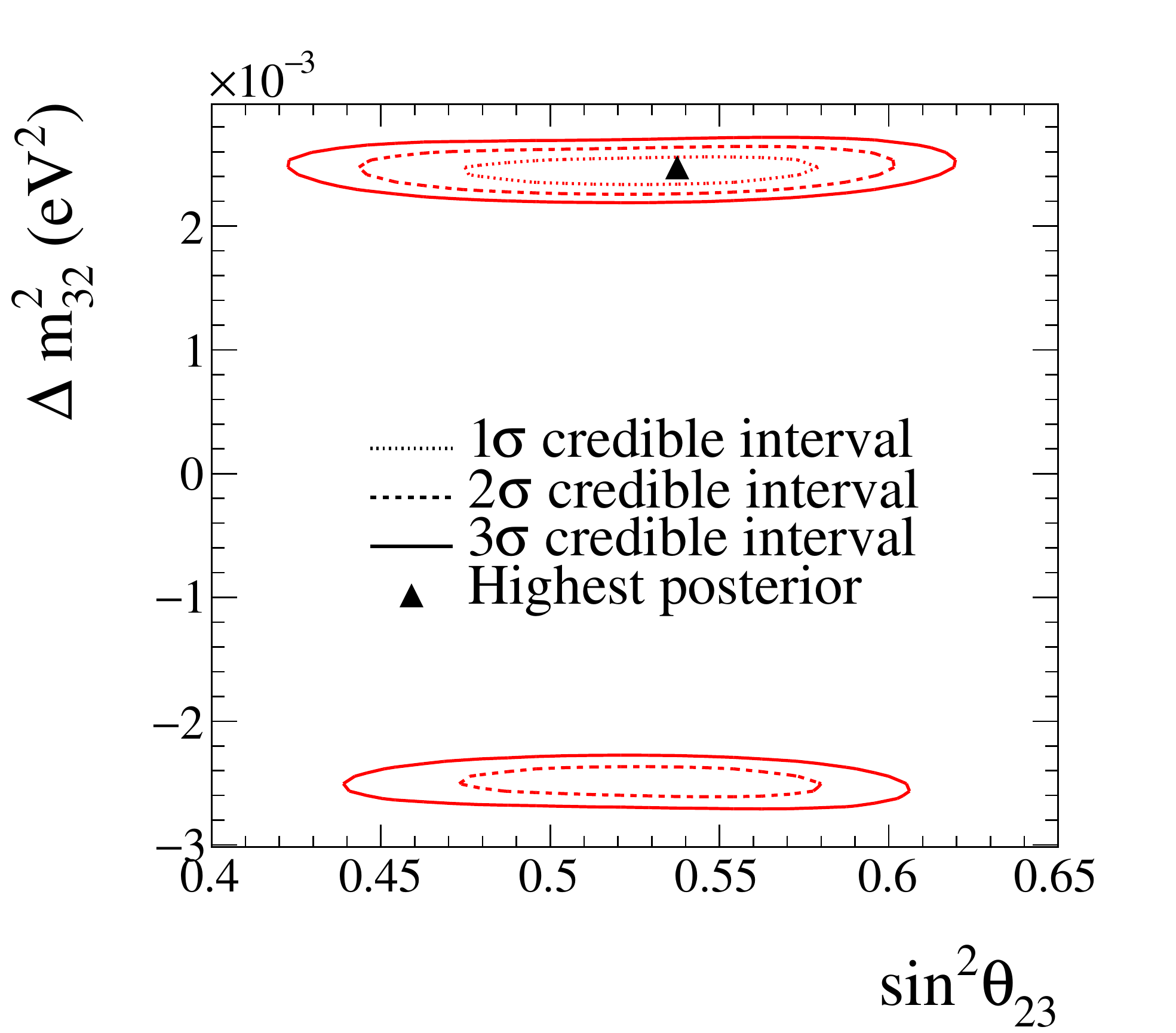}
  \caption{Bayesian credible regions for $\Delta m^2_{32}$ against
  $\sin^2\theta_{23}$ covering both mass orderings.
  }
  \label{fig:bayesian:disappearance}
\end{figure}

\begin{figure*}[htbp]
  \centering
      \includegraphics[width=0.98\textwidth]{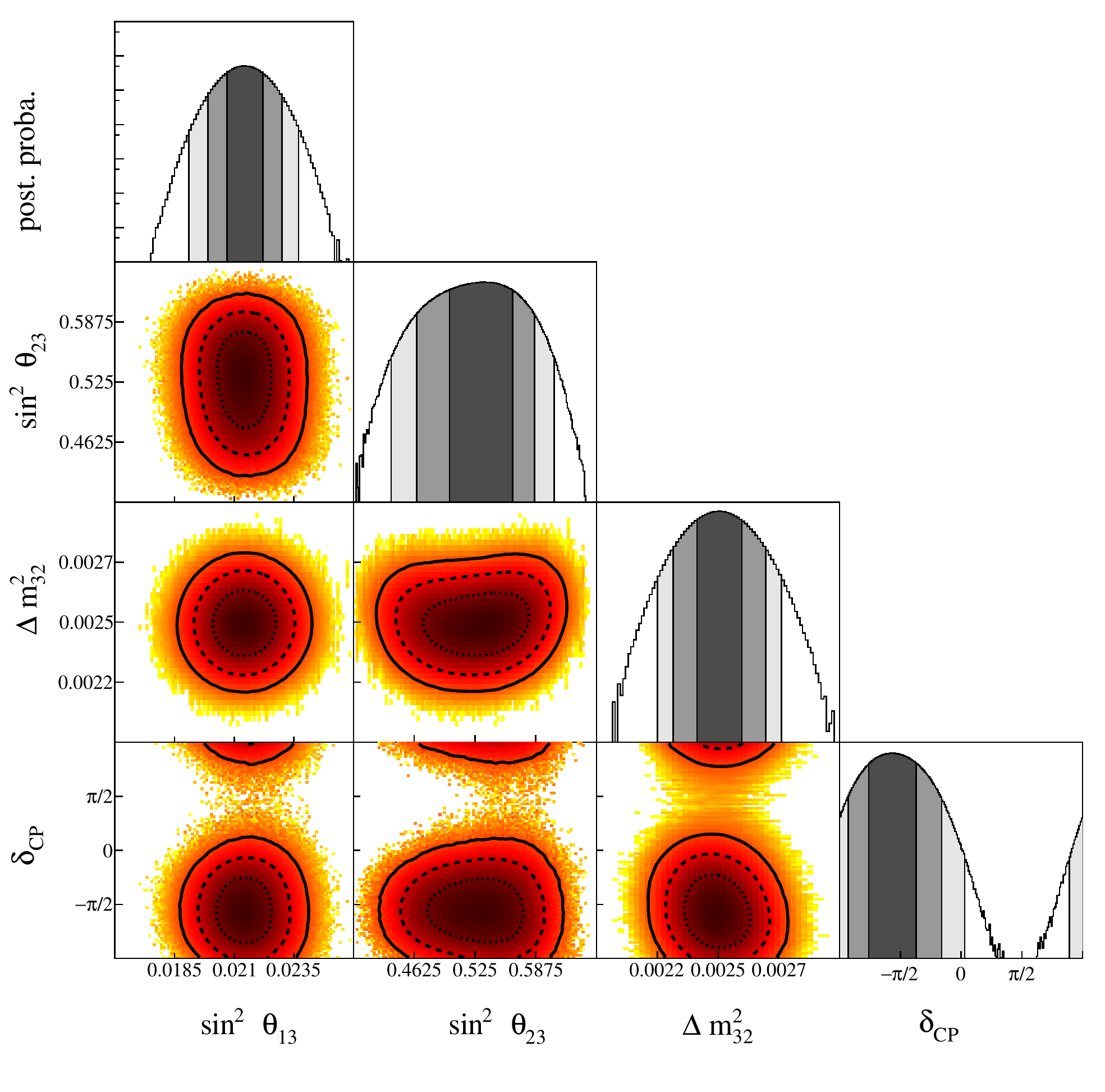}
  \caption{Posteriors probabilities together with 1, 2 and 3~$\sigma$ credible intervals for all the oscillation parameters of interest and their combinations in the normal mass ordering. A logarithmic scale is used for the axis corresponding to the posterior probability density, and darker colors correspond to larger probabilities.}
  \label{fig:bayesian:triangleplot}
\end{figure*}

Figure~\ref{fig:bayesian:triangleplot} shows the posterior
density distributions and credible regions for combinations of all the oscillation parameters, as well as their individual posterior density and credible intervals.
Correlations between the estimates of the different parameters can be seen, as well as the effect of marginalizing over one of them to obtain the one-dimensional posterior distribution of another oscillation parameter. As the \ssqthtwothree posterior distribution has a largely non-Gaussian shape, the marginal likelihood (for the other parameters) used in the analyses described in this paper is expected to have some differences with the profile likelihood commonly used in the neutrino community. Correlations are mostly seen between the atmospheric parameters, \ssqthtwothree and \dmsqtwothree, and between \dmsqtwothree and \deltacp.
The correlation between the atmospheric parameters is clear from Eq.~\ref{equ:disappearance}.
For \dmsqtwothree and $\cos($\deltacp$)$, both parameters shift the energy of the peak (dip) in the electron (muon) neutrino oscillation probability.
Combined with the energy profile of the T2K beam this energy shift produces a change in event rate for the electron-like samples, and a smaller change in the muon-like samples.
This change in both rate and shape introduces a correlation between the two parameters.

\subsection{Additional checks on the validity of \deltacp results}
The main result of the analysis described in this paper is the measurement of \deltacp, with in particular the exclusion of \CP conservation in neutrino oscillations at the $2\sigma$ level, and the fact that some of the possible values of \deltacp are outside of the $3\sigma$ confidence intervals. The observed constraint on \deltacp is considerably stronger than that expected from sensitivity studies, as can be seen by comparing Figs.~\ref{fig:dcp_asimova} and~\ref{fig:dcp_data}. A number of additional studies were done to check the likelihood and robustness of this result.

\subsubsection{Probability of the \deltacp results}

\begin{figure}[htbp]
\centering
\begin{subfigure}[b]{0.47\textwidth}
    \includegraphics[width=0.98\columnwidth]{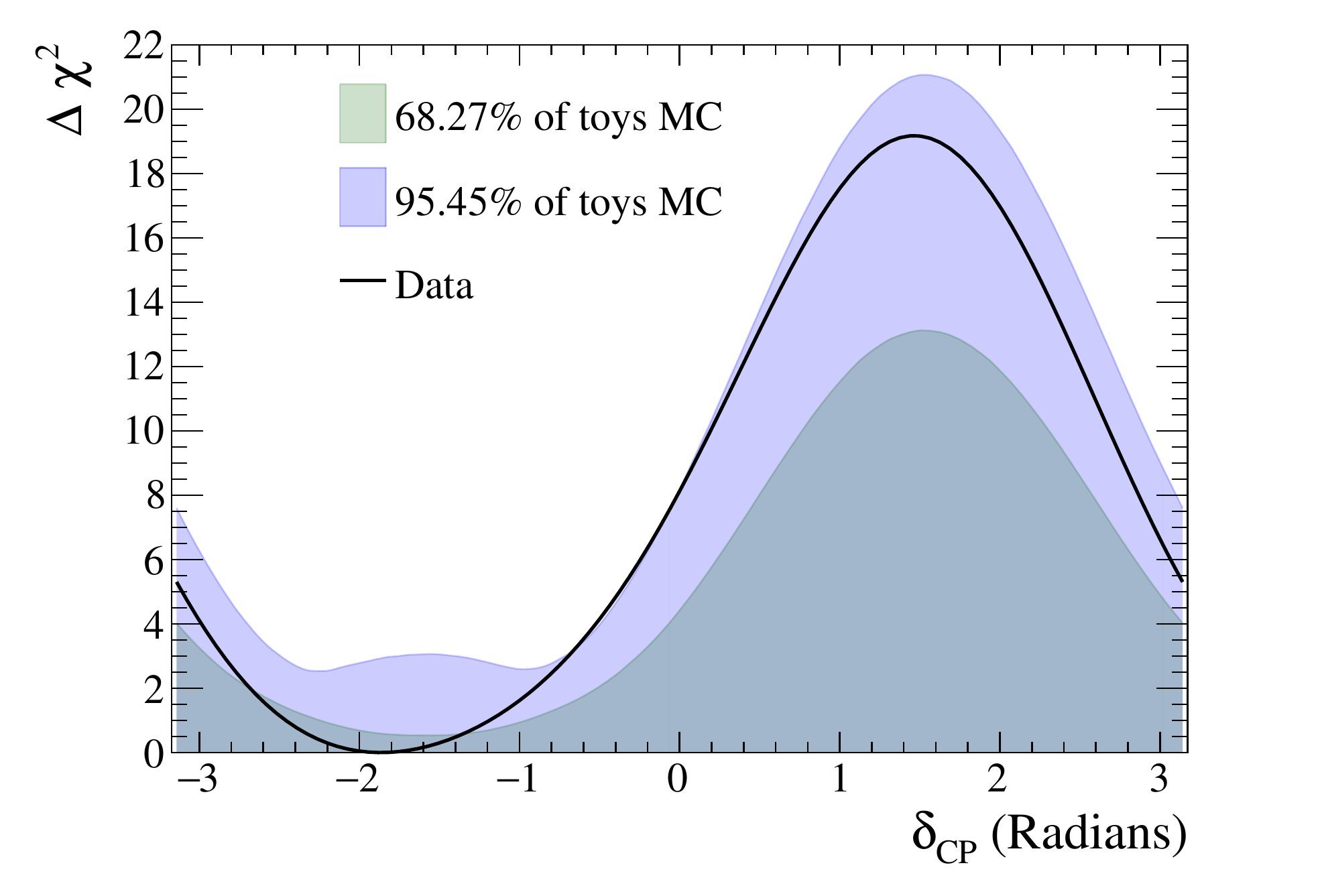}
\end{subfigure}
\begin{subfigure}[b]{0.47\textwidth}
    \includegraphics[width=0.98\columnwidth]{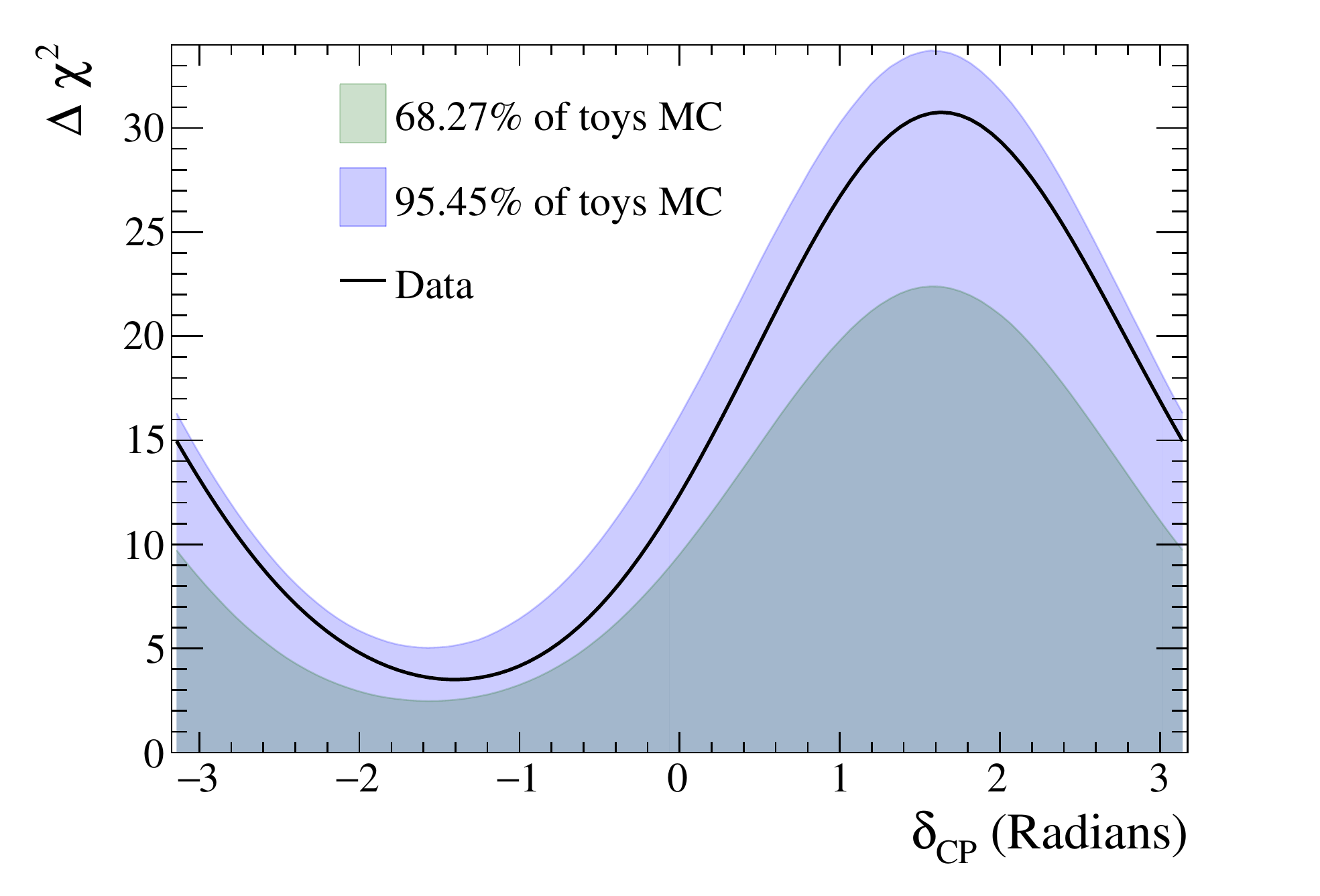}
\end{subfigure}
\caption {
	The observed $\Delta \chi^2=-2 \Delta \ln L$ functions compared with one-sided distributions of $\Delta \chi^2$ values corresponding to 68.27\% and 95.45\% of \num{10000} pseudo-experiments generated for $\deltacp = -\pi/2$ and normal mass ordering. Results for the normal mass ordering (top) and inverted mass ordering (bottom) are shown.
}
\label{fig:brazilian_dcp_mpi2}
\end{figure}

The probability of getting a constraint this strong or stronger considering systematic and statistical uncertainties was evaluated assuming normal ordering and a true value of \deltacp of $-\pi/2$. This case was chosen as it gave the best agreement between data and predictions in Tab.~\ref{tab:events_asimov_a_dcp_var}. \num{10000} pseudo-experiments were generated following the same method as used in the determination of the Feldman--Cousins critical values. The pseudo-experiments were then fitted for \deltacp, and the resulting  $\Delta \chi^2$ distributions were compared to the $\Delta \chi^2$ functions obtained for the data. Figure~\ref{fig:brazilian_dcp_mpi2} shows the observed $\Delta \chi^2=-2 \Delta \ln L$ overlaid on the one-sided $1\sigma$ and $2\sigma$ bands. Those bands are built from the ensemble of $\Delta \chi^2$ values from the different pseudo-experiments at each value of \deltacp, and cover the interval between zero and the  68.27\% ($1\sigma$ band) or 95.45\% ($2\sigma$ band) quantile. The observed $\Delta \chi^2$ function lies within the one-sided $2\sigma$ band for the fits done in the inverted mass ordering scenario, whilst falling just outside the $2\sigma$ band in the normal mass ordering scenario in the region around $\deltacp = 0$. An alternative way to consider the results is to check what fraction of pseudo-experiments exclude one or both of the CP conserving values at a specified confidence level. As shown in Tab.~\ref{tab:exclusion_prob_bf}, for the normal mass ordering, $\deltacp = 0$ and $\deltacp = \pi$ are separately excluded at 90\% confidence level in around 46--48\% of pseudo-experiments and excluded at $2\sigma$ confidence level in around 32--34\%. Both \CP-conserving values are excluded at the $2\sigma$ confidence level in about 25\% of the pseudo-experiments. Overall, a $2\sigma$ exclusion of \CP conservation is not unlikely according to the model used in this analysis.
	
\begin{table}[htp]
	\centering
		\caption{Fraction of the pseudo data sets for which the $\Delta \chi^{2}$ at $\deltacp = 0, \pi$ and both \CP conserving values are above the critical values for 90\% and $2\sigma$\,CL. Pseudo data sets were generated assuming true normal ordering and a value of $\deltacp = -\pi/2$
	}
	\label{tab:exclusion_prob_bf}
	\begin{tabular}{ c  c  c  c }
		\hline \hline
		\deltacp    & 
		\begin{tabular}{@{}c@{}}\rule{0pt}{2.6ex}Mass ordering \\[-0.5ex] scenario \end{tabular}& 90\%  & $2\sigma$ \\ 
		\hline
		0           &          & 0.48 & 0.34     \\
		$\pi$       & Normal   & 0.46 & 0.32     \\
		0 and $\pi$ &          & 0.37 & 0.25     \\[0.5ex]
		0           &          & 0.83 & 0.74     \\
		$\pi$       & Inverted & 0.82 & 0.73     \\
		0 and $\pi$ &          & 0.78 & 0.69     \\
		\hline \hline
	\end{tabular}
\end{table}

\subsubsection{Contributions from individual samples}
\begin{figure}[htbp]
\centering
\includegraphics[width=0.47\textwidth]{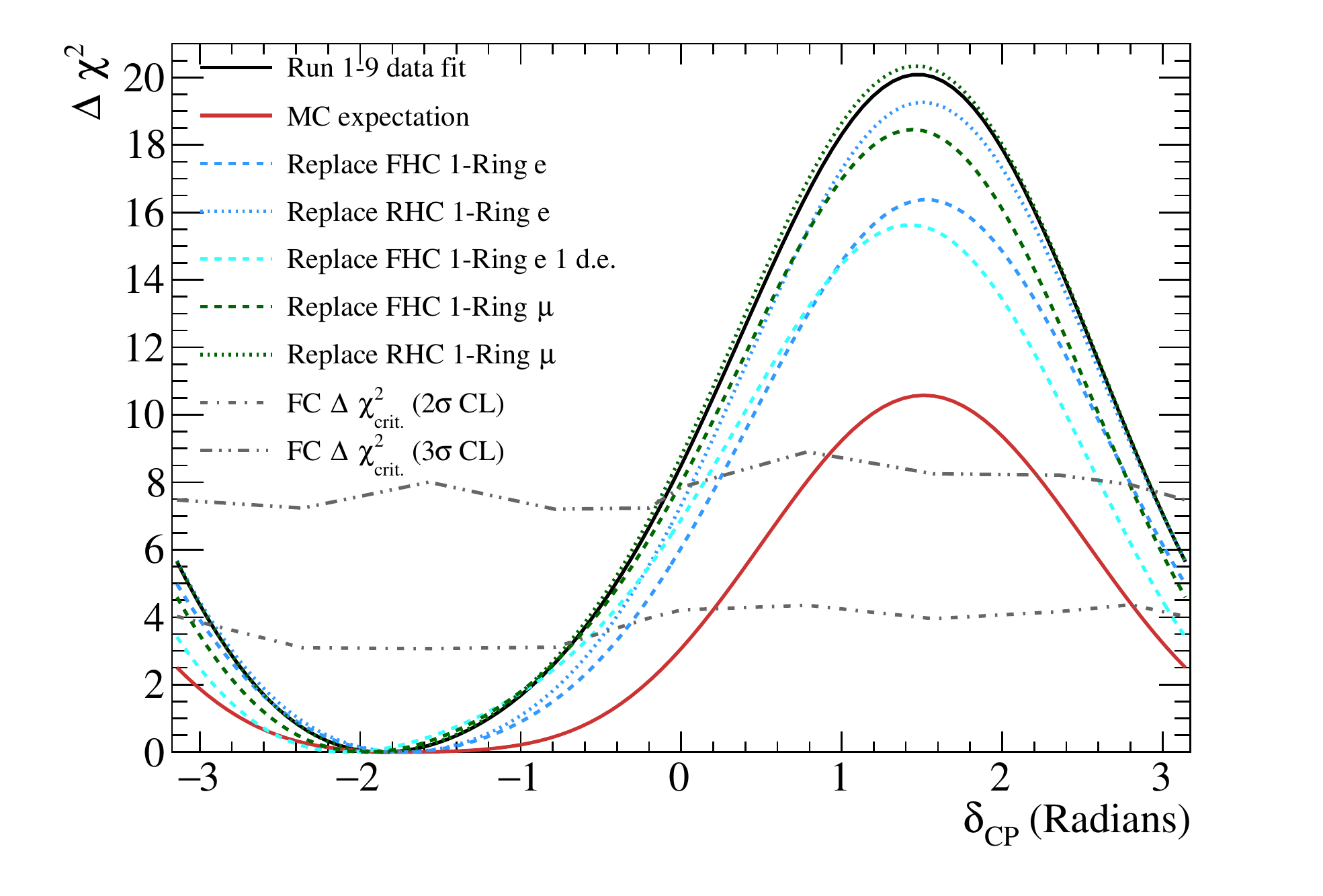}
\caption{
    Change of the results of the fit of the T2K run 1--9 data for $\deltacp$ when observations get replaced by predictions for the different samples. The gray dashed lines correspond to the critical values for 2$\sigma$ and 3$\sigma$ obtained using the Feldman--Cousins method.
}
\label{fig:dcp_hybrid}
\end{figure}

The discrepancies between predictions and observations seen in some of the samples used in the analysis were studied in more detail. In Tab.~\ref{tab:events_asimov_a_dcp_var}, the observed numbers of events in the far detector samples are generally in good agreement with the predicted number of events for $\deltacp=-\pi /2$. Two samples show a difference with the nominal prediction: for the FHC 1R $\mu$-like sample a small deficit of events can be seen in the data, corresponding to 1.33 times the statistical and systematic uncertainty on the predicted number of events. For the FHC $\nue$ CC 1$\pi^{+}$-like sample, the discrepancy is larger with the observed number of events being twice the predicted number. To understand the impact on the \deltacp result of the differences between predictions and observations from each sample, hybrid MC/data fits were performed: the data were replaced by the MC predictions for each sample in turn, as shown in Fig.~\ref{fig:dcp_hybrid}.

It can be seen that in all cases including data, the constraint on \deltacp is stronger than that obtained from the MC predictions alone. For the \CP conserving value $\deltacp=- \pi$, the most significant changes happen when data are replaced by MC predictions for the two samples which showed differences between observations and predictions. In particular, in the case of the FHC $\nue$ CC 1$\pi^{+}$-like sample, the $\Delta \chi^{2}$ at $\deltacp=- \pi$ becomes lower than the 2$\sigma$ critical value obtained with the Feldman--Cousins method, meaning that the conservation of \CP symmetry is no longer excluded with 2$\sigma$ significance in this case.

Additional studies were therefore performed to estimate the likelihood of the observations for the FHC $\nue$ CC 1$\pi^{+}$-like sample. First, the probability to obtain 15 events or more in this sample when taking into account statistical and systematic uncertainties was evaluated. This probability was found to be 2.49\% for true values of the oscillation parameters corresponding to the T2K-only best fit, and 1.34\% for the T2K with reactor constraint best fit point. As there are 5 samples in the analysis, a trial factor should be taken into account. The probability to have such an excess in at least one of the 5 samples (meaning an excess of events in the sample corresponding to a $p$-value smaller or equal to the $p$-value for the FHC $\nue$ CC 1$\pi^{+}$-like sample) was found to be 11.3\% for the T2K only best fit, and 5.8\% for the T2K+reactor best fit point.

\begin{figure}[htbp]
\centering
\includegraphics[width=0.47\textwidth]{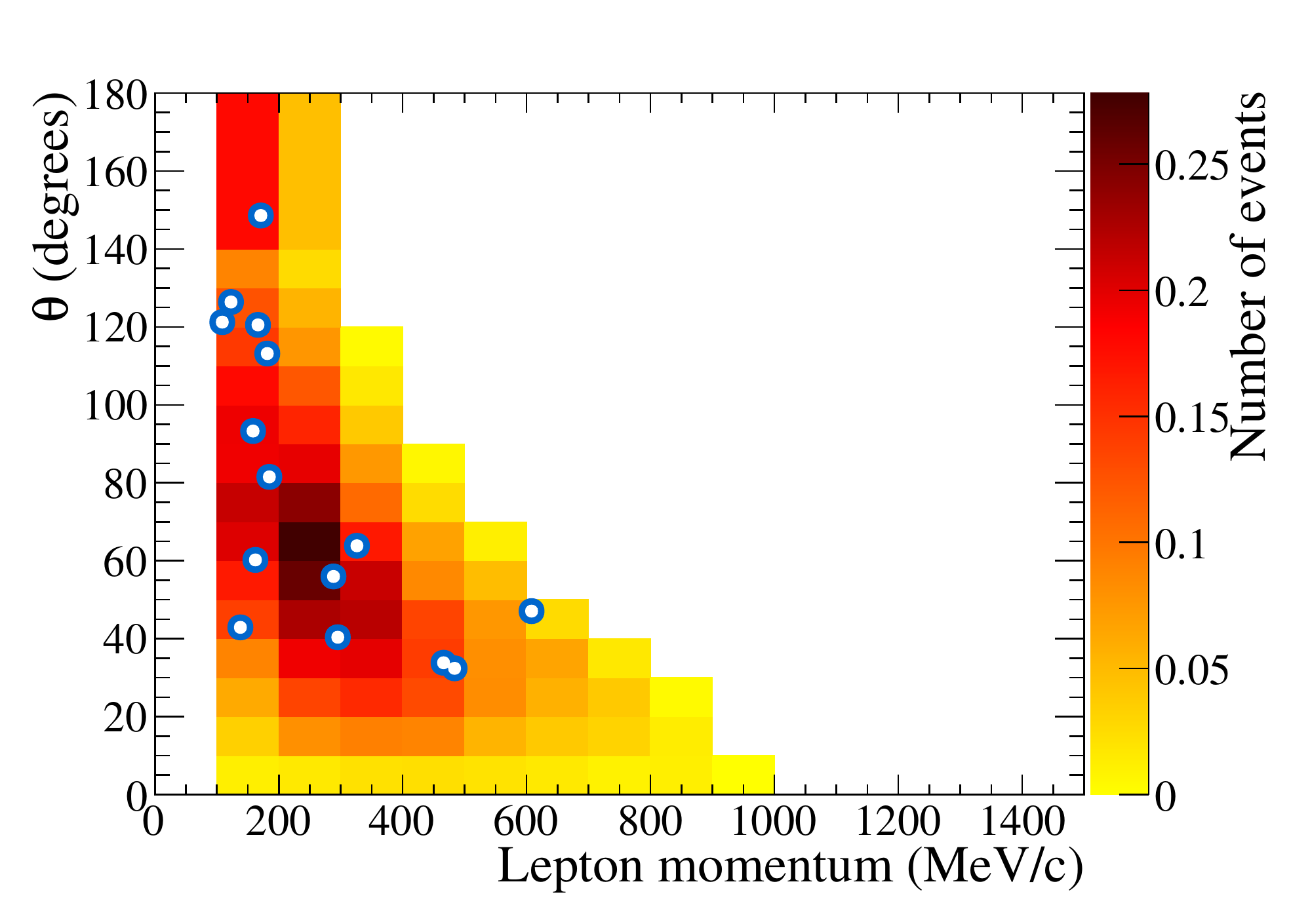}
\caption {
    Lepton momentum and angle with respect to the beam direction for the 15 events observed in the FHC $\nue$ CC 1$\pi^{+}$-like sample, overlaid with the MC predictions.
}
\label{fig:dataMCCC1Pi}
\end{figure}

$p$-values were also calculated when taking into account not only the number of events, but also the distribution of the kinematic variables of the observed events, and were found to be equal to the rate-only $p$-value. As can be seen in Fig.~\ref{fig:dataMCCC1Pi}, the kinematic distributions of the data events are in good agreement with the prediction, and so taking into account the shape information does not increase the disagreement between data and prediction.

\subsubsection{Additional interaction model checks}\label{additonal_model_checks}
Section~\ref{sec:fake_data_studies} described checks for possible biases coming from the choice of interaction model. Those checks were based on comparisons of MC sensitivities obtained with different interaction models. However, some model uncertainties affect primarily the  FHC $\nue$ CC 1$\pi^{+}$ sample, which brings only a small contribution to the \deltacp sensitivity compared to the single ring $e$-like samples, and therefore do not have a significant effect on the simulated data studies. They can nevertheless have a sizable effect on the predictions for the FHC $\nue$ CC 1$\pi^{+}$ sample, and given the large excess of data events in this sample, and its effect on the \deltacp measurement, the possible impact of additional model uncertainties were considered. The additional studies were done in the context of the data fit and not the MC based sensitivity studies. 

The first source of uncertainty studied is the data/MC discrepancy in the pion spectrum for the near detector CC 1$\pi$ sample. Even after tuning using the results of the near detector data fit, the predicted pion momentum spectrum did not reproduce the data for the \emph{near detector} FHC CC 1$\pi^{+}$ sample: the data events had lower pion momentum than predicted by the model.
This discrepancy could have an impact on the prediction for the far detector FHC CC 1$\pi^{+}$  sample: charged pions can appear at SK either as rings if they have high enough momentum ($>156\MeVmom$), or as Michel electrons from the decay of the pion. If the pions produced in CC 1$\pi^{+}$ interactions have lower momentum than our model predicts, a larger fraction of the CC 1$\pi^{+}$ events at SK will enter the FHC CC 1$\pi^{+}$  sample than our MC predicts. Studies showed that a discrepancy of the size observed at the near detector could lead to a 10\% increase in the number of events in the far detector CC 1$\pi^{+}$ sample. To evaluate the possible impact of this uncertainty on our \deltacp results, the fit of the run 1--9 data was redone with two different modifications:
\begin{itemize}
\item adding a 10\% normalization uncertainty on the number of events in the far detector \nueCConepi sample.
\item increasing the number of events in the SK \nueCConepi sample by 10\% in the MC predictions used to fit the data.
\end{itemize}
    
The second model uncertainty concerns the number of hadrons produced in deep inelastic interactions in the low invariant mass region $W<2\GeV$.  Those interactions correspond to a specific interaction mode in the \textsc{NEUT} generator used to produce the MC, and have at least two pions produced at the interaction level. They can nevertheless enter the \nueCConepi sample if some of those pions re-interact in the nucleus (through the final state interactions) or are not detected. The uncertainties on the number of hadrons produced in those interactions produce an uncertainty on the fraction that will enter the \nueCConepi~sample. To assess the potential impact on the \deltacp result, the fit of the run 1--9 data was redone using two alternative models for the hadron multiplicity model in the MC: the first (M1) is based on fits of data from deuterium bubble chamber experiments~\cite{DeuteriumFits}, while the second (M2) is based on the multiplicity part of the AGKY model~\cite{AGKY}. Only the alternative model M1 was found to significantly change the expected number of events, with the biggest effect seen for the \nueCConepi sample ($+13.5\%$). The predicted reconstructed energy spectrum was also affected for this sample, as the increase in the number of events was primarily at low reconstructed energy.

\begin{figure}[htbp]
\centering
\includegraphics[width=0.47\textwidth]{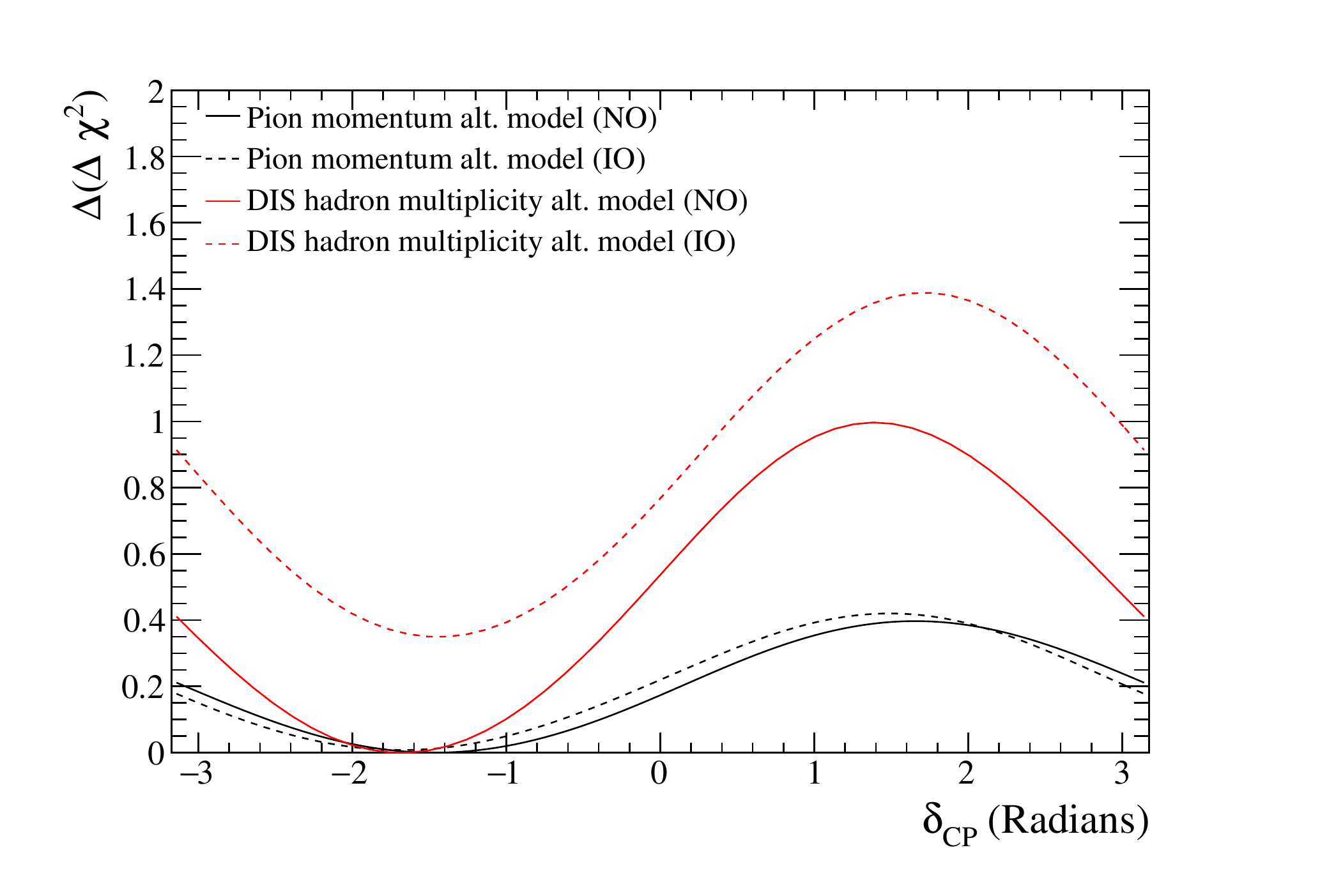}
\caption{
    Change in the $\Delta \chi^{2}$ function for \deltacp when fitting data using an alternative model for DIS hadron multiplicity (red) and pion momentum spectra (black). The $\Delta \chi^{2}$ was \textit{reduced} by the value shown on the plot in each case. The difference is shown for the model change that gave the largest difference in each case.
}
\label{fig:PionSummary}
\end{figure}

Figure~\ref{fig:PionSummary} shows how the $\Delta \chi^{2}$ obtained in the data fit for \deltacp changed when an alternative model was used to fit the data. For each of the two model uncertainties, only the case which gave the largest effect is shown: 10\% increase in the SK \nueCConepi sample MC predictions for the pion spectrum case, and M1 model for the hadron multiplicity model case. Both those changes predict an increase in the number of events in the \nueCConepi sample, reducing the discrepancy with the data, and therefore weakening the constraint on \deltacp. The magnitude of the change in $\Delta \chi^{2}$ is small enough that it does not change the main conclusions obtained in the fit of the data for \deltacp: the \CP-conserving values are excluded at the $2\sigma$ level, and some values of \deltacp are outside of the $3\sigma$ confidence level intervals. It is large enough however to change whether $\deltacp=0$ is excluded at the $3\sigma$ confidence level or not in the results of analyses A and C.

\subsection{Neutrino mass ordering}\label{section_MH}
The question of the mass ordering was studied in a Bayesian framework, by computing posterior probabilities and Bayes factors for each ordering hypothesis. Results shown in this section use the constraint from reactor experiments on $\sin^2 2 \theta_{13}$ unless otherwise stated.

\subsubsection{Posterior probabilities and Bayes factor}
All three analyses evaluated posterior probabilities to estimate the preference for the neutrino mass orderings. Analyses B and C additionally looked at the posterior probabilities for the octant of $\sin^2\theta_{23}$. The practical calculation of the posterior probabilities differs between the analyses, due to differences in the fitting techniques used. Analyses A and C first compute the marginal likelihood for each hypothesis, and compute the posterior probability for hypothesis $H_i$ from:
\begin{equation}
  P(H_i|N^{obs}, \boldsymbol{x}) = \frac{L_{marg}(H_i;N^{obs}, \boldsymbol{x}) P(H_i)}{\sum\limits_{j} L_{marg}(H_j;N^{obs}, \boldsymbol{x}) P(H_j)}.
  \label{eq:pthetamodelselection}
\end{equation}
\noindent Where the denominator sums over all the possible hypothesis combinations (either the two mass orderings, or the 4 combinations of mass ordering and octant), $(N^{obs},\boldsymbol{x})$ is the observed measurement and $P(H)$ is the prior probability, taken to be equal for all the hypotheses. The MCMC-based Analysis B calculates the hypotheses' posterior probabilities by counting the number of MCMC steps in the selected hypothesis against the total number of MCMC steps. That ratio gives a fully marginalized posterior probability
for a given hypothesis. 

Table~\ref{tab:bayesian:bayesfactor} shows the posterior probabilities for the
mass orderings and the octants of $\sin^2\theta_{23}$ obtained with Analysis B.
Most of the posterior probability lies in the upper octant and normal mass
ordering. The obtained values of the Bayes factors, corresponding to the ratio of the marginal likelihoods of the two hypothesis, are of 8.0 for the normal
over the inverted mass orderings, and 3.9 for the upper  over the lower
octant. The commonly-used Jeffreys' scale~\cite{jeffreys1998} classifies both results as
``substantial".
\begin{table}[htbp]
  \centering
  \caption{Model comparison posterior probabilities for normal and inverted
  mass orderings, and for the upper and the lower
  octants of $\sin^2\theta_{23}$, from Analysis B. There is a preference
  for the normal mass ordering and upper octant of $\sin^2\theta_{23}$.
  }
  \label{tab:bayesian:bayesfactor}
  \begin{tabular}{cccc}
    \hline
    \hline
    \multicolumn{1}{c}{\rule{0pt}{2.6ex}}& ~ $\sin^2\theta_{23} < 0.5$ ~  & ~ $\sin^2\theta_{23} >0.5$ ~ & ~ Sum ~ \\
    \hline
    NO ($\Delta m^2_{32} >$ 0) & 0.184  & 0.705 & 0.889 \\
    IO ($\Delta m^2_{32} <$ 0) & 0.021  & 0.090 & 0.111  \\
    \hline
    Sum & 0.205 & 0.795 & 1 \\
    \hline
    \hline
  \end{tabular}
\end{table}

The impact of the details of the analysis methods on the mass ordering results were checked by comparing the results of the different analyses. Analysis C prefers the normal mass ordering and the upper octant of $\sin^2\theta_{23}$ with 91.1\% and 80.4\% posterior probability, respectively, whereas Analysis A finds a posterior probability of 87.7\% for the normal mass ordering. As before, the largest difference is seen between Analysis C on one side and A and B on the other, with the main contribution being the choice of variables used for the kinematic information of the candidate events from the appearance samples.

\begin{figure}[htbp]
\centering
\includegraphics[width=0.47\textwidth]{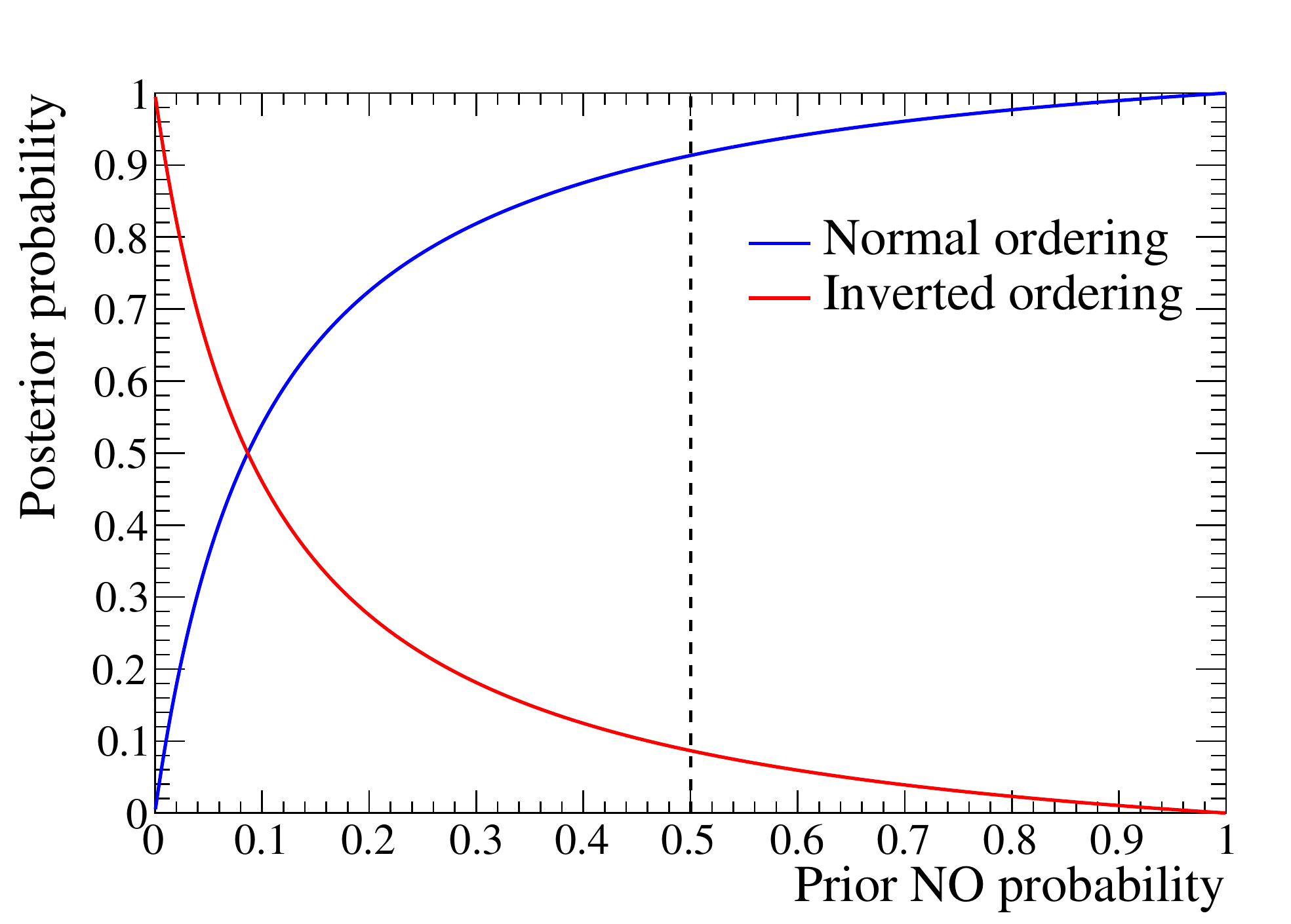}
\caption {
    Mass ordering posterior probabilities as a function of the prior probability assumed for the normal ordering, obtained in the data fit using Analysis C. The dashed black line corresponds to the equal prior probability case used for the main result.  
}
\label{fig:mh_prior}
\end{figure}

The posterior probabilities above assumed equal prior probabilities for the different hypothesis. The effect of the choice of prior probabilities was checked by looking at how the posterior probabilities obtained in the data fit from Analysis C changed as a function of the prior probabilities assumed (Fig.~\ref{fig:mh_prior}). As expected when testing discrete hypotheses using Bayesian methods, the choice of the prior probability has a significant effect on the obtained posterior probabilities. However, the obtained curves are different from $y=\pm x$, demonstrating that the data contain information about the mass ordering.

\subsubsection{Frequentist properties of the Bayesian results for the mass ordering}
As advocated in~\cite{Bayarri2004}, the frequentist properties of the Bayesian mass ordering results were studied. More precisely, it was checked whether excluding a given ordering hypothesis (based on the other ordering having a posterior probability superior or equal to $\alpha\%$) was selecting the true ordering approximately $\alpha\%$ of the time. For this purpose, 20,000 pseudo-experiments were generated, both for each mass ordering hypothesis and for different true values of $\deltacp$. Then, the fraction of the pseudo-experiments which had a posterior probability higher than 95\% for one of the two orderings were studied. 

It was found that this fraction depended strongly on the true value of $\deltacp$ assumed. Only for true values of $\deltacp$ close to $-\pi/2$ could one of the two orderings be excluded a non-negligible part of the time based on this criteria. Additionally, only the normal ordering could have a posterior probability above 95\% in this case, regardless of which of the two hypotheses was assumed to be true when generating the pseudo-experiments. For true $\deltacp=-\pi/2$, the true ordering was excluded 5.67\% of the time the wrong one was. This shows that using the posterior probability to select the mass ordering, although a Bayesian method, nevertheless has reasonable frequentist properties in this case. 

\begin{figure}[htbp]
\centering
\includegraphics[width=0.47\textwidth]{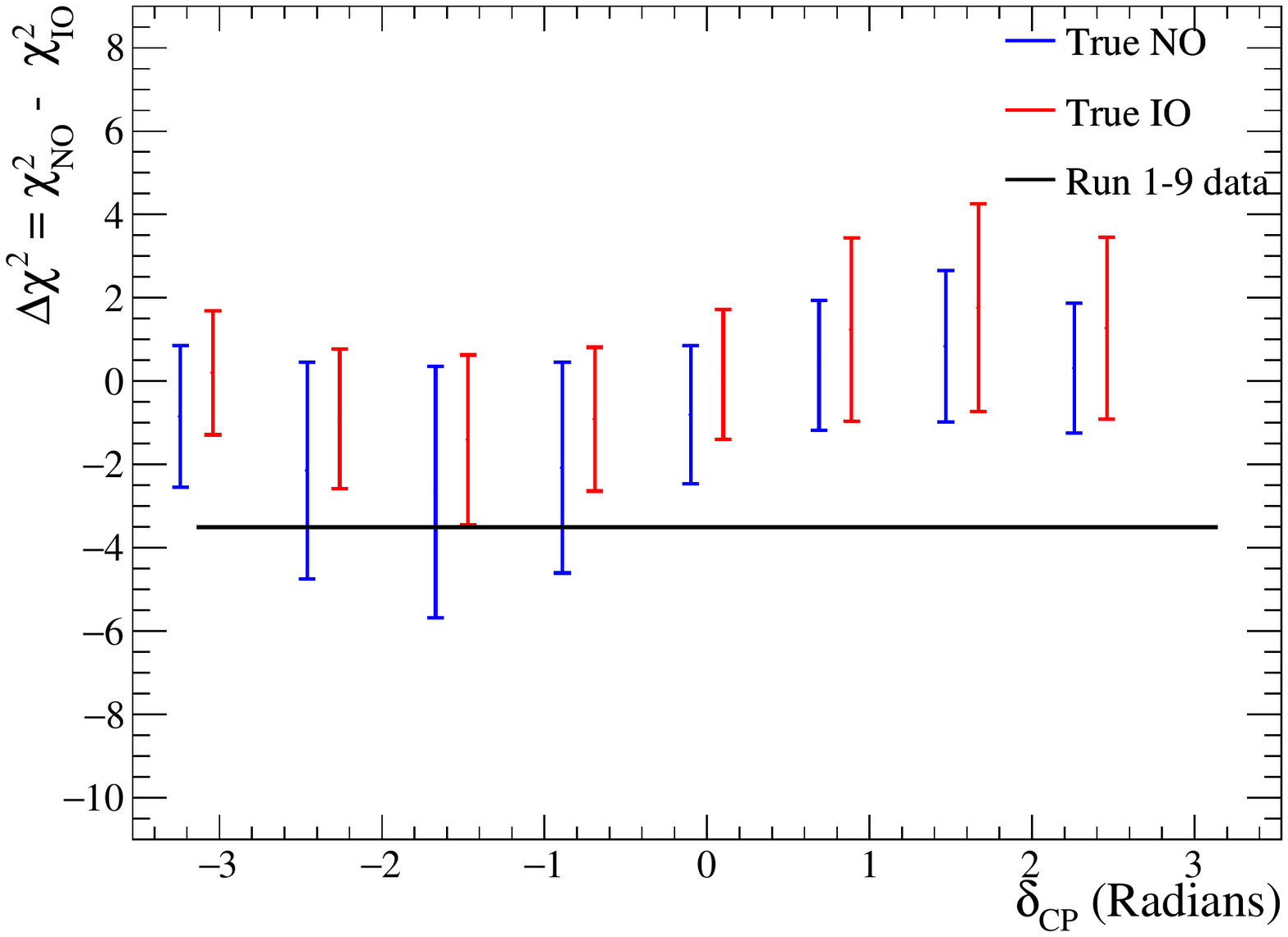}
\caption {
    Ability to distinguish between the two mass orderings as a function of the true value of $\deltacp$. The bar indicates the range of values of $\Delta \chi^{2}=\chi^{2}_{\NO}-\chi^{2}_{\IO}$ which contains 95\% of the values obtained for the pseudo-experiments generated in each case. The black line corresponds to the value obtained in the data fit by Analysis A.
}
\label{fig:mh_sensi}
\end{figure}

The pattern seen highlights the degeneracy between the measurement of $\deltacp$ and the determination of the mass ordering (also visible on Fig. \ref{fig:bievent} of the next section), which have similar effects on the predictions: they both change the two observables (the number of $\nu_{\mu} \rightarrow \nu_{e}$ and $\numub \rightarrow \nueb$ events) in opposite directions. The sensitivity to the mass ordering as a function of $\deltacp$ (Fig.~\ref{fig:mh_sensi}) shows that for $\deltacp<0$ ($\deltacp>0$), only for the true ordering being normal (inverted) does one get some expected values of the test statistics outside of the range predicted for the other ordering hypothesis. One can therefore only hope to exclude with non-negligible statistical significance the inverted (normal) ordering in this case.

\subsubsection{Frequentist results for the mass ordering}
\begin{figure}[htbp]
\centering
\includegraphics[width=0.47\textwidth]{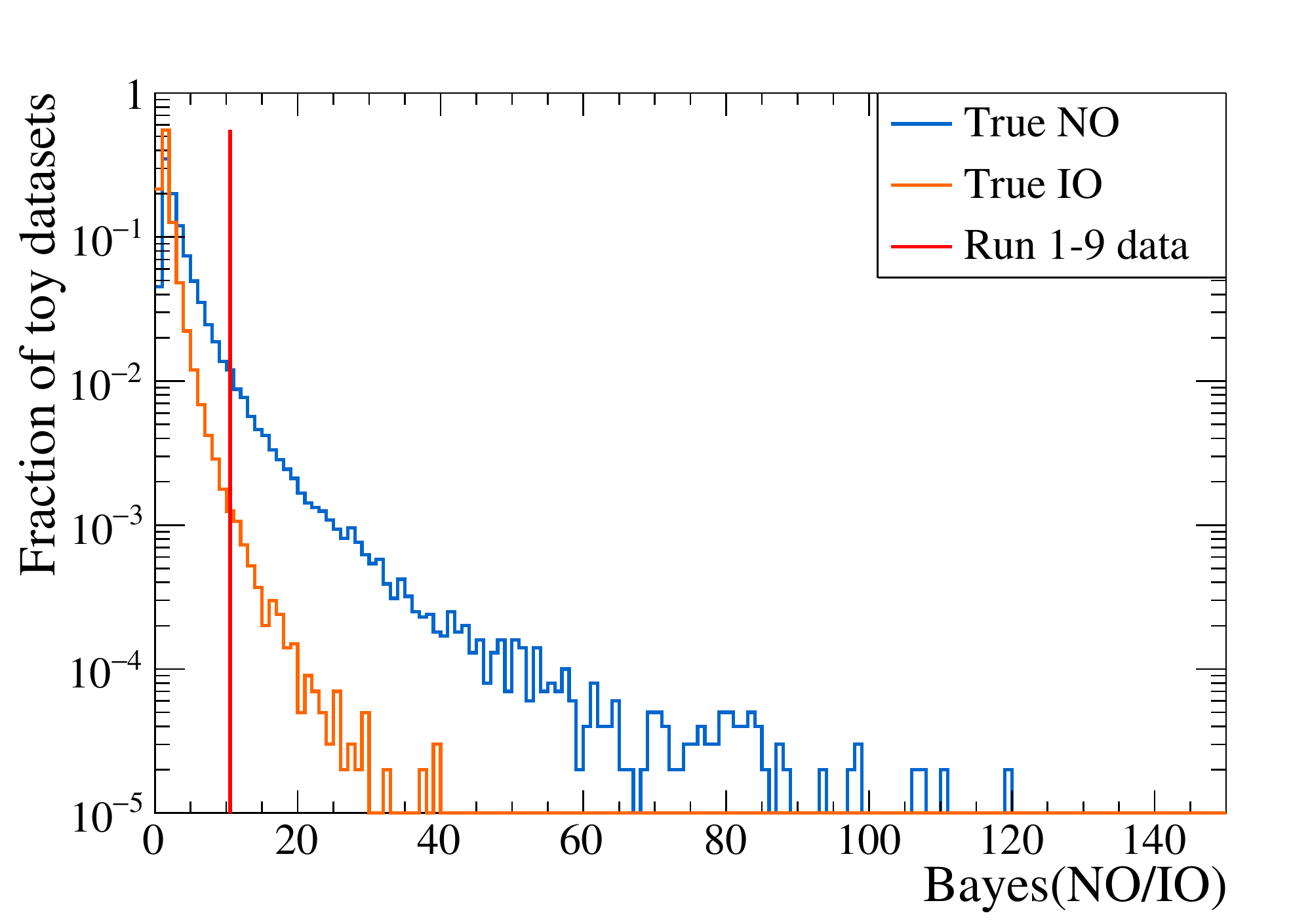}
\caption {
	Expected distributions of the Bayes factor between the mass ordering hypotheses, compared to the value obtained in the data fit.
}
\label{fig:mh_freq}
\end{figure}

For completeness, frequentist results for the mass ordering were derived by computing $p$-values for the Bayes factor. 100,000 pseudo-experiments were generated, randomizing over the nuisance parameters (including $\sin^2\theta_{23}$, $\deltacp$ and $\Delta{}m^2_{32}$, following a similar method as for the frequentist 1D $\deltacp$ results), and the value of the Bayes factor $P(\NO)/P(\IO)$ obtained in the data fit was compared to the values obtained for those pseudo-experiments (Fig.~\ref{fig:mh_freq}) 

The fraction of the pseudo-experiments for which the Bayes factor was larger than what was observed in the data (corresponding to a result more NO-like) was found to be $4.87\times 10^{-3}$ assuming true inverted ordering (inverted ordering $p$-value), and $6.5\times 10^{-2}$ assuming true normal ordering (1 minus normal ordering $p$-value using standard definitions). Those two values are both low, and it would be misleading to claim exclusion of the inverted ordering based solely on the low $p$-value obtained for this hypothesis. An alternative tool sometimes used in the collider community in such cases is \CLs~\cite{Read_2002}, in which the $p$-value obtained for one hypothesis is penalized by one minus the $p$-value for the other hypothesis:
\begin{equation}
\CLs(\IO)=\frac{p_{0}(\IO)}{1-p_{0}(\NO)} 
\end{equation}
where $p_{0}(\IO)$ and $p_{0}(\NO)$ are the $p$-values for respectively the inverted and normal orderings. In this case, $\CLs(\IO)= 0.075$ is obtained.
It should be noted that the more commonly used test statistic $\Delta \chi^{2}=\chi^{2}_\NO-\chi^{2}_\IO$ is equal to $-2\ln(\mathrm{Bayes}(\NO/\IO))$. The $p$-values obtained with this $\Delta \chi^{2}$ test statistics are therefore the same as the ones presented here for the Bayes factor.

\subsection{Summary of results}

To summarize the oscillation parameter constraints, fits to \deltacp, \ssqthonethree, \ssqthtwothree and \dmsqtwothree have been produced using constant $\Delta \chi^{2}$ critical values, and their $1\sigma$ confidence intervals are listed in Tab.~\ref{tab:data_bestfit_t2k_only} and in Tab.~\ref{tab:data_bestfit} with and without using the results of reactor experiments to constrain \ssqthonethree, respectively. Additionally, \deltacp critical values have been calculated using the Feldman--Cousins method, and several confidence intervals are listed in Tab.~\ref{tab:data_deltacp_fc_intervals}.

\begin{figure}[htbp]
\centering
\includegraphics[width=0.47\textwidth]{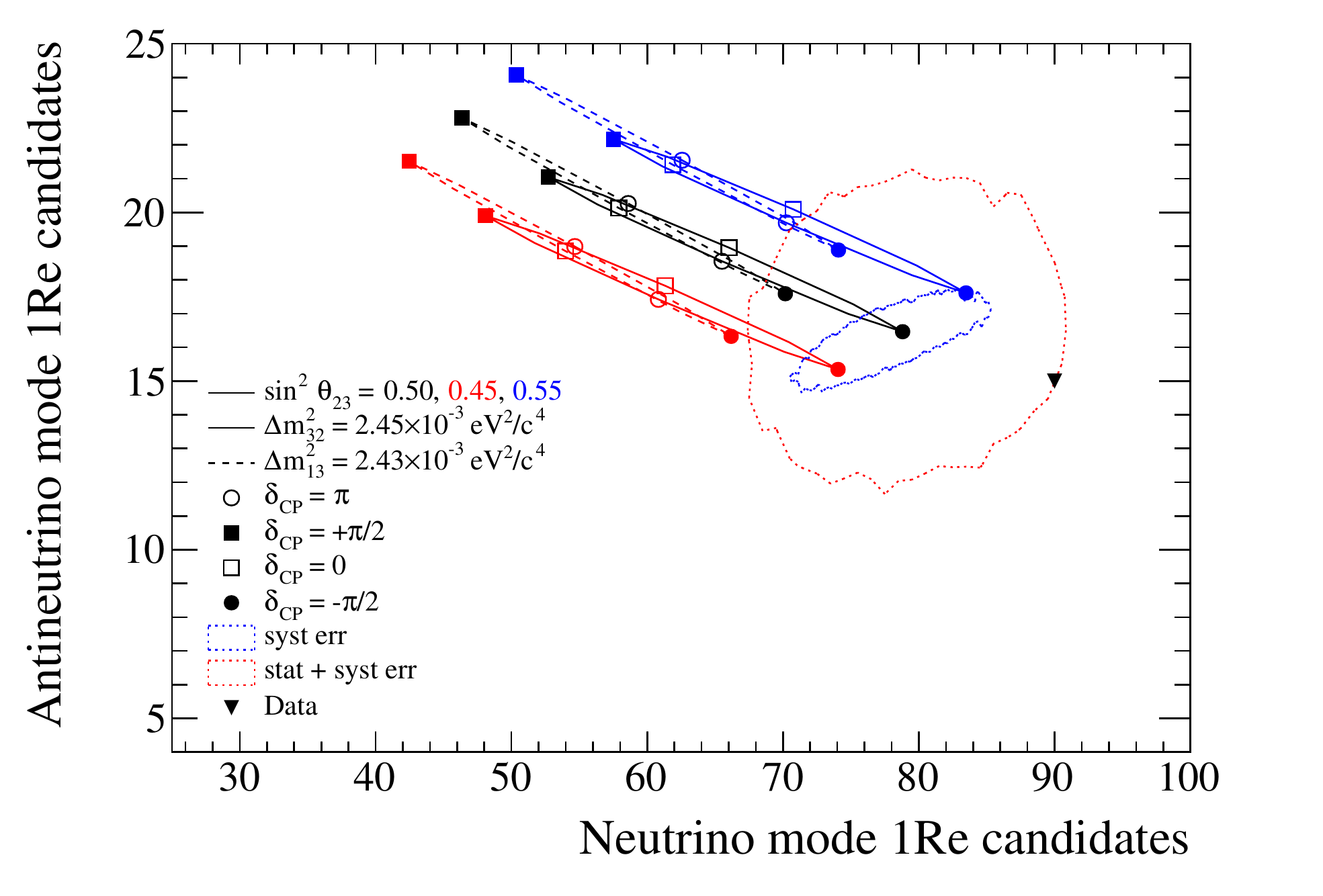}
\caption {
	Candidate RHC one-ring $e$-like event rate vs the candidate FHC one-ring $e$-like event rate (including CC 1$\pi$ events) for a variety of different oscillation parameter values. Predictions are generated for the given values of \deltacp, \ssqthtwothree and mass ordering with remaining oscillation parameters fixed at the central values of the prior probability density functions defined in Sec.~\ref{sec:OA_frquentist_results}. The uncertainty regions are created assuming that \deltacp $=-\pi/2$, \ssqthtwothree $=0.5$, $\dmsqtwothree =2.45 \times 10^{-3}\eVmass$ and normal mass ordering.
}
\label{fig:bievent}
\end{figure}

It is valuable to see how different values of \deltacp, \ssqthtwothree and the mass ordering affect the predicted event rates. Figure~\ref{fig:bievent} shows the predicted $\nueb$ event rate vs $\nue$ event rate for true values of oscillation parameters where \deltacp is varied between \CP conserving and maximally \CP violating values, and \ssqthtwothree is varied around maximal mixing, for both mass orderings. Predicted event rates for a given value of \ssqthtwothree and mass ordering are linearly interpolated between those computed for 9 evenly-spaced values of \deltacp from $-\pi$ to $+\pi$ to indicate the behavior produced by varying \deltacp. The observed number of FHC and RHC 1-ring electron-like candidate events falls on the edge of the $1\sigma$ uncertainty region generated around true $\deltacp =-\pi/2$, $\ssqthtwothree =0.5$, $\dmsqtwothree =2.45 \times 10^{-3}\si{\eV^2\!/\clight^4}$ and normal mass ordering, and is therefore consistent with this point.
	
\newcommand{\Ppmns}{P_{\footnotesize \mathrm{PMNS}}}

\section{$\nueb$ Appearance Analysis}
\label{sec:nuebar}

This section, which differs from the main oscillation analysis reported above, evaluates the significance of the $\numub \rightarrow \nueb$ oscillation under the assumption of two different hypotheses, corresponding to no $\nueb$ appearance, and to $\nueb$ appearance consistent with our current knowledge of the PMNS mixing parameters. To date, the world's best measurement of $\nueb$ appearance has been made by the \nova collaboration \cite{Acero:2019ksn_nova}, which reports an excess of $4.4\sigma$ over the expected background. The T2K measurement of $\nueb$ appearance has been made using the same analysis framework described in Section \ref{sec:fitters} and has been performed by analyses A and C.

The $\nueb$ appearance analysis is performed by multiplying the $\numub \rightarrow \nueb$ PMNS oscillation probability by a factor, $\beta$, i.e. $P\left(\numub \rightarrow \nueb\right) = \beta \times \Ppmns\left(\numub \rightarrow \nueb\right)$.  The parameter $\beta$ is set to either 0 or 1 to select a null hypothesis for two independent tests: when $\beta = 0$, the null hypothesis under consideration is that there is no $\nueb$ appearance, while for $\beta = 1$ the null hypothesis is that $\nueb$ appearance occurs according to the current best knowledge of the PMNS parameters. For each hypothesis, $p$-values are produced from two analyses: a rate-only whose test statistic is the number of candidate events in the RHC one-ring $e$-like sample, and one rate+shape analysis whose test statistic is the difference in marginal negative log-likelihood values between the $\beta = 0$ and $\beta = 1$ cases, denoted $\Delta \chi^2$, as in Eq.~\ref{eq:dchisq01}. 
\begin{equation}
	\Delta \chi^2 = \chi_\mathrm{marg}^2 \left( \beta = 0 \right) - \chi_\mathrm{marg}^2 \left( \beta = 1 \right) 
	\label{eq:dchisq01}
\end{equation}
Here the use of $\chi^2$ is taken to be synonymous with $-2\ln L_{\mathrm{marg}}$. Unlike in the main oscillation analysis, the likelihoods are not only marginalized over the flux, cross-section and detector parameters, but also over all oscillation parameters except $\beta$, including \deltacp and the mass ordering, using \NtoysMargNuebar\ samples of the nuisance parameter space. The number of pseudo-experiments and the number of samples used during marginalization were selected to ensure the stability of the $p$-values. 

To calculate $p$-values, the data are compared to distributions of the test statistics from ensembles of pseudo-experiments produced under the assumption of either $\beta = 0$ or $\beta = 1$. Each pseudo-experiment is produced by randomizing nuisance parameters according to the prior probability density functions (PDF) in Section~\ref{sec:fitters}. Additionally, a uniform PDF in the range $[-\pi, +\pi]$ is used for \deltacp, and a two-point PDF is used for the mass ordering with equal probability for normal and inverted ordering.

T2K data from four ``control samples" (FHC single-ring $e$-like and $\nue \text{CC}1\pi$-like and both neutrino and RHC single-ring $\mu$-like) are used to constrain the oscillation and systematic model parameters. The impact of these four data control samples is estimated differently in the two analyses, A and C. In the latter, each of the \NtoysFakeDataNuebar\ pseudo-experiments used to build the distributions of the rate-only and rate+shape test statistics is weighted by its likelihood over the four control samples, given the T2K data. Analysis A uses an alternative method, applying this constraint by using rejection sampling to select the pseudo-experiments that are most probable according to the data in the control samples.

\begin{figure}[htbp]
	\center
	\begin{subfigure}[b]{0.47\textwidth}
		\includegraphics[width=0.98\columnwidth]{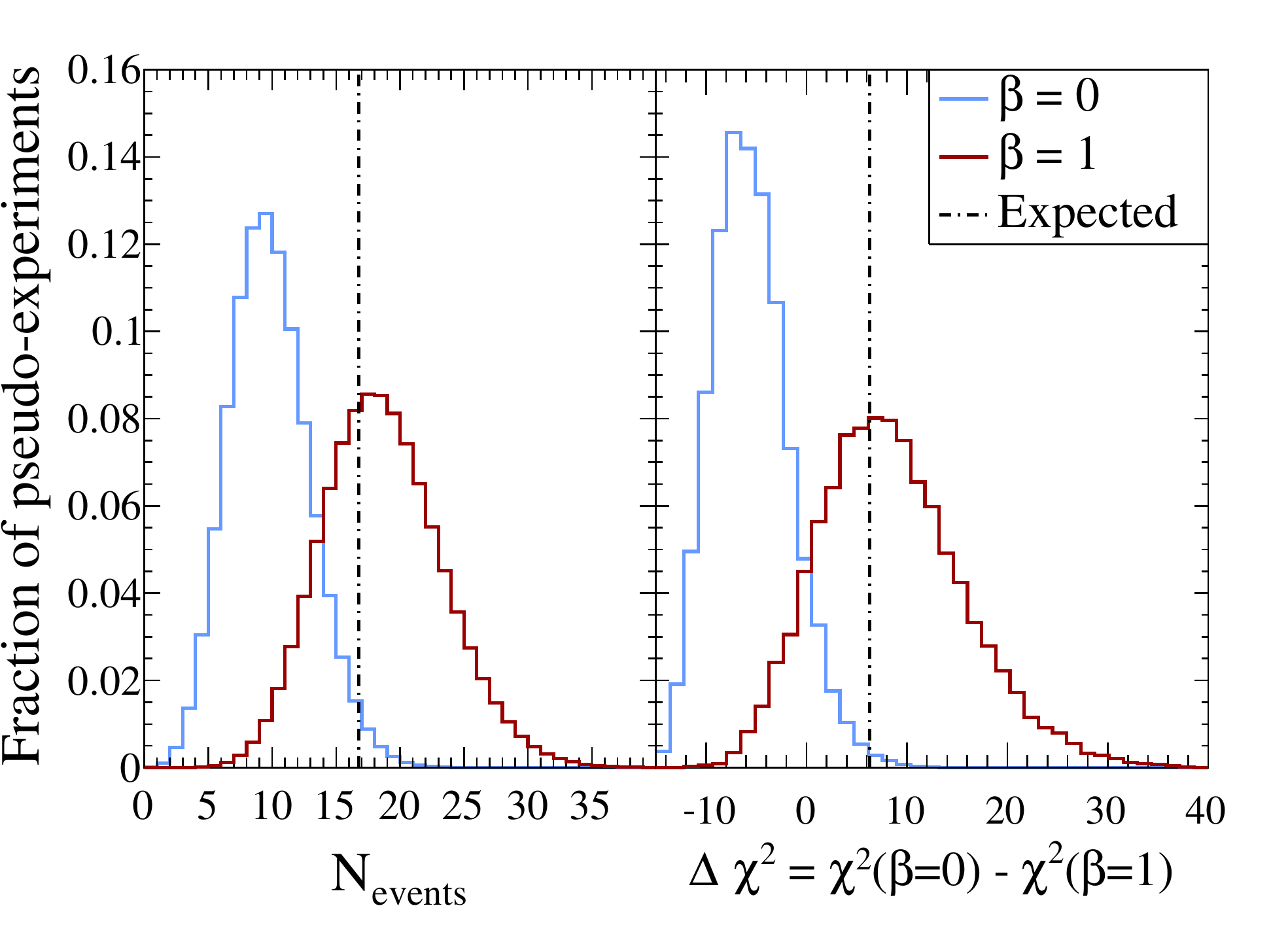}
	\end{subfigure}
	\\
	\begin{subfigure}[b]{0.47\textwidth}
		\includegraphics[width=0.98\columnwidth]{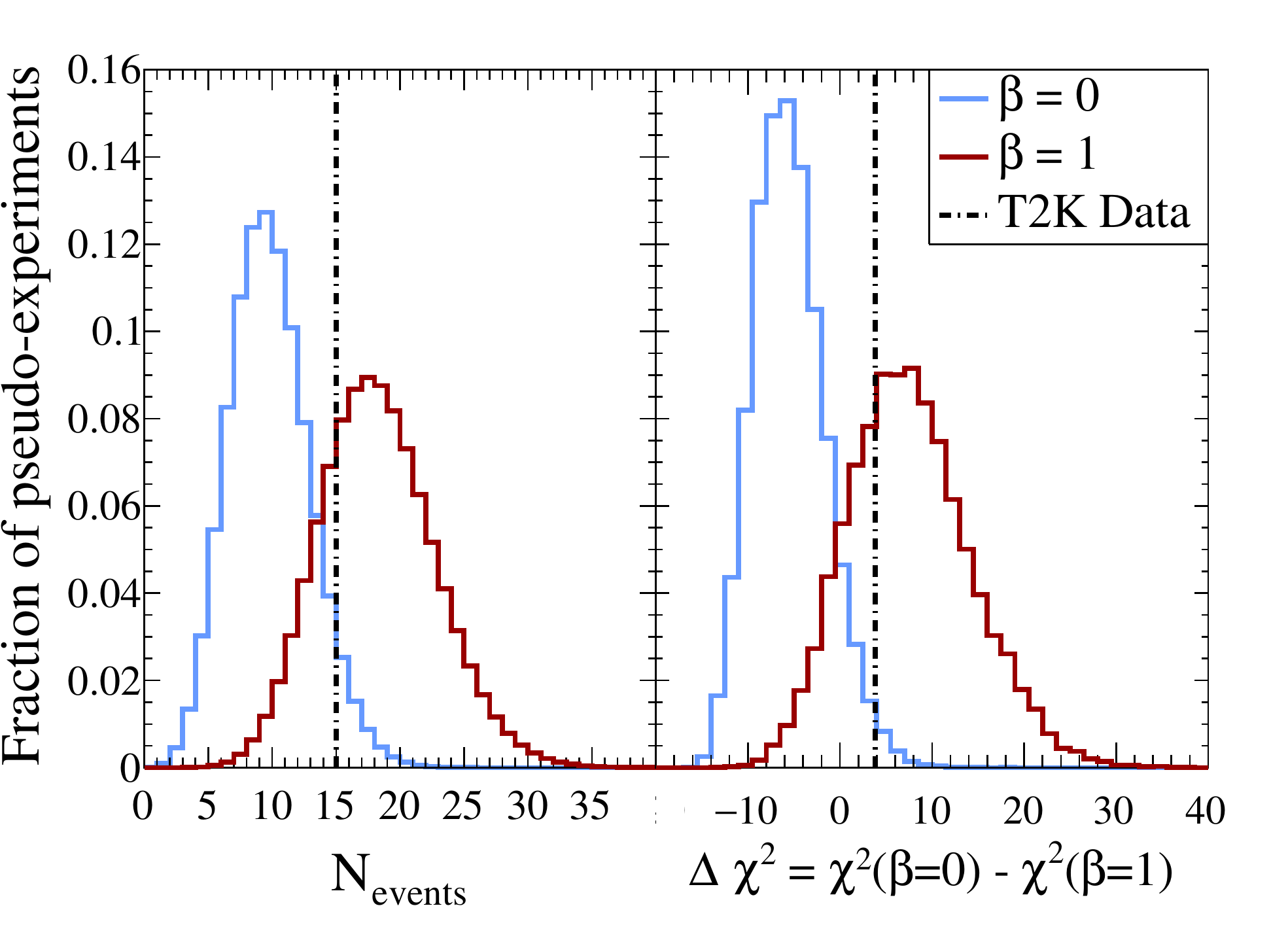}
	\end{subfigure}
	\caption {
		Distributions of the expected (top) and observed (bottom) rate-only (left) and rate+shape (right) test statistics compared to the expected/observed test statistics. Here $\text{N}_{\text{events}}$ denotes the number of observed events in the RHC single-ring $e$-like sample.
	}
	\label{fig:nuebar_distributions}
\end{figure}

Due to the presence of the four control samples, the marginal likelihood used to calculate $\Delta \chi^2$, the rate+shape test statistic, is also constructed differently from the main oscillation analysis. For each pseudo-experiment, the marginal likelihood is constructed based on the prediction in the RHC single-ring $e$-like sample, while its background is constrained using the four control samples. This is reflected in Eq.~\ref{eq:nuebar_chisq_marg}, where $L_{\nueb}$ denotes the likelihood of the RHC single-ring $e$-like sample compared to the pseudo-experiment, $E$, $L_{\textrm{c}}$ is the product of the likelihoods of the four control samples compared to the T2K data, $D$, and $\vec{f}_{j}$ the set of nuisance parameters.
\begin{multline}
\label{eq:nuebar_chisq_marg}
	\chi_{\text{marg}}^2 \left( \beta \right) = \\
	{-}2\ln \bigg{[} 
	\frac{1}{n} \sum_{\smash{j=1}}^{\smash{n}} L_{\nueb} \left( \beta, \vec{f}_{j}; E \right)
	\cdot L_{\textrm{c}} \left( \beta, \vec{f}_{j}; \textrm{D} \right)
	\bigg{]}
\end{multline}
The expected and observed test statistic distributions produced using analysis A are shown in Fig.~\ref{fig:nuebar_distributions} and the corresponding $p$-values are shown in Tab.~\ref{tab:nuebar_pvals}. The hypothesis of no $\nueb$ appearance ($\beta = 0$) is excluded at the $1.9\sigma$ and $2.4\sigma$ levels, respectively using the rate-only and rate+shape analyses. The observed $p$-values provide a weaker exclusion of the $\beta = 0$ case than expected with Asimov data set, in both rate-only and rate+shape analyses. This is primarily due to observing fewer events (15) in the RHC single-ring $e$-like sample than expected (16.8), and strengthened by their relatively background-like spectrum in the rate+shape analysis, as shown in Fig.~\ref{fig:spectra}~(d).
Our data are consistent with the PMNS $\nueb$ appearance hypothesis ($\beta = 1$), with $p$-values of 0.32 and 0.30 for the rate-only and rate+shape analyses, respectively (corresponding to a $1\sigma$ exclusion). Analyses A and C both produce test statistic distributions and $p$-values that are in good agreement with each other, where minor differences between them were determined to result from the difference in kinematic variables used to bin the analysis templates.

\begin{table}[htp]
\centering
\caption{Expected and observed $p$-values and significance of the $\beta = 0$ and $\beta = 1$ hypotheses using both the rate-only and rate+shape analyses}
\label{tab:nuebar_pvals}
  \begin{tabular}{cccccc}
  \hline \hline
  \multirow{2}{*}{$\beta$} 
  & \multirow{2}{*}{Analysis} 
    & \multicolumn{2}{c}{$p$-value} & \multicolumn{2}{c}{Significance ($\sigma$)} \\[-0.25ex]
  & & Expected & Observed & Expected & Observed \\
  \hline
  \multirow{ 2}{*}[0.0ex]{0} 
  & Rate-only 
  & \pvalRateOnlyBetaZeroAsimovARunOneNineD & \pvalRateOnlyBetaZeroDataRunOneNineD 
  & \sigmaRateOnlyBetaZeroAsimovARunOneNineD & \sigmaRateOnlyBetaZeroDataRunOneNineD \\
  & Rate + shape 
  & \pvalRateShapeBetaZeroAsimovARunOneNineD & \pvalRateShapeBetaZeroDataRunOneNineD
  & \sigmaRateShapeBetaZeroAsimovARunOneNineD & \sigmaRateShapeBetaZeroDataRunOneNineD \\[0.5ex]
  \multirow{ 2}{*}[0.0ex]{1}  
  & Rate-only   
  & \pvalRateOnlyBetaOneAsimovARunOneNineD & \pvalRateOnlyBetaOneDataRunOneNineD 
  & \sigmaRateOnlyBetaOneAsimovARunOneNineD & \sigmaRateOnlyBetaOneDataRunOneNineD \\
  & Rate + shape 
  & \pvalRateShapeBetaOneAsimovARunOneNineD & \pvalRateShapeBetaOneDataRunOneNineD 
  & \sigmaRateShapeBetaOneAsimovARunOneNineD & \sigmaRateShapeBetaOneDataRunOneNineD \\
  \hline \hline
  \end{tabular}
\end{table}

It is also desirable to test the robustness of these results to various alternative choices of parts of the interaction model which have a non-negligible effect on the kinematic distributions of the RHC single-ring $e$-like sample. This is done using Analysis C by weighting the nominal near and far detector MC (see Sec.~\ref{sec:fake_data_studies}).  Three simulated data sets are studied: the Kabirnezhad single pion production model, the Nieves LFG model and the data-driven CC $0\pi$ $E_{\nu}-Q^2$ simulated data, described in Sec.~\ref{sec:interaction_model}.
To test their effects on the observed results of the $\nueb$ appearance analysis, the test statistic distributions are weighted by the ratio of the distribution produced using the alternative model to the expected distribution. Similarly, the observed test statistics are shifted by the difference between the median expected pseudo-experiment with and without the use of the alternative model. The change in the observed $p$-values from these alternative models are shown in Tabs.~\ref{tab:nuebar_fake_data_pvals_rate} and~\ref{tab:nuebar_fake_data_pvals_shape} for the rate and rate+shape analyses respectively.
The observed changes are small compared to the nominal $p$-value and the changes do not affect the conclusions, so the analyses are considered robust against alternative model choices.

\begin{table}[htp]
	\centering
	\caption{Nominal and alternative model $p$-values of the $\beta = 0$ and $\beta = 1$ hypotheses using the rate-only analysis.}
	\label{tab:nuebar_fake_data_pvals_rate}
	\begin{tabular}{lcc}
		\hline \hline
		\multirow{2}{*}{\qquad Model} 
		& $\beta=0$	& $\beta=1$ \\
    	& $p$-value ($\sigma$) & $p$-value ($\sigma$)  \\
		\hline
		Nominal                    & 0.0686 (1.82) & 0.246 (1.16) \\
		Kabirnezhad single pion    & 0.0824 (1.74) & 0.176 (1.35) \\
		Nieves LFG 1p1h            & 0.0804 (1.75) & 0.222 (1.22) \\
		Data-driven 2p2h-$\Delta$  & 0.0859 (1.72) & 0.211 (1.25) \\
		\hline \hline
	\end{tabular}
\end{table}

\begin{table}[htp]
	\centering
	\caption{Nominal and alternative model $p$-values of the $\beta = 0$ and $\beta = 1$ hypotheses using the rate+shape analysis.}
	\label{tab:nuebar_fake_data_pvals_shape}
	\begin{tabular}{lcc}
		\hline \hline
		\multirow{2}{*}{\qquad Model} 
		& $\beta=0$ & $\beta=1$ \\
    	& $p$-value ($\sigma$) & $p$-value ($\sigma$)  \\
		\hline
		Nominal                    & 0.0244 (2.25) & 0.261 (1.12) \\
		Kabirnezhad single pion    & 0.0227 (2.28) & 0.225 (1.21) \\
		Nieves LFG 1p1h            & 0.0201 (2.32) & 0.277 (1.09) \\
		Data-driven 2p2h-$\Delta$  & 0.0178 (2.37) & 0.301 (1.03) \\
		\hline \hline
	\end{tabular}
\end{table}

\section{Conclusions}
The T2K collaboration has analyzed the full data set collected by the experiment between 2010 and 2018 to produce measurements of \ssqthtwothree, \dmsqtwothree, \ssqthonethree, \deltacp and the mass ordering.
The parameter values and uncertainties are taken from a simultaneous fit to both muon-like and electron-like event samples at Super-Kamiokande, collected from both neutrino-dominated and antineutrino-dominated beam operation.
This analysis uses a new event selection at SK to increase the effective fiducial volume of the detector, resulting in a 20\% increase in the electron-like sample efficiency and a 40\% reduction in the background contamination in the muon-like samples.

The neutrino interaction model used for this work is improved relative to Ref.~\cite{Abe:2017vif_T2Krun7osc}, incorporating in-medium effects (RPA) and 2p2h shape uncertainties in the charged-current zero-pion signal channel.
A detailed set of simulated data studies were also performed to assess the robustness of the analysis to alternative choices of neutrino interaction model.
These studies demonstrated that the nucleon removal energy uncertainty can have a significant effect on the allowed regions for \ssqthtwothree and \dmsqtwothree and, to a lesser extent, \dmsqtwothree was also sensitive to all of the alternative models studied.
This led to the addition of new systematic uncertainties to the SK samples and oscillation contours to account for these effects.
The results of these simulated data studies are an important outcome of this analysis: long-baseline experiments are now entering the precision era, and changes of a few percent in the reconstructed neutrino energy can have significant effects on the oscillation parameters observed by the experiment.

The T2K oscillation parameter measurements are, however, limited by statistics.
T2K will collect more data in both neutrino and antineutrino beam operation mode.
More complex event topologies will be included in the analyses at both ND280 and Super-Kamiokande, and the collaboration is working towards joint analyses with both the Super-Kamiokande and the \nova experiments.
The combination of new topologies and differing neutrino energies will lift the degeneracies between oscillation parameters present in a single experiment.
This will enable the most sensitive measurements of neutrino oscillations to date, and will provide a template for future analyses at the next generation of long-baseline experiments.
The data related to the measurement and results presented in this paper can be found in \cite{datarelease}.

\renewcommand{\nova}{NOvA\xspace}
\renewcommand{\minerva}{MINERvA\xspace}
\ifnum\sizecheck=0
  \begin{acknowledgments}
We thank the J-PARC staff for superb accelerator performance. We thank the CERN NA61/SHINE Collaboration for providing valuable particle production data. We acknowledge the support of MEXT, JSPS KAKENHI (JP16H06288, JP18K03682, JP18H03701, JP18H05537, JP19J01119, JP19J22440, JP19J22258, JP20H00162, JP20H00149, JP20J20304) and bilateral programs(JPJSBP120204806, JPJSBP120209601), Japan; NSERC, the NRC, and CFI, Canada; the CEA and CNRS/IN2P3, France; the DFG (RO 3625/2), Germany; the INFN, Italy; the Ministry of Education and Science(DIR/WK/2017/05) and the National Science Centre (UMO-2018/30/E/ST2/00441 ), Poland; the RSF (19-12-00325), RFBR(JSPS-RFBR 20-52-50010\\20) and the Ministry of Science and Higher Education(075-15-2020-778), Russia; MICINN (SEV-2016-0588, PID2019-107564GB-I00, PGC2018-099388-BI00) and ERDF funds and CERCA program, Spain; the SNSF and SERI (200021\_185012, 200020\_188533, 20FL21\_186178I), Switzerland; the STFC, UK; and the DOE, USA. We also thank CERN for the UA1/NOMAD magnet, DESY for the HERA-B magnet mover system, NII for SINET5, the WestGrid, SciNet and CalculQuebec consortia in Compute Canada, and GridPP and the Emerald High Performance Computing facility in the United Kingdom. In addition, the participation of individual researchers and institutions has been further supported by funds from the ERC (FP7), “la Caixa” Foundation (ID 100010434, fellowship code LCF/BQ/IN17/11620050), the European Union's Horizon 2020 Research and Innovation Programme under the Marie Sklodowska-Curie grant agreement numbers 713673 and 754496, and H2020 grant numbers RISE-GA822070-JENNIFER2 2020 and RISE-GA872549-SK2HK; the JSPS, Japan; the Royal Society, UK; French ANR grant number ANR-19-CE31-0001; and the DOE Early Career programme, USA.
\end{acknowledgments}

  \bibliographystyle{apsrev4-2}
  \bibliography{2019OA_long}
\fi

\ifnum\PRLsupp=0
  \clearpage
  \appendix*
\section{Post ND280 Fit Parameter Values}
\label{app:appendix_NDFitValues}
The nominal and post-ND280-fit values are shown in Tables~\ref{tab:skfhcflux} and~\ref{tab:skrhcflux} for the flux parameters, and Table~\ref{tab:xsecpostfit} for the cross section parameters.
\begin{table}[ht]
\centering
\caption{Prefit and postfit weights for the SK FHC flux parameters.}
\label{tab:skfhcflux}
\begin{tabular}{l @{\quad} r cl @{\quad} r c l}
\hline \hline
FHC SK flux & &\multirow{2}{0pt}{\clap{Prefit}} & & & \multirow{2}{0pt}{\clap{ND280 Postfit}} &  \\
parameter /GeV & && & && \\
\hline
SK $\numu$ [0.0, 0.4] & 1.00 &$\pm$& 0.10 & 1.01 &$\pm$& 0.06 \\[-0.1ex] 
SK $\numu$ [0.4, 0.5] & 1.00 &$\pm$& 0.10 & 1.03 &$\pm$& 0.05 \\[-0.1ex] 
SK $\numu$ [0.5, 0.6] & 1.00 &$\pm$& 0.09 & 1.02 &$\pm$& 0.05 \\[-0.1ex]
SK $\numu$ [0.6, 0.7] & 1.00 &$\pm$& 0.08 & 0.98 &$\pm$& 0.04 \\[-0.1ex] 
SK $\numu$ [0.7, 1.0] & 1.00 &$\pm$& 0.10 & 0.93 &$\pm$& 0.06 \\[-0.1ex] 
SK $\numu$ [1.0, 1.5] & 1.00 &$\pm$& 0.09 & 0.95 &$\pm$& 0.05 \\[-0.1ex] 
SK $\numu$ [1.5, 2.5] & 1.00 &$\pm$& 0.07 & 1.02 &$\pm$& 0.04 \\[-0.1ex] 
SK $\numu$ [2.5, 3.5] & 1.00 &$\pm$& 0.07 & 1.04 &$\pm$& 0.05 \\[-0.1ex] 
SK $\numu$ [3.5, 5.0] & 1.00 &$\pm$& 0.09 & 1.03 &$\pm$& 0.04 \\[-0.1ex] 
SK $\numu$ [5.0, 7.0] & 1.00 &$\pm$& 0.10 & 0.99 &$\pm$& 0.04 \\[-0.1ex] 
SK $\numu$ [7.0, $\infty$] & 1.00 &$\pm$& 0.11 & 0.97 &$\pm$& 0.05 \\[1ex]
SK $\numub$ [0.0, 0.7] & 1.00 &$\pm$& 0.10 & 0.98 &$\pm$& 0.08 \\[-0.1ex] 
SK $\numub$ [0.7, 1.0] & 1.00 &$\pm$& 0.08 & 0.97 &$\pm$& 0.05 \\[-0.1ex] 
SK $\numub$ [1.0, 1.5] & 1.00 &$\pm$& 0.08 & 0.98 &$\pm$& 0.06 \\[-0.1ex]
SK $\numub$ [1.5, 2.5] & 1.00 &$\pm$& 0.08 & 1.03 &$\pm$& 0.06 \\[-0.1ex] 
SK $\numub$ [2.5, $\infty$] & 1.00 &$\pm$& 0.09 & 1.10 &$\pm$& 0.07 \\[1ex]
SK $\nue$ [0.0, 0.5] & 1.00 &$\pm$& 0.09 & 1.02 &$\pm$& 0.05 \\[-0.1ex] 
SK $\nue$ [0.5, 0.7] & 1.00 &$\pm$& 0.09 & 1.02 &$\pm$& 0.04 \\[-0.1ex] 
SK $\nue$ [0.7, 0.8] & 1.00 &$\pm$& 0.08 & 1.02 &$\pm$& 0.04 \\[-0.1ex] 
SK $\nue$ [0.8, 1.5] & 1.00 &$\pm$& 0.08 & 1.01 &$\pm$& 0.04 \\[-0.1ex] 
SK $\nue$ [1.5, 2.5] & 1.00 &$\pm$& 0.08 & 1.03 &$\pm$& 0.04 \\[-0.1ex] 
SK $\nue$ [2.5, 4.0] & 1.00 &$\pm$& 0.08 & 1.03 &$\pm$& 0.04 \\[-0.1ex] 
SK $\nue$ [4.0, $\infty$] & 1.00 &$\pm$& 0.09 & 1.03 &$\pm$& 0.06 \\[1ex]
SK $\nueb$ [0.0, 2.5] & 1.00 &$\pm$& 0.07 & 1.04 &$\pm$& 0.06 \\[-0.1ex] 
SK $\nueb$ [2.5, $\infty$] & 1.00 &$\pm$& 0.13 & 1.08 &$\pm$& 0.12 \\
\hline \hline
\end{tabular}
\end{table}

\begin{table}[ht!]
\centering
\caption{Prefit and postfit weights for the SK RHC flux parameters.}
\label{tab:skrhcflux}
\begin{tabular}{l@{\quad} r c l@{\quad} r c l}
\hline \hline
RHC SK flux & &\multirow{2}{0pt}{\clap{Prefit}} & & & \multirow{2}{0pt}{\clap{ND280 Postfit}} &  \\
parameter /GeV & && & && \\
\hline
SK $\numu$ [0.0, 0.7] & 1.00 &$\pm$& 0.09 & 0.98 &$\pm$& 0.07 \\[-0.1ex]  
SK $\numu$ [0.7, 1.0] & 1.00 &$\pm$& 0.08 & 0.99 &$\pm$& 0.05 \\[-0.1ex]  
SK $\numu$ [1.0, 1.5] & 1.00 &$\pm$& 0.08 & 1.00 &$\pm$& 0.05 \\[-0.1ex]  
SK $\numu$ [1.5, 2.5] & 1.00 &$\pm$& 0.08 & 1.05 &$\pm$& 0.05 \\[-0.1ex]  
SK $\numu$ [2.5, $\infty$] & 1.00 &$\pm$& 0.08 & 1.04 &$\pm$& 0.05 \\[1ex]
SK $\numub$ [0.0, 0.4] & 1.00 &$\pm$& 0.11 & 1.00 &$\pm$& 0.07 \\[-0.1ex]  
SK $\numub$ [0.4, 0.5] & 1.00 &$\pm$& 0.10 & 1.01 &$\pm$& 0.05 \\[-0.1ex]  
SK $\numub$ [0.5, 0.6] & 1.00 &$\pm$& 0.09 & 0.99 &$\pm$& 0.05 \\[-0.1ex]  
SK $\numub$ [0.6, 0.7] & 1.00 &$\pm$& 0.08 & 0.97 &$\pm$& 0.04 \\[-0.1ex]  
SK $\numub$ [0.7, 1.0] & 1.00 &$\pm$& 0.10 & 0.97 &$\pm$& 0.05 \\[-0.1ex]  
SK $\numub$ [1.0, 1.5] & 1.00 &$\pm$& 0.09 & 0.99 &$\pm$& 0.05 \\[-0.1ex]  
SK $\numub$ [1.5, 2.5] & 1.00 &$\pm$& 0.07 & 1.03 &$\pm$& 0.04 \\[-0.1ex]  
SK $\numub$ [2.5, 3.5] & 1.00 &$\pm$& 0.07 & 1.06 &$\pm$& 0.05 \\[-0.1ex]  
SK $\numub$ [3.5, 5.0] & 1.00 &$\pm$& 0.09 & 1.06 &$\pm$& 0.07 \\[-0.1ex]  
SK $\numub$ [5.0, 7.0] & 1.00 &$\pm$& 0.09 & 1.04 &$\pm$& 0.06 \\[-0.1ex]  
SK $\numub$ [7.0, $\infty$] & 1.00 &$\pm$& 0.12 & 1.00 &$\pm$& 0.09 \\[1ex]
SK $\nue$ [0.0, 2.5] & 1.00 &$\pm$& 0.07 & 1.04 &$\pm$& 0.05 \\[-0.1ex]  
SK $\nue$ [2.5, $\infty$]& 1.00 &$\pm$& 0.08 & 1.04 &$\pm$& 0.07 \\[1ex]
SK $\nueb$ [0.0, 0.5] & 1.00 &$\pm$& 0.10 & 1.01 &$\pm$& 0.05 \\[-0.1ex]  
SK $\nueb$ [0.5, 0.7] & 1.00 &$\pm$& 0.09 & 1.00 &$\pm$& 0.05 \\[-0.1ex]  
SK $\nueb$ [0.7, 0.8] & 1.00 &$\pm$& 0.09 & 1.00 &$\pm$& 0.05 \\[-0.1ex]  
SK $\nueb$ [0.8, 1.5] & 1.00 &$\pm$& 0.08 & 1.01 &$\pm$& 0.04 \\[-0.1ex]  
SK $\nueb$ [1.5, 2.5] & 1.00 &$\pm$& 0.08 & 1.04 &$\pm$& 0.05 \\[-0.1ex]  
SK $\nueb$ [2.5, 4.0] & 1.00 &$\pm$& 0.09 & 1.04 &$\pm$& 0.07 \\[-0.1ex]  
SK $\nueb$ [4.0, $\infty$] & 1.00 &$\pm$& 0.15 & 1.08 &$\pm$& 0.13 \\ 
\hline \hline
\end{tabular}
\end{table}

\begin{table}[ht!]
\centering
\caption{Prefit and postfit values for the cross-section parameters.}
\label{tab:xsecpostfit}
\begin{tabular}{l@{\quad} r c l@{\quad} r c l}
\hline \hline
Cross-section& &\multirow{2}{0pt}{\clap{Prefit}} & & & \multirow{2}{0pt}{\clap{ND280 Postfit}} &  \\
parameter & && & && \\
\hline
$M_{A}^{QE}$(GeV/c$^2$) & 1.20 &$\pm$& 0.03 & 1.13 &$\pm$& 0.08 \\[-0.1ex]
pF $^{12}$C(MeV/c) & 217\phantom{.} &$\pm$& 13\phantom{.0} & 224\phantom{.} &$\pm$& 13\phantom{.3} \\[-0.1ex]
pF $^{16}$O(MeV/c) & 225\phantom{.} &$\pm$& 13\phantom{.0} & 205\phantom{.} &$\pm$& 15\phantom{.0} \\[0.6ex]
2p2h norm $\nu$ & 1.00 &$\pm$& 1.00 & 1.50 &$\pm$& 0.20 \\[-0.1ex]
2p2h norm $\nub$ & 1.00 &$\pm$& 1.00 & 0.73 &$\pm$& 0.23 \\[-0.1ex]
2p2h norm $^{12}$C / $^{16}$O ratio & 1.00 &$\pm$& 0.20 & 0.96 &$\pm$& 0.17 \\[-0.1ex]
2p2h shape $^{12}$C & 1.00 &$\pm$& 3.00 & 2.00 &$\pm$& 0.21 \\[-0.1ex]
2p2h shape $^{16}$O & 1.00 &$\pm$& 3.00 & 2.00 &$\pm$& 0.35 \\[0.6ex]
BeRPA A & 0.59 &$\pm$& 0.12 & 0.69 &$\pm$& 0.06 \\[-0.1ex]
BeRPA B & 1.05 &$\pm$& 0.21 & 1.60 &$\pm$& 0.12 \\[-0.1ex]
BeRPA D & 1.13 &$\pm$& 0.17 & 0.96 &$\pm$& 0.13 \\[-0.1ex]
BeRPA E & 0.88 &$\pm$& 0.35 & 0.87 &$\pm$& 0.35 \\[-0.1ex]
BeRPA U & 1.20 &$\pm$& 0.10 & 1.20 &$\pm$& 0.10 \\[0.6ex]
$C_A^5$ & 0.96 &$\pm$& 0.15 & 0.98 &$\pm$& 0.06 \\[-0.1ex]
$M_A^{RES}$(GeV/c$^2$) & 1.07 &$\pm$& 0.15 & 0.81 &$\pm$& 0.04 \\[-0.1ex]
$I=\frac{1}{2}$ background & 0.96 &$\pm$& 0.40 & 1.31 &$\pm$& 0.26 \\[-0.1ex]
$\nue$/$\numu$ & 1.00 &$\pm$& 0.03 & 1.00 &$\pm$& 0.03 \\[-0.1ex]
$\nueb$/$\numub$ & 1.00 &$\pm$& 0.03 & 1.00 &$\pm$& 0.03 \\[-0.1ex]
CC DIS & 0.00 &$\pm$& 0.40 & 0.39 &$\pm$& 0.21 \\[-0.1ex]
CC coherent $^{12}$C & 1.00 &$\pm$& 0.30 & 0.87 &$\pm$& 0.28 \\[-0.1ex]
CC coherent $^{16}$O & 1.00 &$\pm$& 0.30 & 0.87 &$\pm$& 0.28 \\[0.6ex]
NC coherent & 1.00 &$\pm$& 0.30 & 0.94 &$\pm$& 0.30 \\[-0.1ex]
NC 1$\gamma$ & 1.00 &$\pm$& 1.00 & 1.00 &$\pm$& 1.00 \\[-0.1ex]
NC other ND280 & 1.00 &$\pm$& 0.30 & 1.21 &$\pm$& 0.26 \\[-0.1ex]
NC other SK & 1.00 &$\pm$& 0.30 & 1.00 &$\pm$& 0.30 \\[0.6ex]
FSI inelastic low-E & 0.00 &$\pm$& 0.41 & $-0.32$ &$\pm$& 0.08 \\[-0.1ex]
FSI inelestic high-E & 0.00 &$\pm$& 0.34 & $-0.01$ &$\pm$& 0.13 \\[-0.1ex]
FSI pion production & 0.00 &$\pm$& 0.50 & 0.04 &$\pm$& 0.19 \\[-0.1ex]
FSI pion absorption & 0.00 &$\pm$& 0.41 & $-0.35$ &$\pm$& 0.15 \\[-0.1ex]
FSI charge exch. low-E & 0.00 &$\pm$& 0.57 & $-0.09$ &$\pm$& 0.31 \\[-0.1ex]
FSI charge exch. high-E & 0.00 &$\pm$& 0.28 & 0.02 &$\pm$& 0.10 \\ 
\hline \hline
\end{tabular}
\end{table}

\fi

\end{document}